\documentclass[12pt]{article}
\usepackage{cite}
\usepackage{amsmath, amsthm, amssymb,slashed}
\usepackage{ifpdf}
\ifpdf
  \usepackage[pdftex]{graphicx}
  \usepackage{epstopdf}
\else
  \usepackage[dvips]{graphicx}
\fi
\parskip=\baselineskip
\def\Bbb{\mathbb}
\def\Tr{{\mathrm {Tr}}}
\def\16{{\mathbf 16}}
\def\1{{\bf 1}}
\def\2{{\bf 2}}
\def\3{{\bf 3}}
\def\4{{\bf 4}}
\def\GG{S}
\def\W{{\mathcal W}}
\def\Spin{{\mathrm{Spin}}}
\def\bar{\overline}
\def\tilde{\widetilde}
\def\R{{\Bbb{R}}}\def\Z{{\Bbb{Z}}}
\def\N{{\mathcal N}}
\def\hat{\widehat}
\def\frak{\mathfrak}
\def\cal{\mathcal}
\font\teneurm=eurm10 \font\seveneurm=eurm7 \font\fiveeurm=eurm5
\newfam\eurmfam
\textfont\eurmfam=\teneurm \scriptfont\eurmfam=\seveneurm
\scriptscriptfont\eurmfam=\fiveeurm
\def\eurm#1{{\fam\eurmfam\relax#1}}
\font\teneusm=eusm10 \font\seveneusm=eusm7 \font\fiveeusm=eusm5
\newfam\eusmfam
\textfont\eusmfam=\teneusm \scriptfont\eusmfam=\seveneusm
\scriptscriptfont\eusmfam=\fiveeusm

\font\tencmmib=cmmib10 \skewchar\tencmmib='177
\font\sevencmmib=cmmib7 \skewchar\sevencmmib='177
\font\fivecmmib=cmmib5 \skewchar\fivecmmib='177
\newfam\cmmibfam
\textfont\cmmibfam=\tencmmib \scriptfont\cmmibfam=\sevencmmib
\scriptscriptfont\cmmibfam=\fivecmmib

\numberwithin{equation}{section}
\def\neg{\negthinspace}
\def\d{\mathrm d}
\def\C{{\Bbb C}}
\def\Z{{\Bbb Z}}
\def\A{{\mathcal A}}

\def\bar{\overline}
\DeclareMathAlphabet{\mathpzc}{OT1}{pzc}{m}{it}
\font\teneurm=eurm10 \font\seveneurm=eurm7 \font\fiveeurm=eurm5
\newfam\eurmfam
\textfont\eurmfam=\teneurm \scriptfont\eurmfam=\seveneurm
\scriptscriptfont\eurmfam=\fiveeurm
\def\eurm#1{{\fam\eurmfam\relax#1}}
\font\teneusm=eusm10 \font\seveneusm=eusm7 \font\fiveeusm=eusm5
\newfam\eusmfam
\textfont\eusmfam=\teneusm \scriptfont\eusmfam=\seveneusm
\scriptscriptfont\eusmfam=\fiveeusm

\font\tencmmib=cmmib10 \skewchar\tencmmib='177
\font\sevencmmib=cmmib7 \skewchar\sevencmmib='177
\font\fivecmmib=cmmib5 \skewchar\fivecmmib='177
\newfam\cmmibfam
\textfont\cmmibfam=\tencmmib \scriptfont\cmmibfam=\sevencmmib
\scriptscriptfont\cmmibfam=\fivecmmib

\def\L{{\mathcal L}}
\def\grav{{\mathrm{grav}}}

\begin{document}

\begin{titlepage}

\vskip 1.5in
\begin{center}
{\bf\Large{Fivebranes and Knots}}\vskip
0.5cm {Edward Witten} \vskip 0.05in {\small{ \textit{School of
Natural Sciences, Institute for Advanced Study}\vskip -.4cm
{\textit{Einstein Drive, Princeton, NJ 08540 USA}}}
\vskip.05in{and}
\vskip .05in {\textit{Department of Physics, Stanford University}}
\vskip-.15in {\textit{Palo Alto, CA 94305 USA}}}

\end{center}
\vskip 0.5in
\baselineskip 16pt
\date{September, 2010}

\begin{abstract}
We develop an approach to Khovanov homology of knots via gauge theory (previous physics-based approaches involved other descriptions of the relevant spaces of BPS states).
The starting point is a system of D3-branes ending on an NS5-brane with a nonzero theta-angle.
On the one hand, this system can be related to a Chern-Simons gauge theory on the boundary
of the D3-brane worldvolume; on the other hand, it can be studied by standard techniques of $S$-duality and $T$-duality.
Combining the two approaches leads to a new and manifestly invariant description of the Jones polynomial of knots, and its generalizations, and to
a manifestly invariant description
of Khovanov homology, in terms of certain elliptic partial differential equations in four and five dimensions.
\end{abstract}
\end{titlepage}
\vfill\eject \tableofcontents
\section{Introduction}

\def\CC{\Sigma}
\subsection{Knot Polynomials}
The Jones polynomial \cite{Jones,Jones2} associates to a knot $K$ in Euclidean
three-space $\R^3$ (or in a three-sphere $S^3$) a Laurent
polynomial  $\mathcal J(q;K)$ in a single variable $q$.  The coefficients in this
Laurent polynomial are integers.  Some further details  are  explained below.

The Jones polynomial -- and its many generalizations which are also Laurent polynomials
with integer coefficients  -- can be constructed in a variety
of ways from two-dimensional mathematical physics. The key ingredients include
lattice statistical
mechanics, Yang-Baxter equations,
conformal field theory, and braid group representations
\cite{Homfly,Kauffman,Turaev,PT,BW,TK}.   These constructions are very efficient for
computing the knot polynomials, demonstrating their topological invariance, and
showing that they  indeed are Laurent polynomials with integer coefficients.

However, such constructions do not make manifest the three-dimensional symmetry of
the Jones
polynomial.       For this purpose,
 three-dimensional quantum gauge theory with a Chern-Simons action
\cite{Schwarz,Schonfeld,DJT} turns out to be useful.    The Chern-Simons action for
a gauge theory with
gauge group\footnote{In this paper, $G$ is always a compact Lie group, and all representations
considered are finite-dimensional.}  $G$ and gauge field $A$ on an oriented three-manifold $W$  can be written
\begin{equation}\label{csaction} I = \frac{k}{4\pi}\int_W\Tr\,\left(A\wedge\d
A+\frac{2}{3}A\wedge
A\wedge A\right).\end{equation} Here $k$ is an integer for
topological reasons; up to a choice of orientation, one may take $k$ to be positive. In
this theory, to an oriented embedded loop $K\subset W$ and a representation
$R$ of $G$, one can associate an observable, the trace of the holonomy or Wilson loop
operator:
\begin{equation}\label{zonko} \W(K,R)=\Tr_R\,P\exp\oint_K A.\end{equation}
Reversing the orientation of $K$ has the same effect as replacing $R$ by its complex conjugate.
  It turns out
\cite{witten} that the Jones polynomial and its generalizations
can be computed as expectation values of Wilson loop operators, if
we express the argument $q$ of the knot polynomials in terms of
the Chern-Simons level $k$ by
\begin{equation}\label{morko} q=\exp\left(2\pi i/(k+h)\right),\end{equation}
where $h$ is the dual Coxeter number of $G$.
For example, if we take $G=SU(2)$, $R$ to be the two-dimensional irreducible
representation of $SU(2)$, and $W=S^3$, then the expectation value of $\W(K,R)$ is
equal to the Jones
polynomial:
\begin{equation}\label{orko}\mathcal J(q;K)=\bigl\langle \W(K,R)\bigr\rangle.\end{equation}

\subsubsection{Some Details}\label{details}

We will spell out a few details about the function $\mathcal J(q;K)$.
First of all, the definition extends immediately to an oriented link, that is a union $L$ of $\nu$ disjoint oriented embedded circles $K_i$.  We label the $K_i$ by representations $R_i$ of $G$ and set
\begin{equation}\label{zorko} J(q;K_i,R_i)=\bigl\langle \prod_i \W(K_i,R_i)\bigr\rangle.\end{equation}
For $G=SU(2)$ and all $R_i$ equal to the two-dimensional representation, this function is known as the Jones polynomial of the link $L$. We denote this special case as  $\mathcal J(q;L)$.

In (\ref{orko}) and (\ref{zorko}), the symbol $\langle ~~\rangle$ refers to an expectation value, that is,
a ratio of two path integrals
\begin{equation}\label{omorko} J(q;K_i,R_i) =\frac{\int DA\,\exp(iI)\,\prod_i\W(K_i,R_i)}{\int DA\,\exp(iI)}.
\end{equation}
For $W=S^3$, the denominator is non-trivial (for example, it equals $\sqrt{2/(k+2)}\sin(\pi/(k+2))$
for $G=SU(2)$) and it is necessary to divide by this factor to obtain a function $J(q;K_i,R_i)$ that has the simple properties we will explore in this paper.  However, in our framework, it will be more natural to study a path integral rather than
a ratio of two path integrals.  $J(q;K_i,R_i)$ can be expressed in this form by simply replacing $W=S^3$ with $W=\R^3$. The ratio in (\ref{omorko}) is unaffected, but now the denominator (regularized by the procedure in the present paper to deal with the behavior at infinity) equals 1 and can be omitted.
Taking $W=\R^3$ will also simplify the arguments in this paper by suppressing infrared fluctuations, in a sense
that will be clear later,  and in certain
other technical details.  Accordingly, though we will define an analog of Khovanov homology on any three-manifold, its relation to Chern-Simons theory is most simple for the case of links in $\R^3$.

We should warn the reader of a few differences between our conventions and the ones that are most
common in the mathematical literature.  First,
a very basic case of a link is the empty link $\varnothing$ for which the number of embedded circles is $\nu=0$.
With our definition, $J(q;\varnothing)=1$.  In the mathematical literature, it is customary to normalize the
Jones polynomial so that its value is 1 for the unknot $K_0$ rather than the empty link $\varnothing$,
so the usual mathematical definition corresponds to what we would call  $\tilde{ \mathcal J}(q;L)=\mathcal J(q;L)/\mathcal J(q;K_0)$.  An analogous statement holds for the more general invariants
$J(q;K_i,R_i)$.

The precise sense in which $J(q;K_i,R_i)$ is a Laurent polynomial is as follows. In general, depending on
 the representations $R_i$, $J(q;K_i,R_i))$ is either a Laurent
polynomial in $q$, or $q^{1/2}$ times a Laurent polynomial in $q$. For example, the Jones polynomial
is $q^{\nu/2}$ times a Laurent polynomial,
\begin{equation}\label{odork} \mathcal J(q;L)=\sum_{n\in \Z+\nu/2} a_nq^n,~~a_n\in\Z.\end{equation}
The coefficients $a_n$ are integers and all but finitely many of them vanish.
The half-integral powers
are often suppressed by taking the basic variable to be not our $q$ but
$\tilde q=q^{1/2}$.   In many ways, however, the variable $q$ is more natural.  For example, it will
turn out to be the natural instanton counting factor in a dual gauge theory description.  The fractional
powers of $q$ turn out to have a natural topological interpretation, and it seems unnecessary to suppress them. (In a sense, it is also ultimately fruitless to try to suppress them, since as will become clear, on a general
three-manifold, we meet general fractional powers of $q$, not just half-integral powers.)

One further detail is that, as explained via gauge theory
in \cite{witten}, the invariants $J(q;K_i,R_i)$ are most naturally
defined for framed links.  (A framing of an embedded circle $K\subset W$ is a trivialization of the normal bundle to $K$ in $W$.)  Under a change in framing, $J(q;K_i,R_i)$ is multiplied by a certain (generically
fractional) power
of $q$.  For links in $S^3$ or $\R^3$, one can suppress this phenomenon, since an embedded circle in $S^3$
has a distinguished framing (relative to which its self-linking number is zero).  Standard formulas such as (\ref{odork}) implicitly refer to this standard framing.  Similarly, the Chern-Simons path integral
on a general three-manifold $W$ depends naturally on a framing of $W$ (a trivialization of its tangent bundle $T$) or more generally \cite{Atiyah} on a two-framing (a trivialization of $T\oplus T$).  A change of framing of  $W$ has the same sort of effect as a change in framing of a link: it multiplies
the path integral by a power of $q$.  This power cancels out of the ratio (\ref{omorko}),
but when we assert that the denominator is 1 for $W=\R^3$, this statement refers to the path integral defined with the obvious framing associated to a Euclidean metric on $\R^3$.

\subsubsection{What Chern-Simons Theory Doesn't Explain}\label{doesnt}

The Chern-Simons path integral gives a definition of the invariants $J(q;K_i,R_i)$  with manifest three-dimensional
symmetry, provided that $q$ is a root of unity of the particular
form (\ref{morko}).   Granted that $J(q;K_i,R_i)$ is a Laurent
polynomial, it is determined by its behavior at these
values of $q$. However, the gauge theory path integral does
not shed much light on why these functions are Laurent polynomials. This
is clearer in any of the definitions of the link invariants based on
two-dimensional mathematical physics.  The only known way to
deduce that $J(q;K_i,R_i)$ is a Laurent polynomial starting from three-dimensional
gauge theory is to first reduce to a two-dimensional description, for example via representations
of braid groups, in which this
fact is clear.
The Chern-Simons path integral has been used directly \cite{wittentwo} to explain
the existence of an analytic continuation of Wilson loop
expectation values to complex values of $k$, but not the fact that
the result is a Laurent polynomial.

\def\H{{\mathcal H}}
\def\HH{{\mathpzc H}}
\def\K{{\mathcal K}}
\def\KH{{\mathpzc {K}}}
\def\Kh{\KH}
\def\EH{\mathrm P}
\def\EP{\EH}
\def\EF{\mathrm F}
\subsection{Khovanov Homology}\label{kh}
Moreover, none of the constructions so far mentioned give a really good
explanation of why the coefficients $a_n$ of these Laurent polynomials are integers. This has been accomplished in
Khovanov homology \cite{Khovanov}, in which the $a_n$ are
interpreted as the dimensions (in a $\Z_2$-graded sense)
of finite-dimensional vector spaces. For motivation behind Khovanov homology, see
\cite{CF,FK,BFK}, and for an introduction see \cite{BN}.  In this
theory, one associates to a link $L$ in three-space a
finite-dimensional vector space
 $\KH(L)$, known as its Khovanov homology. The original construction was adapted to the Jones
  polynomial -- or, if you like, to a link  labeled by the two-dimensional representation of $SU(2)$. $\KH(L)$ is defined as the cohomology of
a differential $Q$ (a differential is simply  a linear
 operator $Q$ obeying $Q^2=0$) that acts on a larger vector space $\HH(L)$.
 $\KH(L)$ is natural and depends only on $L$, but there is much arbitrariness in the
construction of $\HH(L)$.  $\HH(L)$ is bigraded, with symmetry
generators that we will call $\EF$ and $\EH$. $Q$ obeys $[\EF,Q]=Q$, $[\EH,Q]=0$;
these relations ensure that $\KH(L)$ is bigraded,
\begin{equation}\label{zolp}\KH(L)=\oplus_{m,n}\KH^{m,n}(L),\end{equation}
where $m,n$  are the eigenvalues of $\EF$, $\EH$.  With the usual normalization, $m$ and $n$
take integer values.  In our formulation in this paper, $m$ is $\Z$-valued and $n$ takes values in $\Z+\nu/2$ (where $\nu$ is the number of components of the link $L$) or more generally in a certain coset
of $\Z$ in $\R$.  Despite the nonintegrality of the eigenvalues of $\EH$, we will loosely refer to the group
generated by $\EF$ and $\EH$ as $U(1)\times U(1)$ and the associated grading as a $\Z\times\Z$ grading.
The relation between Khovanov homology
and the Jones polynomial is
\begin{equation}\label{zonk} \mathcal J(q;L)=\Tr_{\KH(L)}\,(-1)^\EF q^\EH.\end{equation}
This formula makes manifest the fact that $\mathcal J(q;L)$ is a Laurent polynomial
with integer coefficients.  (The half-integral powers in
$\mathcal J$ for a link with an odd number of components
arise from the fact that, with our normalization,
for a link in $\R^3$ with $\nu$ components labeled by the
two-dimensional representation of $SU(2)$,
  the eigenvalues of $\EH$ lie in $\Z+\nu/2$.
  See section \ref{transk}.) We can describe eqn. (\ref{zonk}) by saying that the Jones
polynomial can
be recovered from Khovanov homology by taking an equivariant index or Euler
characteristic.  Since $\EF$ is $\Z$-valued but the right hand side of (\ref{zonk}) only depends on
the value of $\EF$ mod 2, this formula also shows that Khovanov homology potentially
contains more information than the Jones polynomial.
It has turned out that the additional information is
really essential.

The success in recovering the Jones polynomial from a homology theory raises the question of whether
a similar construction is possible if components of $L$ are labeled by arbitrary representations $R_i$ of
a compact Lie group $G$.  In the literature, this has been accomplished for many classes of groups and representations.
Here, we will make a general proposal.

From a physical point of view, a three-dimensional quantum field
theory with loop operators will naturally assign a {\it number} --
the value of the path integral -- to a knot.  To associate to a
knot a {\it vector space} (its Khovanov homology) rather than a
number, we want a four-dimensional quantum field theory with
surface operators rather than loop or line operators.  Thus,\footnote{The actual framework we develop
later is more complicated than the idealized sketch offered here, mainly in the need to introduce a fifth
dimension.}
introduce a fourth ``time'' dimension, parametrized by $\R$,  and
consider a four-dimensional topological field theory on $M=\R\times W$, with a surface operator on 
$\CC=\R\times K$; as before, $K$ is
a knot in a three-manifold $W$.  The space of physical states in
such a theory will be a vector space associated to the pair
$(W,K)$; this vector space will be bigraded -- like the Khovanov
homology of a knot in $W=S^3$ -- if the four-dimensional theory
has an appropriate $U(1)\times U(1)$ symmetry.  What has just been
described was part of the original motivation that led to Khovanov
homology \cite{CF} and these matters have also been discussed from
a physical point of view \cite{Gukov}.  From the point of view of
four-dimensional quantum field theory, the index formula
(\ref{zonk}) has a natural interpretation.  Given a
four-dimensional quantum field theory, one can reduce to a
three-dimensional quantum field theory by compactifying on $S^1$.
The partition function of a four-dimensional theory on a
four-manifold of the form $M=S^1\times W$, where $W$ is a
three-manifold, will give a $\Z_2$-graded trace or index. (Here we
assume that if surface operators are present, they are supported
on $S^1\times K$, for some $K\subset W$, to be compatible with the
product form of $M$.)  In the reduction, if there is a conserved
charge $\EH$ that commutes with $Q$, one can make a twist by $q^\EH$
(for some $q$) in going around the circle.   The partition
function of the reduced theory will then be an equivariant index
as in (\ref{zonk}).

In the mathematical literature, there actually is direct evidence
that Khovanov homology is part of a four-dimensional theory with
surface operators.  The main evidence comes from consideration of
cobordism between knots.  Here, we take $M=I\times S^3$, where
$I=[0,1]$ is the unit interval. In $M$, one considers an embedded
two-manifold $\CC$ whose restriction to one boundary $\{0\}\times
S^3$ is a knot $K$, and whose restriction to the other boundary
$\{1\}\times S^3$ is a knot $K'$.  Physically, one would expect
the path integral on $M$ (with $\CC$ understood as the support of a
surface operator) to define a linear transformation from the space
of physical states associated to the pair $(S^3,K)$ to the
corresponding space for $(S^3,K')$.  Mathematically, it has been
found that one can associate to such a cobordism a natural linear
transformation $\Phi_\CC$ from the Khovanov homology of $K$ to that
of $K'$:
\begin{equation}\label{ocirco}\Phi_\CC:\Kh(K)\to \Kh(K').\end{equation}
If one glues together two knot cobordisms, the corresponding transition amplitudes
multiply,
just as one would expect physically.

The literature on Khovanov homology provides at least one more
clue. In close parallel with the early mathematical constructions
of the Jones polynomial and its cousins, mathematical
constructions of Khovanov homology and its extensions are
frequently based on familiar ingredients in mathematical physics.
But these constructions do not make manifest the topological
invariance of Khovanov homology, potentially creating an
opportunity for physicists.   Actually, a number of mathematical
constructions of Khovanov homology are based on ways of
associating a two-dimensional topological quantum field theory (or
at least the category of branes in such a theory) to a two-sphere
$S^2$ with marked points $p_i$, $i=1,\dots,n$.  A natural
interpretation is that these constructions arise by specializing a
four-dimensional quantum field theory to four-manifolds of the
form  $M=\Sigma\times S^2$, where $\Sigma$ is a Riemann surface
and surface operators are supported on the two-manifolds
$\Sigma\times p_i$. In one construction \cite{KR,GWa}, the
effective theory on $\Sigma$ seems to be a Landau-Ginzburg
$B$-model (so that the branes are matrix factorizations); in a
second construction \cite{CK}, the effective theory is a $B$-model
with target space a certain Kahler manifold; other approaches
\cite{SS,Kam} are based on $A$-models. There have also been
attempts \cite{KM,KM2} to make the three- or four-dimensional
symmetry of Khovanov homology manifest by extracting it from a
special case or analog of Donaldson-Floer theory in four
dimensions.  This of course is related to $\N=2$ super-Yang-Mills
theory in four dimensions.

\subsection{Previous Physics-Based Proposals}

Actually, a proposal for a physical construction of Khovanov homology has been made
some years ago.  An initial clue was that \cite{wittenold}
the knot invariants
associated to Chern-Simons theory can be regarded  as open-string analogs of
the usual  $A$-model invariants for closed strings.  On the other hand,
the topological $A$-model for either closed or open strings can be embedded in Type IIA
superstring theory.  For open strings, this embedding plus a hypothesis of a geometric
transition in string theory has led to powerful results
\cite{GV} about Chern-Simons theory.  In addition, by considering the strong
coupling limit of the Type IIA
model, in which the $M$-theory circle opens up, closed string $A$-model amplitudes
(or Gromov-Witten invariants)
can be fruitfully expressed in terms of Gopakumar-Vafa invariants \cite{GV2}.  The
Gopakumar-Vafa
invariants are simply the dimensions of certain spaces of BPS states of $M$-theory
membranes,
so they are automatically integers, unlike the $A$-model amplitudes themselves
(which in general are
rational numbers).
Expressing the closed topological string amplitudes in terms of Gopakumar-Vafa
invariants
is powerful because purely numerical invariants (the Gromov-Witten invariants) are
expressed
in terms  of vector spaces (the spaces of BPS states).

The Gopakumar-Vafa construction has an analog \cite{OV} for open strings, expressing
$A$-model
observables of open strings in terms of spaces of BPS states in the presence of certain
branes.  For further developments, see  \cite{LM,RS,LMV} and for a review of many of these
topics, see \cite{Marino}.  This approach has been
extended into a proposal
\cite{GSV} to identify
the Khovanov homology for a knot $K$ with the space of BPS states
-- for an $M$-theory configuration that depends on the choice of  $K$.
A substantial amount of evidence for this proposal was given in \cite{GSV}, in part by using
geometric transitions as a tool to compute the spaces of BPS states. Moreover, the
proposal implied some new predictions concerning Khovanov homology and has
led to a better understanding of some aspects of this subject \cite{DGR}.
The relevant brane constructions have been further studied in
 \cite{DVV,AY,CNV}.  For an extension of these ideas involving the topological vertex and the Nekrasov
 partition function for instantons, see  \cite{IKV,GIKV}.

A related road to a  physical interpretation of Khovanov homology has
appeared much more recently in a study of supersymmetric line operators in
four-dimensional
gauge theories with $\N=2$ supersymmetry \cite{GMN}.  It was shown that such line
operators form an ``algebra,'' but with the structure constants being vector spaces
rather
than numbers.  For the case that the four-dimensional theory is obtained by
compactifying the
six-dimensional $(0,2)$ model on a Riemann surface $C$, as analyzed in most detail in
\cite{G}, the algebra in question is closely related to the usual algebra of
multiplication
of Wilson loop operators in quantum Chern-Simons theory on $C$ -- except that the
structure
constants in the algebra are replaced by vector spaces.   (One can recover the usual
loop algebra
of Chern-Simons theory by taking a supertrace, as in (\ref{zonk}), to replace the
vector spaces by
numbers.  This has been pointed out
by the authors of \cite{GMN}.)    These results should be related to a
generalization of Khovanov homology for loops in the three-manifold $\R\times C$ --
more precisely for product loops of the form
$p\times \ell$, with $p$ a point in $\R$ and $\ell$ a loop in $C$.

\subsection{The Present Paper}

In this paper, we will re-examine the relation of Khovanov
homology to the spaces of BPS states in $M$-theory, with three
primary goals.  One goal is to give a gauge theory definition of
Khovanov homology (as opposed to a  definition that requires a
full knowledge of string/$M$-theory). String theory and branes will be used as clues,
 but the results can be expressed as a gauge theory construction.  A second goal is to give a
more transparent -- or at least new -- explanation in this context
of the key property of Khovanov homology: the fact that a
supertrace in the space of BPS states gives the path integral of
Chern-Simons theory. The last goal is to develop an effective
framework to understand generalizations of Khovanov homology in
which one varies the three-manifold $W$ or the boundary conditions
or other details. (This program is not actually achieved in the present paper.)
Along the way, we will
clarify some formal properties of Khovanov homology.

\subsubsection{The Basic Idea}\label{basic}

The basic idea behind this paper is simply explained.  We would like to apply nonperturbative
string theory or field theory dualities to three-dimensional Chern-Simons gauge theory,
but there is no obvious way to do this directly. However, it is possible to express the path
integral of Chern-Simons theory on a three-manifold $W$ as a path integral of $\N=4$ super Yang-Mills
theory on a half-space $V=W\times \R_+$, where $\R_+$ is the ray or half-line  $y\geq 0$. (Knots
in $W$ are represented by Wilson operators in the boundary of $V$.)
Once this is done, one can apply standard gauge theory and  string theory
dualities to the $\N=4$ path integral on the four-manifold $V$, leading to a description by a higher-dimensional theory with the desired properties.

The relation of the Chern-Simons path integral on $W$ to the $\N=4$ path integral on $W\times \R_+$ is one
of the main results of \cite{wittenthree} (and the basic idea is suggested in the conclusions of \cite{wittentwo}).  We will give an alternative explanation in this paper,
partly to keep the paper self-contained, and partly to emphasize the aspects that we need.
In general, in this correspondence, the $\N=4$ path integral on $V=W\times \R_+$ depends on a boundary
condition at $y\to \infty$, and the equivalent Chern-Simons path integral is not the usual one but is
a path integral defined with an exotic integration cycle, in a sense described  in \cite{wittentwo}.  However, for the case of links in $\R^3$ or $S^3$, there is essentially (up to a constant multiple) only one possible integration cycle
and the path integral obtained this way is equivalent to the standard one.  From the vantage point of the present paper, this is one of the reasons that Khovanov homology is simplest in the case of links in $\R^3$.

In order to relate the $\N=4$ path integral on $V=W\times\R_+$ to a Chern-Simons path
integral on $W$, we need to use the right boundary condition on the boundary of $W$.
The requisite boundary condition is not exotic.
 It is simply the boundary
condition of the D3-NS5 system of Type IIB superstring theory in the presence of a theta-angle.
 This boundary condition has been described in \cite{gw,gw2}.

At this point, all we have done is to restate the problem of Chern-Simons theory in terms of
an $\N=4$ path integral on $V$.
To get something like Khovanov homology, we want to re-express the $\N=4$ path integral on $V$ as a path integral of some other theory on $V\times S^1$.
A path integral on $V\times S^1$ can be written as a trace (or, in the presence of fermions, as a
$\Z_2$-graded trace) in a Hilbert space $\H$ associated to quantization on $V$.
Suppose that the path integral on $V\times S^1$ is invariant under a supersymmetry generator $Q$ that obeys
$Q^2=0$.  Then, by a standard argument, the $\Z_2$-graded
trace in $\H$ reduces  to a $\Z_2$-graded trace in $\K$, the cohomology of $Q$. (We will write
$\K$ for cohomology spaces arising in quantum field theory and $\Kh$ for Khovanov homology;
we make this distinction because we do not have a proof that these coincide even in situations where $\Kh$ has been
defined.)  Our strategy to get a formula like (\ref{zonk}) for the Jones polynomial is to first express the Jones polynomial as an $\N=4$
 path
integral on $\R^3\times \R_+$ -- with knots represented by Wilson operators at the boundary
--  and then find a duality to re-express this as a path integral on $\R^3\times\R_+\times S^1$.

\def\TN{{\mathrm {TN}}}
The most naive way to try to do this fails in an instructive way.  We first embed the D3-NS5 system
in Type IIB superstring theory on $\R^9\times S^1$, where the $S^1$ direction is transverse to the branes.
Compactifying one of the transverse directions on a circle does not affect anything that has been said so far.  Then we perform a $T$-duality on the $S^1$.  This replaces $S^1$ by a dual circle $\tilde S^1$.
At first sight, it seems that the $T$-dual of the D3-NS5 path integral will be a path integral on
$\R^3\times \R_+\times \tilde S^1$, leading in the desired fashion to a trace.  However, in the presence
of an NS5-brane wrapped on $\R^6\times p\subset \R^9 \times S^1$ (here $\R^6$ is linearly embedded in $\R^9$ and $p$ is a point in $S^1$), $T$-duality maps us not to $\R^9\times\tilde S^1$ but  \cite{Townsend,GHM} to $\R^6\times \TN$,
where $\TN$ is a Taub-NUT space.  $\TN$ is asymptotic at infinity to a twisted $\tilde S^1$ bundle over $\R^3$, but crucially, $\tilde S^1$ shrinks to a point in the interior of $S^3$.  Because
of this, the path integral in this $T$-dual description cannot be interpreted as a trace.

There is a simple way to avoid this difficulty.  Before $T$-duality, we first perform $S$-duality.
$S$-duality converts the D3-NS5 system to a D3-D5 system.  (A system of D3-branes ending on a D5-brane
has special properties that were investigated in \cite{Dia,myers,CM, cmyerst} and interpreted in field
theory language as a boundary condition in $\N=4$ super Yang-Mills theory in \cite{gw}.)  We embed the
D3-D5 system in $\R^9\times S^1$ and now $T$-duality simply maps this to a D4-D6 system on $\R^9\times \tilde S^1$.  Now the path integral can be straightforwardly interpreted as a trace and this leads to a formula like (\ref{zonk}).  What plays the role of $\KH$ is the cohomology of a certain supercharge $Q$
that is preserved by the construction.  (The proper choice of $Q$ depends on details that we have omitted
here.)  $\EF$ corresponds to an $R$-symmetry of the brane configuration, and $\EH$ is, from the point of view of the D4-brane gauge theory,
the Yang-Mills instanton number integrated over $\R^3\times \R_+$.

Most of these steps have analogs with $\R^3$ replaced by a more general three-manifold $W$, but in trying
to formulate the resulting statements about Chern-Simons theory, one runs into infrared divergences and
  a need to understand how $S$-duality acts on the boundary conditions at $y=\infty$.  The simplest case other than $\R^3$ is
likely to be the case that $W$ is obtained by omitting a point from a rational homology sphere.
In this case, projecting the missing point to infinity and taking a metric on $W$ that looks near infinity
like the flat metric on $\R^3$, there are no infrared divergences and a close analog of Khovanov homology
should exist.  One will still have the problem of understanding the action of $S$-duality on the
boundary conditions at $y=\infty$.

\subsubsection{Organization Of The Paper}\label{organization}

In section \ref{csfun}, we describe in more detail, in the context of the D3-NS5 system, the relation of the Chern-Simons
path integral in three dimensions to an $\mathcal N=4$ path integral in four dimensions.  Then
we apply
standard dualities to this situation, first $S$-duality in section \ref{dualities} followed by $T$-duality (or in gauge theory simply the introduction of a fifth dimension) in section \ref{tdual}.  The first step leads to an essentially new description of knot invariants
related to Chern-Simons theory, and the second leads to Khovanov
homology. The two operations
have different status.  $S$-duality  is natural purely as a field theory operation, but $T$-duality
is not and leads to a description by a five-dimensional super Yang-Mills
theory that is not ultraviolet complete.

A better and conceptually more satisfying formulation is to base our construction not on five-dimensional super Yang-Mills theory but on its familiar ultraviolet completion in the
six-dimensional $(0,2)$ model (for example, see \cite{witsix} for a brief introduction).  In section \ref{fivebranes}, we proceed
in this way: we begin with the
$(0,2)$ theory in six dimensions, and work our way down to five, four, and three dimensions.   This gives the most economical and logically complete
treatment of the topic, and it gives the clearest explanation of a number of questions.  The top-down approach of section \ref{fivebranes} certainly could have been the starting point of the present paper.  We have chosen instead a bottom-up presentation in which the relation to Chern-Simons theory is made as clear as possible at the outset.

In section \ref{morebranes}, we explore a second brane construction, which in
some ways is closer to the setting of \cite{GSV}.
The starting point of the second construction is that Wilson operators of Chern-Simons theory
can be expressed as codimension two monodromy defects.  The two formulations -- via Wilson operators or
monodromy defects -- are related to two different semiclassical limits of Chern-Simons theory.  In one
case, one takes the level $k$ to be large
while keeping fixed the representations $R_i$ labeling
the knots.  This is the most direct framework for describing the Jones polynomial, Khovanov homology,
and their generalizations.  In the other type of semiclassical limit, the monodromies produced by the knots are kept fixed as $k$ becomes large.  This second limit
is related to the volume conjecture of Chern-Simons theory, which has been reviewed with
extensive references in \cite{Mura} and explored physically in \cite{Gu} and \cite{wittentwo}.
The formulation of Chern-Simons theory in terms of monodromy defects can be carried through all the dualities of the present paper, leading to descriptions based on codimension two defects in various dimensions, as we explain briefly in section \ref{morebranes}.  This matter certainly merits much
closer attention.

We probably should mention here  two  important puzzles that we will
{\it not} unravel.  First, Khovanov homology is explicitly calculable for any given link in $\R^3$, though the requisite calculations
may not be easy.  Indeed,
Khovanov homology was originally defined  (see \cite{BN} for an accessible account) by an explicit
  algebraic recipe for computing it, though not one that makes topological invariance manifest.  The description in the present paper has the opposite properties: topological invariance is
manifest, but computability is not.  It would be highly desireable to bridge the gap between the two
types of knowledge by deducing a known definition of Khovanov homology from the quantum field theory construction studied here (or its close cousin studied earlier in \cite{GSV}).   To do this requires understanding
concretely the solutions of the localization equations presented later; one must understand the
four-dimensional version of the equations, presented in (\ref{mexico}), to understand the Jones
polynomial,  and the five-dimensional generalization, presented in (\ref{torm}), to understand Khovanov homology.  
Not much of this is done in the present paper; the only examples of 
actual solutions of the equations presented here 
are in section \ref{thooft}.  However, since the present paper was written, a reasonable understanding of the
four-dimensional equations  has
been obtained in \cite{gwnew} and this indeed has given a concrete understanding of how the Jones  polynomial emerges
in the present framework.   Some interesting special solutions of the four-dimensional equations have also been analyzed in
\cite{henningson}.

Second, our approach
here makes some things clearer than has been the case hitherto, but we fail to make contact with one important insight from \cite{GSV}.  We consider each gauge group as a problem in
its own right, while in \cite{GSV}, the $\sf A$ theories were treated in a unified way,
and this has been generalized to $\sf B$, $\sf C$, and $\sf D$  \cite{SV,MMarino,KPW}.

\subsubsection{Comparison To Other Work}\label{comparison}

Some relations of the present paper to other work, beyond what has already been cited,
are as follows.

Geometric Langlands duality (for a review, see \cite{Fre}) has a generalization, sometimes called quantum
geometric
Langlands in the mathematics literature, involving a parameter that was called $\Psi$
in \cite{KW}.  This generalization  has been related to the theory of quantum groups \cite{Gaitsgory},
suggesting  that
geometric Langlands should be related to
Chern-Simons theory.     Indeed, we show in this paper
that if formulated on a four-manifold $V$ of boundary $W$, the four-dimensional topological
field theory associated to geometric Langlands is related to Chern-Simons theory on $W$,
with $\Psi$ as essentially the Chern-Simons level.  Khovanov homology has previously been defined
\cite{CK} using
moduli spaces of geometric Hecke transformations, which are vital in geometric Langlands and were interpreted
via gauge theory in sections 9 and 10 of \cite{KW}.

On an abstract three-manifold $W$, Chern-Simons gauge theory only makes sense if the level
$k$ is an integer.  But we show in the present paper that if $W$ is the boundary of a given
four-manifold $V$, and we are willing to accept an answer that depends on $V$, then
a theory with many of the properties of Chern-Simons theory  can be formulated as a function
of a complex variable $k$.  Moreover, the theory appears to be unitary in Lorentz signature if
$k$ is real.  All this has a counterpart in contemporary developments in condensed
matter physics.  Topological insulators and superconductors -- see for example \cite{qz} for a
review -- are materials of $d$ dimensions (and therefore $d+1$ spacetime dimensions)  that on their $(d-1)$-dimensional surface realize
physical phenomena that could never occur in a purely $(d-1)$-dimensional material.
The values of $d$ that have been realized experimentally are $3$ (a bulk material with
a two-dimensional surface) and 2 (a thin film with a one-dimensional edge).  The $d=3$
topological insulators are materials that ultimately prove to have a ``forbidden'' Chern-Simons
coupling (for the ordinary electromagnetic field), somewhat like the system we study in the
present paper for non-integer $k$.

Apart from papers already cited, a relation between four-dimensional $\N=4$ super Yang-Mills
and three-dimensional Chern-Simons -- or at least $q$-deformed two-dimensional
Yang-Mills -- has been described in certain geometries in \cite{AOSV}.  And a recent paper
dealing with topics relatively close to that of the present paper is \cite{DGH}.

While the present paper was in gestation, it developed that the five-dimensional gauge theory equations that we present in eqn. (\ref{torm}) have been formulated independently by
A. Haydys \cite{Haydys}.  Haydys's point of view was roughly to study the $A$-model with
target the moduli space of complex-valued flat connections on a three-manifold.  He also
presented the two reductions of the equations that are described in section \ref{zelrud}.  Even more recently, the author has become aware of work by M. Kontsevich and Y. Soibelman that may have a bearing on the present topic.

\section{Chern-Simons From Four Dimensions}\label{csfun}

\subsection{The D3-NS5 System With A Theta-Angle}\label{funx}

\def\gym{g_{\sf{YM}}}
As indicated in section \ref{basic}, our starting point is the D3-NS5 system of Type IIB superstring theory. The local picture is that in Minkowski spacetime $\R^{1,9}$, with coordinates $x^0,\dots,x^9$ (and metric
signature $-++\dots+$), we consider $N$ D3-branes supported at
$x^4=x^5=\dots=x^9=0$.  The D3-branes end on a single NS5-brane that is supported at $x^3=x^7=x^8=x^9=0$.
In the four-dimensional spacetime parametrized by $x^0,\dots,x^3$, the D3-brane world-volume spans the
half-space $x^3>0$.  The gauge theory of the D3-branes is a $U(N)$ gauge theory with $\N=4$ supersymmetry.
In this gauge theory, the NS5-brane provides a half-BPS boundary condition, that is, a boundary condition
that preserves half of the supersymmetry.

When the gauge theory $\theta$-angle vanishes, this boundary condition is simply Neumann boundary conditions for gauge fields, extended to the rest of the vector multiplet in a supersymmetric fashion.  However,
the brane construction implies the existence of a more general half-BPS boundary condition even for $\theta\not=0$.  Indeed, Type IIB superstring theory has a complex coupling parameter $\tau=\theta/2\pi
+i/g_s$ ($\theta$ is the expectation value of a Ramond-Ramond scalar and $g_s$ is the string coupling constant), which in the gauge theory becomes $\tau=\theta/2\pi+4\pi i/\gym^2$, with $\gym$ the gauge
coupling constant and $\theta$ the gauge theory theta-angle.  The D3-NS5 system is half-BPS for any value
of $\tau$, so from a gauge theory point of view, Neumann boundary conditions must have a half-BPS generalization for $\theta\not=0$.

This generalization was described in section 2 of \cite{gw} (a more roundabout construction was also
presented in \cite{gw2}). We will summarize the essential points here, referring for more detail to
\cite{gw}.  Though the initial motivation is the D3-NS5 system, once the half-BPS boundary condition is
expressed in field theory language, it makes sense for any gauge group $G$, and we will present it that way.

The $R$-symmetry group of $\N=4$ boundary condition is $SO(6)$ (or actually its spin double cover), acting by rotation of the normal
bundle to the D3-brane.  The presence of the NS5-brane breaks $SO(6)$ to $SO(3)\times SO(3)$, where one factor
 rotates $x^4,x^5,x^6$ and the second rotates $x^7,x^8,x^9$.    In \cite{gw}, the two $SO(3)$'s are called respectively
$SO(3)_X$ and $SO(3)_Y$ and the corresponding two sets of scalar fields on the D3-brane were called
$\vec X$ and $\vec Y$.  The D3-NS5 boundary condition on $\vec Y$ is
\begin{equation}\label{zurmo}\vec Y|=0 \end{equation}
(for any field $\Phi$, its restriction to $x^3=0$ will be denoted as $\Phi|$), irrespective of $\theta$,
but the other boundary conditions are more subtle.

\def\V{{\sf V}}
\def\2{{\mathbf{2}}}
It is useful to adopt a ten-dimensional notation\footnote{\label{alf} We will attempt to follow conventions of \cite{KW}.  In particular, adjoint-valued fields such as gauge fields are real and anti-hermitian.
(This accounts for some minus signs in formulas such as (\ref{zonno}).)  We define
the Levi-Citiva tensor $\epsilon^{\mu\nu\alpha\beta}$ of $\R^{1,3}$ and the corresponding tensor $\epsilon^{\mu\nu\lambda}$
of the hyperplane $x^3=0$ as antisymmetric tensors obeying $\epsilon^{0123}=1=-\epsilon_{0123}$ and $\epsilon^{012}=1=-\epsilon_{012}$, respectively.} in which $\N=4$ super Yang-Mills theory comes by dimensional
reduction from ten dimensions and
the supersymmetries of the D3-brane transform under $SO(1,9)$ as a spinor $\mathbf{16}$ of definite
chirality; thus a generator $\varepsilon$ of supersymmetry obeys
\begin{equation}\label{zurky}\Gamma_{012\dots 9}\varepsilon = \varepsilon,\end{equation}
where $\Gamma_I$, $I=0,\dots 9$, are the $SO(1,9)$ gamma matrices. (As usual, a symbol such as $\Gamma_{I_1\dots I_k}$ denotes the antisymmetrized product of the corresponding gamma matrices.) The D3-NS5 boundary condition is invariant under $\eurm U=SO(1,2)\times SO(3)_X\times SO(3)_Y$, where $SO(1,2)$ acts on the dimensions $x^0,x^1,x^2$
common to the two types of brane.  Each factor in $\eurm U$ has a two-dimensional representation that we denote as $\mathbf 2$, and the $\mathbf{16}$ transforms as two copies of the tensor product $(\2,\2,\2)$.  This tensor product, which we denote as $\V_8$, is
a real representation of $\eurm U$ of dimension 8.  The supersymmetries transform
as $\mathbf{16}=\V_8\otimes \V_2$, where $\V_2$ is a two-dimensional real vector space. The natural operators
that act on $\V_2$ are the even elements of the $SO(1,9)$ Clifford algebra that commute with $\eurm U$.  They are generated by
\begin{align}\label{thirsty}B_0& = \Gamma_{456789}\notag \\
B_1&=\Gamma_{3456} \\ \notag B_2 & = \Gamma_{3789},\end{align}
and in view of the algebraic relations they obey (such as $B_0^2=-1$, $B_0B_1+B_1B_0=0$, etc.), we can choose a basis for $\V_2$ in which
\begin{equation}\label{dux}B_0=\begin{pmatrix}0 & 1\\ -1 & 0\end{pmatrix},~~
  B_1=\begin{pmatrix}0 & 1\\ 1 & 0\end{pmatrix},~~B_2=\begin{pmatrix}1& 0\\ 0 & -1\end{pmatrix}.\end{equation}

The expression $(\varepsilon,\tilde \varepsilon)=\bar\varepsilon\Gamma_3\tilde\varepsilon$
defines an $SO(1,2)\times SO(6)$-invariant bilinear form on the $\mathbf{16}$ of $SO(1,9)$;
it factors as the tensor product of an antisymmetric $\eurm U$-invariant
form on $\V_8$ and an antisymmetric form on $\V_2$.  If we write $\varepsilon_0\in \V_2$
as a column vector $\begin{pmatrix}s\\ t\end{pmatrix}$ and $\bar\varepsilon_0 $ as the row vector $(t,-s)$,
then we can write the antisymmetric inner product on $\V_2$ as $\langle\varepsilon_0,\tilde\varepsilon_0\rangle =\bar\varepsilon_0\tilde\epsilon_0$.

In any half-BPS boundary condition that is $\eurm U$-invariant, the unbroken supersymmetries must be precisely those of the form $\V_8\otimes \varepsilon_0$, for some nonzero
vector $\varepsilon_0\in \V_2$.  Since scaling of $\varepsilon_0$ is immaterial, the choice of $\varepsilon_0$ depends essentially on a single real parameter.  We can take
\begin{equation}\label{plux}\varepsilon_0=\begin{pmatrix}-a \\ 1 \end{pmatrix}, ~~\bar\varepsilon_0 =
\begin{pmatrix} 1 ~ a \end{pmatrix} \end{equation}
(we include the possibility $a=\infty$, which means that the bottom component of $\varepsilon_0$ vanishes).
It is shown in \cite{gw} that for every  $a\in\R\cup\infty$  there is a unique $\eurm U$-invariant half-BPS
boundary condition that preserves all of the gauge symmetry.  The parameter $a$ corresponds to the gauge
theory $\theta$-angle.\footnote{In the context of the D3-NS5 system, $\theta$ is not really an angle as
a shift $\theta\to\theta+2\pi$ would convert the NS5-brane to a $(1,1)$ fivebrane.  Accordingly, the following
formulas have no periodicity.}

Without repeating the full derivation, we will cite the results that we need.  The fermion fields $\lambda$ of
$\N=4$ super Yang-Mills are adjoint-valued fields that transform as the $\mathbf{16}$ of $SO(1,9)$, like
the supersymmetry generators.  The boundary conditions they obey turn out to be
\begin{equation}\label{pux}\lambda|\in \V_8\otimes\vartheta,\end{equation}
where $\vartheta\in \V_2$ is
\begin{equation}\label{hobnob} \vartheta=\begin{pmatrix} a\\ 1\end{pmatrix}.\end{equation}
The boundary conditions on $\vec X$ at $x^3=0$ are
\begin{equation}\label{zux}D_3X_c-\frac{a}{1+a^2}\epsilon_{cde}[X_d,X_e]=0,\end{equation}
and the boundary conditions on the gauge fields at $x^3=0$ are
\begin{equation}\label{pluxt}F_{3\mu}+\frac{a}{1-a^2}\epsilon_{\mu\nu\lambda}F^{\nu\lambda}=0.\end{equation}

At $a=0$ and $a=\infty$, eqns. (\ref{zux}) and (\ref{pluxt}) reduce to the more obvious Neumann boundary
conditions $D_3X_a=F_{3\mu}=0$ (the two choices actually correspond to the D3-NS5
and D3-$\overline{\mathrm{NS5}}$ systems).  The additional terms in the boundary conditions for
generic $a$ reflect boundary corrections to the familiar $\N=4$ super Yang-Mills action in bulk.
Let us first consider $\vec X$.  The usual bulk action for $\vec X$ is in Lorentz signature
\begin{equation}\label{zonno} I_{\vec X}=\frac{1}{\gym^2}\int_{x^3\geq 0}\d^4x \sum_{\mu=0}^3\sum_{c=1}^3 \Tr\, D_\mu X_cD^\mu X_c.\end{equation}
Let us consider when happens when we vary $\vec X$.  If we place no restriction on the value of $\delta X_c$ at $x^3=0$, we will learn that
to make the boundary term in the variation of $I_{\vec X}$ vanish, the boundary condition must be $D_3 X_c=0$.
Suppose, however, that there is an additional boundary coupling
\begin{equation}\label{ponno}\tilde I_{\vec X}=\frac{2 a}{3\gym^2(1+a^2)}\int_{x^3=0}\d^3x \,\epsilon^{cde}\Tr
X_c[X_d,X_e].\end{equation}
If we now vary $\hat I_{\vec X}=I_{\vec X}+\tilde I_{\vec X}$ with respect to $\vec X$, placing again no
restriction on $\delta X_c|$, we find that setting the boundary variation of $\hat I_{\vec X}$ to zero
gives the boundary condition (\ref{zux}).  So the boundary coupling (\ref{ponno}) underlies the boundary
condition (\ref{zux}).

The boundary coupling $\tilde I_{\vec X}$ is unfamiliar, but it has a more familiar analog for gauge fields. The
analog of (\ref{zonno}) for the gauge field $A$, whose field strength we denote as $F_{\mu\nu}$, is
\begin{equation}\label{onno}I_A=\frac{1}{2\gym^2}\int_{x^3>0}\d^4x \sum_{\mu,\nu=0}^3\Tr\,F_{\mu\nu}F^{\mu\nu}.\end{equation}
If we work just with this action, then setting its boundary variation to zero (with no restriction on $\delta A|$), we learn that the boundary condition on the gauge field must be $F_{3\mu}|=0$.  To arrive at (\ref{pluxt}), we need an additional term in the action.  This extra term is the usual topological
term of four-dimensional gauge theory
\begin{equation}\label{zurox}\tilde I_A=-\frac{\theta}{32\pi^2}\int_{x^3\geq 0}\d^4x \,\epsilon^{\mu\nu\alpha\beta}\,\Tr\,F_{\mu\nu}F_{\alpha\beta}, \end{equation}
with
\begin{equation}\label{polx}  \frac{\theta}{2\pi}= \frac{2a}{1-a^2}\frac{4\pi}{\gym^2}.           \end{equation}
Viewed as an equation for $a$ with $\theta$, $\gym$ fixed, (\ref{polx}) has two roots.  The two roots correspond
to half-BPS boundary conditions of the D3-NS5 and D3-$\bar{\mathrm{NS5}}$ systems, respectively.

Although written as a bulk integral, $\tilde I_A$ has only a boundary variation, simply because
on a manifold $V$ without boundary, $\int_V\,\Tr\,F\wedge F$ is a topological invariant.  In
fact, we can almost write $\tilde I_A$ as a boundary integral, the integral over the surface
$x^3=0$ of the Chern-Simons form:
\begin{equation}\label{zormm}\tilde I_A=-\frac{\theta}{8\pi^2}\int_{x^3=0}\d^3x\,\epsilon^{\mu\nu\lambda}\,
\Tr\,\left(A_\mu\partial_\nu A_\lambda+\frac{2}{3}A_\mu A_\nu A_\lambda\right).\end{equation}

But there is a problem with this last formula: the Chern-Simons integral on a three-manifold is not
quite gauge-invariant.  The right hand side of (\ref{zormm})  is gauge-invariant modulo an integer multiple of
$\theta$.  Since the action of a quantum theory must be well-defined modulo
$2\pi\Z$, $\tilde I_A$ would not make sense as the action of a purely three-dimensional theory
unless $\theta$ is an integer multiple of $2\pi$.  This case is not trivial, since in the presence
of an NS5-brane, there is no symmetry of shifting $\theta$ by $2\pi$; a shift $\theta\to\theta+2\pi k$
would convert the NS5-brane to a $(1,k)$ fivebrane.  However, we do not wish to be limited
to the case $\theta\in 2\pi\Z$.  The reason that we are not so restricted is that we are not doing
gauge theory on an abstract three-manifold; rather, the three-manifold at $x^3=0$ on which we do
the integral (\ref{zormm}) is the boundary of a four-manifold $x^3\geq 0$ on which the gauge
theory is defined; the precise, gauge-invariant definition of $\tilde I_A$ is the original four-dimensional integral (\ref{zurox}).   Still, it can be convenient to informally write $\tilde I_A$ as a Chern-Simons integral
(\ref{zormm}), and we will sometimes do so.

\subsubsection{Wick Rotation}\label{wick}

So far, our formulas have been in Lorentz signature, to make contact with \cite{gw} and to emphasize the
fact that, as long as the parameter $a$ is real, our boundary condition is unitary and physically sensible.
However, to make contact with topological field theory in the rest of this paper, it is helpful
to write the formulas analogous to the above in Euclidean signature.  A Wick rotation
$x^0\to -ix^0$ reverses the sign\footnote{$\tilde I_X$ is free of derivatives and is a contribution
to the potential energy $\eurm V$ of the theory.  As usual, $\eurm V$ appears
in the Lorentz signature action with a minus sign and in the Euclidean signature action with
a plus sign.  Concretely, a contribution $\Delta I_L=-\int \d t \,\eurm V$ to the Lorentz signature
action $I_L$  leads in the path integral  to a factor $\exp(i\,\Delta I_L)=\exp(-i\int \d t \,\eurm V)$.  After
Wick rotation $t\to -i t$, this becomes $\exp(-\int\d t\,\eurm V)$, which is interpreted as a factor
in $\exp(-I_E)$, where $I_E$ is the Euclidean action.  So the contribution to $I_E$ is
$+\int\d t\,\eurm V$.  In the case of the Chern-Simons function,  as it is a topological
invariant, it is not affected directly by the Wick rotation.  The coefficient with which it appears in the action acquires a factor of $-i$ under Wick
rotation purely because of the convention that the integrand of the path integral is $\exp(iI_L)$
in Lorentz signature and $\exp(-I_E)$ in Euclidean signature. } of $\tilde I_X$, and multiplies $\tilde I_A$ by $-i$.  So in Euclidean
signature, combining the terms involving $X$ and $A$, the boundary interactions of the D3-NS5 system are
\begin{equation}\label{duster}I^*=\frac{1}{\gym^2}\int_{x^3=0}\d^3x\left(-\frac{2a}{3(1+a^2)}\epsilon^{abc}\Tr\,X_a[X_b,X_c] +i\frac{2a}{1-a^2}\epsilon^{\mu\nu\lambda}\Tr\,\left(
A_\mu\partial_\nu A_\lambda+\frac{2}{3}A_\mu A_\nu A_\lambda\right)\right).\end{equation}

In a convenient notation in which $\N=4$ super Yang-Mills is obtained by dimensional reduction
from ten dimensions, with the ten dimensions labeled by $x^0,\dots, x^{9}$, the Euclidean signature version of the chirality condition for supersymmetry generators
and fermions is
\begin{equation}\label{pmex}\Gamma_0\Gamma_1\cdots\Gamma_{9}\varepsilon=-i\varepsilon,~~\Gamma_0\Gamma_1\cdots
\Gamma_{9}\lambda = -i\lambda.\end{equation}

\subsection{Comparison To Topological Field Theory}\label{topfield}

So far we have emphasized the half-BPS nature of the boundary condition of interest.  We will also need to
understand this boundary condition from the vantage point of topological field theory.
The background necessary for this analysis can be found in section 3 of \cite{KW}, to which we refer
 for detail (some aspects were
treated originally in \cite{yamron}).  Here we will just summarize some necessary facts.

\subsubsection{Twisting}\label{twisting}

The basic idea is to construct a four-dimensional topological field theory by twisting of $\N=4$ super
Yang-Mills theory.  Postponing the consideration of possible boundary conditions, we consider $\N=4$
super Yang-Mills realized on a system of D3-branes parametrized by $x^0,\dots,x^3$.  The usual rotation
group (in Euclidean signature) is $SO(4)$, rotating these coordinates, while the normal directions $x^4,\dots,x^9$ are rotated by the $SO(6)$ group of $R$-symmetries.  To define a topological field
theory, one defines a group $SO'(4)$ that acts by rotating $x^0,\dots,x^3$ in the usual way, while
simultaneously rotating four normal coordinates $x^4,\dots,x^7$.  We pick a supersymmetry generator
$\varepsilon$
that is $SO'(4)$-invariant, meaning that it obeys
\begin{equation}\label{zurko}\left(\Gamma_{\mu\nu}+\Gamma_{4+\mu,4+\nu}\right)\varepsilon=0,~~\mu,\nu=0,\dots,
3.
\end{equation}
Denoting as $Q$ the supersymmetry generated by such
an $\varepsilon$, arguments of a standard type show that upon restricting to $Q$-invariant operators
and states, one obtains a four-dimensional topological field theory.

From the point of view of $SO'(4)$ symmetry, four of the adjoint-valued scalar fields of $\N=4$ super Yang-Mills theory are reinterpreted
as an adjoint-valued one-form $\phi=\sum_{\mu=0}^3\phi_\mu\,\d x^\mu$, while the other two combine
two an adjoint-valued complex scalar field $\sigma$.  $SO'(4)$ commutes with a group $SO(2)\cong U(1)$ of
$R$-symmetries that rotates $x^8$ and $x^9$.  We normalize its generator $\EF$ so that $\sigma$ has charge
2.

This decomposition of the $R$-symmetry group and of the scalar fields of $\N=4$ super Yang-Mills theory
differs from that made in section \ref{funx}.  In that discussion, the $x^\mu$, $\mu=0,\dots,3$ were split
in tangential coordinates with $\mu\leq 2$ and a normal coordinate $x^3$.  In matching the two
descriptions, we identify the tangential part of $\phi$, that is  $\vec\phi=\sum_{\mu=0}^2\phi_\mu\,\d x^\mu$, with $\vec X$, and we identify the normal part $\phi_3$ with a component of $\vec Y$, say $Y_1$.
(We also set $\sigma=Y_2-iY_3$.) The boundary couplings (\ref{duster}) become in this notation
\begin{equation}\label{fluster}I^*=\frac{1}{\gym^2}\int_{x^3=0}\d^3x\,\epsilon^{\mu\nu\lambda}\Tr\,
\left(-\frac{4a}{3(1+a^2)}\phi_\mu\phi_\nu\phi_\lambda +i\frac{2a}{1-a^2}
\left(A_\mu\partial_\nu A_\lambda+\frac{2}{3}A_\mu A_\nu A_\lambda\right)\right).\end{equation}

\subsubsection{Comparing The Two Descriptions}\label{zongor}

However, rewriting (\ref{duster}) in topological field theory notation is only a reasonable thing to do
if the boundary condition that leads to (\ref{duster}) actually preserves the symmetry of the topological
field theory.  So let us explain why this is true.

First of all, the condition (\ref{zurko}) for $SO'(4)$-invariance of the supersymmetry generator actually
has a two-dimensional space of solutions.  It is possible to pick a basis of solutions $\varepsilon_\ell$,
$\varepsilon_r$ that are chiral in the four-dimensional sense,
\begin{equation}\label{chiral}\Gamma_{0123}\varepsilon_\ell=-\varepsilon_\ell,~~\Gamma_{0123}\varepsilon_r
=\varepsilon_r.\end{equation}
It is possible to normalize $\varepsilon_\ell$ and $\varepsilon_r$ so that,\footnote{
In the following formulas, there is no sum over $\mu$; a covariant version  reads
$(\Gamma_\mu\Gamma_{4+\nu}+\Gamma_\nu\Gamma_{4+\mu})\varepsilon_\ell=-2g_{\mu\nu}\varepsilon_r$,
$(\Gamma_\mu\Gamma_{4+\nu}+\Gamma_\mu\Gamma_{4+\mu})\varepsilon_r=2g_{\mu\nu}\varepsilon_l$.} for  $\mu=0,1,2$, or 3,
\begin{equation}\label{porm} \Gamma_{\mu,4+\mu}\varepsilon_\ell=-\varepsilon_r,~~\Gamma_{\mu,4+\mu}\varepsilon_r=\varepsilon_\ell.
\end{equation}
In constructing a topological field theory, we may take the supersymmetry generator $\varepsilon$ to be
an arbitrary linear combination of $\varepsilon_\ell$ and $\varepsilon_r$.  Up to an inessential scaling,
we take
\begin{equation}\label{zonkotz}\varepsilon=\varepsilon_\ell + t\varepsilon_r.\end{equation}
(We allow $t=\infty$, which corresponds up to scaling to $\varepsilon=\varepsilon_r$.)

So we get a family of topological field theories parametrized by a complex variable $t$.
Now we can make contact with the D3-NS5 system.  From (\ref{pmex}), (\ref{chiral}), and (\ref{thirsty}),
we have \begin{equation}\label{look}B_0\varepsilon_\ell=i\varepsilon_\ell,~~B_0\varepsilon_r=-i\varepsilon_r.\end{equation}  Using also (\ref{porm}) and (\ref{zurko}), one can show, with some gamma matrix algebra, that \begin{equation}\label{ook} B_1\varepsilon_\ell
=-\varepsilon_r, ~~B_1\varepsilon_r=-\varepsilon_\ell.\end{equation}  It follows that
\begin{equation}\label{orxo}\left(1+i\frac{1-t^2}{1+t^2}B_0+\frac{2t}{1+t^2}B_1\right)\left(\varepsilon_\ell
+t\varepsilon_r\right)=0.\end{equation}
On the other hand, with the help of (\ref{dux}), we see that the object $\varepsilon_0$ defined in (\ref{plux}) obeys the same equation
\begin{equation}\label{norxo}\left(1+i\frac{1-t^2}{1+t^2}B_0+\frac{2t}{1+t^2}B_1\right)\varepsilon_0=0\end{equation}
if and only if the parameter $a$ used in describing the D3-NS5 system is related to the parameter $t$ of
the topological field theory by
\begin{equation}\label{flox}a=i\frac{1-it}{1+it}.\end{equation}
The half-BPS boundary condition of the D3-NS5 system preserves every supersymmetry with a generator $\varepsilon=\eta\otimes\varepsilon_0$, with $\eta\in \V_8$.  So in particular, once we impose the relation
(\ref{flox}) between the parameters, this boundary condition preserves the supersymmetry generator of the
twisted topological field theory.   Substituting (\ref{flox}) in (\ref{polx}) and solving for $t^2$, we get the surprisingly simple
result
\begin{equation}\label{doner} t^2=\frac{\bar\tau}{\tau}.\end{equation}
The operation $t\to -t$ corresponds to $a\to -1/a$ and to exchange of the D3-NS5 and D3-$\bar{\mathrm{NS5}}$ systems.\footnote{As long as the gauge theory parameters $g_{\sf{YM}}$ and $\theta$ are real, $\bar\tau$ is the complex conjugate of $\tau$,
so (\ref{doner}) implies that $t$ has modulus 1, and (\ref{flox}) then implies that $a$ is real.  When we get to
topological field theory, we may choose to analytically continue $\tau$ and $\bar\tau$ to independent complex
variables, whereupon $t$ no longer has modulus 1 and $a$ becomes complex.}

With the aid of (\ref{flox}), the boundary couplings (\ref{fluster}) can be rewritten
\begin{equation}\label{guster}I^*=\frac{1}{\gym^2}\int_{x^3=0}\d^3x\,\epsilon^{\mu\nu\lambda}\Tr\,
\left(-\frac{t+t^{-1}}{3}\phi_\mu\phi_\nu\phi_\lambda +\,\frac{t+t^{-1}}{t-t^{-1}}\,
\left(A_\mu\partial_\nu A_\lambda+\frac{2}{3}A_\mu A_\nu A_\lambda\right)\right).\end{equation}

\subsubsection{Global Formulation And Brane Construction}\label{desc}

The topological field theory under discussion  can be defined on any (oriented) four-manifold, possibly with
boundary.  One
can motivate how to do this by generalizing the D3-NS5 system beyond the special geometry that we have considered so far.

Introducing a slightly new nomenclature for a reason that will soon be clear,
let $V_0$ be an arbitrary oriented four-manifold, and consider Type  IIB superstring theory on $T^*V_0\times \R^2$.  For the moment, suppose that $T^*V_0$ admits a complete Calabi-Yau metric.  Consider $N$ D3-branes
wrapped on $V_0\times \{0\}\subset T^*V_0\times \R^2$, where $0$ is a point in $\R^2$ (the ``origin'').
This system is topologically twisted in precisely the way described in section \ref{twisting}.  Type IIB
superstring theory on $T^*V_0\times \R^2$ has four unbroken supersymmetries, of which two are preserved
by the D3-branes wrapped on $V_0$.  The two unbroken supersymmetries precisely correspond to the $SO'(4)$-invariant supersymmetries with generators $\varepsilon_\ell$ and $\varepsilon_r$, as described above.  This
approach to realizing topologically twisted gauge theories via branes was described in \cite{bersh}.
The basic idea is that the twisting of the normal bundle to $V_0\subset T^*V_0$ leads to the $R$-symmetry
twist that is used in defining a topological field theory.

The above remarks are unaffected by possible presence of a  Type IIB theta-angle -- which becomes the theta-angle of the gauge theory along the D3-branes. Now  suppose we are given an oriented three-manifold $W\subset V_0$, such that $T^*W\subset T^*V_0$ is a supersymmetric cycle (a complex submanifold).  Then we can
wrap an NS5-brane on $T^*W\times \{0\}\subset T^*V_0\times \R^2$.  The NS5-brane preserves half the supersymmetry of Type IIB on $T^*V_0\times \R^2$ (that is, in the absence of D3-branes, two supercharges are
conserved, while if one includes D3-branes, there is one conserved supersymmetry).
Moreover, such a $W$, being oriented and of codimension 1 in $V_0$, may potentially
divide $V_0$ into two pieces.  Assuming this is the case, either  one of
the pieces, say $V$, is a four-manifold of boundary $W$.  Now instead of D3-branes wrapped on $V_0$, we can consider
D3-branes wrapped on $V$ and ending on the NS5-brane.  The support of the D3-branes is thus $V\times \{0\}
\subset T^*V_0\times \R^2$.  With both types of brane present, there is now only one conserved supercharge; its generator is a linear combination of $\varepsilon_\ell$ and $\varepsilon_r$, depending on the theta-angle and other parameters.

The geometry assumed above is rather special. For example, a complete Calabi-Yau metric on $T^*V_0$
exists if $V_0$ is $S^4$ or $S^2\times S^2$, but not for most $V_0$.  Actually, the above construction can be
generalized by replacing $T^*V_0$ by any Calabi-Yau four-fold $X$ that admits $V_0$ as a special Lagrangian
four-cycle; similarly, $T^*W$ can be replaced by any divisor in $X$. Moreover, we really only care about
$V$, not $V_0$.  So many cases can be realized, but we probably do not have enough freedom to accomodate an arbitrary $W$ and $V$.  Similarly, the brane construction naturally has a D3-brane gauge group $U(N)$, and though one could accomodate orthogonal or symplectic gauge groups by adding an orientifold plane to the construction, this construction does not naturally lead to exceptional gauge groups.

From our point of view, the most obvious merit of the brane construction is motivational.  It presumably does not literally work, globally, for all oriented four-manifolds $V$ with arbitrary boundary $W$; nor does it work for all gauge groups.  But the brane construction suggests a purely field theoretic construction that does work in general.  The $R$-symmetry twist that was sketched in section \ref{twisting} (and was described in far more detail in section 3 of \cite{KW}) preserves two
supercharges when the theory is formulated on an arbitrary four-manifold $V$; one linear combination
of these two supercharges is preserved when $V$ has a boundary  $W$, with a boundary condition
that is modeled locally on the D3-NS5 system.  All these statements can be verified by infinitesimal calculations on $V$ and $W$, and the fact that they work in the brane construction is enough to ensure
that, as field theoretic statements, they work in general.

Apart from encouragement, what else do we gain from the brane construction?
One answer is that ultimately, we will have to understand the behavior under certain nonperturbative dualities.  For this, the brane construction provides invaluable insight.  A second answer is that to
understand Khovanov homology, we will have to ultimately go to five dimensions, where Yang-Mills quantum
field theory is not ultraviolet-complete.  The most rigorous and general formulation of
our construction will ultimately be given in purely field theoretic terms, but the field theory required is
the six-dimensional $(0,2)$ theory (from which five-dimensional super Yang-Mills theory can be derived), whose existence and properties are known only from its multiple relations to string theory, $M$-theory, and branes.  So the insights that come from brane constructions are again essential.

\subsubsection{Wilson Loops}\label{uzim}

$\N=4$ super Yang-Mills theory in four dimensions admits 1/16-BPS Wilson loop operators \cite{zarembo}.
They are constructed as follows.  The supersymmetry transformation law for the bosonic fields of this theory is
\begin{equation}\label{zar}\delta A_I=i\bar\varepsilon\Gamma_I\lambda=-i\bar\lambda\Gamma_I\varepsilon,~~I=0,\dots,9.
\end{equation}
Here we use a ten-dimensional notation; for $I\leq 3$, $A_I$ is a component of a gauge field,
and for $I\geq 4$, it is a scalar field.  By twisting, we have converted four of the scalar fields
to a one-form $\phi$. Usually, we use Greek letters $\mu,\nu\,\dots$ for four-dimensional indices, so we write $A=\sum_{\mu=0}^3 A_\mu\d x^\mu$, $\phi=\sum_{\mu=0}^3 \phi_\mu\,\d x^\mu=\sum_{\mu=0}^3 A_{4+\mu}\d x^\mu$.

Suppose that $\varepsilon$ is such that
\begin{equation}\label{omar}\left(\Gamma_\mu+i\Gamma_{4+\mu}\right) \varepsilon=0,~~\mu=0,\dots,3.  \end{equation}
Clearly, in this case,  Wilson operators of the form
\begin{equation}\label{qom}\Tr_R\,P\exp\oint_K (A+i\phi) \end{equation}
are invariant, for an arbitrary embedded loop $K$ in spacetime and any representation $R$ of the gauge group.  Similarly, if
\begin{equation}\label{zomar}\left(\Gamma_\mu-i\Gamma_{4+\mu}\right) \varepsilon=0,~~\mu=0,\dots,3,  \end{equation}
then there are supersymmetric Wilson operators of the form
\begin{equation}\label{qaom}\Tr_R\,P\exp\oint_K (A-i\phi). \end{equation}

As explained in \cite{KW}, the supersymmetry generator $\varepsilon=\varepsilon_\ell+t\varepsilon_r$
of interest here obeys (\ref{omar}) or (\ref{zomar}) precisely for $t=i$ or $t=-i$.  Therefore, in general,
supersymmetric Wilson operators appear in this family of topological field theories precisely at those values of $t$.  The occurrence of supersymmetric Wilson operators at $t=\pm i$ is actually important in
geometric Langlands, and played a major role in \cite{KW}.  But in the present paper, we are interested
in other values of $t$.

Therefore, we do not have supersymmetric Wilson operators -- except at the boundary of $V$.
For a Wilson operator supported entirely at the boundary of $V$, we can use the boundary conditions obeyed by $\lambda$, as well as the conditions obeyed by  $\varepsilon$, to establish supersymmetry.
We will explore the conditions that on the boundary of $V$
\begin{equation}\label{lomar}0=\delta(A_\mu+ w \phi_\mu)=-i\bar\lambda(\Gamma_\mu+w\Gamma_{4+\mu})\varepsilon,~~\mu=0,1,2.\end{equation}
The reason that we impose this condition only for $\mu<3$ is that the goal is to construct Wilson operators
that are supersymmetric only on the boundary of $V$, at $x^3=0$.  In (\ref{lomar}), $w$ is a complex
number, to be determined.  If (\ref{lomar}) holds, then
upon setting
\begin{equation}\label{dolbo}\A_w=A+w\phi,\end{equation}
we can construct
supersymmetric Wilson operators
\begin{equation}\label{plom}\Tr_R\,P\exp\oint_K\A_w,\end{equation}
for any knot $K$ in the boundary of $V$.

A preliminary reduction is that $\bar\lambda(\Gamma_\mu+w\Gamma_{4+\mu})\varepsilon=\bar\lambda\Gamma_\mu(
1+w\Gamma_{\mu,4+\mu})\varepsilon=\bar\lambda\Gamma_\mu(1+iwB_0B_1)\varepsilon$.  In the second step,
we used the fact that $\Gamma_{\mu,\mu+4}\varepsilon=iB_0B_1\varepsilon$.  This follows from (\ref{porm}),
(\ref{ook}), (\ref{look}), and the fact that $\varepsilon$ is a linear combination of $\varepsilon_\ell$ and $\varepsilon_r$.  So we need to explore the vanishing of
\begin{equation}\label{tomax}\bar\lambda \Gamma_\mu(1+iwB_0B_1)\varepsilon.\end{equation}
The expression $(\lambda,\varepsilon)=\bar\lambda\Gamma_\mu\varepsilon$, for any $\mu$, gives a symmetric
bilinear form on the $\mathbf{16}$ of $SO(1,9)$.  As before, we decompose $\mathbf{16}=\V_8\otimes \V_2$.
For $\mu\leq 2$, $\bar\lambda\Gamma_\mu\varepsilon$ is the tensor product of a symmetric bilinear form on $\V_8$ (transforming as $(\mathbf 3,\mathbf 1,\mathbf 1)$ under $SO(1,2)\times SO(3)_X\times SO(3)_Y$) with
a symmetric bilinear form on $\V_2$.  If we represent $\vartheta,\varepsilon_0\in \V_2$ as two-component
column vectors, then the form on $\V_2$ can be written as $\vartheta^T\varepsilon_0$.  The fermion
boundary condition of the D3-NS5 system says that $\lambda$, on the boundary, is the tensor product of some
vector in $\V_8$ with $\vartheta\in \V_2$ (where $\vartheta$ was defined in eqn. (\ref{hobnob})), and similarly the generator $\varepsilon$ of any unbroken supersymmetry of the D3-NS5 boundary condition, including the
one of topological interest, is the tensor product of some vector in $\V_8$ with $\varepsilon_0$ (defined
in eqn. (\ref{plux})).  So to justify the definition (\ref{plom}) of supersymmetric Wilson loops, we require
\begin{equation}\label{zomax}\vartheta^T(1+iwB_0B_1)\varepsilon_0=0.\end{equation}
With the definitions of $\vartheta$ and $\varepsilon_0$ and the formulas (\ref{dux}) for $B_0$ and $B_1$,
it is straightforward to compute that (\ref{zomax}) is obeyed precisely if
\begin{equation}\label{potomoz}w=i\frac{a^2-1}{a^2+1}=\frac{t-t^{-1}}{2},\end{equation} where
in the last step, we used the relation (\ref{flox}).  For real $\theta$ and $\gym$, $a$ is always real (by virtue of (\ref{polx})),
so the first formula in (\ref{potomoz}) shows that $w$ is always imaginary.
With the help of (\ref{doner}), we find
\begin{equation}\label{ozox}w=\mp i\frac{\mathrm{Im}\,\tau}{|\tau|},\end{equation}
with the signs corresponding to $t=\pm |\tau|/\tau$.

The action $I$ of $\N=4$ super Yang-Mills theory on a four-manifold $V$ is the sum of a term proportional to $1/\gym^2$, which contains the kinetic energy for all fields, and a term proportional to $\theta$:
\begin{equation}\label{fornox}I=\frac{1}{\gym^2}\int_V\d^4x\sqrt g\L_{\mathrm{kin}}+i\frac{\theta}{32\pi^2}\int_V\d^4x\,\epsilon^{\mu\nu\alpha\beta}\,\Tr\,F_{\mu\nu}F_{\alpha\beta}. \end{equation}
Here, for later reference, the part of $\L_{\mathrm{kin}}$ that involves $A,\phi$ only is
(in Euclidean signature)
\begin{equation}\label{dolfo}\L^{A,\phi}_{\mathrm{kin}}=-\Tr\,\left(\frac{1}{2}F_{\mu\nu}F^{\mu\nu}+D_\mu\phi_\nu D^\mu\phi^\nu +
R_{\mu\nu}\phi^\mu\phi^\nu+\frac{1}{2}[\phi_\mu,\phi_\nu]^2          \right). \end{equation}
($R_{\mu\nu}$ is the Ricci tensor of $V$; when $V$ is not Ricci-flat, the indicated term proportional to $R_{\mu\nu}$ is needed for
$Q$-invariance.)

Let us first consider the case that $V$ has no boundary.  Both terms on the right hand side of (\ref{fornox}) are
$Q$-invariant.  The $\theta$ term is $Q$-invariant because, more generally, it is a topological invariant, unchanged in any continuous
deformations.  It represents a nonzero class in the cohomology of $Q$ (unless $t=\pm i$, as discussed momentarily).  One might suspect that the integral of $\L_{\mathrm{kin}}$ would
vanish in the cohomology of $Q$, as happens in many twisted topological field theories, but this is actually not the case.  Instead, as shown in \cite{KW}, the first term on the right of (\ref{fornox})
is equivalent mod $\{Q,\cdots\}$ to a multiple of the second term.  The precise relation is
\begin{equation}\label{ornox} I=\{Q,\cdots\}+\frac{2 \pi i \Psi}{32\pi^2}\int_V\d^4x\,\epsilon^{\mu\nu\alpha\beta}\,
\Tr\,F_{\mu\nu}F_{\alpha\beta},
  \end{equation}
where
\begin{equation}\label{zornox}\Psi=\frac{\theta}{2\pi}+\frac{4\pi i}{\gym^2}\frac{t-t^{-1}}{t+t^{-1}}\end{equation}
was called in \cite{KW} the canonical parameter.

Before twisting, $\N=4$ super Yang-Mills theory in four dimensions depends on a complex parameter $\tau=\theta/2\pi+4\pi i/\gym^2$,
which is valued in the upper half-plane.  Upon twisting, an additional complex parameter $ t$ appears in the choice of the topological
supercharge.   It was shown in \cite{KW} that the  topological field theory obtained in this way depends on
the two parameters $\tau$ and $t$ only via their combination $\Psi$.   A sketch of this argument is as follows.  For the special cases
$t=\pm i$, which correspond to $\Psi=\infty$, one shows directly that both terms on the right of (\ref{fornox}) are of the form
$\{Q,\cdots\}$, so the parameter $\tau$ is irrelevant if $\Psi=\infty$.  (The case $\Psi=\infty$ is important for geometric Langlands,
but not for the present paper.)  For $t\not=\pm i$, it is shown in \cite{KW} that by including auxiliary fields and making a local
redefinition of the fermion fields, one can make the $Q$-transformation laws of all fields independent of $t$.  After  one thus eliminates
the dependence of the theory on $t$ that is hidden in the definition of $Q$, eqn. (\ref{ornox}) shows that for fixed $\Psi$, $t$ appears only
in a term $\{Q,\cdots\}$ and thus  is irrelevant for the topological field theory.

In \cite{KW}, the transformation of $t$ under electric-magnetic duality was determined.  It was shown that under a general
$S$-duality transformation
\begin{equation}\label{lomus}\tau\to\frac{ a\tau+ b}{c\tau+d},\end{equation}
$t$ transforms by
\begin{equation}\label{zomus}t\to \frac{c\tau+d}{|c\tau+d|}t \end{equation}
and that
$\Psi$ transforms just as $\tau$ does:
\begin{equation}\label{blomus}\Psi\to \frac{a\Psi+b}{c\Psi+d}.\end{equation}
(Unlike $\tau$, $\Psi$ is not restricted to take values in the upper half plane.)
The formula (\ref{zornox}) for $\Psi$ holds for all $\tau,$ $t$.
Imposing the relations (\ref{polx}), (\ref{flox}) that are natural in studying the D3-NS5 system, we can derive several interesting alternative
formulas.  Eliminating $t$  in favor of $\gym$ and $\theta$, we find
\begin{equation}\label{por} \Psi=\frac{|\tau|^2}{\mathrm{Re}\,\tau},\end{equation}
showing that $\Psi$ is always real for the D3-NS5 system with physical values of the parameters (real $\gym$ and $\theta$).
Alternatively, eliminating $\theta$  in favor of $\gym$ and $t$, we get
\begin{equation}\label{tror}\Psi=\frac{4\pi i}{\gym^2}\left(\frac{t-t^{-1}}{t+t^{-1}}-\frac{t+t^{-1}}{t-t^{-1}}\right).\end{equation}

\def\CS{{\mathrm{CS}}}
Now let us discuss what happens when $V$ has a boundary.     The integral $\int_V\d^4x\,\epsilon^{\mu\nu\alpha\beta}\Tr\,F_{\mu\nu}
F_{\alpha\beta}$ is no longer $Q$-invariant, but varies by a boundary term.  It is convenient to replace this integral
by a multiple of the Chern-Simons function.
We define the Chern-Simons function $\CS(\A)$, for  any connection $\A$, possibly complex-valued, by
\begin{equation}\label{boscombo}\mathrm{CS}(\A)=
\frac{1}{4\pi}\int_{\partial V}\d^3x\,\epsilon^{\mu\nu\lambda}\Tr\,\left(\A_\mu\partial_\nu \A_\lambda
+\frac{2}{3}\A_\mu \A_\nu \A_\lambda\right).\end{equation}
In terms of this function, we can make the following substitution on the right hand side of eqn.
(\ref{ornox}):
\begin{equation}\label{bornox}\frac{2 \pi i \Psi}{32\pi^2}\int_V\d^4x\,\epsilon^{\mu\nu\alpha\beta}\,
\Tr\,F_{\mu\nu}F_{\alpha\beta}\to i\Psi\,\mathrm{CS}(A),\end{equation}   As was explained in the context of (\ref{zormm}), the relation (\ref{bornox})
must be treated with care, since $\mathrm{CS}(A)$ is not quite gauge-invariant (but only invariant under topologically trivial gauge
transformations), and the equality suggested in (\ref{bornox}) really holds only modulo an integer multiple of $2\pi i\Psi$.
The substitution  (\ref{bornox}) is a convenient shorthand, which can be used in computing the variation of the integral on the left
under a small change in the connection, such as that generated by $Q$. For future reference, writing $h$ for
the dual Coxeter number of $G$, we can write a formula equivalent to (\ref{boscombo}) in terms of a trace
$\Tr_{\mathrm{ad}}$ in the adjoint representation of $G$:
\begin{equation}\label{boscomb}\mathrm{CS}(\A)=
\frac{1}{8\pi h}\int_{\partial V}\d^3x\,\epsilon^{\mu\nu\lambda}\Tr_{\mathrm{ad}}\,\left(\A_\mu\partial_\nu \A_\lambda
+\frac{2}{3}\A_\mu \A_\nu \A_\lambda\right).\end{equation}

Concretely, when we write $\Psi$ as in (\ref{zornox}), the part of $i\Psi\,\mathrm{CS}(A)$ that is proportional  to $\theta$ is already present in (\ref{guster}).
The part proportional to $1/\gym^2$ appears upon writing the kinetic energy as $\{Q,\cdots\}$
plus a multiple of the theta term, to arrive at (\ref{ornox}). In the derivation of (\ref{ornox}), one can assume that $V$
has no boundary, since the integral $\int_V\Tr\,F\wedge F$ is in general non-zero even
in that case.  In section \ref{zomo}, we will repeat the derivation of eqn. (\ref{ornox}), for the case that $V$ has a non-empty boundary.  When we do this, additional boundary terms will appear; this should come as no surprise, since one such term is already visible in (\ref{guster}) and $Q$-invariance implies that there must be more.  In fact, the boundary couplings must be a function
of $\A_w$ only (modulo $Q$-exact terms), since this is the only non-trivial $Q$-invariant combination of boundary fields.

One can determine the form of the full boundary couplings without any computation, using
gauge invariance and dimensional analysis plus the fact that the boundary coupling is
a function only of $\A_w$.
These conditions imply that it must be simply a multiple of $\CS(\A_w)$; there is no other local, gauge-invariant functional of dimension three.
 For a reason that we will explain momentarily, the coefficient of $\CS(\A_w)$ is precisely $i\Psi$.  So the generalization of (\ref{ornox}) in the presence of a boundary is
\begin{equation}\label{tornox}I=\{Q,\cdots\}+i\Psi\,\mathrm{CS}(\A_w).\end{equation}
When $\CS(\A_w)$  is written explicitly as a function of $A$ and $\phi$, the $\phi$-dependent terms are given by local, gauge-invariant
integrals, since
\begin{align}\label{lornox}\mathrm{CS}(\A_w)=\mathrm{CS}(A)+\frac{1}{4\pi}\int_{\partial V}\d^3x\,\epsilon^{\mu\nu\lambda}\, \Tr\,\left(w\phi_\mu F_{\nu\lambda}
+w^2\phi_\mu D_\nu\phi_\lambda +\frac{2w^3}{3}\phi_\mu\phi_\nu\phi_\lambda\right). \end{align}
Because those terms are local, gauge-invariant integrals over the boundary of $V$, they cannot be detected directly by a computation that assumes that this boundary is empty.

However, because $\mathrm{CS}(A)$ is not completely gauge-invariant, and must really be written as an integral over $V$, its coefficient
is determined by the analysis of the case $\partial V=\varnothing$ in \cite{KW} and can be read off from
(\ref{ornox}), via the substitution (\ref{bornox}).  From this we learn that the coefficient of $\mathrm{CS}(A)$ in the boundary interaction is
$ i\Psi$, and in view of (\ref{lornox}),  the coefficient of $\CS(\A_w)$ must
be the same.  Still, one would naturally like to  generalize (\ref{ornox}) to the
case $\partial V\not=\varnothing$, so as to see explicitly the origin of the $\phi$-dependent
boundary couplings.  This is one of our next goals.

\subsection{Localization And The Boundary Formula}\label{zomo}

\def\cV{{\mathcal V}}
Under favorable conditions, computations in topological field theory can be localized on configurations that obey $\{Q,\zeta\}=0$, for all fermion fields $\zeta$.  Among the fermions of\footnote{The fermion number $\EF$ was
defined in section \ref{twisting}.} $\EF=-1$ in the present model are a selfdual two-form $\chi^+$, an anti-selfdual two-form $\chi^-$, and a scalar
$\eta$ (like all fields of $\N=4$ super Yang-Mills theory, they are adjoint-valued). They have
the property  that $\cV^+=\{Q,\chi^+\}$, $\cV^-=\{Q,\chi^-\}$,
and $\cV^0=\{Q,\eta\}$ depend on $A,\phi$ only:
\begin{align}\label{mexico}\notag  \cV^+ & = \left(F-\phi\wedge\phi+t \d_A\phi\right)^+ \\
                                             \cV^-& = \left(F-\phi\wedge\phi-t^{-1}\d_A\phi\right)^- \\
                                       \notag      \cV^0& = D_\mu\phi^\mu.\end{align}
Here for any two-form $\alpha$, we write $\alpha^+$ and $\alpha^-$ for its selfdual and anti-selfdual projections.
Localization on real fields $A,\phi$ can be achieved for real\footnote{According to eqn. (\ref{doner}),
$t$ is not real for physical values of the parameters; in fact, for weak coupling, it is close to
$\pm i$.  We are here using our freedom to change $t$ as we wish while keeping $\Psi$
fixed.} $t$ by adding a suitable term to the action $I$:
\begin{equation}\label{expfram} I\to I-\frac{1}{\epsilon}\left\{Q,\int_V\Tr\,\left(\chi^+\cV^++\chi^-\cV^-+\chi^0\cV^0\right)\right\}
  = I-\frac{1}{\epsilon}\int_V\Tr\,\left((\cV^+)^2+(\cV^-)^2+(\cV^0)^2+\dots\right),\end{equation}
where $\epsilon$ is a small parameter and
the omitted terms are fermion bilinears.  For  $t$ real, $\cV^+$, $\cV^-$, and $\cV^0$ are real, and the modified action diverges
as $1/\epsilon$ unless the localization equations
\begin{equation}\label{explicitly}\left(F-\phi\wedge\phi+t \d_A\phi\right)^+ = \left(F-\phi\wedge\phi-t^{-1}\d_A\phi\right)^-  = D_\mu\phi^\mu
=0\end{equation}
are satisfied.  So the path integral is supported, for $\epsilon\to 0$, on the space of solutions of those equations.  On the other hand,
the integral is independent of $\epsilon$, since the term we have added to the action is of the form $\{Q,\cdots\}$.  The fact that this sort of argument is most straightforward for real $t$ is  not a major inconvenience, since for any $\Psi$
(other than $\Psi=\infty$) there is always a convenient choice of real $t$.

There are also localization equations that depend on $\sigma$.  For $t\not=\pm i$, they are
\begin{equation}\label{sigmaq} D_\mu\sigma=[\phi_\mu,\sigma] =[\sigma,\bar\sigma]=0. \end{equation}
They say that the gauge transformation generated by $\sigma$ is a symmetry of the whole configuration.  Under favorable
conditions (for instance, if the gauge field is irreducible and has no continuous gauge symmetries, or if a boundary conditions
sets $\sigma$ to zero somewhere), they imply that $\sigma$ is identically zero.

To understand explicitly the origin of the $\phi$-dependent boundary terms in (\ref{lornox}), we have to make more explicit the relation of
the localization procedure of eqn. (\ref{expfram}) to the physical action of $\N=4$ Yang-Mills theory.  The identity we need is actually
the generalization of eqn. (3.33)  of \cite{KW} to the case that $\partial V\not=\varnothing$:
\begin{align}\label{omox}\notag -&\int_V\d^4x\,\Tr\,\left(\frac{t^{-1}}{t+t^{-1}}\cV^+_{\mu\nu}\cV^{+\mu\nu}+\frac{t}{t+t^{-1}}\cV_{\mu\nu}^-\cV^{-\mu\nu}+(\cV^0)^2\right)\\  =&\int_V\d^4x \sqrt g\,\L_{\mathrm{kin}}^{A,\phi}+\frac{t-t^{-1}}{4(t+t^{-1})}
\int_V\d^4x\,\epsilon^{\mu\nu\alpha\beta}\,\Tr\,F_{\mu\nu}F_{\alpha\beta} \\ & \notag+\int_{\partial V}\d^3x \,\epsilon^{\mu\nu\lambda}\,\Tr\,
\left(  -\frac{2}{t+t^{-1}}\phi_\mu F_{\nu\lambda}-\frac{t-t^{-1}}{t+t^{-1}}\phi_\mu D_\nu\phi_\lambda+\frac{4}{3}\frac{1}{t+t^{-1}}\phi_\nu\phi_\nu
\phi_\lambda  \right) .
\end{align}
The left hand side of (\ref{omox}) is of the form $\{Q,\dots\}$ modulo fermion bilinears, by the
same reasoning as in (\ref{expfram}).  One can write a more complete version of the formula
that includes the fermions and also $\sigma$; this makes the formulas longer
without contributing additional boundary terms.
On the right hand side of (\ref{omox}), $\int \L_{\mathrm{kin}}^{A,\phi}$ is (after including fermions and $\sigma$) the part of the bulk action of $\N=4$
super Yang-Mills theory that is proportional to $1/\gym^2$.  The boundary terms that we want are the remaining terms on the right
hand side of (\ref{omox}).  Thus, after multiplying by $1/\gym^2$ and making the substitution (\ref{bornox}) in one term, we can rewrite (\ref{omox}) as follows:
\begin{align}\label{doormat}\notag \frac{1}{\gym^2}\int_V \d^4x \sqrt g \L_{{\mathrm{kin}}} =&\,\{Q,\cdots\}\\ +\frac{1}{\gym^2}\int_{\partial V}\d^3x\,&\epsilon^{\mu\nu\lambda}
\,\Tr\,\left(-\frac{t-t^{-1}}{t+t^{-1}}\left(A_\mu\partial_\nu A_\lambda+\frac{2}{3}A_\mu A_\nu A_\lambda  \right)+\frac{2}{t+t^{-1}}\phi_\mu F_{\nu\lambda}\right. \\\ \notag &~~~~~~~~~~~\left.+\frac{t-t^{-1}}{t+t^{-1}}\phi_\mu D_\nu\phi_\lambda-\frac{4}{3}\frac{1}{t+t^{-1}}\phi_\mu\phi_\nu\phi_\lambda \right) .  \end{align}
When we add the boundary terms that have appeared in (\ref{doormat}) to the boundary terms (\ref{guster}) that are already present in the
physical theory, before twisting, we find that the action has the expected form
\begin{equation}\label{uster}\{Q,\cdots\}+i\Psi \CS(\A_w),\end{equation}
with the expected value
$w=(t-t^{-1})/2$.

\subsection{Relation To Chern-Simons Theory}\label{onecite}

\def\U{{\mathcal U}}
So far we have analyzed this problem starting with the D3-NS5 system.  The coupling parameters
$g_{\sf{YM}}$ and $\theta$ and the parameter $a$ in the boundary condition were all real.  This physical starting point has many advantages,
such as the insight that it will give about the behavior under various nonperturbative dualities.

But let us see what we can say purely from the standpoint of topological field theory. Here we allow ourselves
to continue all parameters to complex values. Keeping $\Psi$ fixed, we may choose, roughly
speaking, any value of $t$ that we wish.  The only restriction is that we may only vary $t$ in such a way that the path integral continues
to converge.    What is convenient is to pick $t$ to be real, for then, as we recalled in section \ref{zomo}, there is a straightforward
procedure to localize the path integral on solutions of the equations $\cV^+= \cV^-=\cV^0=0$.

These are elliptic differential equations, as described in \cite{wittentwo}.  On rather general grounds, given a system of elliptic differential equations on a manifold $V$ with
boundary $\partial V=W$, the space of solutions of the equations gives a  cycle $\Gamma$ in the space of boundary data and this cycle is within a finite amount of
being middle-dimensional.
In the present problem, the boundary data are the fields $\A_w=A+w\phi$ on $W$, and we want to interpret $\Gamma$ as an integration
cycle in the integral over $\A_w$.

We are actually now in a situation that has been analyzed in  detail in section 5 of \cite{wittenthree}.  Localization of the path integral on  the space of solutions of the equations means
that a path integral over bosons and fermions on the four-manifold $V$ reduces to an integral over the purely bosonic fields $\A_w$ on the
three-manifold  $W=\partial V$.
Localization further means that the integral over the boundary fields $\A_w$ reduces to an integral over the cycle $\Gamma$.
In this reduction, the part of the action that is of the form $\{Q,\dots\}$ gets dropped, leaving only -- in the present
context -- the boundary action $i\Psi\,\CS(\A_w)$.

Actually, at this stage we have a problem of index theory.  The classical theory under discussion has the conserved fermion number $\EF$.
This conservation law has an anomaly that is related in the usual way to the index theorem for the Dirac operator of the theory.
This operator and its elliptic boundary condition are described in Appendix A of \cite{wittenthree}. A nonzero index means
that the four-dimensional path integral vanishes unless we insert a suitable operator violating $\EF$ in the appropriate way.  We say that
$\Gamma$ is a middle-dimensional cycle when the index vanishes, and in general that $\Gamma$ departs from being middle-dimensional
by an amount equal to the index.   In the present problem, the index was analyzed\footnote{The operator whose index
we want is the operator $\d_A+\d_A^*$ mapping differential forms of odd degree to those of even degree.  The
requisite boundary conditions, which were described in Appendix A of \cite{wittenthree}), are slightly unusual, but they are homotopic to standard boundary
conditions in which the restriction of a differential form on $V$ to $\partial V$ vanishes.  With these boundary conditions, the index
is $-\chi(V)\mathrm{dim}\,G$.}  in section 4.1.1 of \cite{wittentwo}. It  is independent of the choice of underlying  $G$-bundle $E\to V$, simply because the fermions of given $\EF$ transform
in a real representation of $G$ (namely the adjoint representation), and is proportional to the Euler characteristic of $V$.

We will be interested primarily in the case that the index vanishes. (A typical example of a similar
problem in which the index is nonzero, so that an operator
insertion is needed to get a nonzero path integral, is described in section 2 of \cite{wittenthree}.)  Then $\Gamma$ is a middle-dimensional cycle.  The four-dimensional path integral is generically nonzero and localization means that it
reduces to an integral of the boundary fields over $\Gamma$:
\begin{equation}\label{zomixo}\int_\Gamma D\A_w\exp(-i\Psi\,\CS(\A_w)).\end{equation}
This has been described in section 5.2.2 of \cite{wittenthree}.

At this point, the precise value of $w$ is not important.  All that matters is that it has a nonzero imaginary part, so that $\A_w=A+w\phi$
is a complex-valued connection.   The integral (\ref{zomixo}) has no dependence on $w$ except in the definition of $\A_w$,
so we can eliminate $w$ by simply writing $\A$ for $\A_w$.  (In \cite{wittenthree}, $w$
was set to $i$, but the analysis could have been made in the same way for any $w$ with nonzero imaginary part.)    Accordingly,
we rewrite (\ref{zomixo}) with $w=i$:
\begin{equation}\label{tomixo}\int_\Gamma D\A\exp(-i\Psi\,\CS(\A)).\end{equation}

Now we should address the question of what are the possible values of $\Psi$.
In our derivation starting with the D3-NS5 system, with physically sensible values of the parameters, $\Psi$ has turned out
to be an arbitrary nonzero real number, given according to (\ref{por}) by $\Psi=|\tau|^2/\mathrm{Re}\,\tau$. From a topological
field theory point of view, as in \cite{wittenthree}, one can make a more general choice of the twisting parameter $t$ and then $\Psi$
is an arbitrary nonzero complex number.\footnote{Alternatively, one can reach generic $\Psi$ by analytically continuing to complex
values of the gauge theory theta-angle $\theta$, and otherwise using the formulas of the present paper.  Giving $\theta$ an imaginary
part violates unitarity, and indeed it appears
that reality of $\Psi$ is related to unitarity.}  Both points of view are useful.  The physical one based on the D3-NS5 system
will enable us to understand the role of nonperturbative dualities.  The topological field theory point of view leads among other things
to holomorphy in $\Psi$, which we will make use of momentarily.

The relation of a ``contour'' integral such as (\ref{tomixo})  to ordinary Chern-Simons gauge theory with compact gauge group $G$ has been discussed in \cite{wittentwo}. Let  $\frak g$ and $\frak g_\C$ be the Lie algebras of $G$ and of its complexification $G_\C$, and let $\U$
be the space of all real gauge fields, that is all $\frak g$-valued connections $A$ on some principal $G$-bundle $E\to W$.  And let $\U_\C$
be the  complexification of $\U$, or in other words
the space of all $\frak g_\C$-valued connections on the complexification of $E$. We denote such
a connection as $\A$.   The path integral of ordinary Chern-Simons theory with the
compact gauge group $G$ is
\begin{equation}\label{fordo}\int_\U DA\,\exp(-ik\,\CS(\A)),\end{equation}
and here the ``level'' $k$ has to be an integer, in order to make the integrand of the path integral gauge-invariant.  (There is no such restriction
on $\Psi$ in (\ref{tomixo}),  as explained in \cite{wittenthree}, because of the choice of integration cycle $\Gamma$.)  Usually one says that the path integral does not make sense if $k=0$ (since one needs a nontrivial oscillatory factor $\exp(-ik\, \CS(\A))$ to define a sensible integral
over the space of connections), and one chooses the orientation of $W$ to restrict to the case $k>0$.  We will instead consider both signs
of $k$.

It looks like the ordinary Chern-Simons path integral with gauge group $G$ is the special case of (\ref{tomixo}) with $\Gamma=\U$,
that is, with the integration cycle chosen to be the obvious cycle that parametrizes real gauge fields. To emphasize this, in (\ref{fordo})
we have denoted the argument of the Chern-Simons function as a complex connection $\A$, although the integral is evaluated on the real cycle $\U$, where $\A$
reduces to a real connection $A$. However, before drawing conclusions about the relation of (\ref{tomixo}) to ordinary
Chern-Simons theory, we have to be careful
in comparing the holomorphic volume forms that appear in the two integrals.

The integration form that has been called $DA$ in (\ref{fordo})
arises by analytic continuation to $\U_\C$ of the usual integration
form (which we also call $DA$) of the Feynman integral of the
$\frak g$-valued theory.  The corresponding form $D\A$ is induced
from the four-dimensional path integral on $V$.
Both $DA$ and $D\A$ are Calabi-Yau volume forms on the same space, namely
$\U_\C$.  So {\it a priori}, their ratio is an invertible
holomorphic function on $\U_\C$.  We propose that the relation is
\begin{equation}\label{zondo}DA=
D\A \,\exp(-ih\,\mathrm{sign}(k)\,\CS(\A))\,\frak N_0.\end{equation}
Here $h$ is the dual Coxeter number of $G$, and $\mathrm{sign}(k)$ is the
sign of the integer $k$.  (Formulas
somewhat analogous to (\ref{zondo}) are described in section 2.7.1 of \cite{wittentwo}.)
In (\ref{zondo}), we have included a possible multiplicative constant $\frak N_0$, which
is allowed by holomorphy.
  The constant $\frak N_0$ might depend on the three-manifold
$W$ and the choice of a homomorphism $\rho:\pi_1(W)\to G_\C$ at $y=\infty$ to
define the $\N=4$ path integral, but holomorphy
 in $\Psi$, together with the fact that we have already incorporated the effects of the gauge
 theory theta-angle,  does not allow contributions to $\frak N_0$ beyond one-loop
order.

The relation (\ref{zondo}) should be demonstrated
explicitly -- and the constant $\mathfrak N_0$ calculated -- by comparing the one-loop determinant for $\N=4$ super Yang-Mills theory on $V$
to the one-loop path integral of ordinary Chern-Simons theory on $W$. We will not make such an analysis in the present paper.
Instead, we content ourselves with the following observation.  Suppose that one expands the Chern-Simons path integral (\ref{fordo})
around a critical point,
that is, around a flat connection $\A_0$.  The integrand of the path integral has a phase factor $\exp(-ik\CS(\A_0))$.  As computed in
\cite{witten}, the phase of the one-loop
determinant corrects this to $\exp(-ik'\, \CS(\A_0))$ where
\begin{equation}\label{golly} k' = k +h\,\mathrm{sign}(k).\end{equation}
Usually, $k$ is taken to be positive  so this formula is written $k'=k+h$, but we want to allow both signs of $k$, which requires replacing $h$ with  $h\,\mathrm{sign}(k)$.  (Chern-Simons theory on
a three-manifold $W$ is invariant under a reversal of orientation of $W$ together
with a change of sign of $k$; this means that $k'$ must be an odd function of $k$.  Concretely, the term in $k'$ that is linear in $h$ comes from an $\eta$-invariant
that changes sign if the sign of $k$ is changed.)

Now let us consider the analogous issue for $\N=4$ super Yang-Mills on $V$, with a boundary condition that leads to a ``contour'' integral
(\ref{tomixo})
in the space of $\frak g_\C$-valued connections. The integral is holomorphic in $\Psi$, so a one-loop shift in the phase factor
$\exp(-i\Psi \,\CS(\A_0))$ would have to be holomorphic in $\Psi$.  Since there is no holomorphic function that restricts to
$\mathrm{sign}(\Psi)$ when $\Psi$ is real, such a term cannot arise.

Our proposal is that no such shift arises from the
one-loop determinant of $\N=4$ super Yang-Mills theory.  Instead, the shift is contained in the comparison (\ref{zondo}) between the
path integral measures of the two theories.    There is no problem in holomorphy here, since the left hand side is only defined when
$k$ is a nonzero integer.
According to our proposal, in comparing $\N=4$ super Yang-Mills theory on $V$ to Chern-Simons theory on $W$, we should use not the naive $\Psi=k$ but
\begin{equation}\label{zolt}\Psi = k + h\,\mathrm{sign}(k).  \end{equation}
To be more exact, from $\N=4$ super Yang-Mills theory on $V$, we can generate a theory that works for general (nonzero) complex $\Psi$.
It can be compared to Chern-Simons theory when $\Psi$ is an integer; in making this comparison we should use (\ref{zolt}).

As is clear both from section \ref{uzim} of the present paper and from the analysis in \cite{wittenthree}, we can add knots and Wilson
loop operators to this analysis.  $\N=4$ super Yang-Mills theory with supersymmetric Wilson lines inserted on $W=\partial V$
gives an unusual integration cycle in Chern-Simons theory on $W$ with the same Wilson line insertions.  A more complete microscopic
explanation of the origin of the knots will be presented in section \ref{surfop}.

\subsection{Choice Of $V$}\label{infrared}

Now we will explain the choice of $V$ that will be most useful in the rest of this paper.

Given an oriented three-manifold $W$, we want to pick in a natural and general way an oriented four-manifold $V$ with $\partial V=W$.  There is no
way to do this if $V$ is supposed to be compact.  Instead we will pick $V=W\times \R_+$, where $\R_+$ is a half-line $y\geq 0$.  Thus $y$ corresponds to the normal coordinate to the boundary,
which earlier has been called $x^3$.

Since $V$ is not compact, we need a boundary condition at $y=\infty$.  The boundary condition will be given by a $y$-independent
solution of the localization equations (\ref{explicitly}).  As explained in \cite{wittentwo}, such solutions correspond to conjugacy classes of homomorphism\footnote{To be more precise \cite{corlette}, the solutions correspond to homomorphisms that obey a mild condition of semi-stability: their
monodromies should not be strictly triangular.} from
$\pi_1(W)$, the fundamental group of $W$, to $G_\C$, the complexification of $G$.  We let $\rho:\pi_1(W)\to G_\C$ be such a homomorphism.

Since $V=W\times \R_+$ has two ends -- the boundary at $y=0$ and the end at $y=\infty$ --
we have to be more careful with the formula (\ref{uster}) for the action.  The complete
version of the formula has contributions from both ends:
\begin{equation}\label{muster}I=\{Q,\dots\}+i\Psi\CS(\A)-i\Psi\CS(\A_\infty).\end{equation}
Here we write simply $\A$ for the complex connection at $y=0$, and $\A_\infty$ for its
counterpart at $y=\infty$.   $\A_\infty$ is completely determined by the boundary condition
at $y=\infty$ and in particular by the choice of $\rho$, so the term we have added is simply
a constant.  It is more precise to include the resulting constant in (\ref{tomixo}), so the
$\N=4$ path integral on $W\times \R_+$ is really
\begin{equation}\label{tomixoy}\frak N \int_\Gamma D\A\exp(-i\Psi\,\CS(\A)).\end{equation}
where $\frak N$ is a normalization factor
\begin{equation}\label{omiox}\frak N=\exp(i\Psi\,\CS(\A_\infty)).\end{equation}

\subsection{Some Key Details}\label{technical}

We now run into an important point, which has also been discussed in section 5.2.2 of \cite{wittenthree}.  If $W$ is compact, then
$W\times \R_+$ is macroscopically one-dimensional, and we must worry about infrared divergences.

If $\rho$ is irreducible (which we take to mean that the homomorphism $\rho:\pi_1(W)\to G_\C$ commutes with at most a finite subgroup of $G_\C$),
then our boundary condition at $y=\infty$ makes the theory ``massive'' -- in the effective one-dimensional physics obtained by compactification
on $W$, all bosons and fermions are massive.  Under these conditions, the choice of $\rho$ satisfactorily specifies the boundary conditions.

If instead $\rho$ is reducible -- it leaves unbroken a subgroup of $G$ of positive rank -- then our boundary condition at $y=\infty$ leads to a reduced one-dimensional theory in which the potential energy as a function of scalar fields has flat directions: there are some scalar fields (such as some components
of $\sigma$) that can acquire expectation values, at no cost in energy.  In one dimension, quantum fluctuations of massless scalars are
inevitable and important.  The boundary condition at $y=\infty$ is in this case not adequately specified by the choice of $\rho$; one
also needs a quantum wavefunction describing the initial conditions for the massless scalar fields at $y=\infty$. Here we view $y$ as a Euclidean
time coordinate.

The dependence on $\rho$ presents a number of problems for the constructions
that we will make in the rest of this paper.
Our next step, in section \ref{elmdual}, will be electric-magnetic duality.  At a minimum,
to proceed in a situation in which $\rho$ is important,
we would need to know how $\rho$ transforms under electric-magnetic duality.
Not much is known about this, though a little can
be gleaned (for some special choices of $W$) from \cite{VW,HW}.
The reducible $\rho$'s  are certainly important for understanding the standard Chern-Simons
path integral, since when expressed in terms of cycles associated to flat bundles, it certainly
receives contributions from reducible flat bundles.

What happens to the choice of $\rho$ under electric-magnetic duality is a question that presumably can be answered, in principle.
The infrared divergences that arise in the reducible case pose another problem that may be more serious.  After making electric-magnetic
duality, we will in section \ref{tdual} make a $T$-duality to introduce a new time coordinate, and then we will want to consider quantum states
that propagate in the time direction.  Describing quantum states that propagate in the time direction is, at least at first sight, incompatible with specifying a boundary condition by fixing a quantum state that propagates in the $y$ direction.   One would at least need a better language to describe what
happens here.

Presumably, none of these problems are insuperable, but there clearly is some work to be done to overcome them.

There is actually a straightforward way to circumvent these problems.  This is the approach we will take in most of this paper; it also is the
approach that leads to Khovanov homology.  Instead of taking $W$ to be compact, we will take $W=\R^3$.  (It then is essential to include
knots or Wilson loop operators, since Chern-Simons theory on $\R^3$ is trivial without them.)  For $W=\R^3$, fluctuations of massless
scalar fields  on $V=W\times \R_+$ do not present a problem, because $V$ has four non-compact directions.  Also, as $\R^3$
is simply-connected, when we take $W=\R^3$, there is a unique choice of $\rho$ (corresponding to the trivial flat connection), and this
choice must map to itself under electric-magnetic duality.   So as long as we restrict ourselves to knots in $\R^3$, we avoid all technical
problems related to infrared divergences and the behavior of $\rho$ under electric-magnetic duality.

There are additional technical advantages in taking $W=\R^3$.   Our approach in this paper naturally leads to
an integral (\ref{tomixo}) over a cycle $\Gamma$ defined by solving flow equations on $V=W\times \R_+$. $\Gamma$ depends on the
choice of $\rho$, so we might denote it in more detail as $\Gamma_\rho$. Khovanov homology
is related instead to ordinary real Chern-Simons theory, the integration cycle being the real cycle $\U$.   In general, as described in \cite{wittentwo}, one can expand
$\U$ as an integer linear combination of the $\Gamma_\rho$'s, but it may be hard to determine the coefficients explicitly.  For $W=\R^3$,
as $\rho$ is unique, all integration cycles are integer multiples of a fundamental one, and the relation is simply\footnote{$\U$ is precisely $\Gamma$, rather than a more general integer multiple of $\Gamma$, because in
general when the real integration cycle is expressed in terms of cycles associated to critical points, the cycles associated to real critical points always enter with coefficient 1, as explained in \cite{wittentwo}, eqn. (3.39).} $\Gamma=\U$.   So the integration
cycle that emerges naturally from $\N=4$ super Yang-Mills theory in four dimensions is equivalent to the usual one of ordinary Chern-Simons
theory on the boundary.

Furthermore, the normalization factors $\frak N$ and $\frak N_0$ of (\ref{omiox}) and (\ref{zondo}) equal 1 for $W=\R^3$. We have  $\frak N=1$ because $\A_\infty$ is trivial.  And $\frak N_0=1$
on $\R^3$ because we are studying a topological field theory.  A ``constant'' arising
from a one-loop determinant on $\R^3$ would be a shift in the ground state energy per unit
volume, but such a shift is not possible in a topological field theory.

So there are many advantages to taking $W=\R^3$.
Some but not all of these advantages persist in the following more general case.  Let $W_0$ be a rational homology sphere
and let $W=W_0\backslash p$ be $W_0$ with a point $p$ omitted.  $W$ is not compact and we pick on $W$ a metric that near its
noncompact end looks like the flat metric on $\R^3$.  In this type of example, there are no infrared divergences, but there are in general non-trivial choices of $\rho$, and to proceed one would need to understand how $\rho$ transforms under electric-magnetic duality, and how to expand
$\U$ as a linear combination of the $\Gamma_\rho$'s.

Khovanov homology has been defined in the literature for knots in $\R^3$ (or $S^3$).  It has proved difficult so far to generalize Khovanov
homology to other three-manifolds.  The difficulties may be related to some of the points made above.  We note, however, that results
in \cite{GMN} appear to be part of an analog of Khovanov homology for the case $W=\R\times C$, with $C$ a Riemann surface.

\section{$S$-Duality}\label{dualities}

To learn something new about Chern-Simons gauge theory, we will apply  dualities to the framework
analyzed in section \ref{csfun}.  The relevant dualities are standard.   Here we consider $S$-duality and in section \ref{tdual}, we follow
with $T$-duality.

\subsection{Electric-Magnetic Duality}\label{elmdual}

We begin by applying electric-magnetic duality to $\N=4$ super Yang-Mills theory on $V=W\times \R_+$.

\def\ng{{\mathfrak n_{\mathfrak g}}}
The gauge group $G$ is transformed to the Goddard-Nuyts-Olive or Langlands dual group, which we will denote as $G^\vee$.
The $G^\vee$ gauge theory has a theta-angle and gauge coupling, which we call $\theta^\vee $ and $\gym^\vee$.
As usual, we define
\begin{equation}\label{umon}\tau^\vee=\frac{\theta^\vee}{2\pi}+\frac{4\pi i}{(\gym^\vee)^2}.\end{equation}
The standard relation between $\tau^\vee$ and $\tau$, generalized \cite{AKS} to the case that $G$ is not simply-laced, is
\begin{equation}\label{zoldo}\tau^\vee=-\frac{1}{\ng\tau},\end{equation}
where $\ng$ is the ratio of length squared of long and short roots of $G$, or equivalently of $G^\vee$.  (Thus, $\ng=1$ if $G$ is
simply-laced.)
The formula (\ref{zoldo}) can be written as $\tau^\vee=(a\tau+b)/(c\tau+d)$ where
\begin{equation}\label{priod}\begin{pmatrix} a & b \\ c & d\end{pmatrix}=\pm \begin{pmatrix}0  & -\sqrt{\ng} \\ \sqrt{\ng} & 0\end{pmatrix}.
\end{equation}
The two choices of sign differ by the possibility of combining electric-magnetic duality with a discrete chiral symmetry.
(This symmetry is an element of the center of the $R$-symmetry group $SU(4)_R$; it reverses the sign of the twisting parameter $t$ and
maps $(A,\phi)\to (A,-\phi)$.)   The two choices correspond to duality of the D3-NS5 system
with a D3-D5 or D3-$\bar{\mathrm{D5}}$ system, respectively.  There is no natural choice of which is which.  Either way,
the boundary condition of the D3-NS5 system maps to a dual boundary condition, which we will discuss in section \ref{boundcond}.
Wilson operators supported at $y=0$ map to 't Hooft operators supported at $y=0$; these are described in section \ref{thooft} and modify
the boundary conditions.

The family of twisted topological field theories that is relevant in the present paper is mapped to itself by electric-magnetic duality.
The twisting parameter $t^\vee$ of the dual description with gauge group $G^\vee$ is related to the twisting parameter $t$ in the original
description by
\begin{equation}\label{perry} t^\vee = \pm \frac{\tau}{|\tau|}t.\end{equation}
This formula is a special case of (\ref{zomus}); the sign is the same as the one in (\ref{priod}).
For the D3-NS5 system, we have $t=\pm \sqrt{\bar\tau/\tau}$ according to (\ref{doner}), and this leads
to the amazingly simple
\begin{equation}\label{doofut} t^\vee = \pm 1.\end{equation}
The sign does not matter, as the two choices are exchanged by the discrete chiral symmetry mentioned in the last paragraph.
In this paper, we will take $t^\vee=1$.
The localization equations in the $G^\vee$ gauge theory then take a particularly simple form:
\begin{equation}\label{oofut} F-\phi\wedge\phi +\star \d_A\phi=0=\d_A\star \phi.\end{equation}

\def\adj{{\mathrm{adj}}}
The transformation law (\ref{blomus}) for the canonical parameter $\Psi$ tells us that the parameter $\Psi^\vee$ of the dual theory
is related to $\Psi$ by
\begin{equation}\label{ofus}\Psi^\vee=-\frac{1}{\ng\Psi}.\end{equation}
On the other hand, since $t^\vee =1$, the formula (\ref{zornox}) for $\Psi^\vee$ reduces to
\begin{equation}\label{fus}\Psi^\vee=\frac{\theta^\vee}{2\pi}.\end{equation}
Combining these formulas,
\begin{equation}\label{nofus}\theta^\vee=2\pi\Psi^\vee=-\frac{2\pi}{\ng\Psi}.\end{equation}
For $G^\vee=SU(N)$, we define
 the instanton number of the $G^\vee$ gauge theory by
\begin{equation}\label{toffus}\EH = \frac{1}{32\pi^2}\int_V
\epsilon^{\mu\nu\alpha\beta}\,\Tr\,F_{\mu\nu}F_{\alpha\beta},\end{equation}
where $\Tr$ is the trace in the $N$-dimensional representation.
For any $G^\vee$, we can take
\begin{equation}\label{tofus}\EH = \frac{1}{2h^\vee}\frac{1}{32\pi^2}\int_V
\epsilon^{\mu\nu\alpha\beta}\,\Tr_\adj \,F_{\mu\nu}F_{\alpha\beta},\end{equation}
where\footnote{Thus, in our notation, $h$ is the dual Coxeter number of $G$ and $h^\vee$ is
the dual Coxeter number of $G^\vee$.   (Note that some authors use $h^\vee$ for the dual
Coxeter number of $G$.)}  $h^\vee$ is the dual Coxeter number of $G^\vee$, and $\Tr_{\adj}$ is the trace in
the adjoint representation of $G^\vee$.  The symbol $\Tr$ will be used as an abbreviation for
$\Tr_\adj/2h^\vee$ even if $G^\vee$ is not $SU(N)$.   We will eventually modify the definition (\ref{tofus}) by subtracting
a $c$-number term, that is a term that does not depend on the gauge field $A$ (see eqn. (\ref{lofus}) below).

The role of $\theta^\vee$ in the path integral is simply to weight a field of instanton number $\EH$ by a factor
$\exp(-i\theta^\vee\EH)$.  We set
\begin{equation}\label{flodux}q=\exp(-i\theta^\vee)=\exp(2\pi i/\ng\Psi), \end{equation}
so that the $\theta^\vee$-dependent factor by which we weight a field of instanton number $\EH$ is $q^\EH$.
Recalling (\ref{zolt}), we see that when we compare the $G^\vee$ gauge theory to Chern-Simons theory on $W=\partial V$
with gauge group $G$, we must take
\begin{equation}\label{lodux} q=\exp\left(\frac{2\pi i}{\ng (k+h\,\mathrm{sign}(k))}\right).\end{equation}
At least for simply-laced $G$, this is essentially the standard definition of $q$ in Chern-Simons gauge theory (the formula is usually written
for positive $k$, and what we call $q$ is sometimes called $q^2$ or $q^{-1}$).   Hence,
for example, the Jones polynomial of a knot in $\R^3$ (and its generalizations for other
groups and representations) is essentially a Laurent polynomial in this variable; for a precise
statement, see eqn. (\ref{odork}).

\subsection{Computing The Partition Function}\label{computing}

Now let us discuss how to compute the partition function of the $G^\vee$ gauge theory on $V$.  Because $t^\vee$ is real, the model is
analogous to a two-dimensional $A$-model (or four-dimensional Donaldson theory) and computations can be carried out by  an appropriate procedure of counting of classical
solutions of the localization equations (\ref{oofut}).
The value $t^\vee=1$ makes the
procedure particularly simple.  As $\Psi^\vee$ is independent of $\gym^\vee$, to calculate
the partition function for given $\Psi^\vee$, we can take $\gym^\vee$ to be arbitrarily small.  The partition
function then reduces to a sum over classical solutions of the localization equations.
The expected dimension of the moduli space of solutions of those equations is given
by the index of a certain Dirac-like operator.  As is typical of $A$-type topological field theories,
the operator in question is the fermion kinetic operator of the theory, whose
index equals the anomaly in the fermion number $\EF$. So the expected dimension of
the moduli space  must vanish in order for the twisted
$\N=4$ path integral on $V$ without any operator insertions to be non-vanishing.\footnote{When the index is non-zero, we make a suitable operator insertion to replace the twisted $\N=4$ partition function
with a non-vanishing path integral.  (This can actually only be done when the index is positive,
because the cohomology of $Q$ in the space of local operators vanishes for  $\EF<0$.)  As in other theories
of $A$-model type,  the operator insertions have the effect of constraining
the solutions of the localization equations and reducing to a situation much like what
prevails when the index vanishes.  We omit the details, as we do not need them and they
are standard in topological field theories of this type.}

Let us suppose that this is the case and consider the contribution to the path integral from
a given solution of the localization equations.  For simplicity, assume that in expanding
around such a solution, there are no bosonic or fermionic zero modes and no unbroken gauge symmetries.
This is the generic state of affairs when the index vanishes. In expanding around such a solution,
since we can take $\gym^\vee$ to be arbitrarily small, we can make a one-loop approximation to the path integral, which -- apart from a factor coming from the classical action --
reduces to the
ratio  of fermion and boson determinants.  The determinants
are equal up to sign, because of supersymmetry,  and the boson determinant is always positive.  So the ratio of determinants
is $\pm 1$, depending on the sign of the fermion determinant.    The factor in the
path integral from the classical action is  $q^\EH$, coming from the part of the classical action proportional to $\theta^\vee$.

The sum of the contributions of all solutions with $\EH=n$ is then $a_nq^n$ for some integer $a_n$; $a_n$ is simply  the sum of contributions
$+1$ and $-1$ from classical solutions with $\EH=n$ and positive or negative fermion determinant.
The partition function is the sum of $a_nq^n$ over all values of $n$:
\begin{equation}\label{dunky} Z(q)=\sum_n\, a_nq^n. \end{equation}
As explained in section \ref{infrared}, the $\N=4$ partition function $Z(q)$ will be most
simply related to Chern-Simons theory if $V=\R^3\times\R_+$, in which case $Z(q)$ and the
Chern-Simons path integral on  $\R^3$  should simply coincide.  To make this
case interesting, we include knots in $\R^3$ on the Chern-Simons side and
the corresponding loop operators in the boundary of $V$ in the $\N=4$ description.   The
formula $Z(q)$ has been obtained in a dual description by $G^\vee$ gauge theory, so the
loop operators are 't Hooft operators (rather than the Wilson operators that were introduced
in section \ref{uzim}).  The presence of 't Hooft operators affects the coefficients $a_n$ in the
partition function because it affects  the boundary conditions
along $\partial V$, as we will describe in section \ref{thooft}.

The claim that the sum (\ref{dunky}) reproduces the knot invariants of Chern-Simons theory is one of the main
claims of the present paper.  For a direct verification of this for the special case corresponding to the Jones
polynomial (that is, $G=SU(2)$ with loop operators associated to the two-dimensional representation of $G$) see
\cite{gwnew}.

For future reference, we can rewrite (\ref{dunky}) as follows.  Let $S$ be the set of classical solutions of the localization
equations.  For $s\in S$, let $n_s$ be the value of $\EH$ for the corresponding solution, and denote  the sign
of the fermion determinant obtained when one expands around that solution as $(-1)^{g_s}$.  Then
\begin{equation}\label{unky}Z(q)=\sum_{s\in S}\, q^{n_s}(-1)^{g_s}.\end{equation}

What values of the instanton number $n$ occur in (\ref{dunky})?  Suppose first that $G^\vee$ is simply-connected.
Then $n$ is an integer if $V$ is compact and without boundary, but if $V$ has a boundary
or an end at infinity,
then $n\in \Z+ \delta$, where the constant $\delta$ depends on the boundary conditions and the
behavior at infinity.  (We will
analyze this dependence in section \ref{framan}.)  If $G^\vee$ is not simply-connected but $V$
is compact and without boundary, then $n\in \Z/w$, where the integer $w$ depends only on $G^\vee$ (for example, $w=4$ if $G^\vee =SO(3)$,
since the instanton number of an $SO(3)$ bundle $W\to V$ is congruent to $\int_V w_2(E)^2/4$ mod $\Z$).  If $G^\vee$ is not
simply-connected and $V$ has a boundary or a non-compact end, then $n\in\Z/w+\delta$ for
some constant $\delta$.  Despite these details, we will loosely refer to a sum of the form
(\ref{dunky}) as a Laurent polynomial if $a_n$ vanishes except for finitely many values of $n$.

Given that the Chern-Simons path integral for a knot in $\R^3$  can be expressed as in (\ref{dunky}), can we get a new understanding of the fact that these functions are actually  Laurent polynomials?  This is true if the localization equations have solutions only for finitely many values of $\EH$, since $a_n$ certainly
vanishes if there are no solutions at all with $\EH=n$.   It is shown in \cite{KW}, section 3.3, that if $V$ is a compact four-manifold without boundary,
then the localization equations (for any value of $t$ other than 0 or $\infty$) have no solutions except for $\EH=0$.  Hopefully, for $\partial V\not=\varnothing$, with the boundary conditions of sections \ref{boundcond} and \ref{thooft}, and possibly with a noncompact end,
there is a more general result giving a bound on $|\EH|$ for any solution.    This will ensure that the path integral is a Laurent polynomial.

\subsubsection{Some Further Details}\label{zdetails}

In our simplified explanation of (\ref{dunky}), we have omitted a few details that will be important in some generalizations.

First of all, under electric-magnetic duality, the action may obtain a $c$-number term of the form $\alpha\chi(V)+\beta\sigma(V)$ where
$\chi(V)$ and $\sigma(V)$ are the Euler characteristic and signature of $V$ and $\alpha,\beta$ are universal constants.  Such an effect has been described in \cite{VW} in the context of a different twist of $\N=4$
super Yang-Mills theory. If it occurs in the present context, this would
multiply the right hand side of (\ref{dunky}) by $\exp(\alpha\chi(V)+\beta\sigma(V))$.
 This may be important for some applications,
though not for the case $V=W\times \R_+$ that we focus on in the present paper.

\def\ZZ{{\mathcal Z}}
 Second, we should discuss the role of  unbroken gauge symmetries.  Given a solution of the localization equations, we write $H$ for the subgroup of $G^\vee$ consisting of gauge transformations
that leave fixed the given solution.  We call a solution reducible if $H$ is a Lie group of positive dimension and irreducible if $H$ is
a finite group, in which case we denote the number of its elements as $\# H$.  Reducible solutions (such as the trivial solution with $A=\phi=0$)
are inevitably present if $\partial V=\varnothing$.  In expanding around a reducible solution, there are flat directions in the classical
potential (for example, the potential vanishes for some components of $\sigma$), and one has to learn how to integrate over this space of flat
directions in order to determine the contribution of a reducible solution to the path integral.  This is  a rather delicate question,
and we will not investigate it here.

There is also some subtlety concerning irreducible solutions when $H$ is non-trivial.
For compact $V$, the contribution of
an irreducible solution with non-trivial $H$ is actually not $\pm q^n$ but $\pm q^n/\#H$,
where the factor $1/\#H$ results from the process of dividing by the volume of the gauge group.
Suppose that $V$ has a nonempty boundary and we use the boundary condition described in section
\ref{boundcond}.  This boundary condition explicitly breaks $G^\vee$ down to its center, which we call
$\ZZ(G^\vee)$.  The center  is always a symmetry of any classical solution, so in this situation we always have $H=\ZZ(G^\vee)$.  If in addition $V$ is compact, (\ref{dunky}) should be multiplied by $1/\#\ZZ(G^\vee)$, reflecting
the fact that $\ZZ(G^\vee)$ acts trivially on the space of fields.  However, if $V$
also has a noncompact end (as in our basic example $V=W\times \R_+$), one divides
only by gauge transformations that are trivial at infinity, and hence the factor of $1/\# \ZZ$
does not arise.

For $V=W\times \R_+$, we have to define a boundary condition at infinity.  We do
this just as we did for the original D3-NS5 system: we pick a $y$-independent solution
of the localization equations at infinity.  In the present case, this corresponds to a
homomorphism $\rho^\vee: \pi_1(W)\to G^\vee_\C$.  The partition function (\ref{dunky}) can
be defined for each $\rho^\vee$, so we really get a family of partition functions $Z_{\rho^\vee}(q)$,
labeled by $\rho^\vee$.  Similarly, the integral (\ref{tomixo}) is really a family of path  integrals
$I_\rho$, labeled by homomorphisms $\rho:\pi_1(W)\to G_\C$.  One expects that electric-magnetic
duality will lead to formulas of the general nature
\begin{equation}\label{poly}Z_{\rho^\vee}(q)=\sum_\rho m_{\rho^\vee,\rho} I_\rho(q),\end{equation}
with some matrix $m_{\rho^\vee,\rho}$. But little is clear about the nature of this matrix.
This problem was pointed out in section \ref{infrared}.
Luckily, for the important case $W=\R^3$, we avoid this question.

\subsection{The Dual Boundary Condition}\label{boundcond}

We are mainly interested in the case that the four-manifold $V$ has a boundary, so we need to describe the appropriate
boundary condition in the $G^\vee$ gauge theory.  (We describe here the boundary condition
away from possible 't Hooft operators.  The more elaborate boundary condition that must
be used near an 't Hooft operator is described in section \ref{thooft}.)

For $G^\vee=G=U(N)$, the boundary conditions that we want are those of the D3-D5 system, or equivalently, the D$p$-D$(p+2)$ system for any  $p$.
This boundary condition, which is of a rather surprising nature, was first formulated in \cite{Dia} by comparing to known results
about the Nahm transform of BPS monopoles.  More intuitive explanations have been given in \cite{CM,myers,cmyerst} in terms of the
D$(p+2)$-brane theory and a ``fuzzy funnel.''   A formulation
of the boundary condition purely in field theory terms,  along with a generalization to any $G^\vee$,
has been given in \cite{gw}.

The boundary condition of the D3-D5 system is defined not by imposing a condition on the fields or their normal derivatives,
as in the case of familiar boundary conditions such as Dirichlet and Neumann, but by specifying the singular behavior
that the fields should have near the boundary.  (This is somewhat like the procedure used to define
an 't Hooft operator, or a disorder operator in statistical
mechanics; these are also defined by specifying a desired singularity.)   The desired  behavior is described by giving
a model solution of the equations (\ref{oofut}) that has
the desired singularity.  In the context of topological field theory, the model solution has to obey the equations in order
to preserve the desired topological supersymmetry at $t^\vee=1$.

In fact, the boundary condition of the D3-D5 system has much more symmetry than that; it is half-BPS, and is invariant under
translations and rotations and in fact even  conformal transformations that leave fixed the boundary.
It is convenient to define
the model solution  on the half-space $x^3\geq 0$, and to write $y$ for $x^3$.   In the model
solution, the gauge field $A$ vanishes, as does the normal part of the one-form $\phi$.  We write $\vec \phi=\sum_{i=0}^2\phi_i\,\d x^i$
for the tangential part of $\phi$.  Rotation and translation invariance tell us to look for a model singular solution such
that $\phi$ is a function of $y$ only.  Given all this, the equations (\ref{oofut})
reduce to Nahm's equations
\begin{equation}\label{nahm}\frac{\d \vec \phi}{\d y}+\vec\phi\times\vec\phi= 0.\end{equation}
Here $\vec\phi\times\vec\phi$ is the triple of elements of $\frak g$ defined by
$(\vec\phi\times\vec\phi)_0=[\phi_1,\phi_2]$ plus cyclic permutations of indices,
or equivalently by  $(\vec\phi\times\vec\phi)_i=[\phi_{i+1},\phi_{i-1}]$, where we consider the integer-valued label $i$ to be defined modulo 3.

Conformal invariance of the D3-D5 boundary condition means the boundary condition
is defined by a solution in which  \begin{equation}\label{nelfo}\vec\phi=\vec t/y\end{equation} for some constant elements $\vec t$
of the Lie algebra $\frak g^\vee$.
Nahm's equations then reduce to
\begin{equation} \label{mexo}   [t_i,t_j]=\epsilon_{ijk}t_k,~~i,j,k=0,1,2,\end{equation}
where $\epsilon_{ijk}$ is the antisymmetric tensor with $\epsilon_{012}=1$.
Eqn. (\ref{mexo}) is equivalent to saying that the elements $\vec t$ are the images of a standard set of $SU(2)$ generators under some Lie algebra homomorphism
$\xi:\frak{su}(2)\to \frak g^\vee$.

Having picked $\xi$, the boundary condition on $\vec\phi$ is
\begin{equation}\vec\phi=\frac{\vec t}{y}+\dots,\end{equation}
where the ellipses refer to terms less singular than $1/y$.
The other three scalar fields (the normal part of $\phi$ and the real and imaginary parts of
$\sigma$) vanish at $y=0$, regardless of $\xi$.  This is deduced in \cite{gw} as a consequence
of supersymmetry; in a D3-D5 brane construction, it asserts that scalar fields that describe
motion of the D3-branes normal to the D5-brane must vanish on the boundary.  The gauge field
$A$ obeys a shifted version of Dirichlet boundary conditions, as described in section
\ref{vivisect} below.

The procedure just sketched, with any choice of $\xi$,  leads to a half-BPS boundary condition that preserves the  desired
supersymmetry. However, as explained in
\cite{gw}, the boundary condition we want ($S$-dual to the generalized Neumann boundary conditions that were our starting
point in section \ref{csfun})
corresponds to the case $\xi$ is a
``principal embedding''  \cite{Kostant} of $\frak{su}(2)$ in $\frak g^\vee$.  A principal embedding is unique up to
conjugacy, for any $G^\vee$.

For $G^\vee=SU(N)$ or $U(N)$, a principal embedding is defined by picking an $SU(2)$ subgroup of $G^\vee$ such that  the fundamental
$N$-dimensional  representation of $G^\vee$ restricts to an irreducible representation of $SU(2)$.   For $G^\vee=U(N)$, the principal
embedding arises for $N$ D3-branes ending on a single D5-brane; other choices of $\xi$ can be realized with $N$ D3-branes
ending on multiple D5-branes.

For all other groups, a principal embedding is,
roughly speaking, as close to irreducible as possible.  For example, for $G^\vee=SO(2k+1)$, the fundamental $2k+1$-dimensional representation
is irreducible under a principal $SU(2)$ subgroup.  This is possible because an irreducible $2k+1$-dimensional representation of $SU(2)$
is real, and hence the $SU(2)$ matrices acting in this representation can be embedded in $SO(2k+1)$.  For $G^\vee=SO(2k)$, the best
we can do is to pick an $SU(2)$ subgroup under which the fundamental representation decomposes as ${2k}=(2k-1)+
1$, and this is a principal $SU(2)$ subgroup.  For $G^\vee=Sp(2k)$, a principal $SU(2)$ subgroup is one under which the fundamental
$2k$-dimensional representation of $G$ transforms irreducibly; this is possible because an irreducible $2k$-dimensional representation of
$SU(2)$ is pseudoreal, so the representation matrices can be embedded in $Sp(2k)$.  For all these classical groups, the principal
embedding arises for $N$ D3-branes ending on a single D5-brane in the presence of an orientifold plane.
To give one example involving an exceptional Lie group, for $G^\vee= {\sf G}_2$, the principal $SU(2)$ embedding is characterized by the fact that the $7$-dimensional
representation of $ {\sf G}_2$
transforms irreducibly under a principal $SU(2)$ subgroup of $ {\sf G}_2$.

We will later need to know a few more basic facts about a principal $\frak{su}(2)$ subalgebra of $\frak g$.  If $G$ is a simple Lie group of rank $r$,
then its Lie algebra $\frak g$ decomposes under a principal $\frak{su}(2)$ subalgebra as a direct sum of precisely $r$ irreducible representations of dimensions $2j_i+1$,
$i=1,\dots ,r$.   (The $j_i$ are always integers.) For $G=SU(N)$, the $j_i$ are $1,2,3,\dots,N-1$
and of course in general
\begin{equation}\label{delf}\sum_{i=1}^r(2j_i+1)=\mathrm{dim}\,G.\end{equation}
The ring of invariant polynomials on the Lie algebra $\frak g$ is freely generated by $r$ fundamental
Casimir invariants, which are homogeneous of degrees $d_i=j_i+1$, $i=1,\dots,r$.  For $SU(N)$,
these invariants are the functions $\Tr\,a^d$, $d=2,\dots,N$.

As a point of terminology, we will refer to the singularity that $\vec\phi$ has at the boundary
for the case of a principal $\frak{su}(2)$ embedding as a regular Nahm pole.  Referring to
this singularity as a Nahm pole requires no explanation.  The term ``regular'' refers to the fact
that the raising operator of a principal $\frak{su}(2)$ subalgebra is a regular element of the
complex Lie algebra $\frak g_\C$.  (An element of this Lie algebra is called regular if the subalgebra that
commutes with it has the minimum possible dimension -- the rank of $G$.)  For a fuller
explanation, see the discussion of eqn. (\ref{onkey}).

\subsection{Embedding the Tangent Bundle}\label{vivisect}

\def\ad{{\mathrm{ad}}}
So far we have described the behavior near the boundary for the case that $V=\R^3\times\R_+$,
$\partial V=\R^3$.  Now we want to generalize to the case that the boundary of $ V$ is an arbitrary
three-manifold $W$ with Riemannian metric $g_{ij}$.  We assume that near
its boundary, $V$ looks like a product $W\times\R_+$.

Let us first consider the case that $G^\vee$ is $SU(2)$ or $SO(3)$.  The gauge field $A$,
restricted to $W$, is a connection on a  $G^\vee$ bundle $E\to W$.

In section \ref{boundcond}, for $W=\R^3$, we described the singular part of $\vec\phi$ as $\vec t/y$.  In the context of the twisted topological field theory, since $\vec\phi$ is interpreted
as a one-form, an
identification of the Lie algebra $\frak{su}(2)$ with the tangent space to $\R^3$ is implicit here.  To make it explicit,
we introduce the Kronecker delta $\delta_i^a$ and write, in more detail,
 \begin{equation}\label{nugh}\vec\phi\cdot \d \vec x=\frac{\sum_{i,a}\delta_i^a \,t_a\,\d x^i}{y}+\dots,\end{equation}
where $t_a$ are a standard set of $\frak{su}(2)$ generators, obeying $[t_a,t_b]=\epsilon_{abc}
t_c$ and (therefore) $\Tr\,t_at_b=-\delta_{ab}/2$.  It is convenient to define a quadratic
form on the $\frak{su}(2)$ Lie algebra by $(x,y)=-2\,\Tr\,xy$, so $(t_a,t_b)=\delta_{ab}$.

In the case of a general $W$, the generalization of (\ref{nugh}) can only be
\begin{equation}\label{zugh}\vec\phi=\frac{\sum_{i,a}e_i^a \,t_a\,\d x^i}{y}+\dots,\end{equation}
where now $e_i^a$ is some tensor that, at any point $p\in W$,  reduces to $\delta_i^a$, up to a gauge transformation, in any locally Euclidean coordinate system at $p$.  Such a coordinate
system is one in which the metric at $p$ is $g_{ij}=\delta_{ij}$.  A covariant way to state
the condition on $e_i^a$ without any restriction on the coordinate system or any choice of gauge is to say that
\begin{equation}\label{zobbo}(e_i^at_a,e_j^bt_b)=g_{ij}, \end{equation}
which implies that in a locally Euclidean coordinate system, $e_i^a=\delta_i^a$ up to a gauge transformation.
 An equivalent statement is
\begin{equation}\label{obbo}e_i^ae_j^b\delta_{ab}=g_{ij}.\end{equation}
But this is a familiar condition in Riemannian geometry.  The object $e$ is usually
called the vierbein; it establishes an isomorphism between the bundle $\ad(E)$ with its natural $\frak{su}(2)$-invariant quadratic
form and the
tangent bundle $TW$ of $W$ with the quadratic form determined by the metric tensor of $W$.

Now we have to look more closely at the equations (\ref{oofut}).  As $\vec\phi\sim 1/y$,
the equations have terms of order $1/y^2$.   By taking the $t_i$ to obey the $\frak{su}(2)$
commutation relations, we ensure vanishing of the $1/y^2$ terms in the equations.
We still must consider the terms of order $1/y$ in the equations.  Here we find that
we need \begin{equation}\label{zelda} D_ie_j-D_je_i=0,\end{equation} 
where $D_i=\partial_i+[A_i,~\cdot~]$ is the usual gauge theory connection.  This is another basic equation in Riemannian geometry.  It uniquely determines the restriction of $A$ to $W$ to be the Riemannian connection on $TW$.  In fact, this equation is usually taken as the definition of the Riemannian connection on the tangent
bundle.  We will denote the Riemannian connection on $TW$ as $\omega$.

This is all there is to say if $G^\vee=SO(3)$: the $G^\vee$ bundle $E\to V$, restricted to  the boundary $W=\partial V$, is the tangent
bundle to $W$, and the connection restricted to $W$ is the Riemannian connection.
For $G^\vee=SU(2)$,  the $G^\vee$-bundle $E\to W$ is not completely determined by the above
description of $\ad(E)$; the additional data required is a choice of spin structure.

The extension of this discussion to any $G^\vee$ is straightforward.    The polar part of $\vec\phi$ establishes an isomorphism
between $TW$ and a subbundle of $\ad(E)$, and this subbundle  corresponds to an
$\frak{su}(2)$ subalgebra of $\frak g$. The case we want is that the subalgebra is principal. The equation (\ref{zelda}) says that the gauge field $A$,
restricted to the boundary, is valued in this $\frak{su}(2)$ subalgebra and that its restriction
to $\frak{su}(2)$ is the Riemannian connection.  Differently put, the bundle $\ad(E)$ is
associated to $TW$ by  a principal embedding $\frak{su}(2)\subset\frak g$.
If the center $\ZZ(G^\vee)$ of $G^\vee$ is trivial, then the $G^\vee$ bundle $E\to W$
is completely characterized by this description of $\ad(E)$.
Otherwise, if $W$ is not simply-connected, the global description of $E$
may involve some additional discrete data analogous to a choice of spin structure: the holonomies of $E$ around noncontractible loops in $W$
are not uniquely determined by the Riemannian structure of $W$, but can be modified by tensoring with a homomorphism  $\pi_1(W)\to \ZZ(G^\vee)$.

\subsection{The Framing Anomaly}\label{framan}

\subsubsection{A Gravitational Coupling}\label{gravcoup}

This last result presents us with a quandary.  According to section \ref{computing}, the contribution of
a given classical solution to the partition function is $\pm q^n$, where $n$ is the instanton
number of that solution.   But the boundary conditions of section \ref{vivisect} do not lead to a natural
definition of the instanton number.

\def\fb{{\frak b}}
 The instanton number of a $G^\vee$-bundle $E\to V$ is a topological invariant if $V$ is a four-manifold
without boundary.  It remains a topological invariant if $V$ has a non-empty boundary and we are given a trivialization
of $E$ on $W=\partial V$.

We have just discovered that instead of being trivialized on $W$,
$E$ is identified on $W$ with the tangent bundle $TW$ to $W$; the gauge field $A$ restricted to  $W$ is similarly identified with the Riemannian connection $\omega$ on $TW$, or more precisely with its $G^\vee$-valued image $\xi(\omega)$, where $\xi:\frak{su}(2)\to \frak g^\vee$ is a principal
embedding.  This means that
the instanton number $\EH$ is not invariant under a change of metric of $V$.  In general,
under any change in the gauge field $A$, the change in $\EH$ is given by the change
in the Chern-Simons invariant of the restriction of $A$ to the boundary $W$:
\begin{equation}\label{zonki}\delta \EH=\frac{1}{2\pi}\delta {\mathrm{CS}}(A).\end{equation}
(This  is the content of eqn. (\ref{bornox}), for example.)  Since when restricted to $W$ we have $A=\xi(\omega)$, we can equivalently write
\begin{equation}\label{gollf}\delta\EH=\frac{1}{2\pi}\delta\mathrm{CS}(\xi(\omega)).\end{equation}
In turn, $\mathrm{CS}(\xi(\omega))$ is (modulo the standard $2\pi$ ambiguity) the same as
$\fb\,\mathrm{CS}(\omega)$ where
$\CS(\omega)$ is the Chern-Simons invariant of $\omega$ as an $SU(2)$ connection  (before
embedding it in $G^\vee$), and $\fb$ is an integer, analyzed in section \ref{zelod}, that results from the embedding.  So we can slightly simplify (\ref{gollf}) to
\begin{equation}\label{golf}\delta\EH=\frac{\fb}{2\pi}\delta\mathrm{CS}(\omega).\end{equation}

If $V$ is a compact manifold with boundary, there is a simple cure for this.  We simply modify the definition (\ref{tofus}) of $\EH$
by subtracting the integral over $V$ of a suitable curvature integral.  The curvature integral is a multiple of $\int_V\Tr\,R\wedge R$,
with $R$ the Riemmann tensor of $V$.  This integral is a topological invariant if $\partial V=\varnothing$, and in general its
variation is a multiple of $\delta\mathrm{CS}(\omega)$.  We pick the coefficient to cancel the boundary term in the variation of $\EH$.
Thus, we replace the definition  (\ref{tofus}) with
\begin{equation}\label{lofus}\hat \EH = \frac{1}{2h^\vee}\frac{1}{32\pi^2}\int_V
\epsilon^{\mu\nu\alpha\beta}\,\Tr_\adj \,F_{\mu\nu}F_{\alpha\beta}-\frac{\fb}{4}\frac{1}{32\pi^2}\int_V\epsilon^{\mu\nu\alpha\beta}\,
\Tr_{T\negthinspace V}\,
R_{\mu\nu}R_{\alpha\beta},\end{equation}
where we view the Riemann tensor as a two-form with values in endomorphisms of the tangent bundle $TV$ of $V$ and take the trace
accordingly.\footnote{If $V$ is spin and we pick one of the spin bundles of $V$, say the bundle $S_+$ of spinors of positive chirality,
then we can use in (\ref{lofus}) a trace in $S_+$, rather than $1/4$ of a trace in $TV$.  Even if $V$ has a boundary, but assuming the metric
is a product near the boundary, the two formulas differ by a topological
invariant, a multiple of the Euler characteristic of $V$.}
  With the boundary condition of sections \ref{boundcond} and \ref{vivisect}, $\hat\EH$ is an integer-valued topological invariant.  The modification of $\EH$
amounts to adding to the underlying Lagrangian a coupling of the gauge-theory theta-angle to
$\Tr_{T\negthinspace V}\,R\wedge R$, in addition to its usual coupling to the gauge theory
instanton density.    If $V$ has no boundary,
this modification does not affect the topological invariance of the theory, while
if $V$ has a boundary, it eliminates the dependence on the Riemannian metric of the boundary.

\subsubsection{The Product Case And The Framing Anomaly}\label{prodcase}

What has just been described does not quite work if $V$ is the noncompact four-manifold $W\times \R_+$ that will be essential in
our applications.  Let us discuss this case closely.   We always assume a product metric on $W\times \R_+$; considering more general
metrics does not add anything.

On $V=W\times \R_+$, we should first worry about a possible problem in defining $\EH$ at infinity, as well as the problem at the boundary
of $V$.  At infinity on $\R_+$, we take a boundary condition that is given by a homomorphism $\rho^\vee:\pi_1(W)\to G^\vee_\C$
(as in  the last paragraph of section \ref{zdetails}).  Such a homomorphism is given by a complex-valued connection $\A=A+i\phi$ that
is independent of $y$.   The complex-valued Chern-Simons invariant $\mathrm{CS}(\A)$ is, of course, independent of the metric of $W$,
and, given that $\A$ is flat, the real part of $\mathrm{CS}(\A)$ coincides with $\mathrm{CS}(A)$.   So $\mathrm{CS}(A)$ is independent
of the metric of $W$.  Hence varying the metric of $W$ does not produce a contribution at infinity to the change in $\EP$; the only such
contribution comes at $y=0$, that is, at the boundary of $V$.  Still, if $\rho^\vee$ is non-trivial, the constant value of $\CS(A)$ does
represent a contribution to $\EP$.  Because of this contribution as well as the contribution at $y=0$, the values of $\EP$ are
not integers.  However,  differences in values of $\EP$ continue to be integers.

We pause to explain this last important statement.
The statement is clear if $G^\vee$ is simply-connected, for then
any two bundles that obey the boundary conditions differ by a twist by an element of $\pi_3(G^\vee)$; as usual this twist shifts the instanton
number by an integer.  But even if $G^\vee$ is not simply-connected, differences in the values of $\EP$ are still integers in the special
case of $V=W\times \R_+$.  Let us explain the reason for this for the case $G^\vee=SO(3)$.  In this case, a $G^\vee$ bundle $E\to V$ has
an invariant $w_2(E)\in H^2(V,\Z_2)$, and if $V$ is a compact four-manifold without boundary, the instanton number of the bundle
$E$ is congruent to\footnote{This is a standard topological result. First, let us explain why $\int_V w_2(E)^2$ can
be evaluated mod 4 even though $w_2(E)$ is defined only mod 2.  For simplicity, we make a very mild
assumption that $W_3(M)=0$,
which implies that $w_2(E)$ can be lifted to a class $x\in H^2(M,\Z)$.  Though $x$ is only uniquely determined mod 2,
$\int_M x^2$ is well-defined mod 4. This is so simply because $(x+2y)^2=x^2+4(xy+y^2)$ so $\int_Mx^2$ is
invariant mod 4 under $x\to x+2y$.  So $\frac{1}{4}\int_Mw_2(E)^2$ is well-defined mod $\Z$.  Now
we wish to show that this number coincides with the instanton number of $E$ mod $\Z$.  By obstruction
theory, this is true for all $SO(3)$ bundles $E$ with a given value of $w_2(E)$ if it is true for one such bundle.  (The
basic idea here is that any two such bundles differ by a twist by $\pi_3(SO(3))=\Z$, and such a twist
shifts the instanton number by an integer.)  So it suffices to consider a convenient choice of $E$.  For such
a choice,  let $\L$ be a complex line bundle with $c_1(L)=w_2(E)$ mod 2, and let $E=\R\oplus \L$
where $\R$ is a trivial real line bundle and $\L$ is viewed as a real bundle of rank 2.  Then $w_2(E)=c_1(\L)$ mod 2
and the instanton number of $E$ is $\frac{1}{4}\int_Mc_1(L)^2$.} $\int_Vw_2(E)^2/4$ mod $\Z$.  This is why, potentially, values of $\EP$ might not differ by integers.
However,
for $V=W\times\R_+$, our boundary condition at $y=0$ says that $E|_W=TW$,
and hence (as any oriented three-manifold is spin), the restriction of $w_2(E)$ to $W$ vanishes.  Since $V=W\times \R_+$ is contractible onto
$W$, this ensures that $w_2(E)$ vanishes altogether, so the $G^\vee$ bundle $E$ is liftable to a $\hat G^\vee$ bundle, where $\hat G^\vee =SU(2)$ is the universal cover of $G^\vee$.
This being so, we can replace $G^\vee$ by $\hat G^\vee$ in analyzing the possible values of $\EP$, and these differ by integers just
as if $G^\vee$ is simply connected.  For any $G^\vee$, the argument proceeds in the same way, using the boundary condition at $y=0$
to show that $E$ can be lifted to a bundle with structure group $\hat G^\vee$.

We still have to face the metric dependence of $\EP$ that comes from the behavior at $y=0$.
On $V=W\times \R_+$, we cannot eliminate the metric-dependence of $\EP$ by subtracting a curvature integral, as above.  For a product
metric on $V$, the integral $\int_V\Tr\,R\wedge R$ vanishes.  If we use a more general metric, adding such a term
would merely move the problem from $y=0$ to $y=\infty$.    Instead, we will have to proceed as in \cite{witten}, where a precisely
analogous problem arose in analyzing Chern-Simons theory on a three-manifold $W$.

If $\mathrm{CS}(\omega)$, the Chern-Simons function of the spin connection, were a well-defined real-valued function, we could
eliminate the problem by subtracting from $\EH$ a multiple of this function to define
\begin{equation}\label{tolfox} \hat\EH=\EH - \frac{\fb}{2\pi}\,\mathrm{CS}(\omega).\end{equation}
$\hat\EH$ would then be an integer-valued topological invariant that we would use instead of $\EP$ in the formula for
the partition function.

Actually, $\mathrm{CS}(\omega)$ has the usual $2\pi$ ambiguity, and is not well-defined as a real-valued function unless
we are given more information.  The additional information we need is known as a ``framing,'' a trivialization (up to homotopy) of the bundle in question.  We have defined $\CS(\omega)$ as the Chern-Simons invariant
of the Riemannian connection regarded as an $SU(2)$ connection on the spin bundle, so the information we need to define $\CS(\omega)$
as a real-valued function
is a framing of the spin bundle.  Actually, we will proceed in a slightly different way.  $\CS(\omega)$ has a dependence on the choice of spin structure of $W$,
and this is unnatural in our problem (unless $G^\vee$ is such that the boundary
condition of section \ref{boundcond} entails a choice of spin structure).
Although $\CS(\omega)$ depends on the spin structure, its variation in a change in metric does not (the
dependence of $\CS(\omega)$ on the spin structure is a topological invariant);
this is why eqn. (\ref{golf}) for the metric dependence of $\EP$ does not depend on a spin structure.
In redefining $\EP$ to eliminate its metric-dependence, we want to avoid introducing an unnatural dependence
on spin structure; we can accomplish this by simply rewriting (\ref{tolfox}) in
terms of the Chern-Simons invariant
of the Riemannian connection $\omega$ regarded as an $SO(3)$ connection on $TW$, the tangent bundle of $W$.
In \cite{witten}, the Chern-Simons invariant of $\omega$ as an $SO(3)$ connection was called $\CS_{\mathrm{grav}}$.  The relation between the $\CS(\omega)$ and $\CS_{\mathrm{grav}}$ is simply
\begin{equation}\label{zondox}\CS_{\mathrm{grav}} = 4\, \CS(\omega).\end{equation}
The factor of 4 reflects the fact that the trace of a product of Lie algebra elements (such as $F\wedge F$) in the
three-dimensional representation of $SO(3)$ is four times the trace of the same product in the two-dimensional representation of $SU(2)$.    To define $\CS_{\mathrm{grav}}$ as a real-valued function, the topological data that we need
is a framing of the tangent bundle $TW$.  This is usually called simply a framing of $W$.

 Given  a framing,
$\mathrm{CS}_\grav$ becomes a well-defined real-valued function, and we eliminate the metric-dependence of
$\EP$ by defining, as in (\ref{tolfox}):
\begin{equation}\label{utolfox}  \hat \EH=\EH-\frac{\fb}{2\pi}\CS(\omega)=\EH - \frac{\fb}{8\pi}\,\mathrm{CS}_\grav. \end{equation}
The quantity $\hat\EH$ is an invariant, valued in a coset of $\Z$ in $\R$ that depends on
the choice of $\rho^\vee$ at infinity and on the framing but not on the metric of $W$.

Replacing $\EP$ by $\hat\EP$ introduces in the partition function $Z$  an extra factor
\begin{equation}\label{zolfox}q^{-\fb\,\mathrm{CS}(\omega)/2\pi}=q^{-\fb\,\mathrm{CS}_\grav/8\pi}.\end{equation}
Under a unit change
of framing, with $\mathrm{CS}_\grav\to\mathrm{CS}_\grav+2\pi$, $\hat\EH$ as defined in (\ref{utolfox}) maps to $\hat\EH-v/4$.
So under a unit change of framing, the partition function transforms by
\begin{equation}\label{costly}Z\to Z q^{-\fb/4}.\end{equation}

Precisely such a dependence on a choice of framing appears in Chern-Simons theory.  In section \ref{zelod}, we will compare
the framing anomaly as we have computed it in eqn. (\ref{costly}) in $\N=4$ super Yang-Mills theory to the standard framing
anomaly as found in Chern-Simons theory.

The relation of what has just been said to the treatment in section \ref{gravcoup} is that if one is given a compact $V$
with boundary $W$, then the curvature integral on $V$  gives a natural lift of $\mathrm{CS}_\grav$ (or $\mathrm{CS}(\omega)$) to a real-valued function.
On $V=W\times\R_+$, there is no natural lift and we simply have to pick one.

Actually, something slightly less than a framing of $TW$ is enough.  In comparing
two framings of $TW$, one runs into an integer winding number, associated with
$\pi_3(SO(3))=\Z$,
and, depending on the topology of $W$, one also encounters some two-torsion information derived
from $\pi_1(SO(3))=\Z_2$.  The two-torsion information is not relevant for the framing anomaly
of Chern-Simons theory.  There is a convenient way to eliminate it \cite{Atiyah}.
Two framings of $TW$ that induce the same framing of $TW\oplus TW$ lead to the same
definition of $\CS_\grav$.  One can therefore consider the basic concept needed to define $\CS_\grav$ to be a framing of $TW\oplus TW$.  A framing of $TW\oplus TW$ is called a two-framing.
Globally, by making use of the signature theorem on a four-manifold with boundary,
one can define  a canonical two-framing for any three-manifold $W$.  This canonical two-framing is often used, explicitly or otherwise, in writing formulas for the Chern-Simons partition function.  Because there is no local recipe for constructing it, it is  natural to allow any framing (or two-framing) and determine how the partition function changes in a change of framing.

\subsubsection{Comparison With Chern-Simons Theory}\label{zelod}

\def\sign{{\mathrm{sign}}}
According to \cite{witten}, the framing dependence of Chern-Simons theory on a three-manifold $W$ arises from the fact that to cancel an anomalous
dependence of the partition function $Z$ on the  metric of $W$, we must pick a framing of $W$ and include in the definition of $Z$ a
factor
\begin{equation}\label{pumult}\exp\left(\frac{ic(k) \mathrm{sign}(k) \,\CS_{\mathrm{grav}}}{24}\right).\end{equation}
Here $c(k)$ is the central charge of $G$ current algebra at level $|k|$:
\begin{equation}\label{umult}c(k)=\frac{k\,\dim(G)}{k+h\, \sign(k)},\end{equation}
where $\dim(G)$ is the dimension of the gauge group $G$ and $h$ is its dual Coxeter number.
Both equations (\ref{pumult}) and (\ref{umult}) are usually written for $k>0$; we have included factors of $\sign(k)$ so that
they are valid for any nonzero integer $k$.  (The required factors are determined by the fact that the partition function is invariant
under  $k\to -k$ together with a reversal of the orientation of $W$, which changes the sign of $\CS_{\mathrm{grav}}$.)

It is convenient to expand
\begin{equation}\label{wumult}c(k)=\dim(G)-\frac{h\,\dim(G)\,\sign(k)}{k+h\,\sign(k)}. \end{equation}
Here the first term, $\dim(G)$, arises in the one-loop approximation to Chern-Simons theory.  In fact, it comes from the metric-dependence
of an Atiyah-Patodi-Singer $\eta$-invariant, as explained in \cite{witten}.  When inserted in (\ref{pumult}), this term gives a factor
\begin{equation}\label{rumult} \exp( i \dim(G)\,\sign(k)\,\CS_\grav/24).  \end{equation}
This factor is not analytic in $k$ or $q$ and hence will not match any computation in $\N=4$
super Yang-Mills theory.

Instead, we interpret this factor as part of the constant $\frak N_0$ in  the relation (\ref{zondo})
between two different holomorphic volume forms on the space of complex-valued connections.
One of these, which we call $DA$, arises   by analytic continuation of the path
integral measure of Chern-Simons
theory (with a compact gauge group $G$), while the second, which we call $D\A$,
is induced from $\N=4$ super Yang-Mills theory (together with a boundary condition
defined by a flat connection $\A_\infty$ at $y=\infty$, associated with some homomorphism
$\rho:\pi_1(W)\to G_\C$).
If what we have just found were a complete formula for $\frak N_0$, we would have
\begin{equation}\label{zondoz}DA\cong D\A \,\exp\left(-ih\,\mathrm{sign}(k)\,\CS(\A)+ i \dim(G)\,\sign(k)\,\CS_\grav/24\right).\end{equation}
Unfortunately, this cannot quite be a complete formula.
Because of  the factor of $1/24$ multiplying
$\CS_\grav$, the formula actually leaves unspecified a $24^{th}$  root of unity in the relation between
$DA$ and $D\A$.    There is actually yet another
 root of unity that should be included; this is a fourth root
of unity that arises on the Chern-Simons side from a spectral flow invariant that is described in \cite{gf}. It seems that $\frak N_0$ depends on $\rho$, at least by these roots of unity,
as well as on the metric of $W$. The factor involving $\CS_\grav$ and the roots of unity
all come from the $\eta$ invariant which arises in the one-loop approximation
to Chern-Simons theory evaluated at the flat connection $\A_\infty$.
Perhaps $\frak N_0$ should simply be written in terms of this $\eta$-invariant.
Luckily, in this paper we mostly take $W=\R^3$ and $\A_\infty=0$, enabling us to avoid these issues.

The higher order terms turn out to have a more clear-cut interpretation.
We write $c(k)=\dim(G) +\Delta c$, where $\Delta c=-h\,\sign(k)\,\dim(G)/(k+h\,\sign(k))$ is the  part of $c(k)$ that in Chern-Simons theory comes from diagrams of two or more loops.  The natural perturbative expansion in Chern-Simons
theory is in powers of $1/k$; $\Delta c$ has contributions of all orders in this expansion.
On the other hand, in $\N=4$ super Yang-Mills theory, the natural expansion parameter
is $1/\Psi$ where $\Psi=k+h\,\sign(k)$, so in this expansion, $\Delta c$ is purely a two-loop effect.
This fact remains to be explained.

In any case, the framing anomaly associated to $\Delta c$ has a straightforward interpretation in
the $S$-dual description by $G^\vee$ gauge theory.  The part of (\ref{pumult}) involving
$\Delta c$ is $\exp(-ih \,\dim(G)\,\CS_\grav/24(k+h\,\sign(k)))$.  Under an elementary change of
framing $\CS_\grav\to\CS_\grav+2\pi$, this factor changes by \begin{equation}\label{dunno}\exp\left(-\frac{2\pi i h\,\dim(G)}{24(k+h\,\sign(k))}\right)
=q^{-{h \dim(G) \frak n_{\frak g}/24}},\end{equation} where $q$ was defined in (\ref{flodux}).  For the $S$-dual description, the equivalent formula (\ref{costly}) says that in an elementary change of framing, the partition function changes
 by a factor of $q^{-\fb/4}$.  So obviously to reconcile the two formulas, we need $\fb=\frak n_{\frak g}\,h\,\dim(G)/6$.

 So let us evaluate $\fb$.  We start with an $SU(2)$ gauge field $A$ of instanton number 1.
 Such a gauge field has the property that if $\Tr_{\frak{su}(2)}$ is the trace in the adjoint representation of
 $SU(2)$,  then
 \begin{equation}\label{turnod} 1=\frac{1}{2\cdot 2}\cdot \frac{1}{32\pi^2}\int_V
\epsilon^{\mu\nu\alpha\beta}\,\Tr_{\frak{su}(2)}\, F_{\mu\nu} F_{\alpha\beta}.\end{equation}
In the denominator, we have replaced $2 h^\vee$ in the definition of the instanton number by
$2\cdot 2$, since $h^\vee=2$ for $SU(2)$.
Now $\fb$ is defined as the instanton number of the $G^\vee$ gauge field $\xi(A)$,
where $\xi$ is a principal embedding $\frak{su}(2)\to\frak g$. Hence
\begin{equation}\label{urnod}\fb=\frac{1}{2\cdot h^\vee}\frac{1}{32\pi^2}\int_V
\epsilon^{\mu\nu\alpha\beta}\,\Tr_{\frak g} \,\xi(F_{\mu\nu})\xi(F_{\alpha\beta}).\end{equation}
The trace is now taken in the adjoint representation of $G^\vee$, and to be pedantic, we have
written $\xi(F)$ for the $\frak g$-valued image of $F$.    The ratio of traces in (\ref{urnod})
and (\ref{turnod}) is the same as the ratio of the traces of the quadratic Casimir operator of
$\frak{su}(2)$ in the two representations (namely $\frak g$ and $\frak{su}(2)$).  The value
of the Casimir operator in an irreducible representation of $\frak{su}(2)$ of dimension $2j+1$
is $j(j+1)$, and its trace is $j(j+1)(2j+1)$.  So the ratio of the two traces is
$\sum_{i=1}^rj_i(j_i+1)(2j_i+1)/6$, where (as discussed at the end of section \ref{boundcond}) $\frak g$ is a direct sum of $\frak{su}(2)$ modules
of dimensions $2j_i+1$.  So finally
\begin{equation}\label{morkey}\fb= \sum_{i=1}^r\frac{j_i(j_i+1)(2j_i+1)}{3h^\vee}.\end{equation}

The desired relation $\fb=\frak n_{\frak g}\,\dim(G)\,h/6$ hence becomes
\begin{equation}\label{orkey}\sum_{i=1}^rj_i(j_i+1)(2j_i+1)=\frac{1}{2}\frak n_{\frak g}\,\dim(G)\,hh^\vee.\end{equation}
As a check, this relation holds for $G$ if and only if it holds for $G^\vee$.  Indeed,
the $j_i$,  $\frak n_{\frak g}$, and $\mathrm{dim}\,G$ are invariant under the exchange $G\leftrightarrow G^\vee$,
while $h$ and $h^\vee$ are exchanged.

For a proof of this relation, see \cite{Panyushev}, Proposition 3.1.
It is actually not difficult to verify the relation by hand for all simple Lie groups,
whether of type $\sf {A,\,B,\,C,\,D,\,E,\,F,}$ or $\sf G$.  As an example, if $G$ and therefore
also $G^\vee$ are of type ${\sf G}_2$, then the $j_i$ are $1$ and 5, while $\frak n_g=3$, $\mathrm{dim}(G)=14$,
and $h=h^\vee=4$.  The left and right of (\ref{orkey}) both
equal 336.

\subsection{'t Hooft Operators In The Boundary}\label{thooft}

\subsubsection{Preliminaries}\label{prelims}

In section \ref{uzim}, we showed that, when the gauge theory theta-angle  is nonzero, the D3-NS5 system admits supersymmetric
Wilson line operators at, and only at, the boundary of a four-manifold $V$.  Dually, the same must be true for the D3-D5 system, but now with supersymmetric
't Hooft operators rather than Wilson operators.  Our goal in the present section will be to concretely explain how to define these
't Hooft operators.

In general, 't Hooft operators are analogous to disorder operators in statistical mechanics -- and also analogous to the D3-D5 boundary
condition that we have described in section \ref{boundcond}.  Just as our boundary condition was described by specifying the singularity
that fields must have along the boundary of $V$, so an 't Hooft operator is
defined, as explained in \cite{Kapustin}, by describing the singular behavior that four-dimensional fields should have
along a chosen one-manifold $S$, which usually is taken to lie in the interior of $V$. To explain what singular behavior one wants, one selects a local model solution of the supersymmetric Yang-Mills equations
on $\R^4\backslash \R$ (i.e., $\R^4$ with $\R$ removed) with a singularity of some desired type along $\R$.  Normally, one picks a solution that is invariant
under rotations and translations (and possibly conformal motions) of $\R^4$ that map $\R$ to itself, and possibly
under some supersymmetries.
Concretely, for the usual
half-BPS 't Hooft operators, the requisite singular solutions are very simple: they are obtained by embedding an abelian Dirac monopole
in the nonabelian Yang-Mills gauge group.  Once a singularity type is chosen, one calculates in the presence of an 't Hooft operator
supported on a one-manifold $S\subset V$ by doing gauge theory on $V\backslash S$ with fields that have a singularity
along $S$ of the chosen type.

In our problem, we want to follow the same general ideas, with one important difference: $V$ is a four-manifold with boundary $W$,
and $S$ is contained in $W$.  (We expect from duality that $S$ must be contained in $W$, but we can also see this directly by following the analysis of Wilson-'t Hooft operators in section 6.2
of \cite{KW}.\footnote{It is shown there that 't Hooft operators away from the boundary preserve the topological symmetry
only if $\Psi=0$.  It is also shown, however, that for any rational value of $\Psi$, there are
combined
Wilson-'t Hooft operators in bulk (as one would expect from $S$-duality).
These are undoubtedly important for understanding
special properties of the theory at rational values of $\Psi$.})  But the basic idea of defining an 't Hooft operator by specifying a model solution
still applies.

For the model solution, we now take $V$ to be a half-space, say the space $x^3\geq 0$ in a Euclidean space with coordinates $x^0,\dots,x^3$.
And we take $S$ to be a straight line in the boundary of $V$, say the line $x^1=x^2=x^3=0$.  We look for a solution of the Yang-Mills
equations on $V$ that is invariant under symmetries that map $S$ to itself, that is, under translations of $x^0$ and rotations of the
$x^1-x^2$ plane.  In addition, as we want an 't Hooft operator that preserves the supersymmetry $Q$ of our topological field theory,
the singular solution should obey the supersymmetric equations (\ref{oofut}).  (Actually our 't Hooft operator will preserve more supersymmetry
than just the one supercharge $Q$, which it will accomplish by obeying a stronger system of equations, as described later.)  The solution
should become trivial for $x^3\to\infty$, far from the position of the 't Hooft operator.   At a generic boundary point, it must
have the boundary behavior of the regular Nahm pole as described in section \ref{boundcond}.  This in particular means that the desired singular solution cannot be a simple
abelian one, like the singular solution used to describe an 't  Hooft operator away from the boundary.  At a boundary point that is located
on the line $S$, the singular behavior is more complicated.   That more complicated behavior is exactly what we wish to determine.

We will carry out this program in full for $G=SU(2)$. For $G$ of higher rank, we carry out some of the steps but the precise singular
solution of relevance is not yet known.

\subsubsection{First Reduction Of The Equations}\label{doxon}

As just explained, we want to find on the half-space $V$ given by $x^3\geq 0$ a special type of solution of the supersymmetric equations
\begin{equation}\label{torzo} F-\phi\wedge\phi +\star \d_A\phi =0 = \d_A\star \phi.\end{equation}
The solution should be invariant under translations in $x^0$,  should become trivial for $x^3\to\infty$, and away from the line
$S$ given by $x^1=x^2=x^3=0$, its boundary behavior should coincide with the regular Nahm pole described in section \ref{boundcond}.

A drastic simplification comes from the fact that in solving the equations, we can set $A_0=\phi_3=0$.  The reader may choose to
view this as a lucky ansatz that can be used to simplify the equations.  However, there are also several ways to predict {\it a priori}
that the solution we want has $A_0=\phi_3=0$.  For one thing, one can use a vanishing argument similar to that discussed in eqn. (4.13)
of \cite{wittentwo} to prove that a solution on $V$ with the desired asymptotic behavior has $A_0=\phi_3=0$.  (The proof is standard:
one squares the equations (\ref{torzo}), integrates over $V$, and then integrates by parts, showing that in any solution, $A_0$ and $\phi_3$ are
annihilated by strictly positive linear differential operators.)  Alternatively, one can use supersymmetry.   Obeying (\ref{torzo}) ensures invariance under one supersymmetry, but duality with the boundary Wilson lines studied in section
\ref{uzim} indicates that the 't Hooft operators of interest should preserve four global supercharges (half of the supercharges preserved
by the half-BPS boundary condition).  The extra supersymmetry puts additional constraints on the solution, leading to the
structure that we describe momentarily.

\def\D{{\mathcal D}}
The equations obtained from (\ref{torzo}) after setting $A_0=\phi_3=0$ can be described as follows.
Define the three operators
\begin{align}\label{thrice} \D_1 & = \frac{D}{D x^1}+i\frac{D}{Dx^2}=\frac{\partial}{\partial x^1}
+i\frac{\partial}{\partial x^2}+[A_1+iA_2,\,\cdot\,]\cr
                                                     \D_2 & = D_3-i[\phi_0,\cdot]=\frac{\partial}{\partial x^3}+[A_3-i\phi_0,
                                                     \,\cdot\,] \cr    \D_3 & = [\phi_1-i\phi_2,\,\cdot\,] .\cr \end{align}
Thus, $\D_1$ and $\D_2$ are first order differential operators, while $\D_3$ is of order zero.
In (\ref{thrice}), for an adjoint-valued field $\Lambda$,
the symbol $[\Lambda,\cdot ]$ represents the commutator with $\Lambda$.

With this understood, the equations (\ref{torzo}) take the form
\begin{equation}\label{rice}[\D_i,\D_j]= 0,~~i,j=1,\dots,3\end{equation}
together with
\begin{equation}\label{ice}\sum_{i=1}^3[\D_i,\D_i^\dagger]=0.\end{equation}
Here $\D_i^\dagger$ is the adjoint of the differential operator $\D_i$.  Concretely, (\ref{ice})
takes the form
\begin{equation}\label{mice} F_{12}-[\phi_1,\phi_2]-D_3\phi_0=0.\end{equation}
To similarly make (\ref{rice}) explicit is immediate from the definitions of the $\D_i$.

Before trying to understand these equations, let us describe some special cases.   If we set $A_1=A_2=0$ and take the fields to be independent of
$x^1$ and $x^2$, we get Nahm's equations.  If we set $A_3=\phi_0=0$ and take the fields to be independent of $y=x^3$, we get Hitchin's
equations.  Finally, if we set $\phi_1=\phi_2=0$, we get the Bogomolny equations. So our
system is a hybrid of all those equations. This
hybrid was encountered in \cite{KW} and called the extended Bogomolny
equations (see eqn. (10.36) of that paper, where the  equations are
written in the gauge $A_y=0$).  The main interest there was the role in these equations
of 't Hooft operators in the bulk (and their interpretation in terms of Hecke modifications of Higgs
bundles).  Our concern here will instead be the more subtle case of 't Hooft operators in the
boundary.

It is also helpful to consider some analogous equations.
For an interesting analogy, consider gauge theory of a connection $A$ on $\R^6\cong \C^3$.  We endow $\C^3$ with complex
coordinates $z^i$, $i=1,\dots,3$, and define
\begin{equation}\label{tice}\D_i=\frac{\partial}{\partial\bar {z^i}}+A_{\bar i}.\end{equation}
In other words, the $(0,1)$ part of the connection is $\sum_i \d\bar{z^i}\D_i$.  The equations $[\D_i,\D_j]=0$ assert that the
$(0,2)$ part of the curvature vanishes, so that the connection defines a holomorphic bundle, while the remaining equation $\sum_i [\D_i,\D_i^\dagger]=0$ can be solved only if the holomorphic bundle is semi-stable, and, according to a theorem of Donaldson and
of Uhlenbeck and Yau, it has a unique solution in that case.   The combined equations are known as the hermitian Yang-Mills
equations, and can be formulated on a general complex manifold, not necessarily $\C^3$.  Physically, the hermitian Yang-Mills
equations are
 familiar in the context of the heterotic string on a Calabi-Yau threefold.  In that context, solutions of those equations preserve four supercharges, and the same is true for the
equations (\ref{rice}) and (\ref{ice}), though we will not demonstrate this here.

As in the other cases that we have just mentioned, the key to understanding the equations (\ref{rice}) and (\ref{ice}), is to first observe that equations
(\ref{rice}) have a larger gauge symmetry than the full system. The full system of equations is invariant under an ordinary gauge transformation
\begin{equation}\label{trex}\D_i\to g\D_i g^{-1},~~i=1,\dots,3,\end{equation}
where $g$ is  $G^\vee$-valued.  But eqns. (\ref{rice}), since they involve only the operators $\D_i$
and not their adjoints, are invariant under complex-valued gauge transformations, that is gauge transformations in which we allow
$g$ to be valued in $G^\vee_\C$, the complexification of $G^\vee$.  The space of solutions of eqns. (\ref{rice}), modulo complex-valued
gauge transformations, is naturally a complex manifold.  In all the problems that we have mentioned -- including Nahm's equations, Hitchin's equations, the Bogomolny equations, the hermitian Yang-Mills equations,
and also our present problem -- the remaining equation (\ref{ice}) can be interpreted
as an equation for vanishing of the moment map.  In other words, in each case, one can define a symplectic structure on the space of fields
such that the moment map for the action of the compact gauge group ($G^\vee$ in our problem)  is the left hand side of eqn. (\ref{ice}).  One then aims to compare {\it (i)} the space
of solutions of the full system of equations, modulo $G^\vee$-valued gauge transformations, to {\it (ii)} the solutions of the holomorphic
equations modulo $G^\vee_\C$-valued gauge transformations.  Typically, one aims to show (as in the result of Donaldson and Uhlenbeck-Yau
concerning the hermitian Yang-Mills equations) that {\it (i)} and {\it (ii)} coincide after correcting {\it (ii)} to incorporate a certain condition
of stability.  In our present problem, the desired boundary condition at $y=0$ ensures
that the gauge group acts freely on the space of solutions, and one may hope that in
a proper formulation -- which will have to take into account the boundary behavior
in an essential way -- {\it (i)} and {\it (ii)} -- will simply coincide.

\subsubsection{The Holomorphic Data}\label{hol}

The holomorphic data in this problem are easily described.  Since a holomorphic
$G^\vee_\C$-bundle over the complex $z$-plane is trivial, we can make a complex
gauge transformation to go to a gauge in which $A_1+iA_2=0$, so that $\D_1$
reduces to $\partial_1+i\partial_2=2\partial_{\bar z}$.  But actually, since $[\D_1,\D_2]=0$,
we can do better: we can make a complex gauge transformation setting $A_1+iA_2=A_3-i\phi_0
=0$.  In this gauge, $\D_1=2\,\partial/\partial\bar z$ and $\D_2=\partial/\partial x^3$.
The equations $[\D_1,\D_3]=[\D_2,\D_3]=0$ then say that $\varphi=\phi_1-i\phi_2$ is
holomorphic in $z$ and independent of $y=x^3$.  We are still free to make a gauge
transformation by a holomorphic map $g(z):\C\to G^\vee_\C$.

In short, the holomorphic data consist of a $\frak g^\vee_\C$-valued holomorphic function
$\varphi(z)$, modulo conjugation by a $G^\vee_\C$-valued holomorphic function $g(z)$.
What sort of function $\varphi(z)$ we should consider depends on what behavior we want
at infinity.  Let us remember that vacuum states of $\N=4$ super Yang-Mills theory are specified
by the asymptotic values of the scalar fields (which moreover must commute with each other
to ensure the vanishing of the classical potential energy).  In particular, a choice of vacuum
state at infinity determines the conjugacy class of $\varphi=\phi_1-i\phi_2$ at $y=\infty$.
For the present paper, the most convenient vacuum to consider is the one in which the scalar
fields simply vanish at infinity.  So we will look for solutions of the extended Bogomolny equations
in which $\varphi\to 0 $ at infinity.  In any event, the real interest in the present section is
in the singular behavior of the solution near special boundary points where 't Hooft operators
are inserted, and  we do not care too much about what happens far away. For our immediate purposes, asking for $\varphi$ to vanish at infinity is just
a convenient auxiliary condition that will make it easier to find a solution with the singularity
we want.

\def\P{{\mathcal P}}
The equation $[\D_2,\D_3]=0$ is equivalent to  $\partial_3\varphi=-[A_3-i\phi_0,\varphi]$.  It
says that the $x^3$ derivative of $\varphi$ is a commutator of $\varphi$ with some matrix,
so that the conjugacy class of $\varphi$ is independent of
$y=x^3$.  It is not correct to conclude from this and the fact that $\varphi$ vanishes at $y=\infty$
that $\varphi$ is identically zero.  The correct conclusion is only that $\varphi$ is nilpotent.  To prove nilpotency, let $\P$ be a homogeneous  invariant polynomial of positive degree on the complex Lie algebra $\frak g_\C$.
Since the conjugacy class of $\varphi$ is independent of $y$, we have  $\partial_y \P(\varphi)=0$.  So if $\varphi$ vanishes
at infinity, then $\P(\varphi)$ vanishes for all $y$.  An element $\varphi\in\frak g_\C$
such that $\P(\varphi)=0$ for all $\P$ of the assumed kind is nilpotent.  So $\varphi$ is nilpotent for all $y$ (and $z$).

A simple example of a solution in which $\varphi$ is everywhere nilpotent but not zero and approaches zero at
infinity
is  the basic Nahm pole solution
(\ref{nelfo}) with $\vec\phi=\vec t/y$, where $\vec t$ are images of a standard set of $\frak{su}(2)$
generators under an embedding $\xi:\frak{su}(2)\to\frak{g}$.  In this
solution, $\varphi=(t_1-it_2)/y$ is indeed nilpotent (it is a lowering operator with respect to $t_0$).  Its conjugacy
class is independent of $y$ (this is proved by conjugating by $t_0$) and it vanishes for $y\to\infty$.  

We are actually interested in the case that $\xi$ is a principal embedding, which is equivalent to the condition
that  $\varphi$ is   a regular nilpotent element of $\frak g_\C$.
We pause to explain this concept.
Every complex simple Lie algebra has a finite set of nilpotent conjugacy classes.  For
example, a nilpotent element $\varphi\in\frak{sl}(n,\C)$ can be conjugated to a Jordan canonical  form in which
all matrix elements vanish except just above the main diagonal:
\begin{equation}\label{onkey}\varphi=\begin{pmatrix}0 & * & 0 & \dots & 0 \cr
                                                                                  0 & 0 & * & \dots & 0 \cr
                                                                                    & & & \ddots &   \cr
                                                                                     0 & 0 & 0 &\dots & *\cr
                                                                                      0 & 0 & 0 & \dots & 0\cr\end{pmatrix},\end{equation}
and moreover the matrix elements just above the main diagonal are all 1 or 0.
The conjugacy classes of nilpotent elements of $\frak{sl}(n,\C)$ are classified by the pattern of 1's and 0's, up to
obvious permutations of blocks.       An element of a complex
Lie algebra $\frak g_\C$ is called regular if the subalgebra of $\frak g_\C$ that commutes
with it is as small as possible, that is if its dimension equals $r$, the rank of the algebra.
There is always a unique nilpotent conjugacy class of maximal dimension, known
as the regular nilpotent conjugacy class.
This is the class containing the
raising and lowering operators for a principal $\frak{su}(2)$ subalgebra.  For $\frak{sl}(n,\C)$,
the regular nilpotent conjugacy class is the one with a single Jordan block (all elements
labeled $*$ in (\ref{onkey}) actually equal 1).  A generic nilpotent element is contained
in this regular nilpotent conjugacy class.  In particular, in the solution associated to the
principal $\frak{su}(2)$ embedding, $\varphi$ is a regular nilpotent element.

Finally, we can describe the solutions that are relevant for boundary 't Hooft operators.
We look for a solution in which $\varphi(z)$ is holomorphic in $z$ and everywhere nilpotent.
Moreover, for a generic value of $z$, the behavior for $y\to 0$ must coincide with the model
solution (\ref{nelfo}), so $\varphi$ is a regular nilpotent.  At isolated points $z=z_j$, $j=1,\dots,s$,
$\varphi$ is in a more special nilpotent conjugacy class.  These are the points at which
't Hooft operators are inserted.

For example, for the case that $G^\vee=SU(2)$, any everywhere nilpotent $\varphi(z)$ is conjugate to
\begin{equation}\label{nobbo}\varphi(z)=\begin{pmatrix}0 & f(z) \cr 0 & 0 \end{pmatrix},\end{equation}
for some holomorphic function $f(z)$.  Only the zeroes of $f$ and the degrees of
their zeroes have an invariant meaning,
since where $f(z)$ is not zero, 	we can  set $\varphi = g\varphi_1g^{-1}$,  with
\begin{equation}\label{onk}\varphi_1=\begin{pmatrix} 0 & 1\cr 0 & 0 \end{pmatrix}\end{equation} and
\begin{equation}g(z)= \begin{pmatrix} f(z)^{1/2} & 0 \\ 0 & f(z)^{-1/2}\end{pmatrix}.\end{equation}
The case of a single 't Hooft operator is the case that the function $f(z)$ has only one zero,
say of order ${\frak r}$:
\begin{equation}\label{orobbo}\varphi=\begin{pmatrix} 0 & z^{\frak r} \cr 0 & 0 \end{pmatrix}\end{equation}
In section \ref{sutwo}, we will find for each positive integer ${\frak r}$ a unique solution of the extended Bogomolny equations  with
this $\varphi$ and the appropriate asymptotic behavior at the boundary $y=0$ and at infinity.

For a more systematic explanation of the above formula, let us recall that GNO or Langlands duality associates to
a representation of $G$ a dual magnetic weight of $G^\vee$.  This magnetic weight is
a conjugacy class of  homomorphisms from $\C^*$ to $G^\vee_\C$.  For $G=SO(3)$,
the homomorphism to $G^\vee_\C=SL(2,\C)$ associated to the spin $j$ representation of $G$ is
\begin{equation}z\to g(z)=\begin{pmatrix} z^j & 0 \cr 0 & z^{-j}\end{pmatrix}.\end{equation}
For $G=SU(2)$, $j$ may be half-integral and then the formula should be written in the spin 1 representation; $g(z)$ is well-defined as a homomorphism
from $\C^*$ to $G^\vee_\C=SO(3)_\C$.
   In all cases, the relation between $\varphi$ and $g$
is $\varphi=g\varphi_1 g^{-1}$, so that in the notation of eqn. (\ref{orobbo}), ${\frak r}=2j$.

The analog of this for $G=SU(n)$ is hopefully clear.  Instead of (\ref{orobbo}), we look for a solution
with
\begin{equation}  \varphi  =\begin{pmatrix}0 & z^{{\frak r}_1} & 0 & \dots & 0 \cr
                                                                                  0 & 0 & z^{{\frak r}_2} & \dots & 0 \cr
                                                                                    & & & \ddots &   \cr
                                                                                     0 & 0 & 0 &\dots & z^{{\frak r}_{n-1}}\cr
                                                                                      0 & 0 & 0 & \dots & 0\cr\end{pmatrix},      \end{equation}
where the ${\frak r}_i$ are non-negative integers, not all zero, representing the highest weight
of a representation of $G$.  More generally, for any $G^\vee$, the corresponding formula
is obtained as follows.  Pick a principal $\frak{su}(2)$ embedding and within it a Cartan subalgebra.
 Relative to this choice, let $\varphi_1$
be a raising operator of the chosen $\frak{su}(2)$ subalgebra, and let $T^\vee_\C$ be the maximal
torus of $G^\vee_\C$ that commutes with the chosen Cartan subalgebra of $\frak{su}(2)$.
Pick a homomorphism $g(z):\C^*\to T^\vee_\C$ such that $\varphi= g\varphi_1 g^{-1}$ has
no pole at $z=0$.  The choices for $g(z)$ are in natural correspondence with the highest
weights of $G$ representations, and therefore with Wilson operators of $G$ gauge theory.  By solving the extended Bogomolny equations with the
corresponding $\varphi$ and identifying the singular behavior at $y=z=0$, we get our
candidate for the definition of the boundary 't Hooft operator in $G^\vee$ gauge theory
that is dual to a given Wilson operator of $G$.

In section \ref{sutwo}, we will explicitly find the relevant solutions of the extended Bogomolny equations for $G=SU(2)$.  For $G$
of higher rank, this  remains open.

\subsubsection{Solving The Equations For $SU(2)$}\label{sutwo}

Starting with the holomorphic data (\ref{orobbo}), with all other fields vanishing, we want to make a complex gauge transformation
$\D_i\to g\D_i g^{-1}$ so as to obey the extended Bogomolny equations.  Since the $\D_i$ will obey $[\D_i,\D_j]=0$ for any choice
of $g$, we really need only chose $g$ to obey the remaining condition $\sum_i[\D_i,\D_i^\dagger]=0$.

The extended Bogomolny equations are invariant under $\varphi\to e^{i\alpha}\varphi$ with $\alpha$ a real constant.  The holomorphic
data (\ref{orobbo}) are invariant under this symmetry, up to a diagonal gauge transformation.  So it is natural to choose $g$ so as to preserve
the symmetry.  This means that $g$ must be diagonal:
\begin{equation}\label{onzo}g = \begin{pmatrix} e^{v/2} & 0 \cr 0 & e^{-v/2}\end{pmatrix}.\end{equation}
Moreover, using the invariance of the extended Bogomolny equations under unitary gauge transformations (those valued in $G^\vee$
rather than its complexification), we can take $v$ to be real.   After transforming $\D_i\to g \D_i g^{-1}$, we find
\begin{align}\label{mathco} A_1+i A_2 & = -\frac{(\partial_1+i\partial_2)v}{2} \begin{pmatrix} 1 & 0 \cr 0 & -1 \end{pmatrix} \cr
                                                   F_{12} & = \frac{i(\partial_1^2+\partial_2^2)v}{2} \begin{pmatrix} 1 & 0 \cr 0 & -1 \end{pmatrix} \cr
                                                      \phi_0 & = -\frac{i\,\partial_3 v }{2} \begin{pmatrix} 1 & 0 \cr 0 & -1 \end{pmatrix} \cr
                                                     \varphi & = z^{\frak r} e^v\begin{pmatrix} 0 & 1\cr 0 & 0 \end{pmatrix}.\end{align}
And finally, the ``moment map'' equation $\sum_i[\D_i,\D_i^\dagger]=0$ becomes
\begin{equation}\label{zon}-\left(\frac{\partial^2}{\partial x_1^2}+\frac{\partial^2}{\partial x_2^2}+\frac{\partial^2}{\partial y^2}\right)v
+|z|^{2{\frak r}} \exp(2v) = 0,\end{equation}
where we write $y$ for $x_3$ and $z$ for $x_1+ix_2$.

This equation has the simple exact solution
\begin{equation}\label{elx} v=-{\frak r}\log|z|-\log y,\end{equation}
corresponding to
\begin{equation}\label{melx} \varphi = \frac{(z/\bar z)^{{\frak r}/2}}{y}\begin{pmatrix} 0 & 1\cr 0 & 0 \end{pmatrix}.\end{equation}
This solution is singular at $z=0$, but the singularity can actually be removed by a unitary gauge transformation $\varphi\to h\varphi h^{-1}$
with
\begin{equation}\label{yelx} h =\begin{pmatrix}(z/\bar z)^{-{\frak r}/4} & 0 \cr 0 & (z/\bar z)^{{\frak r}/4}\end{pmatrix}.\end{equation}
After this gauge transformation, we arrive at the basic solution (\ref{nelfo}) in which the gauge field $A$ vanishes while $\varphi$
is $1/y$ times a raising operator.  This is the solution that defines the boundary condition we want at boundary points with $z\not=0$,
that is, anywhere away from the insertion of the 't Hooft operator.

To describe an 't Hooft operator at the boundary, we want a solution with the same behavior as (\ref{elx}) for $y\to 0$ with $z\not=0$, but regular along the open ray $z=0$,
$y\not=0$.  Exactly what will happen near $z=y=0$ will be determined by the equations.  That will be the answer to our question: the 't Hooft
operator of charge ${\frak r}$ will be defined by the singularity that the equation forces upon us at $z=y=0$.

It is useful to make a small change of variables:
\begin{equation}\label{pelx} v = -({\frak r}+1)\log|z|+u.\end{equation}
The desired behavior of $u$ is hence
\begin{equation}\label{tomx} \begin{cases} u\sim \log|z|-\log y
 & \mathrm{for} ~y\to 0~ \mathrm{with} ~z\not=0\\
                                                                      u\sim ({\frak r}+1)\log |z|
                                                                       & \mathrm{for}~ z\to 0~\mathrm{with}
                                                                        ~y\not=0. \end{cases}\end{equation}
(The second condition ensures that $v$ is regular at $z=0$, $y>0$.)
In terms of $u$, the equation becomes
\begin{equation}\label{zonzo}-\left(\frac{\partial^2}{\partial x_1^2}
+\frac{\partial^2}{\partial x_2^2}+\frac{\partial^2}{\partial y^2}\right)u
+|z|^{-2} \exp(2u) = 0.\end{equation}
Writing the equation this way makes visible a scaling symmetry
$z\to\lambda z$, $y\to\lambda y$.  There is also an obvious symmetry
of rotation of the $z$-plane.

\def\NN{N}
It is natural to expect the fields produced by an 't Hooft operator at $y=z=0$ to be scale-invariant and rotation-symmetric.
For a rotation-symmetric solution, writing $r=|z|$, the equation becomes
\begin{equation}\label{onzox}-\left((r\partial_r)^2+(r\partial_y)^2)\right) u +\exp(2u)=0.\end{equation}  Scale-invariance means that $u$ is a function only of  $s=r/y$.  Acting on a function with this property,
we can substitute $r\partial_r\to s\,\partial_s$, $r\partial_y\to -s^2\,\partial_s$, so the equation becomes
\begin{equation}\label{nzo} -\left(\left(s\frac{\d}{\d s}\right)^2+\left(s^2\frac{\d}{\d s}\right)^2\right)u+e^{2u}=0.\end{equation}
This equation can be neatly solved by transforming from $s$ to another coordinate $\tau(s)$ with the property that
\begin{equation}\label{nzone} \left(s\frac{\d}{\d s}\right)^2+\left(s^2\frac{\d}{\d s}\right)^2=\frac{\d^2}{\d\tau^2}. \end{equation}
This equation is conveniently equivalent to
\begin{equation}\label{pzonne}\left(\sqrt{s^2+s^4}\frac{\d}{\d s}\right)^2=\frac{\d^2}{\d\tau^2},\end{equation}
leading to
\begin{equation}\label{qzone} \frac{\d s}{\sqrt{s^2+s^4}}=\d\tau. \end{equation}
This equation can be integrated, but for the moment let us refrain from doing so.
In terms of $\tau$, our equation (\ref{nzo}) becomes
\begin{equation}\label{rzone}\frac{\d^2 u}{\d \tau^2} =\exp(2 u).\end{equation}
This implies that
\begin{equation}\label{kzone} \frac{\d u}{\sqrt{e^{2u}+b^2}}=\pm \,\d\tau,\end{equation}
with an integration constant $b^2$.  Setting
\begin{equation}\label{mozone} e^{u(\tau)}=b\,p(\tau),\end{equation}
we get
\begin{equation}\label{pzone} \frac{1}{b}\frac{\d p}{\sqrt {p^4+p^2}}=\pm \d \tau,\end{equation}
and comparing to (\ref{qzone}), we see that we can eliminate $\tau$:
\begin{equation}\label{mzone}\frac{1}{b}\frac{\d p}{\sqrt{p^4+p^2}}=\pm \frac{\d s}{\sqrt{s^4+s^2}}.\end{equation}
Using now the indefinite integral
\begin{equation}\label{tzone} \int \frac{\d t}{\sqrt{t^4+t^2}}=-\log\left(\frac{t}{\sqrt{1+t^2}-1}\right)+C,\end{equation}
we find that
\begin{equation}\label{lzone}\frac{p}{\sqrt{1+p^2}-1}=\NN \left(\frac{ s}{\sqrt{1+s^2}-1}\right)^{\pm b},\end{equation}
for a constant $\NN$.    For $y\to 0$ with fixed $z\not=0$, we have $s\to\infty$, and according to (\ref{tomx}), we want $u\to\infty$ in this
limit, and hence also $p\to\infty$.  It then follows from (\ref{lzone}) that we must set $\NN=1$.
Compatibility with  (\ref{tomx}) for $s\to 0$ (that is, for $z\to 0$ with fixed $y\not=0$) gives $b={\frak r}+1$
(and also tells us to use the plus sign in the exponent in (\ref{lzone})).  Taking these values and
solving for $p$, we get
\begin{equation}\label{pliny}p(s)=\frac{2s^{{\frak r}+1}}{\left(\sqrt{1+s^2}+1\right)^{{\frak r}+1}-\left(\sqrt{1+s^2}-1\right)^{{\frak r}+1}}.
\end{equation}
The original variable $v(s)$ is
\begin{equation}\label{liny}e^{v(s)}=\frac{({\frak r}+1)p(s)}{|z|^{{\frak r}+1}}.\end{equation}

This is the solution in the presence of a single  't Hooft operator that is dual to a Wilson operator with $j={\frak r}/2$.  More generally, the singularity of this solution at $y=z=0$ defines what
we mean by a boundary 't Hooft operator of this magnetic charge.

To understand the solution a little better, let us evaluate the gauge field on the boundary plane $y=0$.  From (\ref{liny}), we have
$v=-\log y -{\frak r}\,\log z+\mathrm{constant}+\mathcal O(y)$, so from (\ref{mathco}) we get
\begin{equation}\label{dorf}A_i=\frac{\epsilon_{ij}x_j}{x_1^2+x_2^2}\frac{{\frak r}}{2}\begin{pmatrix} i & 0 \cr 0 & -i \end{pmatrix}+
\mathcal O(y).\end{equation}
This is a familiar type of two-dimensional $U(1)$ gauge field, except that here it is embedded in $SU(2)$.  It describes a point
vortex with ${\frak r}/2$ magnetic flux quanta, located at $z=0$.
The gauge field is flat in the boundary, away from $z=0$.  The monodromy around the point $z=0$ is
\begin{equation}\label{zongo} \begin{pmatrix} e^{i\pi {\frak r}} & 0 \cr 0 & e^{-i\pi {\frak r}}\end{pmatrix}.\end{equation}
As long as ${\frak r}$ is an integer, the monodromy is $\pm 1$, and in fact it is always 1 when regarded as an element of $G^\vee$.
(We recall that odd ${\frak r}$ corresponds to half-integral $j={\frak r}/2$, and hence to $G=SU(2)$, $G^\vee=SO(3)$.)

\subsubsection{Solutions With A Line Singularity}\label{anothersol}

In section \ref{morebranes}, we will actually want some additional solutions of the
 same equations that have a singularity not just at $z=y=0$, but along the whole
 ray $z=0$, $y\geq 0$.  We call this ray $\ell$.

Some new solutions correspond to the case ${\frak r}=-1$ of the ansatz (\ref{mathco}).
Thus,  the holomorphic data are given by $\varphi=g\varphi_1 g^{-1}$, with $g$ as in (\ref{onzo}) and
\begin{equation}\label{zombo}\varphi_1=\begin{pmatrix}0 & z^{-1}\cr 0 & 0 \end{pmatrix}.\end{equation}
 For ${\frak r}=-1$, $v$ and $u$
coincide.  As for the asymptotic behavior of the solution, for $y\to 0$ or $s\to\infty$, we want
the usual behavior
\begin{equation}\label{rombo}v\sim \log|z|-\log y=\log s,~~ s\to\infty,\end{equation}
so as to agree at a generic point on the boundary  with the usual solution with a regular Nahm pole.
Along the line $\ell$, we look first for a solution that is singular but less singular than $1/|z|$.
For $\varphi$ to be less singular than $1/|z|$ means that we need $v\to-\infty$ for $|z|\to 0$,
but for $A$ to be less singular than $1/|z|$ means that $|v|$ should diverge more slowly than
$\log |z|$.  These conditions force us to take $b=0$, which is not a surprise since in general
we had $b={\frak r}+1$.
 For $b=0$, the substitution (\ref{mozone}) is not useful, but we can directly
 combine (\ref{kzone}) and (\ref{qzone}) to get (with $v=u$)
 \begin{equation}\label{lomxo} \frac{\d v}{e^v}= \frac{\d s}{\sqrt{s^2+s^4}}.\end{equation}
 Using (\ref{tzone}) and adjusting the integration constant to match what we want for $s\to\infty$, we find
 the unique solution
 \begin{equation}\label{polzom} e^v=\frac{1}{\log\left(s/\left(\sqrt{1+s^2}-1\right)\right)}.\end{equation}

A slightly more general solution in which we do not take $b=0$ is also of interest.  To find this solution,
we simply combine (\ref{mozone}) and (\ref{lzone}).  We set $v=u$ as we still assume ${\frak r}=-1$, and we keep $\NN=1$
to leave the behavior unchanged for $y\to 0$ or $s\to\infty$.  The solution is
\begin{equation}\label{omigo} e^v=\frac{2bs^{b}}{\left(\sqrt{1+s^2}+1\right)^{b}-\left(\sqrt{1+s^2}-1\right)^{b}}.\end{equation}
The asymptotic behavior is
\begin{equation}\label{wombix} \begin{cases} v\sim \log s   & \mathrm{for} ~s\to\infty\\
                                                                                v\sim~b\log s& \mathrm{for}~ s\to 0. \end{cases}\end{equation}
The Nahm pole for $y\to 0$ or $s\to \infty$ is unchanged, and in particular, if we restrict to the boundary
plane at $y=0$, then  the monodromy around the point $z=0$  remains trivial (as an element of\footnote{For
$G^\vee=SU(2)$, to make the monodromy in the boundary plane trivial, we modify the solution by twisting by
a flat line bundle on the complement of $\ell$ whose monodromy around $\ell$ is $-1$.  Differently put,
we modify the solution by the gauge transformation (\ref{yelx}), with ${\frak r}=-1$.}
 $G^\vee=SO(3)$), just as in (\ref{zongo}).  However, the singularity along $\ell$ at a point with $y>0$ is controlled by the behavior for $z\to 0$ with fixed $y$, or in other words for $s\to 0$.  This monodromy  can be determined by the same computation that led to (\ref{zongo}), simply replacing the behavior
$v\sim -{\frak r}\log |z|$ assumed there by $v\sim b\log |z|$.  So the monodromy is
\begin{equation}\label{kombix} \begin{pmatrix}e^{-i\pi b} & 0 \cr 0 & e^{i\pi b}\end{pmatrix}.\end{equation}

For a further generalization, we continue to require that the singularity in the holomorphic data corresponds to a simple pole at $z=0$, but we drop the assumption that $\varphi$ is nilpotent.  So we take $\varphi
= g \varphi_1 g^{-1}$, with
\begin{equation}\label{yombo} \varphi_1=\frac{\lambda}{z}\begin{pmatrix} 0 & 1\cr 1 & 0 \end{pmatrix},\end{equation}
where $\lambda$ is an arbitrary nonzero complex number.  (Equivalently, we could take $\varphi_1=M/z$,
where $M$ is any $2\times 2$ matrix of determinant $-\lambda^2$, but then we would have to slightly alter the
rest of the ansatz.)
So
\begin{equation}\label{oyombo}\varphi = g\varphi_1 g^{-1}=\frac{\lambda}{z}\begin{pmatrix} 0 & e^{v}\cr e^{-v}&0\end{pmatrix}. \end{equation}  Keeping the rest of the ansatz (\ref{mathco}) unchanged,
the equation (\ref{zon}) is replaced by
\begin{equation}\label{zonif}-\left(\frac{\partial^2}{\partial x_1^2}+\frac{\partial^2}{\partial x_2^2}+\frac{\partial^2}{\partial y^2}\right)v+\frac{|\lambda|^2}{|z|^2}
\left(e^{2v}-e^{-2v}\right)=0. \end{equation}
We assume that $v$ is a function only of $s=|z|/y$ with
\begin{equation}\label{tomix} \begin{cases} v\sim \log s  & \mathrm{for} ~s\to\infty\\
                                                                                v~\mathrm{bounded}& \mathrm{for}~ s\to 0. \end{cases}\end{equation}
Eqn. (\ref{nzo}) is replaced by
\begin{equation}\label{fronz} -\left(\left(s\frac{\d}{\d s}\right)^2+\left(s^2\frac{\d}{\d s}\right)^2\right)v+|\lambda|^2\left(e^{2v}+e^{-2v}\right)=0.\end{equation}                                                                               Introducing $\tau$ as in (\ref{qzone}), we get now
\begin{equation}\label{sophox}\frac{\d v}{\sqrt{e^{2v}+e^{-2v}+2E}}=|\lambda|\, \d\tau = |\lambda| \frac{\d s}{\sqrt{s^2+s^4}},
\end{equation} where $E$ is an integration constant.
For $v$ to be regular for all $s\geq 0$, we have to take $E=-1$, whereupon we get
\begin{equation}\label{ophox}\frac{\d v}{e^v-e^{-v}}  = |\lambda|\frac{\d s}{\sqrt{s^2+s^4}}, \end{equation}
leading to
\begin{equation}\label{gophox} \frac{e^v-1}{e^v+1}=\left(\frac{\sqrt{s^2+1}-1}{s}\right)^{2|\lambda|},\end{equation}
so that
\begin{equation}\label{tombix} \begin{cases} v\sim \log s -\log|\lambda|+\dots  & \mathrm{for} ~s\to\infty\\
                                                                                v\sim~2\left({s}/{2}\right)
                                                                                ^{2|\lambda|}& \mathrm{for}~ s\to 0. \end{cases}\end{equation}

Eqn. (\ref{gophox}) is equivalent to
\begin{equation}\label{nuphox} e^v=\frac{1+\left((\sqrt{s^2+1}-1)/s\right)^{2|\lambda|}}{1-\left((\sqrt{s^2+1}-1)/s\right)^{2|\lambda|}}.
\end{equation}
Taking $\lambda\to 0$, we get
\begin{equation}\label{uphox}e^v\sim \frac{1}{|\lambda|\log(s/(\sqrt{s^2+1}-1))}.\end{equation}
Thus, even though the form of the differential equation (\ref{zonif}) suggests that the solution might become regular in the limit $\lambda\to 0$, this is not the case.
However, if we shift $v$ by $-\log|\lambda|$, then (\ref{uphox}) coincides with the solution (\ref{polzom}) in which $\varphi$ is
nilpotent.  Modulo the shift in $v$ (and an ordinary gauge transformation that depends on the argument of $\lambda$), the ansatz (\ref{oyombo}) converges for $\lambda\to 0$ to the ansatz (\ref{zombo}) with a nilpotent pole.  Thus,
starting with the solution (\ref{gophox}) in which $\varphi$ has a pole at $z=0$ with distinct eigenvalues
$\pm \lambda$, and taking the limit $\lambda\to 0$, we get  the solution (\ref{polzom}) in which $\varphi$ has a pole with nilpotent residue.  An analogous phenomenon
is known for solutions of Hitchin's equations with a regular singularity \cite{Simpson}.

In the language of section \ref{zelfus}, the solution (\ref{omigo}) has $\alpha^\vee\not=0$ with $\beta^\vee=
\gamma ^\vee=0$,
while the solution (\ref{nuphox}) has $\beta^\vee,\gamma^\vee\not=0$ with $\alpha^\vee=0$.
The solution (\ref{polzom})
is the limit for $\alpha^\vee,\beta^\vee,\gamma^\vee\to 0$.  It would be desireable to find a solution with generic
values of $\alpha^\vee,\beta^\vee,\gamma^\vee$
(that is, a solution in which $\varphi$ has a pole at $z=0$ whose residue has distinct
eigenvalues and the monodromy around the ray $\ell$ is generic).
This appears to require a more complicated ansatz than the one we have used.

\subsubsection{Two-Sided Solutions}\label{twos}

The solutions that we have studied so far have been motivated by the problem of D3-branes on $\R^3\times\R_+$, with D3-D5 boundary conditions and 't Hooft operators in the boundary.  It is also of interest to consider a two-sided problem\footnote{This problem is related to Chern-Simons theory on the boundary with a complex gauge group, as will be described elsewhere.}  of D3-branes on $\R^3\times I$, where $I$ is a compact
interval, for instance the unit interval $0\leq y\leq 1$, and we assume that the D3-branes end on D5-branes
both at $y=0$ and at $y=1$.  A time-independent configuration of 't Hooft operators is still described
by the three-dimensional equations (\ref{rice}), (\ref{ice}). Now we want a solution that describes 't Hooft operators on both components of the boundary.

A simple modification of the above ansatz gives  examples of solutions of that type.  (It does not
give the most general such solutions.)  We set
\begin{equation}\label{uton}\varphi_1= \begin{pmatrix}0 & f(z)\cr h(z) & 0 \end{pmatrix}\end{equation}
where $f(z)$ and $h(z)$ are two polynomials.  Zeroes of $f$ and of $h$ will be, respectively, the positions
of 't Hooft operators at $y=0$ and at $y=1$.  We take $\varphi=g\varphi_1g^{-1}$ with $g$ as in (\ref{onzo}), and we leave the rest of the ansatz (\ref{mathco}) unchanged.  Eqn. (\ref{zon}) for $v$
becomes
\begin{equation}\label{opelf}-\left(\frac{\partial^2}{\partial x_1^2}+ \frac{\partial^2}{\partial x_2^2}+
\frac{\partial^2}{\partial y^2}\right)v+|f|^2e^{2v}-|h|^2e^{-2v}=0.\end{equation}

To understand what sort of solution to look for, first consider the case that $f$ and $h$ are constants,
so that no 't Hooft operators are present.  Then one can look for a solution\footnote{This solution
is related to one of the original solutions of Nahm's equation.}  that depends only on $y$. An elementary integration gives an implicit form of the solution
\begin{equation}\label{zuton} y= C -\int_0^v \frac{\d w}{\sqrt{|f|^2e^{2w}+|h|^2e^{-2w}+E}}, \end{equation}
with constants $C,E$.  These constants can be adjusted in a  unique way to ensure that $v\to +\infty$ for
$y\to 0$ and $v\to -\infty$ for $y\to 1$.  Then one has $v\sim -\log y-\log |f|$ for $y\to 0$, and $v\sim \log(1-y)+\log |h|$
for $y\to 1$.  At both $y=0$ and $y=1$, the solution has a regular Nahm pole.  Looking at the way $v$ was
introduced in eqn. (\ref{onzo}), we see that a sign change of $v$ can be compensated by a Weyl transformation that exchanges the two eigenvalues of a diagonal matrix; the structures at $y=1$ and $y=0$
are related in this way.

In general, for any polynomials $f,h$, we look for a solution such that $v\to +\infty$ for $y\to 0$ and
$v\to -\infty$ for $y\to 1$.  Then near $y=0$, the term $-|h|^2e^{-2v}$ is unimportant in (\ref{opelf}).
The analysis of the boundary behavior is the same as in the one-sided case; near a boundary point at which
$f$ is not zero, we have $v\sim -\log y-\log |f|$, while near a point at which $f$ is zero, the boundary
behavior is given by the appropriate model solution with an 't Hooft operator.  Similarly, near $y=1$,
the term $|f|^2e^{2v}$ is unimportant.  The behavior near $y=1$ is the same as the behavior near $y=0$
with the substitutions $v\to -v$, $f\to h$, $y\to 1-y$.

\subsection{The Framing Anomaly For Knots}\label{framknots}

\def\j{{(j)}}
We have described the singularity associated to an 't Hooft operator supported on a knot
$K$ for the idealized case that $K$ is a copy of $\R$ linearly embedded in $W=\R^3$.
For the general case, we simply require that there should be a singularity along $K$
that in the directions normal to $K$ looks like this ideal solution.  Away from $K$, the structure
must be what we have already described in sections \ref{boundcond} and \ref{vivisect}.

An important consequence of this is the framing anomaly for knots.  We will describe
this for $G^\vee=SO(3)$, which in any event is the case that we understand the 't Hooft operator
in most detail.  We consider an 't Hooft operator of spin $j$ supported on  $K$.  In the absence of the 't Hooft operator, the restriction $E|_W$ of $E$ to $W$ coincides
with $TW$, the tangent bundle to $W$, as we have seen in section \ref{vivisect}.  In what follows,
we are only concerned with the behavior along $W$, so we write simply $E$ for $E|_W$.  In
the presence of the 't Hooft operator, $E$ is modified along $K$ and we denote this modification
as $E_\j$.  The Riemannian connection $\omega$ on $E$ is modified to a connection on $E_\j$
that we will call $\omega_\j$.   In the absence of the 't Hooft operator, a step in defining the
partition function was to define a real-valued Chern-Simons function $\CS(\omega)$ (or
$\CS_\grav$, but this refinement is not relevant in discussing the framing anomaly for knots).
Similarly, to define the partition function in the absence of the 't Hooft operator, we need
to be able to define a real-valued Chern-Simons function $\CS(\omega_\j)$.  A framing of $W$ makes it possible to define a
lift of $\CS(\omega)$ to a real-valued function, but does not suffice for defining a natural real-valued $\CS(\omega_\j)$.

The additional information we need turns out to be
a framing of $K$. For $K\subset W$ a knot, let $N\neg K$ be the normal bundle to $K$ in $W$.
The fibration $N\neg K\to K$ has structure group $SO(2)$ (we have taken $W$ orientable from
the beginning, since this is required in the definition of Chern-Simons theory, and $K$ is certainly orientable, so $N\neg K$ is orientable).  Since $K$ is  a one-manifold and $SO(2)$ is connected, it follows that the fibration $N\neg K\to K$ is trivial.   But it
has different homotopy classes of trivializations; given any one trivialization, any other can be found by twisting the first by a map from $K\cong S^1$ to $SO(2)$.  In other words, two
trivializations differ  by an element of $\pi_1(SO(2))\cong\Z$.
A framing of $K$ is a trivialization of $N\neg K$ up to homotopy.  As we will see below, a real-valued
function $\CS(\omega_j)$ can be defined if we are given framings of both $W$ and $K$.
Thus, the knot invariants that we obtain in the $G^\vee$ description can be naturally
understood as invariants of framed knots in a framed three-manifold.\footnote{\label{dolfus}
Here we can make
a remark that parallels what was said about framings of three-manifolds at the end of section \ref{prodcase}.  A knot $K\subset \R^3$ has a canonical framing (relative to which its
self-linking number vanishes).  Formulas for the Jones polynomial and related invariants are usually written
relative to this canonical framing.  Because the canonical framing cannot be found locally,
it is natural to define the invariants for an arbitrary framing.  In any event, in a general three-manifold $W$, a knot does not have a canonical framing.}

Similarly, the knot invariants of Chern-Simons theory are most naturally defined for
framed knots.  Let us recall some details of this that will help in understanding what to look for
on the $G^\vee$ side.    The tangent bundle $TW$, when restricted to a knot $K$, is a direct
sum $T\neg K\oplus N\neg K$, where $T\neg K$ is the tangent bundle to $K$.  Unless $K$ is
a geodesic, this decomposition is not invariant under parallel transport along $K$.
However, the Riemannian connection $\omega$ on $TW$ induces a natural $SO(2)$ connection $\varpi$ on
$N\neg K$.  Parallel transport of a vector in $N\neg K$ with respect to $\varpi$ is defined as transport with respect to $\omega$ with a projection back to $N\neg K$.  Concretely, with respect to the decomposition $TW|_K=T\neg K\oplus N\neg K$, $\varpi$ is the lower right block of $\omega$:
\begin{equation}\label{noggo} \omega=\begin{pmatrix} 0  & * \cr * & \varpi\end{pmatrix}.\end{equation}

The holonomy of the connection $\varpi$ is an element of $SO(2)$ that we can write $\exp({\tau I})$
with
\begin{equation}\label{retro}I=\begin{pmatrix}0 & 1\cr -1 & 0  \end{pmatrix}.
\end{equation}
For a ``bare'' knot,  $\tau$ takes values in $\R/2\pi\Z$, but for a framed knot, $\tau$ is
$\R$-valued.  Indeed, once a framing is picked,
 the connection $\varpi$ becomes
$\varpi=\lambda I$,
where now $\lambda$ is an ordinary one-form, and $\tau$ is simply  $\oint_K\lambda$.
If the framing of $K$ is shifted by one unit (by making an $SO(2)$-valued gauge transformation of $N\neg K\to K$ with
winding number 1 around $K$), $\tau$ transforms by $\tau\to\tau+2\pi$.

As essentially  found for abelian Chern-Simons theory in
\cite{Polyakov} and more generally in \cite{witten}, in computing the expectation value of a Wilson
loop operator $\W_R(K)$ in Chern-Simons theory on $W$ with gauge group $G$, one runs into an analog of what was described for three-manifolds
in section \ref{zelod}.  The expectation value of $\W_R(K)$ is not independent of the metric of $W$
unless one modifies its classical definition by including a factor that depends on $\tau$:
\begin{equation}\label{torno} \W_R(K)\to \W_R(K)\exp(i d_R\tau).\end{equation}
Here $d_R$ is a constant that can be usefully characterized using the relation of three-dimensional Chern-Simons theory to conformal field theory in two dimensions.  For $k>0$, $d_R$ is  the dimension of the primary field associated to the representation $R$ in two-dimensional current algebra
with symmetry group $G$ at level $k$.  Thus
\begin{equation}\label{zorno}d_R=\frac{c_2(R)}{k+h\,\sign(k)},\end{equation}
where $c_2(R)$ is the value  in the representation
$R$ of the quadratic Casimir operator  of $G$ (normalized to equal $h$
in the adjoint representation).   This formula is usually written only for $k>0$; we have extended it to
all nonzero integers $k$ so that $d_R$ is an odd function of $k$ (this reflects the fact that for $k<0$, Chern-Simons theory
is related to an antiholomorphic rather than holomorphic current algebra in two dimensions). It follows from (\ref{torno}), (\ref{zorno}), and the definition of $q$ in (\ref{lodux}) that under
a unit change in framing of $K$, the Wilson loop operator transforms by
\begin{equation}\label{bonzo} \W_R(K)\to \W_R(K) q^{\frak n_{\frak g} c_2(R)}.\end{equation}
For example, if $G=SU(2)$ and $R$ is the spin $j$ representation, then
\begin{equation}\label{onzort}\W_R(K)\to \W_R(K)q^{j(j+1)}.\end{equation}

The difference between $E$ and $E_\j$ is local
along $K$,  so to understand what happens in the dual $G^\vee$ description,
it suffices to consider a local model of
the neighborhood of $K\subset W$. We take such a neighborhood to be  $W_0=S^1\times D$ where $D$ is a disc of radius $R$.  We assume that $W$ is the union of two pieces $W_0$ and $W_1$, glued along their common
boundary $\Xi=S^1\times \tilde S^1$, where $\tilde S^1$ is the boundary of $D$.  $W_1$ may be arbitrarily complicated, but $W_0$ will be very simple.  To describe $W_0$, we introduce an angular
coordinate $\alpha$ on $S^1$ and polar coordinates $r,\beta$ ($0\leq r\leq R$) on $D$, and
we  take the obvious flat metric:
\begin{equation}\label{dunfo} \d s^2 =\d \alpha^2 +\d r^2 + r^2\d \beta^2,\end{equation}
but with a twist of the following sort.  We take $\beta$ to be an ordinary angular variable, \begin{equation}\label{bunfo}\beta\cong
\beta+2\pi,\end{equation} while under a $2\pi$ shift of $\alpha$, we rotate $\R^2$ by an angle $\tau$:
\begin{equation}\label{unfo}
\alpha\to \alpha+2\pi,~~\beta\to \beta-\tau.\end{equation}
The definition of $W_0$ only depends on $\tau$ mod $2\pi$, since $\beta\to\beta+2\pi$ is an equivalence anyway.  We take the knot $K$ to be located at $r=0$.  Relative to the obvious orthonormal frame field
\begin{equation}\label{frames}e_1=\d \alpha, ~~e_2=\d(r\cos\beta),~~e_3=\d(r\sin\beta),\end{equation}
the Riemannian connection $\omega$ simply vanishes.  However, it has a nontrivial monodromy around $S^1$
because the orthonormal frame used in (\ref{frames}) has a monodromy under (\ref{unfo}):
\begin{equation}\label{rames}\begin{pmatrix}e_2\cr e_3\end{pmatrix}\to \exp\left(\tau I\right)\begin{pmatrix}e_2\cr e_3\end{pmatrix}. \end{equation}
It is convenient to work with a single-valued orthonormal frame consisting of $e_1$ and
\begin{equation}\label{ames}\begin{pmatrix} \tilde e_2\cr \tilde e_3\end{pmatrix}
=\exp\left(-\frac{\tau \alpha}{2\pi} I\right)\begin{pmatrix} e_2\cr e_3\end{pmatrix}.\end{equation}
Unlike all the previous formulas, this one depends on $\tau$ as a real number, not just
an angle.  In fact, when
restricted to $K$, $\tilde e_2$ and $\tilde e_3$ define a framing of $K$.  This framing is shifted by $n$
units if we modify (\ref{ames}) by $\tau\to\tau+2\pi n$.  The orthonormal frame
$e_1,\tilde e_2,\tilde e_3$ also defines a framing of $W_0$, but this framing contains no
relevant topological information.\footnote{Because $\pi_1(SO(3))=\Z_2$, the topological
class of the framing of $W_0$ depends on $n$ precisely mod 2.  But the two-torsion
information contained in a framing
is not relevant in Chern-Simons theory.  A convenient
way to eliminate it \cite{Atiyah} is to pass from a framing of $TW$ to the corresponding
framing of $TW\oplus TW$.}  We assume that the framing of $W_0$ given by
$e_1,\tilde e_2,\tilde e_3$ (or at least the corresponding two-framing) is somehow
matched to a framing of $W_1$, giving a framing of $W$.  We want to see what happens to $\CS(\omega_\j)$ when we vary the framing
of $K$ while keeping fixed the framing or two-framing of $W$.

Relative to the orthornormal frame $e_1,\tilde e_2,\tilde e_3$, the Riemannian connection is
\begin{equation}\label{orft}\omega =\frac{\tau\,\d \alpha}{2\pi}\begin{pmatrix} 0 & 0 & 0 \cr 0 & 0 & 1\cr
        0 & -1 & 0 \end{pmatrix}.   \end{equation}
It is clumsy to write such a formula with a first row and column of zeroes.  Everything of interest will happen in the lower right $2\times 2$ block, and the $2\times 2$ matrices will all
be easily constructed from the $SO(2)$ generator $I$ of eqn. (\ref{retro}).  So we will abbreviate a formula
such as this one as
\begin{equation}\label{norft} \omega=\frac{\tau\,\d\alpha}{2\pi}I.\end{equation}

Now we want to include the 't Hooft operator.  As in eqn. (\ref{dorf}) (which however was written
in the two-dimensional representation while now we are in the adjoint representation),
 this means that the Riemannian connection $\omega$ is replaced by a connection $\omega^*$ that is obtained from the Riemannian
connection by adding a singular vortex of flux $2j$ acting on the normal bundle.  In the same
abbreviated notation as in (\ref{norft}), we take
\begin{equation}\label{romfo}\omega^*=2j\,\left(\d\beta+\frac{\tau}{2\pi}\d\alpha\right)I+ \frac{\tau\,\d\alpha}{2\pi}I.\end{equation}  This formula was chosen so that for fixed $\alpha$ it agrees
with the singular vortex connection (\ref{dorf}), and also so that $\omega^*$ is gauge-equivalent to $\omega$ for $r\not=0$.
The gauge transformation between them is
\begin{equation}\label{donkey}\d+\omega=\exp(-s)(\d+\omega^*)\exp(s),\end{equation}
with
\begin{equation}\label{monkey}s=-2j\left(\beta+\frac{\tau\alpha}{2\pi}\right) I.\end{equation}
$s$ has been defined so that $\exp(s)$ is single-valued on the complement of the knot $K$.

We want to modify $\omega^*$ slightly near $r=0$ to remove its singularity.  We introduce a cutoff function
 $g(r)$ such that $g(r)=1$ for $r>\epsilon$ (with some very small $\epsilon<<R$) but
$g(r)\sim r^2$ for $r\to 0$.  We modify $\omega^*$ to
\begin{equation}\label{zomf}\hat\omega =  2j\left(g(r)\d\beta+\frac{\tau\,\d\alpha}{2\pi}\right)I
+\frac{\tau\,\d\alpha}{2\pi}I.\end{equation}
(One can think of this modification as meaning that instead of restricting the bundle $E$ literally to the
boundary $W$ of $V=W\times \R_+$, we restrict it to a three-cycle that coincides with the boundary away from knots, but near a knot $K$ bends slightly into the interior of $V$ to avoid the singularity along $K$.)

Now we can describe the desired bundle $E_\j\to W$ and the connection $\omega_\j$ on
this bundle whose Chern-Simons function we want.
On $W_1$, $E_{(j)}$ coincides with $TW_1$, and the connection is  the Riemannian connection
$\omega$.  On $W_0$,
$E_{(j)}$ is a trivial bundle with connection $\hat\omega$ defined in eqn. (\ref{zomf}).  On the common
boundary $\Xi$ of $W_0$ and $W_1$, the bundles and connections are glued together with the gauge transformation
(\ref{donkey}).  The framing (or more exactly the two-framing) of $TW_0$ that is given by $e_1,\tilde e_2,\tilde e_3$ has an extension over $W$ that will be kept fixed while varying the framing of $K$.  Everything is
in place to compute a real-valued Chern-Simons function $\CS(\omega_\j)$ and determine its dependence on the framing of $K$.  We use eqn. (\ref{boscomb}), in which $\CS(A)$ is defined for any connection $A$
using a trace in the adjoint representation (and we set $h=2$).
In the present context, it is convenient to evaluate the
right hand side of (\ref{boscomb}) as the sum of an integral over $W_1$ with the connection $\omega$,
an integral over $W_0$ with the connection $\hat\omega$, and a correction term on the common boundary
$\Xi$ of $W_0$ and $W_1$ that involves the gauge transformation between $\omega$ and $\hat\omega$:
\begin{align}\label{yelf}\CS(\omega_\j)=&\frac{1}{16\pi}\int_{W_1}\Tr_\ad \left(\omega\wedge\d
\omega+\frac{2}{3}\omega\wedge\omega\wedge\omega\right)\cr +&
\frac{1}{16\pi}\int_{W_0}\Tr_\ad\,\hat\omega
\wedge \d\hat\omega -\frac{1}{16\pi}\int_\Xi\, \Tr_\ad \,\d s \wedge \hat\omega.\end{align}
($\Tr_\ad$ is the trace in the adjoint representation of $SO(3)$; some minor simplifications in (\ref{yelf}) reflect the fact that $\hat\omega$ and the gauge transformation relating
it to $\omega$ are actually abelian, taking values in an $SO(2)$ subgroup. Evaluation of (\ref{yelf})
uses $\Tr_\ad\,I^2=-2$ and the orientation of $W_0$ given by $e_1\wedge e_2\wedge e_3$.)  The terms in (\ref{yelf}) that depend on the framing of $K$ are the integrals
over $W_0$ and $\Xi$.  A straightforward evaluation gives
\begin{equation}\label{melmo}\CS(\omega_\j)=-\tau j(j+1)+\dots \end{equation}
where the ellipses come from the integral over $W_1$ and do not depend on the framing of $K$.
Using (\ref{zolfox}) (with $v=1$ for $G^\vee=SO(3)$), the dependence of the partition function on $\CS(\omega_\j)$ is
a factor of $q^{-\CS(\omega_\j)/2\pi}$.  So finally, under a unit change in framing, $\tau\to\tau+2\pi$,
the partition function is multiplied by $q^{j(j+1)}$, just as in Chern-Simons theory.

There is another issue that could be treated here using these ideas.  This is to show that, for $W=\R^3$,
with a knot $K$ labeled by the spin $j$ representation of $SU(2)$, and using our boundary conditions, the instanton number $\EP$ takes values in $\Z+j$.  Setting $j=1/2$, this accounts for the fact that the Jones
polynomial is actually $q^{1/2}$ times a Laurent polynomial in $q$.  More generally, for  $W=\R^3$
with a link $L$ with $\nu$  components labeled by $j_1,\dots,j_\nu$, $\EP$ takes values in $\Z+\sum_{s=1}^\nu  j_s$.
 We will postpone these issues and consider them in section \ref{fourfram} from a
 higher-dimensional perspective.  Similarly, in section \ref{fourfram},
 we will give a new and possibly
more transparent computation of the framing anomaly for knots.

\section{$T$-Duality And Khovanov Homology}\label{tdual}

\subsection{Lift To Five Dimensions}\label{lift}

\subsubsection{Five-Dimensional Super Yang-Mills And $T$-Duality}\label{td}

So far we have found a new way to calculate the partition function of three-dimensional Chern-Simons gauge theory with gauge group $G$, using $G^\vee$ gauge theory
in four dimensions.  To get to Khovanov homology takes an additional step: we need a fifth dimension.

From a field theory point of view, we can try to proceed by claiming that four-dimensional
maximally supersymmetric Yang-Mills theory is the theory obtained at low energies by
compactifying five-dimensional maximally supersymmetric Yang-Mills theory on a circle.
Thus, instead of considering four-dimensional $\N=4$ super Yang-Mills theory on a four-manifold
$V$,
we consider the corresponding five-dimensional theory on $V\times S^1$ (with supersymetry-preserving boundary conditions in going around $S^1$).   The twisting along $V$ and the
boundary conditions at the boundary of $V$ preserve the same supersymmetry that they
did in the purely four-dimensional formulation of the theory.  (The boundary condition
of section \ref{boundcond} can be lifted to five dimensions in an obvious way; three of the
scalar fields have the singular behavior at the boundary described there.)  In particular, the topological supercharge $Q$ that is familiar in four dimensions is still a symmetry when the model is lifted to five dimensions.

Once the model is lifted to $V\times S^1$, we can pick a point $p\in S^1$ and construct a physical Hilbert space $\H(V)$
associated to quantization on the codimension one submanifold $V\times p$.  The path integral
on $V\times S^1$ can then be written as a trace in $\H(V)$.
In the present approach, $\H(V)$ plays the role of the space that was called by that name in our introductory
sketch of Khovanov homology in section \ref{kh}.    $Q$ automatically acts on $\H(V)$,
as it generates a symmetry of the theory. We write $\K(V)$ for the cohomology of $Q$, acting on $\H(V)$.  Then $\K(V)$ is
our candidate for the generalization to this situation of Khovanov homology.  (Since we do not have
a proof that the cohomology of $Q$ is equivalent to Khovanov homology as defined in the literature, even if one specializes to the situation of knots in $\R^3$ where Khovanov
homology has been defined, we denote the cohomology of $Q$ as $\K$ and write $\Kh$ for
Khovanov homology.)

From a D-brane point of view, the lift from four to five dimensions amounts to $T$-duality.  Thus, for the case that the gauge group
is $G^\vee=U(N)$, consider a system of $N$ D3-branes wrapped on
$V$, with some twisting of the normal bundle to $V$ to preserve supersymmetry.  This picture
was described in section \ref{desc}. Without changing anything essential in that discussion,
we can take one of the spacetime directions transverse to $V$ to be compactified on a circle $\tilde S^1$.  Explicitly, we replace what in section \ref{desc} was $T^*V_0\times \R^2$
by $T^*V_0\times \R\times \tilde S^1$. Then we perform $T$-duality
on $\tilde S^1$, converting the spacetime to $T^*V_0\times \R\times S^1$.  The D3-branes wrapped on $V\subset V_0$ are converted to D4-branes wrapped on $V\times S^1$.  If as in section \ref{desc}, the D3-branes end on a D5-brane (wrapped on $T^*W$ with $W=\partial V$),  then  $T$-duality converts the D3-branes to D4-branes that end on a D6-brane (wrapped on $T^*W\times S^1$).   So, when the appropriate geometry exists, the lift to five dimensions
simply amounts to $T$-duality from  the D3-D5 system that we have studied so far  to a D4-D6 system.

None of the approaches just mentioned is entirely satisfactory.
The disadvantage of the description by five-dimensional super Yang-Mills theory is that  this theory is not ultraviolet complete.
The brane construction also has a few drawbacks, which were described in section \ref{desc}.  The appropriate Calabi-Yau geometry
may not exist for generic $V$, and even if it exists, it may entail unnatural choices. The brane construction does not help very much with exceptional gauge groups. Also,  the brane construction and the full string theory have many degrees of freedom that are not relevant
to the problem of defining an analog of Khovanov homology and relating it to Chern-Simons theory.

There is a completely satisfactory alternative to the approaches that we have summarized so far.  Five-dimensional maximally super Yang-Mills
theory has a canonical ultraviolet completion in the six-dimensional (0,2) superconformal field theory.  This gives a general and economical
framework for the topic considered in the present paper,
and for many purposes it is probably the most powerful framework.
In section \ref{fivebranes}, we will develop a top-down approach to the subject with this starting point.  As an illustration of the power of this
viewpoint, we will show that in the six-dimensional picture, the existence of supersymmetric Wilson and 't Hooft operators precisely at
the boundary of $V$ follows from standard facts, while
in the four and five-dimensional pictures,
this seems to require the  detailed computations in sections \ref{uzim} and \ref{thooft}.

But some important points, especially the representation (\ref{turkox}) of the Chern-Simons partition function as a trace in
Khovanov homology, do not require the six-dimensional machinery.
So it seems reasonable to begin with an explanation in five dimensions.

\subsubsection{The Bigrading}\label{bigrading}

To agree with Khovanov homology, $\K(V)$ should admit a $U(1)\times U(1)$ action, so that it will be $\Z\times \Z$ graded.\footnote{This
is a slight simplification as in general the eigenvalues of the symmetry generators $\EF$ and $\EH$ may lie in a coset of $\Z\times \Z\subset
\R\times \R$. The most important consequence of this was described in section \ref{framan}. More generally, if $G^\vee$ is not simply-connected, the eigenvalues of $\EH$ may lie in a coset of $\Z/w\subset\R$ for
some integer $w$, rather than in a coset of $\Z$.  This last effect, which was discussed in relation to eqn. (\ref{dunky}), is not directly
relevant to Khovanov homology, because it does not arise
for $V=W\times \R_+$ with the sort of boundary conditions that we impose on $\partial V$.}
One generator of $U(1)\times U(1)$ is the instanton number, evaluated on the four-cycle $V$.  The definition is the same as it was in section \ref{elmdual}:
\begin{equation}\label{toffusl}\EH = \frac{1}{32\pi^2}\int_V
\epsilon^{\mu\nu\alpha\beta}\,\Tr\,F_{\mu\nu}F_{\alpha\beta}.\end{equation}
However, the physical interpretation is different: in the five-dimensional interpretation,
$\EH$ is an operator acting on quantum states that are obtained by quantizing
fields on $V$, while in the four-dimensional interpretation, $\EH$ was a term in the classical action.

The other generator of $U(1)\times U(1)$ is an $R$-symmetry generator $\EF$ that is left unbroken by the twisting procedure that is used
to define a topological field theory. In the four-dimensional analysis of section \ref{topfield}, we began with the $R$-symmetry group
$SO(6)$ of $\N=4$ super Yang-Mills theory in four dimensions, and twisted by identifying an $SO(4)$ subgroup of $SO(6)$ with the Riemannian
holonomy of $V$.  This left an unbroken subgroup $SO(2)\subset SO(6)$, and we defined the generator of this $SO(2)\cong U(1)$ to be $\EF$.
When we lift to five dimensions, the $R$-symmetry group is reduced to $SO(5)$, so embedding an $SO(4)$ holonomy group in the $R$-symmetry
group would not leave an unbroken $SO(2)$.  To compensate for this, we specialize to $V=W\times \R_+$ (or $V=W\times S$ for
any one-manifold $S$), with $W$ a three-manifold.  This ensures that the holonomy group of $V$ reduces to $SO(3)$, so that its embedding in the $R$-symmetry group,
which is now $SO(5)$, again leaves
an unbroken $SO(2)$.  We again call the generator of this symmetry $\EF$.  For general $V$, we do not get a $\Z$-grading by $\EF$, but
there is always a $\Z_2$-grading that distinguishes bosonic states from fermionic ones.   When $\EF$ can be defined, the $\Z_2$-grading by
statistics is the mod 2 reduction of the $\Z$-grading by $\EF$.  It turns out, however,
that the lift to five dimensions is useful primarily when the conserved charge $\EF$ can be defined,
so we will be mainly interested in that case.

Of course, when $V$ has a boundary, to define $\EF$, the boundary condition must be $\EF$-invariant.  But there is no problem with this.
We use the boundary condition of section \ref{boundcond}, lifted to five dimensions.
Three of the five scalar fields of five-dimensional maximally supersymmetric Yang-Mills theory have  expectation values that diverge
at the boundary,
leaving an unbroken $SO(2)$ symmetry that rotates the other two.  The two scalars that are rotated by $\EF$ play the role of the complex field
$\sigma$ of section \ref{twisting}.  In any supersymmetric classical solution, $\sigma$ vanishes and the value of $\EF$ also vanishes.
Quantum mechanically,  for a quantum state associated to a given classical solution, the eigenvalue of $\EF$ is computed by summing over
the $\EF$ quantum numbers of all fermions in the filled Dirac sea.  In that sense, it makes sense to refer to $\EF$ as a fermion number.

A more detailed and complete explanation of many of these matters is given in section \ref{fivebranes} in the context of an ultraviolet
completion of five-dimensional super Yang-Mills theory in six dimensions.
For now, it is enough to know that, not for all $V$, but for $V$ of the form $W\times \R_+$, $\K(V)$ is bigraded, like Khovanov homology.

Since Khovanov homology has been defined in the literature only for links in $\R^3$, to make a precise conjecture about the relation
of $\K(V)$ to Khovanov homology, we must restrict to  $V=\R^3\times\R_+$.
For Khovanov homology, we consider a link $L\subset \R^3$ consisting of a disjoint union of embedded circles $K_i\subset \R^3$.
We label each $K_i$ by an irreducible representation $R_i$ of a compact Lie group $G$.  In the four-dimensional description of section
\ref{csfun} via $G$ gauge theory, we include supersymmetric Wilson operators of the representations $R_i$, supported on $K_i\times \{0\}$, where $\{0\}$ is the endpoint of $\R_+$.   In the $S$-dual description in
section \ref{dualities}, the gauge group is $G^\vee$, the Goddard-Nuyts-Olive or Langlands dual of $G$, and the Wilson operators
in the boundary of $V$ are converted to the dual 't Hooft operators of $G^\vee$ gauge theory.  The description of 't Hooft operators in the boundary of $V$ is somewhat subtle and was described in section \ref{thooft}. In this situation,  $\mathcal K(V)$ is a candidate for Khovanov homology.

\subsubsection{Notation}\label{notation}

As we move to five dimensions, the cast of characters will get longer.  To make the arguments
easier to follow, in the rest of the paper we write $V_4$ and $W_3$ for the four-manifold
and three-manifold that earlier we have called simply $V$ and $W$.   Thus $W_3$ is always
the boundary of $V_4$.

\subsection{Procedure For Computing $\K$}\label{york}

\def\po{{0}}
Now we would like to sketch the concrete procedure for computing $\K(V_4)$, for a four-manifold
$V_4$,  via five-dimensional supersymmetric Yang-Mills theory.
This procedure is in no way novel; it is a standard procedure in topological applications of supersymmetric theories; typical
examples involve Morse theory \cite{wittenmorse}
or Floer cohomology \cite{floer}.  We sketch the procedure here for completeness.

We want to describe a procedure to determine the space of quantum ground states of twisted super Yang-Mills theory on the five-manifold
$M_5=\R\times V_4$.  For comparison to Chern-Simons theory (or Khovanov homology), we take $V_4=W_3\times\R_+$ for some $W_3$, but
the general procedure to describe the space of ground states holds for any $V_4$.

First of all, the condition for a five-dimensional field configuration to preserve the $Q$ symmetry gives a system of elliptic
differential equations in five dimensions.   It is straightforward to derive these equations, and we will do so in section \ref{gaugedesc} (see
eqn. (\ref{torm})  for the final result).
But for now, we do not need the details.   All we need to know is that these are elliptic differential equations that, in the time-independent case, specialize to the familiar  four-dimensional equations
\begin{equation}\label{zmosc}F-\phi\wedge\phi+\star \d_A\phi=0=\d_A\star\phi.\end{equation}

The first approximation to finding the space of quantum ground states is to find the space of classical ground states.
A classical ground state is a time-independent classical solution of the five-dimensional equations for unbroken supersymmetry.
So in other words, a classical ground state is a solution of the equations (\ref{zmosc}) on the four-manifold $V_4$.  For simplicity
we are going to assume that this equation has a finite set of solutions, up to gauge transformation, and further that these solutions
are all nondegenerate (there are no bosonic zero modes in expanding around a given solution).
 Let $S$ be the set of these solutions.  If $V_4$ has a non-empty boundary,
then on $\partial V_4$ we impose the boundary conditions of section \ref{boundcond}; with these boundary conditions, the solutions
are automatically all irreducible (they leave unbroken only a finite group of gauge symmetries, in fact the center of $G^\vee$).  If $V_4$
has no boundary, we assume for simplicity that the solutions are all irreducible.

Nondegeneracy means that the expansion around a given classical solution gives, at least perturbatively, a single quantum state of zero
energy.  We will let $\K_\po$ be the space of quantum ground states in the classical approximation; it has a basis consisting of a single
state $\psi_s$ for each $s\in S$.  We let $n_s$ be the instanton number $\EH$  for the $s^{th}$ classical
solution, as defined in eqn. (\ref{toffusl}).
Assuming that $V_4=W_3\times\R_+$ for some $W_3$, we let $f_s$ be the fermion number $\EF$ of the $s^{th}$ classical
solution.  (It equals  the value of $\EF$ for the filled Dirac sea that
one obtains in expanding around the $s^{th}$ solution.)  For any $V_4$,
$\K_\po$ is $\Z\times\Z_2$-graded, where the $\Z$-grading is by the eigenvalue  of $\EH$, and the $\Z_2$ distinguishes
fermionic states from bosonic ones.  For $V_4=W_3\times \R_+$, $\K_\po$ is $\Z\times\Z$ graded by the eigenvalues of $\EH$ and $\EF$.

Now we want to consider quantum corrections to this spectrum.  Once one has an asymptotic approximation to the space of supersymmetric states -- in this
case $\K_\po$ -- states can only disappear from the supersymmetric spectrum in bose-fermi pairs. The reason for this is familiar:
eigenstates of the supersymmetric Hamiltonian with a nonzero energy occur in pairs, corresponding to a bosonic state and a fermionic state
of the same energy.   In the $\Z\times\Z_2$-graded case,
a pair of states that are going to disappear must have the same $\EH$ eigenvalue (since $\EH$ commutes with $Q$) and opposite
statistics.  In the $\Z\times \Z$-graded case, a pair of states that are going to disappear from the supersymmetric spectrum must
have the same eigenvalue of $\EH$ and eigenvalues of $\EF$ that differ by 1.  (The last statement is a consequence of the commutation
relation $[\EF,Q]=Q$, which implies that a supermultiplet of energy eigenstates with nonzero energy consists of a pair of states
with values of $\EF$ differing by $\pm 1$.)

In perturbation theory, nothing happens to the supersymmetric spectrum.  Indeed, perturbation theory around a given classical solution only
``knows'' about a single approximate supersymmetric state, namely the one obtained by quantizing that classical solution.  In perturbation theory, there is no way for that approximate supersymmetric
ground state to pair up with another one and disappear.   However, just as in
supersymmetric quantum mechanics or Floer cohomology,  instanton effects involving
tunneling from one classical solution to another can lift a pair of supersymmetric states away from zero
energy.  In the present context, instantons are solutions of the five-dimensional supersymmetric equations, the ones that are presented in
eqn. (\ref{torm})  and whose reduction to the time-independent case agrees with eqn. (\ref{zmosc}).  An instanton that
interpolates between one solution of (\ref{zmosc}) in the past and another in the future can lift away from zero energy the supersymmetric quantum states that correspond to the two solutions.

Let $\K$ be the exact supersymmetric spectrum that we get after allowing for the effects of instantons.  A precise and general recipe for
computing $\K$ is that it is the cohomology of a certain operator acting on $\K_\po$.  This operator is simply $Q$ evaluated in the
space $\K_\po$ generated by the approximate supersymmetric states $\psi_s$.  A precise formula for $Q$, up to conjugation, is
\begin{equation}\label{omex}Q\psi_s=\sum_{\{t\in S|f_t-f_s=1\}}\,n_{st}\psi_t,\end{equation}
where $n_{st}$ is computed by summing over instantons that begin at the $s^{th}$ solution in the past and end on
the $t^{th}$ solution in the future.  Such solutions come in one-parameter families
generated by time translation invariance; each such family contributes 1 or $-1$ to $n_{st}$, depending on the sign of
the fermion determinant that arises in linearizing around the given solution, after removing the zero mode that comes from time-translation
invariance.   The details are standard in Floer cohomology and related theories, and will not be described here.

\subsubsection{Relation To Chern-Simons Theory}\label{ondo}

Now we want to explain how $\K(V_4)$, as just described, is related to the $S$-dual four-dimensional construction of section \ref{dualities}.
For brevity, we  focus on the $\Z\times\Z$-graded case $V_4=W_3\times \R_+$, so that we also will get a link to Chern-Simons theory on $W_3$.
The general case is similar, except that the function $L(q,y)$ that is introduced shortly is only defined for $y=-1$ since the grading is only by $\Z\times\Z_2$.

\def\LL{L}
First of all, if we know $\K(V_4)$, then we can compute the function
\begin{equation}\label{nobble}\LL(q,y)=\Tr_{\K(V_4)}\,q^\EH y^\EF.\end{equation}
For $V_4=W_3\times \R_+$, this function is an invariant of $W_3$, or of $W_3$ together with the knot or link it may contain, if any.
However, there is no convenient way to represent this function by a path integral.

To get a trace associated to $V_4$, we should consider
a path integral on the five-manifold $M_5=V_4\times S^1$.  If $\H$ is the Hilbert space of all physical states of five-dimensional super Yang-Mills
theory (not necessarily annihilated by $Q$), $H$ is the Hamiltonian acting on $\H$, and $\beta$ is the circumference of $S^1$, then
a path integral on $M_5$ with an insertion of the operator $q^{\EH}y^{\EF}$ can compute
\begin{equation}\label{obble}G(q,y)=\Tr_{\H}\,q^\EH y^\EF\exp(-\beta H).\end{equation}
However, this trace receives contributions from states of nonzero energy. A pair of states with $H=E$, $\EH=n$, and $\EF=f,f+1$
contribute
\begin{equation}\label{obo} q^n\exp(-\beta E)\left(y^f+y^{f+1}\right)\end{equation}
to $G(q,y)$.  To make this contribution vanish, we must choose $y$ so that $y^f+y^{f+1}=0$; in other words, we need to
take $y=-1$.   Otherwise, $G(q,y)$ is not a topological invariant.  If we set $y=-1$, $G(q,y)$ reduces
to $L(q,y)$.

The study of Khovanov homology has shown that the function $\LL(q,y)$ contains quite a lot of information that we lose if we set $y=-1$.
However, the case $y=-1$ is the case that can be represented by a path integral on $M_5$.  For this value of $y$, the trace in (\ref{nobble}) or
(\ref{obble}) computes
what is usually called the {\it index} of the operator $Q$, or more precisely the equivariant generalization of this index to take account of
the symmetry generated by $\EH$.  (We get the ordinary index of $Q$ if we set $q=1$.)   As is usual, the index of an operator is more readily computed by a path integral than are other topological invariants.

Not only can $\LL(q,-1)$ be represented by a five-dimensional path integral on $M_5$; it can more simply be represented by a path integral on $V_4$.
The reason for this is as follows.  Approximate supersymmetric states that are lifted from the spectrum by instanton effects do not
contribute to $\LL(q,-1)$ (since they have the same value of $\EH$ and have $\EF$ differing by 1).  So we can calculate $\LL(q,-1)$ in the
space $\K_\po(V_4)$ of approximate supersymmetric ground states, instead of  the space $\K(V_4)$ of states of exactly zero energy:
\begin{equation}\label{tobble}\LL(q,-1)=\Tr_{\K_\po(V_4)}\,q^\EH(-1)^\EF. \end{equation}

Before looking at this formula more closely, let us note as an aside that we could also, of course, define a more general trace in $\K_\po(V_4)$:
\begin{equation}\label{qobble}\tilde \LL(q,y)=\Tr_{\K_{\po}(V_4)}\,q^\EH y^\EF.  \end{equation}
But in general, one should not expect $\tilde \LL(q,y)$ to be a topological invariant.  The reason is that, unlike $\K(V_4)$, $\K_\po(V_4)$ is not,
in general, a topological invariant.  In general, one should expect supersymmetric classical solutions to appear and disappear in pairs as
the metric on $V_4$ is varied; when this occurs, $\tilde \LL(q,y)$ will jump with no change in $\LL(q,y)$.  Concretely, when one varies the metric of $V_4$ so that a pair
of time-independent classical solutions appears, there also appears a time-dependent instanton solution that interpolates between them and ensures that the extra two states that have appeared in $\K_\po(V_4)$ do not contribute to $\K(V_4)$.

Since we want to study topological invariants, we set $y=-1$.  Now let us go back to the formula (\ref{tobble}) for $\LL(q,-1)$.
This trace is a sum over classical solutions of the time-independent equations (\ref{zmosc}); as before, we assume that the solutions
are nondegenerate and parametrized by a finite set $S$.  For each $s\in S$, we write $n_s$  and $f_s$ for the $\EH$ and $\EF$
eigenvalues of the approximate ground state $\psi_s$.    The explicit formula for $\LL(q,-1)$ is then
\begin{equation}\label{tomky} \LL(q,-1)=\sum_{s\in S}\,q^{n_s}(-1)^{f_s}.\end{equation}
But this coincides with the formula (\ref{unky}) for the purely four-dimensional path integral on $V_4$ provided the sign $(-1)^{g_s}$
of the four-dimensional fermion determinant coincides with $(-1)^{f_s}$.  The justification for that last statement is 
that as
one varies the metric of $V_4$ or the background fields $A,\phi$ in the Dirac operator, the sign of the four-dimensional fermion determinant
is reversed whenever it has a zero mode; but these are precisely the points at which, from a five-dimensional point of view, the value of $f_s$
jumps by $\pm 1$.   (This argument does not fix an additive constant in $g_s$; this constant depends on a choice of
trivialization of the determinant line bundle in four dimensions.  We fix the constant to reconcile the four- and
five-dimensional formulas.)

In turn, we know that for $V_4=W_3\times \R_+$, the four-dimensional path integral (\ref{unky}) equals the Chern-Simons path integral
$Z_{W_3}^{\mathrm{CS}}(q)$ on $W_3$.  Putting everything together, we have obtained the relation
\begin{equation}\label{turkox} Z^{\mathrm{CS}}_{W_3}(q)=\Tr_{\K(W_3\times\R_+)}\,q^\EH (-1)^\EF \end{equation}
between Chern-Simons theory on $W_3$ and our candidate $\K(W_3\times \R_+)$ for the generalized Khovanov homology.  But in general, something is hidden
in the way we have written this formula.

On the left hand side of this formula, the possible integration cycles of the Chern-Simons
theory on $W_3$ that must be used for computing $Z_{W_3}^{\mathrm{CS}}$ are
associated to  critical points of the $G_\C$-valued  Chern-Simons function on $W_3$ --
in other words, to homomorphisms $\rho:\pi_1(W_3)\to G_\C$.  On the right
hand side, $\K(W_3\times\R_+)$ is defined using a homomorphism
$\rho^\vee:\pi_1(W_3)\to G^\vee_\C$ to set the boundary condition at infinity.
To use the formula in general, we would have to understand the relation between
$\rho$ and $\rho^\vee$ determined by $S$-duality.  A more precise version of the
formula would involve a sum as in (\ref{poly}) with an unknown matrix $m_{\rho^\vee,\rho}$.
We can avoid this problem if
we specialize to $W_3=\R^3$ with a link whose components are labeled by
Wilson operators on the left hand side of (\ref{turkox})
or by the dual 't Hooft operators on the right hand side.  Then $\rho$ and
$\rho^\vee$ are both trivial, so we do not need to analyze
an $S$-duality transformation between them.  The  relation (\ref{turkox})
becomes -- conjecturally --  the classical relation between Khovanov homology (and its
generalization to arbitrary representations of compact Lie groups)
and the Jones polynomial  (and more general knot invariants derived from Chern-Simons theory), as described in eqn. (\ref{zonk}) of the introduction.

\subsection{Lie Groups That Are Not Simply-Laced}\label{zorky}

We are now going to explain a possibly surprising fact: when the gauge group $G$ of Chern-Simons theory is not
simply-laced, there is a perfectly good alternative to what has just been explained.

Although this is a general fact, we will, to be concrete, explain it first for the case that $G=Sp(2n)$ for some $n$.  The GNO or Langlands
dual group is then $G^\vee=SO(2n+1)$.  And this is a subgroup of the simply-laced Lie group $G^*=SO(2n+2)$.  $G^*$ admits
an outer automorphism that we will call $\zeta$ that leaves fixed $G^\vee$.  In the $2n+2$-dimensional representation of $G^*$,
$\zeta$ acts by the matrix $\mathrm{diag}(1,1,\dots,1,-1)$.

As is clear from the explicit description in section \ref{boundcond}, a principal $\frak{su}(2)$ subalgebra of $SO(2n+2)$ can actually be conjugated into the Lie algebra of
 $SO(2n+1)$.  With this choice, it commutes with $\zeta$.  This means that the boundary condition of the D3-D5 system, as described in section \ref{boundcond},
or its $T$-dual, the boundary condition of the D4-D6 system, as studied in this section, is $\zeta$-invariant.

Hence, taking the gauge group to be $G^*$, $\zeta$ acts on the set $S^*$ of solutions of the four-dimensional equations (\ref{zmosc}).
We denote this space as $S^*$, rather than $S$ (as before), to emphasize that we are taking the gauge group to be $G^*$ rather than $G^\vee$.
The set $S$ of solutions of the equations (\ref{zmosc}) with gauge group $G^\vee$ is simply the set of fixed points of $\zeta$ acting on $S$.
We will likewise write $\K^*_\po(V_4)$ and $\K^*(V_4)$ for the spaces of approximate and exact quantum ground states in the $G^*$ theory,
while $\K_\po(V_4)$ and $\K(V_4)$ will be the corresponding spaces for gauge group $G^\vee$.

Since $\zeta$ acts on the set $S^*$, it also acts on the vector space $\K^*_\po(V_4)$, which is simply constructed to have one basis vector $\psi_s$
for every $s\in S^*$.
  $\zeta$ is also a symmetry of the five-dimensional ``instanton'' equations that lift some states in $\K^*_{\po}(V_4)$ (this is hopefully natural
even though we will not actually construct those equations until section \ref{fivebranes}), so it acts on $\K^*(V_4)$ as well.

Using the $\zeta$ action on $\K^*(V_4)$, we can now define a new trace that generalizes (\ref{nobble}):
\begin{equation}\label{zelnick}\LL^*_\zeta(q,y)=\Tr_{\K^*(V_4)}\,q^\EH y^\EF\zeta.\end{equation}
Here for brevity, but also because it is the most interesting case, we assume that $V_4=W_3\times \R_+$ so that we can define the
$\EF$ symmetry.    Note that $\zeta$ commutes with $\EH$ and with $\EF$, as well as with $Q$.

Just as in the discussion of (\ref{nobble}), to represent $\LL^*_\zeta(q,y)$ by a path integral in a simple way is only possible if $y=-1$.
So let us consider the relation of $\LL^*_\zeta(q,-1)$ to Chern-Simons theory.  Just as in (\ref{tobble}), in computing $\LL^*_\zeta(q,-1)$,
we can replace the trace in $\K^*(V_4)$ by a trace in $\K_\po^*(V_4)$:
\begin{equation} \label{elnick} \LL^*_\zeta(q,-1)= \Tr_{\K^*_\po(V_4)}\,q^\EH (-1)^\EF\zeta.\end{equation}
We can evaluate the trace in (\ref{elnick}) by summing over the basis of $\K^*_\po$ given
by the vectors $\psi_s,$ $s \in S^*$.  In this basis, we evaluate the trace by summing
over the diagonal matrix elements of $q^\EH(-1)^\EF\zeta$.  Since $\EH$ and $\EF$ are diagonal
in the chosen basis, the trace receives contributions only from diagonal matrix elements
of $\zeta$.  The action of $\zeta$ in this basis is easily described. $\zeta$ is a permutation
matrix determined by the action of $\zeta$  on the set $S^*$.   $\zeta$ either leaves fixed a given $s\in S^*$ or
exchanges a pair of elements.  Nonzero diagonal matrix  elements of $\zeta$ are all 1 and
correspond
to $\zeta$-invariant elements of $S^*$.  But the $\zeta$-invariant elements of $S^*$ make up
precisely the set $S$ of $G^\vee$-valued solutions of the four-dimensional localization equations.
Hence
\begin{equation}\LL^*_\zeta(q,-1)=
\sum_{s\in S}\,q^\EH(-1)^\EF=\Tr_{\K(V_4)}\,q^\EH(-1)^\EF.\end{equation}
Since we got the same result for $\LL(q,-1)$ in (\ref{tomky}), we learn that $\LL^*_\zeta(q,-1)=
\LL(q,-1)$. Since we have already identified $\LL(q,-1)$ with the Chern-Simons partition
function of $G=Sp(2n)$, we actually now have two alternative formulas for this function:
\begin{equation}\label{elbow}Z^{\mathrm{CS}}_{W_3}(q)= \LL^*_\zeta(q,-1)=\LL(q,-1).\end{equation}
Both of these formulas amount to ways of writing the Chern-Simons partition function as a trace:
\begin{equation}\label{zelbow} Z^{\mathrm{CS}}_{W_3}(q)=\Tr_{\K(W_3\times\R_+)}\,q^\EH(-1)^\EF=\Tr_{\K^*(W_3\times\R_+)}\,q^\EH(-1)^\EF\zeta.\end{equation}

Actually, the attentive reader may notice a small gap in this derivation:
we have assumed that for a given $G^\vee$-valued classical solution, the values of $\EH$
and $(-1)^\EF$ are the same whether calculated in $G^\vee$ or after embedding of the solution in
$G^*$.
For $\EH$, this is a classical fact about the instanton number, but a proof of what we want for $(-1)^\EF$ is not
clear at the moment\footnote{This actually is clear for the case $G^\vee= {\sf G}_2$, $G^*=\mathrm{Spin}(8)$.  The complement
of the $G^\vee$ Lie algebra in that of $G^*$ is two copies of the irreducible seven-dimensional representation of $G^\vee$.
When we embed $G^\vee$ in $G^*$, the fermion determinant is multiplied by the square of a real determinant associated to the
seven-dimensional representation of $G^\vee$, so its sign does not change.} and this is a gap in our explanation.  A proof may follow from a vanishing theorem for the five-dimensional
Dirac operator.

We have treated the case of $G=Sp(2n)$, but a similar derivation works for any
gauge group that is not simply-laced.  For $G=SO(2n+1)$, we have $G^\vee=Sp(2n)$.
We can take $G^*$ to be the simply-laced Lie group $SU(2n)$, which admits an outer
automorphism $\zeta$ that leaves fixed $G^\vee$.  Once again, a principal $\frak{su}(2)$ subalgebra
of $G^\vee$ embeds as a principal $\frak{su}(2)$ subalgebra of $G^*$.  This is clear from
the description of the principal subgroups in section \ref{boundcond}.  So we can repeat
all steps in the above derivation, arriving again at (\ref{elbow}) and (\ref{zelbow}).

The other cases of non-simply-laced Lie groups are similar, though less obvious.
If $G={\sf G}_2$ or ${\sf F}_4$, then again $G^\vee={\sf G}_2$ or ${\sf F}_4$.  For $G^\vee={\sf G}_2$,
we take $G^*=\Spin(8)$ with $\zeta$ a triality automorphism, which is of order 3.  We can
pick $\zeta$ to leave fixed ${\sf G}_2\subset G^*$, and a principal $\frak{su}(2)$ subalgebra
of ${\sf G}_2$ embeds as one of $G^*$.  For $G^\vee={\sf F}_4$, we take $G^*={\sf E}_6$.  ${\sf E}_6$ admits an outer automorphism $\zeta$ of order 2, which we can choose to leave ${\sf F}_4$ fixed.
Again a principal $\frak{su}(2)$ subalgebra of ${\sf F}_4$ embeds as one of ${\sf E}_6$.
(Proofs of the statements in this paragraph about principal $\frak{su}(2)$ subalgebras
have been sketched by B. Kostant.)   So we can repeat the above derivation, leading
to the same conclusions (\ref{elbow}) and (\ref{zelbow}).

\subsection{Ultraviolet Completion}\label{orkit}

Mathematically, the approach to this subject via five-dimensional gauge theory has  the
great advantage of relying on five-dimensional elliptic differential equations, without
needing the full machinery of quantum field theory and string theory.  (We have not yet
described explicitly the relevant five-dimensional equations and their essential properties;
this will be done starting in section \ref{gaugedesc}.)  Indeed, this fact is the main reason that the
present paper may have some mathematical impact in the short term.

Physicists will generally prefer a starting point based on an ultraviolet-complete quantum
field theory.  This we will present in section \ref{fivebranes}.  Some of the drawbacks
of relying on five-dimensional supersymmetric Yang-Mills theory were described at the
end of section \ref{td}.

 The alternative
formulas of eqn. (\ref{zelbow}) for the Chern-Simons partition function when $G$ is not simply-laced give an
interesting challenge for the six-dimensional approach.  In section \ref{zeldow}, we will suggest two slightly different six-dimensional
starting points that lead to the two formulas.

\section{Top-Down Approach}\label{fivebranes}

So far in this paper, we have worked our way up from three to four and then five dimensions.
The logical end of this process is the six-dimensional superconformal field theory that
provides an ultraviolet completion of five-dimensional super Yang-Mills theory.

In the present section, we begin in six dimensions and deduce the five-dimensional
picture that was used in  section \ref{tdual}.  We also fill in many key gaps
in section \ref{tdual}, mainly by deriving the explicit form of the relevant elliptic differential
equations and describing their key properties.

The six-dimensional starting point in the present section will also bring us
closer to the brane constructions that have been used previously in
related work \cite{OV,GSV,DVV,AY,CNV}.

We began our analysis in section \ref{csfun} on a fairly
general four-manifold $V_4$ with boundary $W_3$.   In section \ref{tdual}, we lifted
the analysis to the five-manifold $S^1\times V_4$.  In
that context, as was explained in section \ref{bigrading}, to maintain the bigrading that
gives Khovanov homology much of its power, one must
specialize\footnote{More generally, one could replace
 $\R_+$ by another one-manifold, notably a circle, real line,
 or compact unit interval.} to $V_4=W_3\times\R_+$, for some $W_3$, so that the five-dimensional
 description is based on
$M_5=S^1\times W_3\times \R_+$.  However, it turns
out that this can be generalized. The five-dimensional
version of the construction makes sense on
$M_5=M_4\times \R_+$, with any oriented four-manifold $M_4$  without boundary,
not necessarily of the form $W_3\times S^1$.  (Note that in the important case that
$M_5=S^1\times W_3\times \R_+$, $M_4$ is not the same as $V_4$; $V_4$ is
$W_3\times \R_+$ while $M_4$  is $S^1\times W_3$.)
 We will define a four-dimensional topological field theory that will work
 for an arbitrary $M_4$. Moreover, $M_4$ can be endowed
with ``surface operators,'' supported on a two-manifold $\CC\subset M_4$.
Though any $M_4$ is allowed,  this theory is most interesting  (for a reason explained in section
\ref{symgroup} and again involving the bigrading), if the third Betti number of $M_4$ is
positive -- a fairly typical example being $M_4=S^1\times W_3$.  We will also write $
M_6$ for a fairly general six-manifold, although we will soon concentrate
on the case $M_6=M_4\times D$ for a two-manifold $D$.

We make one change in notation from
the earlier part of this paper.  In section \ref{csfun}, to emphasize
that the starting point was a physically sensible,
unitary boundary condition for the D3-NS5 system,
we started in Lorentz signature and labeled the
coordinates of the D3 world-volume as $x^0,\dots,x^3$.
After establishing some basics, we then Wick rotated to
Euclidean signature (section \ref{wick}), still
labeling the coordinates the same way.  But in section
\ref{tdual}, we introduced a new coordinate by
$T$-duality, and it is natural to think of this as the time coordinate.  To make ``room'' for labeling
the new time coordinate as $x^0$, we relabel the four ``old'' coordinates
by $x^\mu\to x^{\mu+1}$.  The main consequence is that when
we do gauge theory on a five-dimensional half-space, starting in
section \ref{gaugedesc}, the coordinate
normal to the boundary of the half-space will be $y=x^4$, and not $x^3$ as earlier in this paper.

\subsection{Four-Dimensional Topological Field Theory From Six Dimensions}\label{wim}

\subsubsection{Basics}\label{twim}

The basic idea is to construct a four-dimensional topological
field theory by twisting of the six-dimensional $(0,2)$
superconformal field theory associated to a simple and
simply-laced Lie group\footnote{\label{feathers}  To be more precise, the
six-dimensional theory is associated to the Dynkin diagram of $G$
rather than to the choice of a specific global form of the group $G$ (such as the adjoint
group or its simply-connected cover).  In particular, the six-dimensional theory does not distinguish $G$ from $G^\vee$;
in the simply-laced case, they are two global forms of the same group.
On a six-manifold $X$, this theory has a family of partition functions
labeled by the quantization of a finite Heisenberg group associated to
$H^3(X,\mathcal Z)$; here $\mathcal Z=\Gamma^\vee/\Gamma$, with $\Gamma$ the root lattice of $G$ and
$\Gamma^\vee$ its dual.  Within this family, one can make a choice that
on reduction to five dimensions leads to
a desired  global form of $G$; on further reduction to four dimensions, the choices
that lead to $G$ or $G^\vee$ are exchanged by $S$-duality.
The details, which are described
in \cite{wittentech}, will not be important in the present paper.} $G$. (The idea of twisting was briefly described in section \ref{twisting}.)
The $R$-symmetry group of this theory
is $SO(5)_R$ or more precisely its double cover $\Spin(5)_R$. As
there is no non-trivial homomorphism from $\Spin(6)$ (the
structure group of the spin bundle of a generic six-manifold) to
$\Spin(5)_R$, there is no way to construct a six-dimensional
topological field theory by twisting of the six-dimensional
$(0,2)$ model.  However,  it is possible to construct topological
field theories in dimension five or less.

\def\uU{{\mathrm U}}
\def\vV{{\mathrm V}}
\def\vv{{\mathrm v}}
The specific construction that we want gives a four-dimensional
topological field theory.  We use the fact that  $\Spin(5)_R$
contains a subgroup
\begin{equation}\label{belg}\uU=(\Spin(3)\times \Spin(2))/\Z_2\subset
\Spin(5)_R.\end{equation} We specialize to six-manifolds of the
form $M_6=M_4\times D$, where $M_4$ is an oriented four-manifold
and $D$ is an oriented\footnote{The orientation of $M_4$ is
necessary to enable us to make a consistent choice of $\Spin(3)_r$
in eqn. (\ref{polm}).  Given this, $D$ must be oriented because the $(0,2)$ model is only
defined on an oriented six-manifold.} two-manifold. The structure group of the
Riemannian (spin) connection of $M_6$ reduces to the subgroup
\begin{equation}\label{elg}\vV=(\Spin(4)\times \Spin(2))/\Z_2\subset
\Spin(6).\end{equation} Furthermore, we have the exceptional isomorphism
\begin{equation}\label{polm}\Spin(4)\cong
\Spin(3)_\ell\times\Spin(3)_r.\end{equation} So it is possible to define a homomorphism
\begin{equation}\label{olmbo}\upsilon:\vV\to \Spin(5)\end{equation}
that annihilates $\Spin(3)_\ell$ and maps
$(\Spin(3)_r\times\Spin(2))/\Z_2$ isomorphically onto $\uU$. We
define a subgroup $\vV'$ of $\Spin(6)\times \Spin(5)_R$, isomorphic
to $\vV$:
\begin{equation}\label{zolm}\vV'=(1\times\upsilon)(\vV).\end{equation}
(In the action of $\vV'$, a spacetime rotation by a group element $\vv\in
\vV$ is combined with an $R$-symmetry transformation $\upsilon(\vv)$.)

In a standard fashion, we can define a twisted version of the
$(0,2)$ model on $M_4\times D$ in which the spin connection
couples to the currents that generate $\vV'$, rather than $\vV$. For
generic $M_4$, the unbroken supersymmetries of the twisted model
correspond to the $\vV'$-invariant supersymmetries that the model
has if formulated on $\R^6$.  A standard group-theoretic exercise,
starting with the fact that the global supersymmetries of the
$(0,2)$ model transform under $\Spin(6)\times \Spin(5)_R$ as
$\mathbf 4_+\otimes \mathbf 4_R$ (where $\mathbf 4_+$ is a
positive chirality spinor of $\Spin(6)$, and $\mathbf 4_R$ is a
spinor of $\Spin(5)_R$), shows that there is just one
$\vV'$-invariant supersymmetry generator, which we will call $Q$.
$Q$ transforms as a non-trivial character of $\Spin(2)_R$, and we
normalize the generator $\EF$ of $\Spin(2)_R$ so that
\begin{equation}\label{zork} [\EF,Q]=Q.\end{equation}
$Q$ also obeys
\begin{equation}\label{ork} Q^2=0; \end{equation}
indeed, if not zero, $Q^2$ would be a universally defined Killing
vector field on $M_4\times D$.

Once we restrict to the cohomology of $Q$, the theory obtained
this way is a topological field theory on $M_4$, but varies
holomorphically with the complex moduli of $D$. One can understand
this without detailed computation as follows. First, compactify
from  six to four dimensions on $D$, making a $\Spin(2)_R$ twist
to preserve supersymmetry.  This leads to a four-dimensional
theory with $\N=2$ supersymmetry.  The remaining $R$-symmetry
group is the subgroup of $\Spin(5)_R$ that commutes with its
$\Spin(2)_R$ subgroup; this is precisely $\uU$, which is isomorphic
to $(SU(2)\times U(1))/\Z_2=U(2)$, the usual $R$-symmetry group of
an $\N=2$ superconformal field theory in four dimensions.  Indeed,
if $D$ is a compact Riemann surface without boundary (possibly
with  punctures),
compactification from six dimensions on $D$ with a supersymmetric
twist gives a four-dimensional superconformal gauge theory \cite{G}; the
gauge group is semi-simple and the coupling parameters $\tau_i$ of
its simple factors are the moduli of $D$.

Now that we are in four dimensions with $\N=2$ supersymmetry,
there is an essentially unique $R$-symmetry twist, resulting from
the identification of $\Spin(3)_r$ with the corresponding subgroup
of $\uU$.  This leads to a four-dimensional topological field
theory by the same reasoning as in \cite{wittendon}.  The
observables of this theory are computed by counting instanton
solutions and hence they depend holomorphically on the instanton
counting factors $q_i =\exp(2\pi i\tau_i)$, that is, on the moduli
of $D$.  Thus, reduction of the six-dimensional theory on
$M_4\times D$ with an $R$-symmetry twist that preserves
supersymmetry gives a theory that is topological on $M_4$ but
varies holomorphically with the moduli of $D$.

\subsubsection{Brane Construction}\label{altcon}

For the
case that $G$ is of  $\sf A$ or $\sf D$ type, and with favorable choices of $M_4$ and $D$,
this construction has a realization via M5-branes.   Just as in section \ref{desc}, this brane realization is highly informative though  not completely general.

We use the fact that the $(0,2)$-model of type ${\sf A}_{r-1}$ arises at low energies on a system of $r$ parallel M5-branes supported on
$\R^6\subset \R^{11}$.  In this description, the $R$-symmetry
group $\Spin(5)_R$ acts by rotations of the normal bundle to
$\R^6$.   To construct a topological field theory, we simply replace $\R^6$ by $M_4\times D$, twisting
the normal bundle to maintain supersymmetry. To get the model of
type $\sf D_r$, we make an orbifold version of the same construction, starting with $2r$ M5-branes and dividing
by a $\Z_2$ symmetry that acts as $-1$ on the normal bundle to the M5-branes.

We let $X$ be the total space of the bundle $\Omega^{2,+}(M_4)$ of
self-dual two-forms on $M_4$, and let $Y=T^*D$ be the cotangent
bundle of $D$.  Ideally, we would like to endow $X$ and $Y$ with
complete metrics of holonomy, respectively, ${\sf G}_2$ and $SU(2)$ --
conditions that will maintain supersymmetry. Having done so, we consider
$M$-theory on the product $\mathcal X=X\times Y$. Then the low
energy limit\footnote{One reaches this low energy limit by scaling
up the metric of $\mathcal X$ so that the radius of curvature
becomes much greater than the natural $M$-theory length scale.} of
$r$ M5-branes wrapped on $M_4\times D$ will give a realization of
the $(0,2)$ model of type\footnote{Taking account of the center of mass motion of the M5-branes,
one actually gets a $U(r)$ rather than ${\sf A}_{r-1}=SU(r)$ theory; that is, one gets a theory that upon compactification
on a circle reduces at low energy to $U(r)$ gauge theory.} ${\sf A}_{r}$ on that manifold with
the $R$-symmetry twist described above. In this description,  the
$R$-symmetry twist of section \ref{twim} arises geometrically from
the twisting of the normal bundle to $M_4\times D$ in $\mathcal X$.

Alternatively, we consider $M$-theory on $\mathcal X/\Z_2=(X\times
Y)/\Z_2$, where the non-trivial element of $\Z_2$ leaves fixed
$M_4\times D$ and acts as $-1$ on the normal bundle to this space.
Wrapping $2r$ M5-branes on $M_4\times D$ and taking the low
energy limit, we get now a realization of the $(0,2)$ model of type
${\sf D}_r$.

What has just been described is less than a general construction
because the desired complete metrics of special holonomy only
exist for special choices of $M_4$ and $D$.  For example, the
requisite metrics of ${\sf G}_2$ holonomy exist \cite{brysal,gp} if
$M_4$ is $S^4$ or $\Bbb{CP}^2$, while for $D=S^2$, the
Eguchi-Hansen hyper-Kahler metric is suitable.  (In the main
example of this paper, $D$ is an open disc with a cigar-like metric and
the Taub-NUT metric has the right properties.) Actually, existence
of such complete metrics is convenient, but is not necessary for
any construction we will make. For one thing, in the $M$-theory
context, all we really care about is the local structure of
$\mathcal X=X\times Y$ near $M_6=M_4\times D$ and any $M$-theory
solution with the appropriate local structure will do. For many
choices of $M_4$ and $D$, $M_4\times D$ can be embedded as a
supersymmetric cycle in some $X\times Y$ where $X$ and $Y$ are as
described above locally near $M_4$ and $D$ but not globally.

More fundamentally, what we will really study is the
six-dimensional $(0,2)$ model on $M_6$ with the $R$-symmetry twist
described in section \ref{twim}; this has its own life
independently of how it can be embedded in $M$-theory.

The utility of the $M$-theory embedding for the present paper is
largely that it  helps to motivate some  constructions and to make obvious
the outcome of some field theory computations.  We will
not consider results that depend on actual existence of an
$M$-theory embedding of $M_4\times D$. (We do make some arguments that are local
on $M_4$ and use the fact that $D$ can be embedded in a Taub-NUT or Eguchi-Hansen
space.)  When an $M$-theory embedding exists, it can lead
to further results, as shown strikingly in \cite{OV,LMV,GSV}
by analysis of geometric transitions that do follow from a
string/$M$-theory embedding.

\subsubsection{Surface Operators}\label{surfop}

In the twisted $(0,2)$ model described in section \ref{twim},  we
want to include surface operators while preserving the topological
symmetry.

The six-dimensional $(0,2)$ theory has half-BPS surface operators.
The simplest example \cite{bcfm,grw} arises from the fact that an
M2-brane can end on a system of parallel M5-branes
\cite{strominger}.  (For generalizations, see section
\ref{gencon}.) As above, we write $M_6$ for the world-volume of
the M5-branes. $M_6$ is contained in an $M$-theory spacetime $M_{11}$. We consider an M2-brane whose worldvolume is a
three-manifold $P_3\subset M_{11}$; we assume that the boundary of $P_3$ is a
two-manifold $\CC_2\subset M_6$. $P_3$ is oriented, so $\CC_2$ is also.
Taking the low energy limit of such a configuration gives us the
$(0,2)$ model of type $\sf A$ or $\sf D$ in the presence of a
surface operator. This surface operator depends on the
``direction'' with which $P_3$ ends on $\CC_2$.

Let us specialize to the case $M_6=M_4\times D$, embedded in the $M$-theory
spacetime   $\cal X=X\times Y$ as
described in section \ref{altcon}.  For a generic choice of
$\CC_2\subset M_6=M_4\times D$, the topological supersymmetry of the
model is broken.  However, it is preserved if we pick $\CC_2=\CC_2'\times
p$, with $\CC_2'$ an oriented two-manifold in $M_4$ and $p$ a point in
$D$, and also pick $P_3$ correctly.

To pick $P_3$, we proceed as follows (in analogy with the
construction in \cite{OV} of a Lagrangian brane associated to a
knot).  Consider a point $q\in \CC_2'$. The oriented tangent plane to
$\CC_2'$  at $q$ determines a non-zero two-form on $M_4$ at $q$, which we
can take to be normalized in a natural metric. Projecting
this two-form to its self-dual part, we get a non-zero unit vector $v\in
\Omega^{2,+}(M_4)|_q$ (that is, in the fiber at $q$ of the bundle
$\Omega^{2,+}(M_4)$ of self-dual two-forms on $M_4$).  But $\Omega^{2,+}(M_4)$
is the normal bundle to $M_4$ in $X$, so $v$ determines a ray in
the fiber at $q$ of that normal bundle.  (What we have just done is to identify the trivial
summand $\varepsilon$ of eqn. (\ref{restc}).) The union of all these
rays for $q\in \CC_2'$ gives a three-manifold $P_3'\subset X$, with
boundary $\CC_2'$.  We take the support of our M2-brane to be
$P_3=P_3'\times p$.

The key point is that an M2-brane supported on $P_3$ does preserve
the same supersymmetry as an M5-brane supported on $M_4\times D$.
This can be understood as an exercise in ${\sf G}_2$ structures.  The
tangent space to $X$ at the point $q$ is a copy of $\R^7$, with a
${\sf G}_2$ structure defined by a three-form $\Upsilon$.  Choosing on
$\R^7$ suitable coordinates  $x^a,$ $a=1,\dots , 7$ and setting
$x^{a+7}=x^a$, we have
\begin{equation}\label{zelf}\Upsilon=\sum_{a=1}^7 \d
x^a\wedge \d x^{a+1}\wedge \d x^{a+3}=\d x^1\wedge \d x^2\wedge \d
x^4+\dots+\d x^3\wedge \d x^4\wedge \d x^6+\dots .\end{equation} A
supersymmetric three-cycle $U_3\subset \R^7$ is a three-cycle whose
volume form coincides with the restriction of $\Upsilon$;
similarly, a supersymmetric four-cycle $R_4\subset\R^7$ is one whose
volume form coincides with the restriction of $\star \Upsilon$.
For example, the three-manifold $U_3$ defined by vanishing of
$x^3,x^5,x^6,x^7$, and so parametrized by $x^1,x^2,x^4$, is a
supersymmetric three-cycle. Similarly, the four-manifold $R_4$
defined by vanishing of $x^3,x^4,x^6$, and so parametrized by
$x^1,x^2,x^5,x^7$,  is a supersymmetric four-cycle. So branes
wrapped on $U_3$ and $R_4$ both preserve the supersymmetry that is
associated to the ${\sf G}_2$ structure.    The geometrical relation
between $U_3$ and $R_4$ is essentially that between $P_3'$ and $M_4$ as
defined earlier. Indeed, setting $M_4=R_4$, we can identify $X=\R^7$
as $\Omega^{2,+}(M_4)$, and then the ${\sf G}_2$ structure coming from
$\Upsilon$ coincides with the natural one on
$\Omega^{2,+}(M_4)$.  In this picture,
$\CC_2'$ corresponds to the intersection $U_3\cap R_4$, and is the
subspace of $M_4$ parametrized by $x^1$ and $x^2$. Finally, $P_3'$
is the half-space in $U_3$ defined by $x^4\geq 0$.

This ensures that, for any choice of $p\in D$, an M2-brane
supported on $P_3=P_3'\times p$ preserves the same supersymmetry as a
system of M5-branes on $M_4\times D$.

\def\X{{\mathcal X}}
We have presented this construction as if $M_4\times D$ has an
$M$-theory embedding in $\X=X\times T^*D$.  The construction of
the half-BPS surface operator does not really depend on this, but
only on the section $v$ of $\Omega^{2,+}(M_4)|_{\CC_2'}$ that is
described above. It is helpful to recall the simplest construction
of supersymmetric Wilson operators in $\N=4$ super Yang-Mills
theory.  The most simple such operator for a loop $K$ and
representation $R$  is
\begin{equation}\label{rusty}\Tr_R\,P\exp\oint_K\left(A+i\vec
n\cdot\vec \phi\,\d s\right), \end{equation} where $\vec\phi$ are the
adjoint-valued scalar fields of the $\N=4$ theory, $\vec n$ is
a unit vector in the space of these scalar fields, and $\d s$ is the geodesic length element along $K$. The section $v$
is the analog of $\vec n$ in the six-dimensional $(0,2)$ theory,
though in this theory one does not have a description by classical
fields that would make it possible to write a formula analogous to
(\ref{rusty}).

\subsubsection{General Construction Of Surface Operators}\label{gencon}

What we  considered in section \ref{surfop} is the most obvious
example of a surface operator in the $(0,2)$ model, associated
with the boundary of an M2-brane that ends on M5-branes.  This
gives a surface operator in the $(0,2)$ model of type $\sf A$.
Upon compactification on a circle, if the support of the surface
operator wraps the circle, such a surface operator will turn into
a Wilson line operator in the fundamental representation of the
appropriate $\sf A$ group; in the opposite case, it turns into
an 't Hooft operator with minimal nonzero magnetic charge, supported on a two-dimensional
surface.\footnote{
In any dimension, a Wilson operator is defined by the holonomy of a gauge field, integrated
along a curve.  So Wilson operators are always supported on curves.  By contrast,
't Hooft operators in gauge theory are always supported in codimension three, since an 't Hooft
operator is defined, as sketched in section \ref{prelims}, by a codimension three singularity.
The codimension three singularity is that of a singular Dirac magnetic monopole in the three dimensions
normal to the support of the 't Hooft operator.
So
an 't Hooft operator is supported on a point in three dimensions, a curve in four dimensions, or
a two-dimensional surface in five dimensions.}  

 For
our applications, we would like to know which Wilson and 't Hooft
operators in five-dimensional super Yang-Mills theory (associated with what representations or magnetic
charges) arise in this way  by compactifying a half-BPS surface operator in six dimensions.  In this
paper, we will assume that all Wilson and 't
Hooft operators arise like that, though this statement goes
somewhat beyond what has been established in the literature.  In
what follows, we indicate some of the known facts.

Large classes of surface operators have been
constructed\footnote{I thank J. Gomis for a guide to this
literature and for sharing some of his insights.} \cite{gomis,wilsonsurface,ol,degk}, in some cases
somewhat implicitly, for the models of
type ${\sf A}_{N-1}$, using the realization of these models via
$M$-theory on $\mathrm{AdS}_7\times S^4$, with $N$ units of flux
on $S^4$:
\begin{equation}\label{izork}\int_{S^4}\frac{G}{2\pi}=N.\end{equation}
Here $G=\d C$ is the curvature of the $M$-theory three-form field
$C$.   These constructions  all
have better understood and more  extensively studied analogs  \cite{GP,Yam,GMOT}
for line operators in $\N=4$ super Yang-Mills theory that are
derived from branes in $\mathrm{AdS}_5\times S^5$.

One basic construction \cite{gomis} uses an M5-brane
supported on $\Theta=\mathrm{AdS}_3\times S^3\subset
\mathrm{AdS}_7\times S^4$. (The M5-brane can be regarded as a
bound state of several parallel M2-branes, which polarize to an
M5-brane via a Myers effect \cite{myers}.  The support of the
surface operator is, as usual, given by the asymptotic behavior of
$\Theta$ at the boundary of $\mathrm{AdS}_7$.) Here $\mathrm{AdS}_3$ is
linearly embedded in $\mathrm{AdS}_7$ in an obvious sense.  And
$S^3$ is embedded in $S^4$ as follows.  We view $S^4$ as the unit
sphere in $\R^5$. Then for some unit vector $v\in \R^5$ ($v$ corresponds to the object
that was denoted by the same name in
section \ref{surfop}), we parametrize $S^3$ by a point $x\in S^4$
that obeys $(v,x)=\kappa$, where $(~,~)$ is the natural inner product
in $\R^5$ and $\kappa$ is a constant.

The constant $\kappa$  is not arbitrary for the following reason. The
M5-brane supports a two-form field whose curvature $T$ equals the
restriction to the fivebrane world-volume of $C$; differently put,
$C$ is trivialized when restricted to the fivebrane worldvolume.
This means that $\int_{S^3}C/2\pi$ must equal an integer, a
condition that allows only finitely many choices of $\kappa$.  Instead
of discussing the gauge-dependent field $C$, it is convenient to
let $B$ be a closed four-ball in $S^4$ of boundary $S^3$;
concretely, we define $B$ by the inequality $(v,x)\leq \kappa$. The
condition on $C$ and $\kappa$ is equivalent to integrality of
\begin{equation}\label{udolf}t=
\int_B\frac{G}{2\pi}.\end{equation} In $\mathrm{AdS}_7\times S^4$ compactification, $G/2\pi$ is
the volume form of $S^4$, normalized so its integral over $S^4$ is $N$.  Its integral over $B$ is positive
but less than $N$.  Hence the possible values of $t$ are
$1,2,3,\dots,N-1$.

The interpretation \cite{gomis,wilsonsurface} is that upon
compactification on a circle, the surface operator just described
reduces to a Wilson operator associated to the $t^{th}$
antisymmetric tensor power of the defining $N$-dimensional
representation. We denote this representation as ${\mathcal R}_t$.  The
${\mathcal R}_t$ are known as the fundamental representations of $SU(N)$.
In general, every simple Lie group $G$ of rank $r$ has $r$
fundamental representations, associated to the nodes of the Dynkin
diagram of $G$; the highest weights of these representations are
called fundamental weights. The highest weight of any irreducible
representation is a positive integer linear combination of the
fundamental weights.  Related to this, every irreducible
representation of $G$ appears in the algebra of tensor products of
fundamental representations provided that we are willing to allow
integer linear combinations with coefficients that are not
necessarily positive.\footnote{For example, let $R$ be an
irreducible representation of $SU(N)$ described as a third rank
tensor that is neither completely symmetric nor completely
antisymmetric.  Then $R$ can be expressed as ${\mathcal R}_1\otimes{\mathcal R}_2-
{\mathcal R}_3$, since it can be constructed as ${\mathcal R}_1\otimes{\mathcal R}_2$
with the completely antisymmetric part subtracted out.}

For applications to Khovanov homology, one would like to know if
the $(0,2)$ model has additional surface operators such that
negative coefficients can be avoided.  This will determine whether
Khovanov homology groups can be defined for a knot labeled by an
arbitrary representation of $G$, or only for those representations that
appear in the tensor algebra of the fundamental representations
without negative coefficients.  In fact, for the $(0,2)$ model of
type $\sf A$, there is \cite{wilsonsurface,ol} a second
construction of half-BPS surface operators with precisely the same
half-BPS properties that again is based on M5-branes. The M5-brane
world-volume is again $\mathrm{AdS}_3\times S^3$, but this time
$\mathrm{AdS}_3\times S^3$ is embedded in $\mathrm{AdS}_7$ (as the
locus of all points a fixed distance $d$ from an $\mathrm{AdS}_3$
subspace of $\mathrm{AdS}_7$) and is supported at a single point
$v\in S^4$ (the same point $v$ that entered the first
construction). Surface operators of this type are believed to
correspond after compactification on a circle to symmetric tensors
of $SU(N)$, with a rank determined by\footnote{The $\mathrm{AdS}_3\times S^3$ solution for the M5-brane has a nonzero value of $\int_{S^3}T/2\pi$, where $T$ is the selfdual three-form curvature  that
propagates on the M5-brane worldvolume.  One expects that Dirac quantization of the flux of $T$  leads
to a quantization condition on the possible values of $d$.  This is somewhat analogous to
quantization of the parameter $t$ in (\ref{udolf}).
} the distance $d$.

More generally, a supergravity analysis \cite{degk} of half-BPS
solutions of $M$-theory with $\mathrm{AdS}_7\times S^4$
asymptotics indicates that surface operators exist that are
associated to an arbitrary Young tableau (fig. 2 of the paper appears to
show the data of a Young tableau), or in other words (after
reduction on a circle) to an arbitrary irreducible representation of
$SU(N)$.

For the $(0,2)$ model of type $\sf D_r$, all of these
constructions have analogs, starting with the realization of the
model via $M$-theory on $\mathrm{AdS}_7\times \Bbb{RP}^4$.  This
may give surface operators that correspond after reduction on a
circle to an arbitrary irreducible representation of $\sf D_r$.
Unfortunately, this sort of construction has no close analog for
groups of type $\sf E$.

\subsubsection{$U(1)_D$ Symmetry}\label{uonesym}

Now we return to our six-dimensional theory on $M_6=M_4\times D$.
For what follows, we require an action of $U(1)$ on the
two-manifold $D$. Moreover, the theory is ultimately more
interesting if the $U(1)$ action on $D$ has a fixed point. If $D$
is to be a complete Riemannian manifold, there are two possible
choices.  We can take $D=\R^2$, with $U(1)$ acting by rotation
around a single fixed point, which we can think of as the origin
in $\R^2$.  Or we can take $D=S^2$, which admits a $U(1)$ action
with two fixed points. We write $U(1)_D$ for the $U(1)$ action on
$D$. We denote its generator as $\EP$.  (When we reduce back to five dimensions
in section \ref{reducing}, $\EP$ will turn into instanton number.)

We can define $\EP$ in the quantum theory so that it commutes
with the unbroken supersymmetry $Q$.  (This condition is needed to define
the quantum operator $\EP$ uniquely; without it, one could add
to $\EP$ a multiple of $\EF$.)  Thus, recalling
(\ref{zork}) and (\ref{ork}), we have
\begin{equation}\label{polyst} Q^2=0, ~~[\EF,Q]=Q,~~[\EP,Q]=0.\end{equation} Vanishing of $Q^2$ implies that one can
define a cohomology of $Q$ (on either operators or states).  The
commutation relations imply that $\EF$ and $\EP$ act on
this cohomology, so the cohomology of $Q$ is $\Z\times \Z$-graded
by the eigenvalues of $\EF$ and $\EP$.

In view of \cite{GSV} or of arguments given earlier in this paper, we anticipate that  Khovanov homology arises from the case
$D=\R^2$.  (The other choice $D=S^2$ apparently leads to a
close relative of Khovanov homology, related to Chern-Simons theory with a complex
gauge group; we will not explore this in the present paper.)  For $D=\R^2$, it is
convenient to endow $D$ with a ``cigar-like'' metric
\begin{equation}\label{hormo}\d s^2 = \d y^2 + f(y)^2\,
\d\psi^2,\end{equation} where $\psi$ is an angular variable of
period $2\pi$ and $f(y)$ is a smooth, increasing function with
$f(r)\sim r$ for $r$ small and $f(r)\to\mathrm{constant}$ for $r\to\infty$.
With a suitable choice of $f$, the cotangent bundle of $D$
can be endowed with a complete hyper-Kahler metric, namely the
Taub-NUT metric.  This is convenient for the $M$-theory
construction of section \ref{altcon}.  More importantly, the
cigar-like nature of the metric will enable us to reduce to a
gauge theory description in section \ref{gaugedesc}.  For $D=S^2$,
one can similarly regard $D$ as a supersymmetric cycle in a
hyper-Kahler manifold (the Eguchi-Hansen manifold).

The remarks of the last paragraph mean that although we cannot
use the  brane construction of section \ref{altcon} globally
along $M_4$ for arbitrary $M_4$ (as a general $M_4$ is not a
supersymmetric cycle in a manifold of ${\sf G}_2$ holonomy), we can do
so globally along $D$ and locally along $M_4$. Indeed, locally, we approximate $M_4$ by
$\R^4$, which we embed in the flat manifold $\R^7$, whose holonomy
(being trivial) is certainly contained in ${\sf G}_2$.  Thus, to get the
model of type $\sf A$, we consider $M$-theory on
\begin{equation}\label{zolk} \mathcal X = \R^7\times  Y,\end{equation}
where $Y$ is a hyper-Kahler manifold (Taub-NUT or Eguchi-Hansen if $D$ is $\R^2$ or $S^2$), with M5-branes wrapped on
\begin{equation}\label{ozolk}M_6=\R^4\times D,\end{equation}
$D$ being a supersymmetric cycle in $Y$.  For the model of type
$\sf D$, we similarly wrap M5-branes on $\mathcal X/\Z_2$, where $\Z_2$ acts
as $-1$ on the normal bundle to $M_6$.

If surface operators are present, then as described in section
\ref{surfop}, we wish to choose them so as to preserve the
$U(1)_D$ symmetry as well as supersymmetry.  We do this by taking
the support $\CC_2$ of the surface operator to be $\CC_2'\times p$, where
$\CC_2'$ is a two-manifold in $M_4$ and $p\in D$ is a fixed point of
the $U(1)$ action.  For example, for the case $D= \R^2$, surface
operators are required to live at the unique fixed point of
$U(1)_D$, the origin in $\R^2$.

\subsubsection{Hamiltonian Description}\label{hamdesc}

To get Khovanov homology, we go to a Hamiltonian description. For
this, we take $M_4=\R\times W_3$, for some three-manifold $W_3$.
Here $\R$ parametrizes the ``time.''  The overall six-manifold is
therefore now $M_6=\R\times W_3\times D$.

We write $\H$ for the (infinite-dimensional) physical Hilbert space of the twisted $(0,2)$ model
in this geometry.   Actually, we want to consider a generalization with a surface operator included.
In order to be able to construct a space of physical states in the
presence of a surface operator, we wish the surface operator to
have time-independent support.
 So in the case of a surface operator supported
on $\CC_2=\CC_2'\times p$, as in section \ref{uonesym}, we want $\CC_2'=\R\times
K$, where $K\subset W_3$ is a knot (as usual, one can generalize to a link, that is, a disjoint union of knots) and $\R$ parametrizes the time.  The space of physical
states in this situation we designate as $\H_K$.    We take $p$ to be the fixed point of the $U(1)_D$ action on $D$.
In this case, $\H_K$ is $\Z\times
\Z$-graded, because of the $U(1)_R\times U(1)_D$ symmetry.

The operator $Q$ acts on $\H_K$.  We write $\K(K)$ or simply $\K$ for the cohomology of $Q$,
acting on $\H_K$. $\K(K)$ inherits the $\Z\times\Z$ grading of $\H_K$.  This is the candidate for the Khovanov homology of $K$.
In section \ref{gaugedesc}, we relate the present six-dimensional description to the gauge theory description that was the basis for
section \ref{tdual}.

Of course, we are not limited to the case that the two-dimensional
surface $\CC_2'\subset \R\times W_3$ is of the form $\R\times K$ with
$K$ a knot or link.   A more general case,
known mathematically as a link cobordism, was already mentioned in section
\ref{kh}.  We pick two
links  $L$ and $L'$ in $\R\times W_3$,
and pick $\CC_2'$ to coincide with $\R\times L$ in the past and with
$\R\times L'$ in the future.  Then we consider the $(0,2)$ model
on $M_6=\R\times W_3\times D$ with a surface operator on
$\CC_2=\CC_2'\times p$.  This determines a $U(1)\times U(1)$-invariant
quantum transition operator from $\K(L)$ to $\K(L')$. In other
words, we get a $\Z\times\Z$-graded linear transformation
\begin{equation}\label{zurx}\Phi_{\CC_2}:\K(L)\to \K(L').\end{equation}
Link cobordisms can be glued together in an obvious way, and the
corresponding linear transformations multiply.

 Actually, the sense in which $\Phi_{\CC_2}$ is
$\Z\times \Z$ graded is a little subtle.  It shifts the $q$-grading in a way that depends on the
topology and normal bundle of $\CC_2$.  This is a known result in Khovanov homology, and
will be explained from the present point of view in section \ref{fourfram}.

\subsection{Gauge Theory Description}\label{gaugedesc}

\def\done{{\dot 1}}
\def\dtwo{{\dot 2}}
\def\dthree{{\dot 3}}
\def\dfour{{\dot 4}}
\def\dfive{{\dot 5}}
\def\ADE{\sf{A-D-E}}

\subsubsection{Reducing To Five Dimensions}\label{reducing}

Our next task is to reduce this six-dimensional description, which rests upon the
mysteries of the $(0,2)$ model, to the five-dimensional gauge theory description of section
\ref{tdual}.

The basic idea is simply to use the $U(1)_D$ symmetry of the Riemann surface $D$.
By standard arguments, if the metric on $M_4\times D$ is scaled in a way that we
describe momentarily, the $(0,2)$ model
on $M_4\times D$ has a low energy description via maximally supersymmetric gauge theory
on $M_4\times D/U(1)_D$.

We consider the case that $D$ is $\R^2$, endowed with the cigar-like metric of
eqn. (\ref{hormo}):
\begin{equation}\label{kelf}\d s^2=\d y^2+f(y) \d\psi^2,~~0\leq y<\infty,
~0\leq\psi\leq 2\pi.\end{equation}
The $U(1)_D$ symmetry of $D$ acts by constant shifts of the angular variable $\psi$.

While keeping fixed the metric on $M_4$, we
multiply the metric of $D$ by a small constant so that
 the asymptotic
value of $f(y)$ for $y\to\infty$ becomes small.
In the limit, the $(0,2)$ model
on $M_4\times D$ has a low energy description in terms of maximally supersymmetric
Yang-Mills theory on $M_4\times \R_+$.  Here $\R_+$ is the half-line $D/U(1)_D$,
parametrized
by $y$.

\def\TN{{\mathrm{TN}}}
This five-dimensional gauge theory description is actually the same one that we used in section \ref{tdual}.
To see this, consider the description in terms of M5-branes wrapped on $M_4\times D\subset X\times \TN$, where $\TN$
is a Taub-NUT manifold in which $D$ is embedded.  $U(1)_D$ acts on $\TN$, with a unique fixed point $p$ (which coincides with the fixed point
at $y=0$ in the action of $U(1)_D$ on $D\subset \TN$).
In the limit that the $U(1)_D$ orbits are small, $M$-theory on $X\times \TN$ reduces to Type IIA superstring theory on
$X\times \TN/U(1)_D$.  The quotient $\TN/U(1)_D$ is simply a copy of $\R^3$, but with a key subtlety \cite{Townsend,GHM}: in the Type IIA description based on this quotient, there is a D6-brane supported on $X\times p$.

Additionally, when we reduce from $M$-theory to Type IIA,  the M5-branes wrapped on $M_4\times D$ become D4-branes wrapped on $M_4\times \R_+$, where $\R_+=D/U(1)_D$ is
a half-line in $\R^3$ that ends at $p$.     What we have arrived at is a D4-D6 system, with D4-branes supported on $M_4\times \R_+$
and ending on a D6-brane.
  But this is precisely the system that was investigated in section \ref{tdual}.    The advantage of deducing this description from a reduction
of the $(0,2)$ model in six dimensions is that the latter provides an ultraviolet completion of five-dimensional super Yang-Mills theory.

To be consistent with the notation used in section \ref{td} and earlier in this paper, we will denote as $G^\vee$ the gauge group of
the five-dimensional description that arises by reducing on the $U(1)_D$ orbits.  As explained in footnote \ref{feathers}, it is a little
subtle how the global form of $G^\vee$ (as opposed to its Lie algebra) is encoded in the six-dimensional theory.  The details of this will not be important in the present paper.

\subsubsection{The Symmetry Group}\label{symgroup}

Now we have to ask how the $U(1)\times U(1)$ symmetry generated by $\EP$ and $\EF$
is realized in the gauge theory description.

Let us first consider the generator $\EP$ of
rotations of $D$.  In general, when the $(0,2)$ model is reduced on a circle, the momentum
around the circle becomes instanton number in the description by five-dimensional
gauge theory.  (This is clear in the $M$-theory description.  Momentum around the circle
turns into D0-brane charge in Type IIA superstring theory.  But, in the gauge theory of a system
of Type IIA D4-branes, D0-brane charge is carried by instantons.)  So $\EP$ corresponds
in the gauge theory description to instanton number.  In the earlier part of this paper, this result was found in another way (in this other
approach, the coupling of the theta-angle to instanton number in the D3-NS5 system was converted after some dualities to instanton
number as a conserved charge in the D4-D6 system).

Perhaps we should clarify the precise meaning of the statement that $\EP$ corresponds to
instanton number.  Instanton number is associated to the closed four-form $\Tr\,F\wedge F$,
which in five dimensions is dual to a conserved current.  The claim is that this is the conserved
current that generates $U(1)_D$ symmetry.   Its integral over an initial value surface, such
as a surface of fixed time in $M_5=\R\times W_3\times \R_+$, is a conserved quantity $\EP$. Actually in making this claim, we have to be careful, just as in section \ref{framan},  with the behavior at both $y=0$ and $y=\infty$.   That behavior will be analyzed in section \ref{fourfram}, and has some significant consequences.  But the conserved instanton number current does lead to a $\Z$-grading that hopefully corresponds to the $q$-grading of Khovanov homology.

 The topological field theory derived from twisting  the $(0,2)$ model on $M_4\times D$  can be defined 
 on any (oriented) $M_4$, but it is probably more interesting if $M_4$ has a suitable\footnote{This three-cycle may be non-compact,
 as in our main example $M_4=\R\times\R^3$, in which the three-cycle is $\{0\}\times \R^3$, with
 $\{0\}$ a point in $\R$.} three-cycle, leading to a four-cycle in $M_4\times \R_+$.
 In the absence
of such a four-cycle, we effectively lose the grading associated with instanton number.  But Khovanov 
homology loses much of its power if we forget the $q$-grading;
this would be analogous roughly to taking the classical limit $q=1$ in Chern-Simons theory.

The other conserved quantity $\EF$ of the $(0,2)$ model is the generator of an $R$ symmetry
that is left unbroken by the twisting procedure.  It has the same type of interpretation
in the description by five-dimensional gauge theory.

\def\TN{\mathrm{TN}}
\subsubsection{Details of Notation}\label{dfourdsix}

Our next goal is to fill a major gap from section \ref{tdual} and
identify  the elliptic partial differential equations that
are associated with supersymmetry in this problem.

Some notational preliminaries will be helpful.
It is convenient to formulate maximally supersymmetric Yang-Mills
theory in five dimensions via dimensional reduction from ten
dimensions.  This means that we combine the five components of the
five-dimensional gauge field, together with five scalars in the
adjoint representation, and regard them as components of a ten
component ``gauge field'' $A_I$.  ($A_I$ has ten components, but
they depend only on the five coordinates of $M_5=M_4\times \R_+$.)
We label the five coordinates of $M_5=M_4\times \R_+$ as
$x^0,x^1,\dots,x^4$, where $x^0,\dots,x^3$ parametrize $M_4$ and
$x^4=y$.  When we specialize to $M_4=\R\times W_3$, with a
three-manifold $W_3$, we will take $x^0$ to parametrize $\R$ and
call it the ``time'' coordinate.  As for the scalars, we call them
$\phi_{I}$ where $I=\done,\dtwo,\dthree,\dfour,\dfive$.  (We do
not label any of the scalars as $\dot 0$, since none will have
``timelike'' properties.)   The curvature is defined as
$F_{IJ}=[D_I,D_J]$, where $D_I$ is a covariant derivative if
$I=0,1,2,3,4$ and otherwise $D_I$ is one of the scalar fields
$\phi_I$.

The fermions fields $\lambda$ of maximally supersymmetric Yang-Mills
theory can be regarded as a positive chirality spinor field of
$SO(1,9)$ with values in the adjoint representation.  We write
$\Gamma^I$ for the gamma matrices of $SO(1,9)$; again $I$ takes
values $0,1,2,3,4$ and $\done,\dtwo,\dthree,\dfour,\dfive$. Both
$\lambda$ and the supersymmetry generator $\varepsilon$ obey a
chirality condition.  In Euclidean signature, we can take this
condition to be
\begin{equation}\label{purf}\bar\Gamma\lambda=-i\lambda,~~\bar\Gamma\varepsilon=-i\varepsilon,\end{equation}
with
$\bar\Gamma=\Gamma_0\Gamma_1\cdots\Gamma_4\Gamma_{\done}\Gamma_{\dtwo}\cdots\Gamma_{\dfive}$.

\subsubsection{The Boundary Condition}\label{bc}

Now we want to consider this theory on a half-space $\R^4\times \R_+$, where $\R_+$ is the half-line $y\geq 0$, and we want the boundary
condition at $y=0$ that corresponds to reduction on $U(1)_D$ orbits of a system of M5-branes on $\R^4\times D$.  In particular,
this boundary condition will break the $R$-symmetry group $\Spin(5)_R$ to $(\Spin(3)\times \Spin(2))/\Z_2$.  The $\Spin(3)$ symmetry
will later be used in maintaining some supersymmetry when $\R^4$ is replaced by an arbitrary four-manifold $M_4$.

The scalar fields $\phi_I$ represent normal fluctuations in the D4-brane position.  In the context of the D4-D6 system,
they play quite different roles.  Three scalars, which we will call $\phi_\done,\,\phi_\dtwo,\,\phi_\dthree$, describe fluctuations in
the D4-brane position along the D6-brane.  And the remaining two scalars, which we will call $\phi_\dfour$ and $\phi_\dfive$,
describe fluctuations normal to the D6-brane.

The normal fluctuations must vanish at $y=0$ where the D4-brane ends on the D6-brane, so the boundary conditions for the last
two scalars at $y=0$ are $\phi_\dfour=\phi_\dfive=0$.
We combine these two fields to a complex scalar field
\begin{equation}\label{turnoco}
\sigma=\frac{\phi_{\dfour}-i\phi_{\dfive}}{\sqrt 2}.\end{equation}
We define a $\Spin(2)_R$ subgroup of the $\Spin(5)$ $R$-symmetry group of the theory that rotates $\phi_\dfour $ and $\phi_\dfive$
and acts trivially on the other scalars.
We define the generator $\EF$ of $\Spin(2)_R=U(1)_R$
so that $\sigma$  has $\EF=2$; the fermions then
have $U(1)$ charges $\pm  1$.  When we eventually define a topological field theory by picking a supercharge $Q$ that obeys
$Q^2=0$, $Q$ will have $\EF=1$. The field $\sigma$ will then inevitably be $Q$-invariant:
\begin{equation}\label{urnoco}[Q,\sigma]=0.\end{equation}
Indeed, the quantum numbers of $[Q,\sigma]$ (it has spin $1/2$, $\mathrm F=3$, and dimension
3/2, and transforms in the adjoint representation of the gauge group) do not coincide with those of any elementary or composite fermion field of five-dimensional
super Yang-Mills theory.

The three scalar fields that describe the motion of the D4-branes along the D6-brane have a polar
behavior at $y=0$.  This polar behavior is a general property of the D$p$-D$(p+2)$ system for any $p$ and was described in the
context of the D3-D5 system in section \ref{boundcond}.  The polar behavior is that
\begin{equation}\label{normbo}\phi_{\dot k}=\frac{{\xi}(t_{k})}{y}+\cdots,~~ k = 1,2,3,\end{equation}
where the $t_{k}$ are a standard set of $\frak{su}(2)$ generators and ${\xi}:\frak{su}(2)\to\frak g$ is a principal embedding.
We will combine the $\phi_{\dot k }$, $k =1,2,3$ to a three-vector $\vec \phi$. (For the moment, this
three-vector lives in an abstract space; it will be reinterpreted in eqn. (\ref{really}).)  One can define a subgroup $\Spin(3)$ of the
$R$-symmetry group that rotates $\vec\phi$.  It preserves the boundary condition when combined with a gauge transformation.
As expected, the boundary condition has reduced the $R$-symmetry group from $\Spin(5)$ to $(\Spin(3)\times \Spin(2))/\Z_2$.

The polar behavior of $\vec \phi$ preserves half of the supersymmetry
of the model.  To describe which half, we recall that the
supersymmetry transformation law for fermions is
\begin{equation} \label{suptrans}\delta\lambda
=\frac{1}{2}\Gamma^{IJ}F_{IJ}\varepsilon,\end{equation} where
$\varepsilon$ is the supersymmetry generator. (As usual a symbol
such as $\Gamma_{I_1\dots I_k}$ vanishes if two indices are equal
and otherwise equals the product of the indicated gamma matrices.)
Nahm's equations (\ref{nahm}) for the scalar fields $\phi_\done,$ $\phi_\dtwo$, $\phi_\dthree$ can be regarded as a selfduality
condition in the four-dimensional subspace corresponding to directions
$4\done\dtwo\dthree$. Writing $\Gamma_y$ for
$\Gamma_4$, Nahm's equations preserve those supersymmetries whose
generator obeys
\begin{equation}\label{zym}\Gamma_{y\done\dtwo\dthree}\varepsilon=\varepsilon.\end{equation}

The solution (\ref{normbo}) of Nahm's equations preserves the
supersymmetry of eqn. (\ref{zym}) for any choice of homomorphism
${\xi}:\frak{su}(2)\to\frak g$.  However, in the case of D4-branes
ending on a single D6-brane, the appropriate choice is that ${\xi}$
is a principal embedding.  More general choices of ${\xi}$ correspond to D4-D6
systems with multiple D6-branes; this has been described in detail in \cite{gw}.
In terms of the six-dimensional $(0,2)$ theory, these more general choices correspond to
formulating that theory on $M_4\times D$ with a suitable defect operator (of a type considered in
\cite{G}) supported on $M_4\times p$.    These  more general choices can be analyzed by methods similar
to those of the present paper; they do not lead precisely to Khovanov homology, but to an interesting
analog of it.

\subsubsection{Twisting Along $M_4$}\label{twalong}

So far we have described the boundary condition at $y=0$ that
breaks half of the supersymmetry and reduces the $R$-symmetry
group to $(\Spin(3)\times\Spin(2))/\Z_2$.  As explained in section
\ref{twim}, the next step is to twist along $M_4$, making a
$\Spin(3)$ twist so that one supersymmetry remains unbroken for an
arbitrary $M_4$.

It is straightforward to describe this one unbroken supersymmetry.
The $\Spin(4)$ symmetry of $\R^4$ is generated by operators $\Gamma_{\mu\nu}=\frac{1}{2}[\Gamma_\mu,\Gamma_\nu]$
acting on spinors.  When we decompose $\Spin(4)=\Spin(3)_\ell\times \Spin(3)_r$, the two factors are generated by the
anti-selfdual and selfdual parts of $\Gamma_{\mu\nu}$, respectively. According to section \ref{twim}, the desired supersymmetry
generator $\varepsilon$ is invariant under $\Spin(3)_\ell$ and under a diagonal combination of $\Spin(3)_r$ and a group of $R$-symmetries;
we will denote this combination as $\Spin(3)'_r$.
The condition that
$\varepsilon$ is invariant under $\Spin(3)_\ell$ is that
\begin{equation}\label{turf}\left(\Gamma_{01}-\Gamma_{23}\right)\varepsilon=0,\end{equation}
along with similar statements that follow by cyclic permutation of
indices $123$.
The condition that $\varepsilon$ is also invariant under $\Spin(3)'_r$   is
\begin{equation}\label{zurf}\left(\Gamma_{12}+\Gamma_{\done\dtwo}\right)\varepsilon=0,\end{equation}
again with similar statements obtained by simultaneous cyclic permutations of
indices $123$ and $\done\dtwo\dthree$.

The conditions (\ref{zym}), (\ref{turf}), and (\ref{zurf}) have a
one-dimensional space of solutions, which corresponds to the
unbroken supersymmetry of the twisted model on a general $M_4$.
For practice, let us use these conditions to determine how
$\varepsilon$ transforms under the $U(1)_R=\Spin(2)_R$ group that
commutes with the Nahm pole.  Taking the generator of this
symmetry to be $\EF= i\Gamma_{\dfour\dfive}$, we use
(\ref{turf}), (\ref{purf}), and (\ref{zym}) to deduce that
\begin{equation}\label{poj}
\EF\varepsilon=-\varepsilon,\end{equation} implying that the
corresponding supercharge $Q$ has $\EF=+1$.

By standard arguments, any quantum computation in the twisted
model can be localized on fields that are invariant under the
topological supersymmetry.  As in other models of this type, such
as the twisted version of $\N=2$ super Yang-Mills theory that is
related to Donaldson theory, there will be equations --
generalizing the instanton equations of Yang-Mills theory -- that
characterize what fields are invariant under this supersymmetry.
The necessary condition is that the supersymmetry variations of
the fermions -- given in eqn. (\ref{suptrans}) -- should vanish. In
other words, we want
\begin{equation}\label{nurky}0=\Gamma^{IJ}F_{IJ}\varepsilon.\end{equation}
 Having
characterized $\varepsilon$, we can work out the consequences of
this condition.

The  analysis will lead to differential equations on $M_5=M_4\times\R_+$
that will have only
four-dimensional symmetry.  Because of this, we introduce some notation that uses
the product structure
of $M_5$.  It will be
convenient to  write $\Omega^{2,+}(M_4)$ for the bundle of
self-dual two-forms on $M_4$, pulled back to $M_5$.   An important
preliminary point is that in the twisted theory, the scalar fields
$\phi_{\done},\phi_{\dtwo},\phi_{\dthree}$ are best understood as
a section of $\Omega^{2,+}(M_4)$, with values in the adjoint
bundle $\mathrm{ad}(E)$ (derived from the underlying $G^\vee$
bundle $E\to M_5$). Thus, we define a self-dual antisymmetric
tensor field $B$ by \begin{equation}\label{really}B_{0i}=\phi_{\dot i},~
B_{ij}=\epsilon_{ijk}\phi_{\dot k},~i,j,k=1,\dots,3.\end{equation}   We
regard $B$ as a section of $\Omega^{2,+}(M_4)\otimes
\mathrm{ad}(E)$.  A useful fact is that $\Omega^{2,+}(M_4)$ is of rank
3, which ensures that there is a ``cross product'' operation on
sections of $\Omega^{2,+}(M_4)\otimes \mathrm{ad}(E)$; this operation is inherited
from the usual cross product
for vectors in $\R^3$, along with the Lie algebra structure of $\mathrm{ad}(E)$. Explicitly, given
$B$, we define  a new section $B\times B$ of
$\Omega^{2,+}(M_4)\otimes \mathrm{ad}(E)$ by
\begin{equation}\label{pesky}(B\times
B)_{\mu\nu}=\sum_\tau[B_{\mu\tau},B_{\nu\tau}],\end{equation}
where on the right hand side $[~,~]$ is the commutator in the Lie
algebra.
The right hand side of (\ref{pesky}) is selfdual if $B$ is, so in particular $B\times B$ is valued in $\Omega^{2,+}(M_4)\otimes \mathrm{ad}(E)$,
as promised.  One final preliminary is that given a two-form $F$ on
$M_4$ -- such as the gauge curvature $F$ -- we define its selfdual
projection  $F^+=(1+\star)F/2$, with $\star$ the Hodge star (defined so $\star(\d x^0\wedge \d x^1)=\d x^2\wedge \d x^3$).

We consider first the part of eqn. (\ref{nurky}) with $
\EF=-1$.  It is convenient to observe that the spinors with $\EF=-1$ transform
under $\Spin(3)_\ell\times\Spin(3)'_r$ as
$(1/2,1/2)\oplus (0,1)\oplus (0,0)$. The
$(0,0)$ part of the equation is satisfied identically.  The
$(0,1)$ part of the equation is
\begin{equation}\label{plonk}\left(\sum_{\mu,\nu=0}^3\Gamma^{\mu\nu}F_{\mu\nu}+2\sum_{\dot
i=1}^3 \Gamma^y\Gamma^{\dot i}D_y\phi_{\dot
i}+\sum_{\dot i,\dot j=1}^3\Gamma^{\dot i\dot j}[\phi_{\dot i},\phi_{\dot
j}]\right)\varepsilon=0.\end{equation} Using the conditions obeyed
by $\varepsilon$, the condition for this to vanish is
\begin{equation}\label{lonk}F^+-\frac{1}{4}B\times B-\frac{1}{2}D_yB=0.\end{equation}
To derive this formula, it is convenient to look at a particular component, say the 01 component.
A part of equation (\ref{plonk}) is
\begin{equation}\label{zilonk}\left(\Gamma^{01}F_{01}+\Gamma^{23}F_{23}+\Gamma^{y\dot 1}D_y\phi_{\dot 1}+\Gamma^{\dtwo,\dthree}
[\phi_\dtwo,\phi_\dthree]\right)\varepsilon=0.\end{equation}
Using (\ref{turf}), we can replace $\Gamma^{01}$ by $\Gamma^{23}$; using (\ref{zym}), we can replace
$\Gamma^{y\done}$ by $-\Gamma^{\dtwo\dthree}$; and using (\ref{zurf}), we can replace $\Gamma^{\dtwo\dthree}$ by $-\Gamma^{23}$.
At this stage the gamma matrices drop out and we find the equation $F_{01}+F_{23}-D_y\phi_\done -[\phi_\dtwo,\phi_\dthree]=0$.
Using the definitions of $B$ and $B\times B$, this is equivalent to $F^+_{01}-\frac{1}{2}D_yB_{01}-\frac{1}{4}(B\times B)_{01}=0$, which is a component of
(\ref{lonk}).
The equation of type $(1/2,1/2)$ can be written
\begin{equation}\label{ponk}\left(\Gamma^y\Gamma^\mu F_{y\mu}
+\Gamma^\mu\sum_{k=1,2,3}\Gamma^{\dot k}D_\mu\phi_{\dot
k}\right)\varepsilon=0.\end{equation} Reducing this equation in a similar
way to what has just been described, we arrive at
\begin{equation}\label{stork} F_{y\mu}+\sum_{\nu=0}^3 D^\nu B_{\nu\mu}=0,~~ \mu=0,\dots,3.\end{equation}

We also need to analyze the part of eqn. (\ref{nurky}) with
$\EF=1$.  This, however, is more straightforward.  We simply
learn that
\begin{equation}\label{ilik}D_\mu \sigma = D_y\sigma =
[B,\sigma]=0, \end{equation} where $\sigma$ was defined in
(\ref{turnoco}). Eqn. (\ref{ilik}) says that a gauge
transformation generated by the adjoint-valued field $\sigma$ is a
symmetry of the solution. Since our boundary condition at $y=0$
forces the solution to be irreducible (and even if we relax the
assumption that ${\xi}:\frak{su}(2)\to\frak{g}$ is a regular
embedding, supersymmetry requires that $\sigma=0$ at $y=0$), these
conditions force $\sigma$ to vanish.

Now that we have reinterpreted $\vec\phi$ in the twisted theory as a section $B$ of $\Omega^{2,+}(M)\otimes \mathrm{ad}(E)$, we should reconsider the boundary conditions at $y=0$ that were described in section
\ref{bc}.  This will be done in section \ref{boundtwist}.

\subsubsection{What Are These Equations Good For?}\label{goodfor}

According to (\ref{lonk}) and (\ref{stork}),
the equations for a supersymmetric field configuration in this theory read
\begin{align}\label{torm}\notag F^+-\frac{1}{4}B\times B -\frac{1}{2}D_y B & = 0 \\
                                            F_{y\mu}+\,D^\nu B_{\nu\mu} &= 0, \end{align}
along with $\sigma=0$.  We will call these simply the supersymmetric equations.

What is one supposed to do with these equations?  This question was answered in section \ref{york}.
Time-independent solutions of these equations supply a basis for a space $\K_{\po}$ of approximate supersymmetric
ground states.  The actual space $\K$ of supersymmetric ground states is found by constructing the supercharge $Q$ as
a linear transformation of $\K_{\po}$ and computing its cohomology.  Concretely, $Q$ is constructed as in eqn. (\ref{omex})
by counting time-dependent solutions of the equations (\ref{torm}) that interpolate between specified limits in the far past and
future.  Both $\K_{\po}$ and $\K$ are $\Z\times\Z$-graded by the action of $\EP$ and $\EF$.  The eigenvalue of $\EP$ is given
by the classical instanton number; that of $\EF$ is found by computing the charge of the filled Dirac sea of negative energy states.  That is why we refer to $\EF$ as fermion number,
though in the full supersymmetric gauge theory it is carried by some bosons (notably $\sigma$)
as well as fermions.

In section \ref{properties}, we will describe some useful properties of the supersymmetric equations (\ref{torm}).  But it may be well to mention here their most basic property, without which the counting
of solutions outlined in section \ref{york} would not make sense: they are elliptic modulo the action of the gauge group.  This actually follows from the relation of these equations to the underlying super Yang-Mills
theory, as we will explain in section \ref{thaction}.

\subsection{Some Properties Of The Equations}\label{properties}

\subsubsection{Reductions To Four Dimensions}\label{zelrud}

Another basic property of the equations is that they can be specialized to more
familiar
equations in lower dimensions.

We begin with the most obvious specialization.  We can look for solutions on $M_4\times \R_+$  that are
independent of $y$.  We do not assume that the solution is a pullback from $M_4$; rather, we replace the covariant
derivative $D/Dy$ with the commutator with an adjoint-valued scalar field $C$.  So the equations become
\begin{align}\label{zorm} F^+-\frac{1}{4}B\times B-\frac{1}{2}[C,B] & = 0 \\ \notag
                                     -D_\mu C+ \,D^\nu B_{\nu\mu}  & = 0 .\end{align}
These equations have been obtained previously \cite{VW} by topological twisting of four-dimensional
$\N=4$ super Yang-Mills theory.   For our purposes, we do not want to study solutions that are independent of $y$ everywhere,  because our boundary condition at $y=0$ does not allow this.  However, it is natural on $M_4\times\R_+$ to  consider solutions that are $y$-independent
for $y\to\infty$, and thus we define our boundary condition at $y=\infty$ by specifying a solution of the equations (\ref{zorm}).  In
the important case that $M_4=\R\times W_3$, we are primarily interested in  boundary conditions at $y=\infty$
that are invariant under time translations.  In the time-independent
case, the equations (\ref{zorm}) describe complex-valued flat connections $\A=\sum_i(A_i+iB_{0i})\d x^i$.  (This will be clear from
another reduction that we describe momentarily.) So, as in most of this
paper, we define the boundary condition by specifying a complex-valued flat connection at infinity.

It is not hard to see why our supersymmetric equations (\ref{torm}), for fields that are independent of $y$, give an equation that
can be derived from $\N=4$ super Yang-Mills theory.
Suppose that for our starting point, we had taken the $(0,2)$ model on $M_6=M_4\times D$, with now $D$ equal to a two-torus
$\tilde S^1\times S^1$ rather than $\R^2$.  Then, upon reducing on $S^1$, the same derivation would lead to the same supersymmetric
equations (\ref{torm}) on $M_4\times \tilde S^1$, with $y$ now an angular variable parametrizing $\tilde S^1$.  It now makes sense to take the solutions to
be independent of $y$, and this leads to (\ref{zorm}).  The two-step process of reducing on first one circle and then the other amounts to
the usual two-torus compactification from the $(0,2)$ model in six  dimensions to $\N=4$ super Yang-Mills in four dimensions.
So naturally it leads
to equations that can be obtained by a topological twist of the $\N=4$ theory.

There is another reduction of the eqns. (\ref{torm}) that is  more surprising if
one simply starts with those equations, though it is obvious from the derivation we have given.  This  comes if we specialize to the case $M_4=\R\times W_3$, for some $W_3$, and
ask for a solution of eqns. (\ref{torm}) on $M_4\times \R_+$ that is time-independent, that is invariant under translations of $\R$.  This process amounts to undoing the lift
from four to five dimensions which was the first step in section \ref{tdual}.  Starting with the supersymmetric equations of the D4-D6
system, if we drop the dependence on time we will get the corresponding  supersymmetric equations of the D3-D5 system.
We already know what these equations are, from eqn. (\ref{oofut}).  They are the familiar equations
\begin{equation}\label{dorky} F-\phi\wedge\phi+\star\d_A\phi = 0 = \d_A\star\phi\end{equation}
for a pair $(A,\phi)$ where $A$ is a connection
on a $G$-bundle $E\to W_3\times \R_+$ and $\phi$ is an $\mathrm{ad}(E)$-valued one-form on $W_3\times\R_+$.

To actually get these equations by a time-independent reduction of our five-dimensional ones, we proceed as follows.  First
of all, parametrize $\R$ by a time coordinate $x^0$ and $W_3$ by local coordinates $x^i$, $i=1,\dots,3$.  As in the case already considered, we look for a solution
on $\R\times W_3\times \R_+$ that is invariant under translations of $x^0$, but we do not assume that the solution is a pullback from $W_3\times
\R_+$.   In particular, we do not assume that $A_0$, the component of the connnection in the $x^0$ direction, vanishes.  Now we define
an adjoint-valued one-form on $W_3\times \R_+$ by
\begin{equation}\label{omoxo} \phi=\sum_{k=1}^3B_{0k}\,\d x^k - A_0\,\d y.\end{equation}
Notice that $A_0$, which was the component
of the connection $A$ in the $x^0$ direction, has been reinterpreted (apart from a minus sign)
as what we might call $\phi_y$, the component of the one-form $\phi$ in the $y$ direction.  Of
course, this only makes sense because both the $x^0$ direction and the $y$ direction have been factored out in $M_5=M_4\times \R_+=
\R\times W_3\times\R_+$.

It is a short calculation, starting with the five-dimensional supersymmetric equations (\ref{torm}) and the definition (\ref{omoxo}),
to arrive at the four-dimensional supersymmetric equations (\ref{dorky}).    The reason that this result is important is that, as explained
in section (\ref{york}), the time-independent solutions of the supersymmetric equations (\ref{torm}) are the basis for the classical
approximation $\K_{\po}$ to the space $\K$ of supersymmetric ground states.  Understanding these time-independent solutions is the
starting point in studying Khovanov homology via five-dimensional gauge theory in the way described here.

Even though we had a good reason to expect the above results and they are not difficult to prove, they should give us a renewed appreciation for the fact that the five-dimensional equations (\ref{torm})
actually are elliptic.  These equations can be obtained in either of two ways from an elliptic equation in four dimensions by replacing a field
with a covariant derivative.  We start with (\ref{zorm}) and substitute $C\to D/Dy$, or we start with (\ref{dorky}) and substitute
$\phi_y\to -D/Dx^0$.  It is quite exceptional that starting with an elliptic differential equation and replacing one of the fields
by the derivative with respect to a new variable, one arrives at an elliptic differential equation in one dimension more.  However,
equations (\ref{zorm}) and (\ref{dorky}) both have this property.
From the point of view developed in the present paper, the fact that the four-dimensional equations (\ref{dorky}) can be ``lifted'' in this
sense to five dimensions is part of the reason that Chern-Simons gauge theory can be ``categorified,'' which is just a fancy way to say
that it can be derived from a theory in one dimension higher.   Similarly, the fact
that the four-dimensional equations (\ref{zorm}) can be lifted to five dimensions means that the four-dimensional
invariant given by counting solutions of those equations
can be categorified.  Modulo a certain vanishing theorem, this four-dimensional
invariant is the Euler characteristic of instanton moduli space \cite{VW}, and its
categorification is, modulo the vanishing theorem and various technicalities involving
the noncompactness of the moduli space, the
cohomology of instanton
moduli space.

\subsubsection{Relation To Morse Theory}\label{normox}

The twisted version of super Yang-Mills theory that we are studying here has in general one supercharge $Q$ when
formulated on $M_5=M_4\times \R_+$.  However, when we specialize to $M_4=\R\times W_3$, for some $W_3$, the theory becomes unitary and
a second supercharge appears, namely the adjoint of $Q$.  Supersymmetric quantum mechanics with two supercharges is commonly
related to Morse theory \cite{wittenmorse}, and as we will now show, this is the case here.

In general, on a manifold $Z$, with local coordinates $u^i$, a metric tensor $\gamma_{ij}$, and a Morse function $\GG$, the flow equations of
Morse theory read
\begin{equation}\label{doxo}\frac{\d u^i}{\d t}=-\gamma^{ij}\frac{\partial \GG}{\partial u^j}.\end{equation}
We wish to show that in the gauge $A_0=0$, our supersymmetric equations (\ref{torm}) can be written as such flow
equations, if we pick a suitable metric on the space of fields and a suitable Morse function.

This is actually a straightforward exercise.  We endow $W_3\times\R_+$ with a metric $g_{ij}\d x^i\,\d x^j+\d y^2$.  On the space of fields on $W_3\times \R_+$, we define the metric
\begin{equation}\label{tonny}\d s^2=-\int_{W_3\times\R_+}\d^3x\,\d y\sqrt g \,\Tr \left(g^{ij}\delta A_i\delta A_j+\delta A_y\delta A_y
+g^{ij}\delta B_{0i}\delta B_{0j} \right).\end{equation}
And then we define the Morse function
\begin{equation}\label{onny}\GG=-\int_{W_3\times\R_+}\d^3x\,\d y\,\Tr\left(\sqrt gg^{ij}F_{yi}B_{0j}+\frac{1}{2}\epsilon^{ijk}\left(A_i\partial_j
A_k+\frac{2}{3}A_iA_jA_k-B_{0i}D_jB_{0k}\right) +\sqrt g w\right),\end{equation}
with $w$ a constant chosen so that the integral converges for $y\to\infty$.
 (The
required constant of course  depends on which $G^\vee_\C$-valued flat connection $\A=(A_i+iB_{0i})\d x^i$ is used to define the
boundary conditions at $y=\infty$.)
A straightforward computation shows that the supersymmetric equations (\ref{torm}), in the gauge $A_0=0$,
  are indeed the flow equations with
$\GG$ as a Morse function.

What we have just described is really the proper input for section \ref{york}, in which we sketched the use of
Morse theory (as extended to field theory problems in \cite{floer}) to describe the space $\K$ of supersymmetric ground states.  The starting point is a knowledge of the time-independent solutions of the supersymmetric equations.    These correspond to critical points of the Morse function $\GG$, and
they furnish a basis of a space $\K_{\po}$ of approximate quantum ground states.  One then realizes $Q$ as a linear transformation
of $\K_{\po}$ via the formula (\ref{omex}); the main step in constructing this formula is to
count, with appropriate signs, the solutions  of
the Morse theory flow equations (\ref{doxo}) that connect two given critical points.  The cohomology of $Q$ gives then the space $\K$ of
exact supersymmetric ground states.

Because of the connection with Morse theory, the value of $\EF$ associated to a given critical point  has an interesting interpretation:
it is the regularized Morse index of that critical point.  In the case of two critical points on bundles of the same topological type (that is, two
critical points with the same value of $\EP$), the difference of $\EF$ at the two critical points can be computed by spectral flow.
To evaluate this spectral flow, one counts the fermion states of $\EF=1$ or $\EF=-1$ that pass through zero energy when one interpolates
between the two critical points.

The attentive reader might notice an apparent clash between what we have said in section
\ref{zelrud}  about time-independent solutions of the supersymmetric equations
and what we have just described.  In interpreting the time-independent solutions as Morse theory flow equations,
the first step was to go to the gauge $A_0=0$.  On the other hand, in section \ref{zelrud}, we carefully did not set $A_0$ to zero,
and instead gave it a new name $-\phi_y$.  The resolution of this puzzle is that eqn. (\ref{dorky}) is actually subject to a vanishing
theorem: in a solution on $W_3\times \R_+$ with the boundary conditions of interest to us, $\phi_y$ vanishes (see\footnote{\label{theproof}  In brief,  after squaring the equations, integrating,  and integrating by parts, one finds that $\phi_y$ is annihilated by
a strictly positive operator.  This implies vanishing of $\phi_y$, a result that was also used in section \ref{doxon} above. Note that our
$\phi_y$ is called $\phi_t$ in \cite{wittentwo}.} the analysis of
eqn. (4.13) in \cite{wittentwo}).  The claim that time-independent solutions of our supersymmetric equations (\ref{torm}) correspond to critical
points depends on this vanishing theorem.   The equations (\ref{dorky}) are covariant and elliptic with $\phi_y$ included. If one uses the
vanishing theorem to set $\phi_y$ to zero, the equations are of course no longer covariant in four dimensions; they also are not elliptic modulo
the gauge group  (but, assuming that one is expanding around a classical solution, they can be embedded in a larger elliptic complex).   However, setting $\phi_y$ to zero makes the Morse
theory interpretation of these equations clearer.  This is so both for the five-dimensional equations
(\ref{torm}) and for the four-dimensional   equations (\ref{dorky}) that were related to Morse theory in a similar way in \cite{wittentwo}.

The vanishing theorem that we just encountered has a perhaps more familiar analog for Floer theory of the space of connections
on a three-manifold.  If on a four-manifold of the form $\R\times W_3$, one
looks for time-independent solutions of the instanton equation $F^+=0$, one gets in three dimensions the Bogomolny equations $F+\star D A_0=0$.
These equations are the analog of (\ref{dorky}); they are elliptic modulo the action of the gauge group, and they do not correspond
directly to the critical points of any Morse function.  However, assuming that $W_3$ is compact and we want  nonsingular and irreducible solutions,
one can deduce from the Bogomolny equations a vanishing theorem $A_0=0$. (The proof is made by the same sort of  argument as  in footnote \ref{theproof}.)   From this vanishing theorem, one learns that
the time-independent solutions of the instanton equation actually correspond to flat connections on $W_3$.  These are the critical points of a Morse
function, namely the Chern-Simons function $\CS(A)$.  The equation $F=0$ that we get after using the vanishing theorem is not elliptic
modulo the gauge group, but it is part of a larger elliptic complex.

The Chern-Simons function $\CS(A)$ of standard Floer theory is not quite well-defined as a real-valued function on the space of gauge
fields modulo gauge transformations (but only as a circle-valued function); because of this, Floer theory is ultimately not $\Z$-graded by the Morse index of a critical
point, but $\Z/4h\Z$-graded, where $h$ is the dual Coxeter number of the gauge group.  By contrast, in our present problem,
the Morse function $\GG$ is actually a well-defined real-valued function, and hence the grading by the fermion number $\EF$ is
an actual  $\Z$-grading, as we have asserted throughout this paper.  To verify that $\GG$ is well-defined, a slightly subtle point is the following. One contribution  in the definition (\ref{onny}) of $\GG$ is the integral over $W_3\times\R_+$ of a Chern-Simons
three-form (times $\d y$). This contribution may look dangerous since the Chern-Simons integral
is not quite well-defined
as a real number, but we pick the constant $w$ to cancel the limiting value of the Chern-Simons integral at $y=\infty$, and then that
integral  causes no further
difficulties.

\subsubsection{The Action}\label{thaction}

By analogy with familiar facts about the equations for Yang-Mills instantons, we anticipate that the first-order supersymmetric
equations (\ref{torm}) imply the second order Euler-Lagrange equations of supersymmetric Yang-Mills theory.   In many examples,
an efficient way to establish such a result is to square the first-order equations, integrate over spacetime, and compare the result to
the action of the underlying physical theory.

In the case at hand, setting \begin{equation}\label{ylab}Y_{\mu\nu}=(F^+-\frac{1}{4} B\times B -\frac{1}{2}D_yB)_{\mu\nu},~~Z_\mu=F_{y\mu}+D^\sigma B_{\sigma\mu},\end{equation} so that
the supersymmetric equations are $Y=Z=0$,  we find the following identity
\begin{equation}\label{nmop}\begin{split} -\int_{M_4\times \R_+}&
\d^4x \,\d y\,\sqrt g\,\Tr\left(Y_{\mu\nu}Y^{\mu\nu}+Z_\mu Z^\mu  \right) \cr =-\int_{M_4\times \R_+}&
\d^4x \,\d y\,\sqrt g\,\Tr\left(\frac{1}{2}F_{\mu\nu}F^{\mu\nu}
+F_{y\mu}F^{y\mu}+\frac{1}{4}(D_yB_{\mu\nu})^2
+\frac{1}{4}(D_\alpha B_{\mu\nu})^2\right. \cr &\biggl. ~~~~~~~~~~~~~~~
+\frac{1}{16}(B\times B)_{\mu\nu}(B\times B)^{\mu\nu}
+\frac{R}{8}B_{\mu\nu}B^{\mu\nu}
-\frac{1}{4}R_{\lambda\nu\mu\tau}B^{\lambda\nu}B^{\mu\tau}\biggr)
+\dots .\end{split}\end{equation}
Here $R_{\lambda\nu\mu\tau}$ and $R$ are the Riemann tensor
and Ricci scalar  of $M_4$; these curvature
couplings are dictated by supersymmetry when $M_4$ becomes curved.  In
(\ref{nmop}), the ellipses represent the omission of certain terms
 whose local variations vanish -- both surface
terms and a multiple of the instanton number evaluated on $M_4$.
 In fact, with our boundary conditions,
both the volume integral on the right hand side of (\ref{nmop}) and the omitted terms are divergent.
Because their local variations vanish, the omitted terms do not affect the argument below.

The right hand side of (\ref{nmop}) is essentially the bosonic
part of the action of maximally supersymmetric Yang-Mills theory in five dimensions.\footnote{To be more precise,
(\ref{nmop}) can be obtained from the super Yang-Mills action by setting two of the five scalar fields
to zero, twisting the other three to a selfdual two-form $B$, and adding some curvature couplings that
are needed to preserve some supersymmetry when $M_4$ is curved.} What do we learn from this relationship?
If $Y=Z=0$, then the left hand side of (\ref{nmop}) is certainly stationary.  So the right hand side is also.  It follows, then, that the Euler-Lagrange
equations derived from the right hand side of (\ref{nmop}) are consequences of the first order supersymmetric equations.  Those
Euler-Lagrange equations are essentially the usual field equations of super Yang-Mills theory (with some scalar fields twisted to the two-form $B$, with fermions and $\sigma$
omitted, and with some curvature couplings added).

We can use this relation between the first order and second order equations to show that the first order equations
in question are elliptic.  Linearization and gauge-fixing\footnote{\label{onkdy} It is convenient to use a ``background field'' version of
Landau gauge, in which the fluctation $\delta A$ of the gauge field $A$ is constrained to obey $\d_A\star\delta A=0$.} of the equations $Y=Z=0$ gives
a linear differential operator that we may call $\mathcal D$.  The ``leading symbol'' of $\mathcal D$ is given by the highest order
 part of $\mathcal D$, written in momentum space.  Let us call this leading symbol $\sigma$. In the present example, $\mathcal D$ is a
first order operator and $\sigma$ is a matrix-valued linear function of the momentum.  Ellipticity of
a system of equations means that the leading symbol of the linearization is invertible for any nonzero (real) momentum.   Letting $\sigma^t$ denote the transpose of $\sigma$,
certainly $\sigma$ is invertible if $\sigma^t\sigma$ is.  But the relation (\ref{nmop}), or more exactly the relation between first order and
second order equations that it implies, means that   $\sigma^t\sigma$ is the leading symbol
of the equations obtained by linearizing the second order equations of super Yang-Mills theory.  Those equations are certainly elliptic;
indeed (in the gauge mentioned in footnote \ref{onkdy}), their leading symbol is the identity matrix multiplied by the leading symbol of the Laplace
operator on scalars.  That symbol is simply the function of a momentum vector $p$ given by $f(p)=p^2$; it is nonzero for real
nonzero $p$.

\subsubsection{The Boundary Condition After Twisting}\label{boundtwist}

Finally, we should reconsider the boundary conditions at $y=0$ for the supersymmetric equations (\ref{torm}) on $M_4\times\R_+$.  For the special case $M_4=\R^4$ without surface operators, these boundary conditions
have already been described in section \ref{bc}: $\vec\phi$ has a regular Nahm pole at $y=0$.
What happens now that we have reinterpreted $\vec\phi$ in the twisted theory as a section $B$ of $\Omega^{2,+}(M_4)\otimes \mathrm{ad}(E)$?

In fact, what happens is quite similar to what we have already described in one dimension
less in section \ref{vivisect}.  The field $\vec\phi$, which was a section of $TW_3$ for a three-manifold $W_3$, has been promoted to a self-dual two-form $B$ on a four-manifold $M_4$. With
this change, all of the previous statements have close analogs.

Since we are interested in what happens at $y=0$, let us write
simply $E$ for the restriction of the gauge bundle
$E$ to $M_4\times\{y=0\}$. Suppose first that $G^\vee=SO(3)$.
Let us write $B=b/y+\dots$ near $y=0$.  Then, by virtue of the vanishing
of the terms of order $1/y^2$ in the supersymmetric equations (\ref{torm}), $b$ establishes
an isomorphism between $\Omega^{2,+}(M_4)$ and $\mathrm{ad}(E)$, and
this isomorphism identifies the metric on $\Omega^{2,+}(M_4)$ with that of $\mathrm{ad}(E)$.
In section \ref{vivisect}, we used analogous statements, which were deduced in the same way,
to identify the polar residue of $\vec\phi$ with the vierbein $e$.
Here the analogous statement is that $b$ can be identified with the selfdual part of $e\wedge e$.
Moreover, the vanishing of the term of order $1/y$ in the supersymmetric equations implies
that $\d_Ab=0$. And this in turn implies\footnote{Once one knows that $b$ is the selfdual part of $e\wedge e$, the analysis of the condition $\d_Ab=0$ to show that $A$ is the Riemannian connection on $\Omega^{2,+}(M_4)$
is a problem that has been considered in the context of canonical quantum gravity \cite{AT}.} that the identification between $\Omega^{2,+}(M_4)$ and
$\mathrm{ad}(E)$ given by $b$ is covariantly constant, meaning that
the restriction to $M_4$ of the $G^\vee$ connection $A$ is simply the Riemannian
connection on $\Omega^{2,+}(M_4)$.
  So just as
in section \ref{vivisect}, the restriction to the boundary of the bundle $E$ and the connection $A$
are directly determined by the Riemannian geometry.

For any $G^\vee$, there is a similar story making use of a principal $\frak{su}(2)$ embedding
${\xi}:\frak{su}(2)\to\frak g^\vee$.  The restrictions of $\mathrm{ad}(E)$ and $A$ to the boundary are
obtained from $\Omega^{2,+}(M_4)$ and the Riemannian connection on it via the homomorphism
${\xi}$.  (In general, depending on the global form of $G^\vee$, the construction of $E$ itself
as opposed to its adjoint form may require a lift of the structure group of $\Omega^{2,+}(M_4)$ from $SO(3)$ to $\Spin(3)$.) Similarly the polar part of $B$ establishes an isomorphism between
$\Omega^{2,+}(M_4)$ and a subbundle of $\mathrm{ad}(E)$ corresponding to
${\xi}(\frak{su}(2))\subset\frak g^\vee$.

It is illuminating to consider the case that $M_4=S^1\times W_3$ (or $\R\times W_3$) with a product metric, and to look for
solutions on $M_4\times \R_+$ that are pulled back from $W_3\times\R_+$.  The equations (\ref{torm}) then
reduce, according to section \ref{zelrud}, to the four-dimensional equations whose boundary conditions were considered in section \ref{vivisect}.  And, as $\Omega^{2,+}(M_4)$ is the pullback to $M_4$ of $TW_3$,
the boundary conditions that we have just described in the five-dimensional case do reduce to the four-dimensional boundary conditions of section \ref{vivisect}.

So far we have described the appropriate boundary condition away from surface operators.   In the presence
of surface operators, we proceed just as we did in section \ref{thooft}.  We first look at a local problem
with a surface operator supported on $\CC_2=\R^2$ linearly embedded in $M_4=\R^4$.  For this local problem,
we find a model solution on $M_4\times \R_+$ that is invariant under translations along $\CC_2$  and has a singularity in the normal plane to $\CC_2$   that is associated to a given
irreducible representation $R$ of $G$.  Since the solution is invariant under translations of $\CC_2$, it is
the pullback to $M_4\times \R_+$ of a solution of reduced three-dimensional equations on $\R^2_\perp\times \R_+$,
where $\R^2_\perp$ is the normal plane.  But in fact,
the relevant reduced equations coincide with the ones already analyzed in section \ref{thooft}. This again follows from the statements in section \ref{zelrud} about dimensional reduction.  So in particular, for $G^\vee=SO(3)$ or $SU(2)$, the relevant model solutions have been fully described in section \ref{sutwo}.

Once the model solutions are known, a surface operator supported on a general embedded oriented two-manifold $\CC_2\subset M_4$ and labeled by a representation $R$  is defined rather as in section \ref{thooft}: we define a boundary condition for the supersymmetric equations such that near a generic boundary point,
the singular behavior is that of the regular Nahm pole,
while along $\CC_2$ the singular behavior is that of the relevant model solution.

There is one important phenomenon that does not quite have an analog in one dimension less: the topology of $\CC_2$ and of its normal bundle influence the $q$-grading of Khovanov homology.  This we consider next.

\subsection{Surface Operators And $q$-Grading}\label{fourfram}

In general, suppose that in five dimensions one is given a conserved current $J$.  Then the four-form
$\star J$ is a conserved charge density, and given an initial value surface $\Omega$, we define
the conserved charge
\begin{equation}\label{utz} q=\int_\Omega \star J. \end{equation}
We are interested in the case that $\star J$ is the instanton current:
\begin{equation}\label{starj}\star J=\frac{1}{32\pi^2}
\epsilon^{\mu\nu\alpha\beta}\,\Tr\,F_{\mu\nu}F_{\alpha\beta}.\end{equation}
We have normalized the instanton  current so  that, for any simply-connected $G^\vee$, the conserved
charge $q$ takes integer values if $\Omega$ is compact and without boundary.

Let us now specialize to $M_4=\R\times W_3$ and thus $M_5=\R\times W_3\times \R_+$.  Given
a conserved current $J$,
we define a charge at time $t\in\R$ by integration
of this four-form over the initial value surface $\{t\}\times W_3\times \R_+$:
\begin{equation}\label{goofy} q(t) =\int_{\{t\}\times W_3\times\R_+} \star J.\end{equation}
Is $q(t)$ independent of time?  Conservation of $J$ is not quite enough to ensure this, since current might disappear
at the ends $y=0$ and $y=\infty$.  In general the change in $q$ between initial and final times $t_i$ and $t_f$ is
\begin{equation}\label{oofy} q(t_f)-q(t_i)=\int_{\Delta_0(t_f,t_i)}\star J -\int_{\Delta_\infty(t_f,t_i)}\star J.\end{equation}
Here $\Delta_0(t_f,t_i)$ is defined by $y=0$, $t_f\geq t\geq t_i$, and $\Delta_\infty(t_f,t_i)$
by $y=\infty$, $t_f\geq t\geq t_i$.  Taking $t_f\to +\infty$, $t_i\to -\infty$
and writing just $\Delta_0$ and $\Delta_\infty$ for the boundaries at $y=0$ and $y=\infty$, the total
change in the charge is
\begin{equation}\label{omfy}\Delta q = \int_{\Delta_0} \star J -\int_{\Delta_\infty}\star J. \end{equation}

In the case of the instanton current, naively the conserved charge is the instanton number
\begin{equation}\label{noffusl}\EH(t) = \frac{1}{32\pi^2}\int_{\{t\}\times W_3\times\R_+}
\epsilon^{\mu\nu\alpha\beta}\,\Tr\,F_{\mu\nu}F_{\alpha\beta}.\end{equation}
Actually, as in eqn. (\ref{utolfox}), to eliminate a dependence on the metric
of $W_3$ (replacing it with a dependence on a framing of $W_3$), we should subtract from $\EH$ a multiple
of the gravitational Chern-Simons function $\CS_\grav$, replacing $\EH$ with
\begin{equation}\label{zomfy}\hat \EH=\EH- \frac{v\,\CS_\grav}{8\pi}.\end{equation}
Since we will take the metric on $W_3$ to be time-independent, this correction term is time-independent.
So the total change in $\hat\EH$ between the far past and the far future is the same as the change in $\EH$. From (\ref{omfy}), it is the sum of two contributions given by the fluxes of the conserved current at $y=0$
and $y=\infty$.  In the present context, those two terms are the instantons numbers of the $G^\vee$ bundle
$E$, restricted to $y=0$ or $y=\infty$.  We write $\EH(y=0)$ and $\EH(y=\infty)$ for the instanton number
evaluated at $y=0$ or at $y=\infty$, so
\begin{equation}\label{yomf}\Delta\hat\EH=\Delta\EH=\EH(y=0)-\EH(y=\infty).\end{equation}

We want to apply this to Khovanov homology, meaning that the boundary condition at $y=\infty$ is that the
connection $A$ approaches a fixed, time-independent flat connection.  This ensures that $\EH(y=\infty)=0$.
Likewise, $\EH(y=0)$ will vanish if the boundary condition at $y=0$ is time-independent.  This will
happen if there are no knots at $y=0$, since then the boundary condition says that the restriction of the connection to $y=0$ is the pullback of the Riemannian connection on $W_3$.  More generally, this will
 happen if all knots are static and time-independent, for then the boundary condition still identifies the
restriction of the connection to $y=0$ with a pullback from $W_3$.

We want to allow time-dependence by including a surface operator supported on a possibly time-dependent two-manifold $\CC_2\subset
\R\times W_3$.  Such surface operators are associated to the knot cobordisms of Khovanov homology.
To describe a transition from the Khovanov homology of a link $L$ in the far past to the Khovanov
homology of another link $L'$ in the far future, we require that in the past $\CC_2$ looks like $\R\times L$
and in the future it looks like $\R\times L'$.  We assume in addition that $\CC_2$ is an oriented, embedded
surface without boundary and with no other ends apart from the ones just described.  Otherwise, $\CC_2$ may
have an arbitrary time-dependence.  The quantum transition amplitude in this situation from an initial state in $\K(L)$
to a final state in $\K(L')$ will give a linear map $\Phi_{\CC_2}:\K(L)\to\K(L')$.  This linear map is,
in mathematical language, the morphism of Khovanov homology associated to the link cobordism $\CC_2$.

Including $\CC_2$ makes the boundary condition at $y=0$ time-dependent, so there is no reason for $\Delta\hat\EH$ to vanish.  Instead,  $\Delta\hat\EH$ will simply equal $\EH(y=0)$, the instanton number of the bundle $E$ restricted to $y=0$.
$\Delta \hat\EH$ is equal to the amount by which the quantum transition amplitude $\Phi_{\CC_2}$ shifts the
$q$-grading of Khovanov homology.

The fundamental case to understand is the case that
$\CC_2$ is compact and $L$ and $L'$ are empty.  After treating this case in section \ref{compc}, we will
reintroduce the knots in section \ref{transk}.

The problem we consider in section \ref{compc} is somewhat like the one studied for framing of knots
in section \ref{framknots}, but it  is simpler because we will be
computing a characteristic class (the instanton number) rather than a secondary characteristic class
(the Chern-Simons function).  We will see in section \ref{transk} that the simpler computation we do here
actually implies the result of section \ref{framknots}.

\subsubsection{Compactly Supported Surface Operator}\label{compc}

In the following, we consider a surface operator of compact support  in an arbitrary four-manifold
$M_4$, which we regard as the boundary at $y=0$ of $M_5=M_4\times\R_+$.
We write simply $\CC$, rather than $\CC_2$, for the support of the surface operator, and we
 write simply $E$ for the restriction of the gauge bundle $E$ to $M_4$,
that is, to $y=0$.
As in our study of knot framings, we will do this analysis for $G^\vee=SO(3)$.  The instanton number of $E$ is 1/4 times
the first Pontryagin class of $\ad(E)$:
\begin{equation}\label{olm} \EP(y=0)=\frac{1}{4}\int_{M_4}p_1(\ad(E)).\end{equation}
(The factor of $1/4$, which corresponds to $1/2h^\vee$ in eqn. (\ref{tofus}), comes from the ratio of the trace of the four-form $F\wedge F$
in the two-dimensional and three-dimensional representations of $SU(2)$.)

In the absence
of a surface operator, $\ad(E)$ is simply $\Omega^{2,+}(M_4)$, so $\EP(y=0)$ can be expressed in terms of the
Euler characteristic and signature of $M_4$.  We want to determine the shift in $\EP(y=0)$ due to the
presence of the surface operator:
\begin{equation}\label{zolf}\Delta \EP(y=0)=\frac{1}{4}\int_{M_4}\left(p_1(E)-p_1(\Omega^{2,+}(M))\right).
\end{equation}

Let us first describe the restriction to $\CC$ of $\Omega^{2,+}(M_4)$. At a point $p\in \CC$, we pick an orthonormal basis of one-forms $e_1,e_2$ and $f_1,f_2$, such that the $e_i$ are tangent to $\CC$ and the $f_j$
are normal to $\CC$.  Also we orient them so that $e_1\wedge e_2$ and $f_1\wedge f_2$ determine the orientations of the tangent bundle $T  \CC$ to $\CC$ and its normal bundle $N  \CC$, respectively, and
hence the orientation of $M_4$ corresponds to $e_1\wedge e_2\wedge f_1\wedge f_2$.

\def\L{{\mathcal L}}
\def\S{{\mathcal S}}
\def\T{{\mathcal T}}
\def\N{{\mathcal N}}
Now let us simply write down an orthonormal basis of self-dual two-forms at $p$.
We can take one such form to be $w_1=e_1\wedge e_2+f_1\wedge f_2$.  For the other two such forms, we write
\begin{equation}\label{thelf} w_2+iw_3=(e_1+ie_2)\wedge (f_1+if_2)\end{equation}
or
\begin{equation}\label{melf} w_2=e_1\wedge f_1-e_2\wedge f_2,~~w_3=e_1\wedge f_2+e_2\wedge f_1.\end{equation}
Clearly, $w_1,$ $w_2,$ and $w_3$ are indeed selfdual and (in a natural inner product) orthonormal.

The definition of $w_1$ was completely natural, so $\Omega^{2,+}(M_4)|_{\CC}$ contains a one-dimensional trivial real summand that we will call $\varepsilon$.  As for $w_2+iw_3$, it is best understood as lying in the fiber at $p\in \CC$ of a complex line bundle $\mathcal M\to \CC$.  To construct this line bundle, we view $T^*\CC$
and $N^*\CC$ (the duals of $T  \CC$ and $N  \CC$) as rank one complex line bundles, placing on them the
complex structures that act by
\begin{equation}\label{polyp}I(e_1+ie_2)=i(e_1+ie_2),~~J(f_1+if_2)=i(f_1+if_2).\end{equation}
Evidently, $\mathcal M\cong T^*\CC\otimes_\C N^*\CC$, since $e_1+ie_2$ takes values in $T^*\CC$ and $f_1+if_2$ in $N^*\CC$.
So the restriction of $\Omega^{2,+}(M_4)$ to $\CC$ is
\begin{equation}\label{zonkor}\Omega^{2,+}(M_4)|_{\CC}=\varepsilon\oplus \mathcal M,\end{equation} where $\mathcal M$
is regarded as a real vector bundle of rank 2.

As a real bundle of rank 2, $\mathcal M$
is equivalent to its dual.  (In fact, the Riemannian metric on $M_4$ gives
a natural identification between them.)  This means that in (\ref{zonkor}), we can replace $\mathcal M$ by  $\L=\mathcal M^{-1}$. Here $\L=T  \CC\otimes_\C N  \CC=\T\otimes\N$, where  we write
 simply $\T$ and $\N$ for $T \CC$ and $N \CC$ regarded as complex line bundles.
Thus (\ref{zonkor}) is equivalent to
$\Omega^{2,+}(M_4)|_{\CC}=\varepsilon\oplus\L.$
A small neighborhood $\mathcal U$ of $\CC$ is contractible onto $\CC$, and this isomorphism
automatically extends over $\mathcal U$:
\begin{equation}\label{restc}\Omega^{2,+}(M_4)|_{\mathcal U}\cong \varepsilon\oplus \L.\end{equation}

Now we want to modify $\Omega^{2,+}(M_4)$ along $\CC$ by gluing in along $\CC$ an 't Hooft operator supported on $\CC$ and dual to the spin $j$ representation of $G=SU(2)$.  In the full five-dimensional description, the  support of the 't Hooft operator is
on $\CC\times \{y=0\}\subset M_4\times \R_+$, so it is of codimension three as expected for 't Hooft operators.  We denote the modified bundle as $E_\j$.  We can understand the structure of $E_\j$ from the model solution described in section \ref{sutwo} -- lifted now to five dimensions rather than
to four as assumed in section \ref{thooft}.  The gauge field of the model solution is $\frak u(1)$-valued (though the full model solution including the other fields is irreducible).  In the context of a knot $K$ in
a three-manifold $W_3$, the $U(1)$ in question acts on the normal bundle to $K$.  When we lift to a surface $\CC$ in a four-manifold $M_4$, the $U(1)$ in question acts on the subbundle of $\Omega^{2,+}(M_4)|_\CC$ that is orthogonal to $\varepsilon$.  In other words, it acts on $\L$.

To construct $E_\j$, we are supposed to glue in $2j$ units of flux in this $U(1)$ subgroup.  This
means that $E_\j$ restricted to $\CC$ will have the form $\varepsilon \oplus \S$ where $\S$ is a complex
line bundle with the following properties:
{\it (1)} Away from $\CC$, $\S$ is isomorphic to $\L$, ensuring that $E_\j$ is equivalent to $\Omega^{2,+}(M_4)$.
{\it (2)} The isomorphism between $\L$ and $\S$ has a zero along $\CC$ of degree $2j$.
This second condition captures the idea that $E_j$ is obtained from $\Omega^{2,+}(M_4)$ by adding $2j$ units of flux in the normal direction.

The two conditions have a simple and unique solution.
In general, if $\CC$ is a Riemann surface, there is no natural way to pick a section of a complex line bundle $\S\to \CC$.  But let $X$ be the total space of the line bundle  $\S\to \CC$
and let $\pi:X\to \CC$ be the natural projection, and pull back $\S$ to a line bundle $\pi^*\S\to X$.
Then $\pi^*\S$ does have a natural section, which moreover has a simple zero along
$\CC\subset X$.  This section is defined as follows: for $q\in X$, define $p\in \CC$ by $p=\pi(q)$.
Then $q$ lies in $\S_p$, the fiber of $\S$ over $p$.  But by the definition of pullback, $\S_p$ is
naturally isomorphic to the fiber of $\pi^*\S$ over $q$.  This isomorphism maps $q$ to an element
$s(q)$ of this fiber, and the map $q\to s(q)$ is the desired section of $\pi^*\S\to X$.

The most familiar example of this construction is  the case that $\S$ is the canonical bundle
$K_\CC$ of $\CC$; $K_\CC$ has no natural section, but its pullback to the total space of the
fibration $K_\CC\to \CC$ does have a natural section, usually written as $p\,\d x$, where $x$ is
a local coordinate on $\CC$ and $p$ is a fiber coordinate.  We note that $p\,\d x$ has indeed a simple
zero at $p=0$, that is, along $\CC$, and is nonzero for $p\not=0$.

If $s$ is a section of $\pi^*\S\to X$ with a simple zero along $\CC$, then $s^{2j}$ is a section of
$(\pi^*\S)^{2j}\to X$ with a zero along $\CC$ of degree $2j$ and no other zeroes.
Moreover, up to isomorphism, $(\pi^*\S)^{2j}$ and $s^{2j}$ are the unique line bundle
and section with these properties.

To apply this to our problem, we observe that a small neighborhood $\mathcal U$ of $\CC\subset M_4$ can
be identified, in a way that is unique up to homotopy, with a neighborhood of the zero
section in the total space of the fibration $\pi^*\N\to \CC$.  So a line bundle over
$\mathcal U$ that has a section vanishing in degree $2j$ along $\CC$ and nowhere else
is the pullback to $\mathcal U$ of $(\pi^*\N)^{2j}$.  More informally, we call this line
bundle simply $\N^{2j}$.

So a line bundle that is isomorphic to $\L$ away from $\CC$ by an isomorphism that has a zero
of degree $2j$ along $\CC$ is simply $\L\otimes \N^{2j}$.  We thus arrive at a description of $E_\j$.
In a neighborhood of $\CC$ it is
\begin{equation}\label{ofunk} E_\j|_{\mathcal U} =\varepsilon \oplus \L\otimes \N^{2j}=\varepsilon\oplus
\T\otimes \N^{2j+1}.\end{equation}
In general, if $E$ is a rank three real vector bundle that is given globally as $\varepsilon\oplus
\mathcal R$, where $\varepsilon$ is a trivial real line bundle and $\mathcal R$ is a complex
line bundle that we view as a real vector bundle of rank two, then $p_1(E)=c_1(\mathcal R)^2$.  So
from (\ref{zolf}), if the formulas (\ref{restc}) and (\ref{ofunk}) are valid globally on $M_4$,
not just in a neighborhood of $\CC$, then the change in the instanton number due to the
surface operator is
\begin{equation}\label{zoronk}\Delta \EP(y=0)=\frac{1}{4}\int_{M_4}\left(c_1(\T\otimes \N^{2j+1})^2
-c_1(\T\otimes \N)^2\right). \end{equation}

It is possible for (\ref{restc}) and (\ref{ofunk}) to be valid globally, if $\T$ and $\N$ are suitably
extended over $M_4$.  This happens
if $M_4$ is a complex manifold and $\CC$ is a complex submanifold.  In this
case, $\Omega^{2,+}(M_4)=\varepsilon\oplus K_{M_4}$, where $K_{M_4}$ is the canonical line
bundle of $M_4$.  As a real bundle of rank two, $K_{M_4}$ is equivalent to the anticanonical
bundle $K_{M_4}^{-1}$.  When restricted to $\CC$, $K_{M_4}^{-1}\cong \T\otimes \N$, showing that
(\ref{restc}) holds globally.  Similarly (\ref{ofunk}) holds, with $\N$ interpreted as the line
bundle $\mathcal O(\CC)$ whose holomorphic sections are meromorphic functions that
may have a simple pole along $\CC$.  Not only is it possible for (\ref{restc}) and (\ref{ofunk}) to
hold globally, but this can be the case with no restriction on the topology of $\CC$ or its normal
bundle.  So cases of this type must suffice to determine the general result.

Actually, one can justify (\ref{zoronk})  more directly without reference to the question of whether
(\ref{restc}) and (\ref{ofunk}) may hold globally.
The formal difference $E\ominus \Omega^{2,+}(M_4)$
represents a class in the $K$-theory of $\U$ with compact support (since $E$ and $\Omega^{2,+}(M_4)$ are isomorphic on the complement of $\CC$).  The difference between the
formulas (\ref{restc}) and (\ref{ofunk}) is a valid formula in this $K$-theory with
compact support, and this is enough to justify (\ref{zoronk}), which involves only
the first Pontryagin class of $E\ominus \Omega^{2,+}(M_4)$.

As for the actual evaluation of the right hand side of (\ref{zoronk}), all that one needs to know is that
 the integral of $c_1(\N)^2$ is
$\CC\cap \CC$, the self-intersection number of $\CC$, and that the integral of $c_1(\N)\cdot c_1(\T)$
is $\chi(\CC)$, the Euler characteristic of $\CC$.  Both statements follow from the fact that
$\N$ has a section with a simple zero along $\CC$.
So finally the shift in the $q$-grading
due to the surface operator is
\begin{equation}\label{dorfo}\Delta \EP(y=0)=j\,\chi(\CC)+j(j+1)\,\CC\cap \CC.\end{equation}

\subsubsection{Transitions Between Knots}\label{transk}

Now let us consider link cobordisms.  For brevity
in the exposition, let us assume that there are no knots in the past and there is a single knot
$K$ in the future.  The generalization to arbitrary links in the past and future does not change
much; the remarks that follow apply to each boundary component separately. So we take $\CC$ to be compact toward the past and to have an end
toward the future that looks like $K\times \R_+$.  (This $\R_+$ is future-pointing and does not coincide with the usual
$\R_+$ that is parametrized by $y$.)

Nothing changes in the above derivation provided the line bundles $\T$ and $\N$ are
trivialized near the noncompact end of $\CC$.  $\T$ has a natural trivialization near $t=\infty$
associated with a vector field that generates time translations along $K\times\R_+$.  One
can think of this as the reason that there is no problem to define the Euler characteristic
of a noncompact Riemann surface like $\CC$.  However, a time-independent trivialization of $\N$
near $t=\infty$ corresponds to a framing of $\CC$.  If the framing of $K$ is shifted by 1 unit,
then $\CC\cap \CC$, defined relative to this trivialization, shifts by 1 unit.
This shifts $\Delta\EP(y=0)$ by $j(j+1)$, so the $q$-grading of the final state in $\K(K)$
is also shifted by $j(j+1)$.   This is consistent with the fact that the expectation value of
a Wilson operator supported on $K$ in Chern-Simons theory is multiplied by $q^{j(j+1)}$ under a unit
shift in framing of $K$, a fact that we have also explained in another way in section
\ref{framknots}.

Another interesting effect results from the  term in (\ref{dorfo}) proportional to $\chi(\CC)$.
For a closed Riemann surface $\CC$, $\chi$ is even, but for a Riemann surface ending on
a single knot, $\chi$ is odd.  It follows then that if $j$ is half-integral, $\Delta\EP(y=0)$ is
also half-integral and the shift in $q$-grading in a transition from the vacuum (no knots)
to a state in the Khovanov homology of a single knot is half-integral.  This gives a new
explanation of why the Jones polynomial  of a knot (the invariant  associated to $j=1/2$) is
$q^{1/2}$ times a series in (positive and negative) integer powers of $q$.  More generally,
by the same reasoning, the Jones polynomial of a link with $\nu$ components is $q^{\nu/2}$
times a series in integer powers of $q$.

\subsection{Gauge Groups That Are Not Simply-Laced}\label{zeldow}

\subsubsection{Preliminaries}\label{meld}

Starting with section \ref{twim}, the groups $G$ and $G^\vee$ have been simply-laced, for the simple reason that our main tools, the $(0,2)$ models
in six dimensions,  are associated to simply-laced groups.  Nonetheless, it is possible
to deduce $S$-duality in four dimensions for a gauge group $G$ that is not simply-laced  by starting \cite{geom}
 with the six-dimensional model of a simply-laced
group $G^*$.  The relation between $G$ and $G^*$ is the same as it was in section \ref{zorky}: $G^*$ has an outer automorphism $\zeta$,
such that the subgroup of $G^*$ that commutes with $\zeta$ is $G^\vee$, the dual of $G$.  As we have seen in section \ref{zorky}, when
$G$ is not simply-laced, there are two different Khovanov-like formulas, both presented in eqn. (\ref{zelbow}), that express the knot invariants of $G$ Chern-Simons theory  as traces in some space akin to Khovanov homology.  Our goal here is to identify two six-dimensional constructions,
starting with
 the $(0,2)$ theory of type $G^*$, that lead to these two formulas.

The first basic fact that one needs to know is that for every pair $(G^*,\zeta)$ that appeared in section \ref{zorky}, the $(0,2)$ model of type
$G^*$ has $\zeta$ as a global symmetry.  One way to see this is to use the unified description \cite{comments} of $(0,2)$ models for all $\sf {A-D-E}$ groups
in terms of Type IIB superstring theory at the corresponding $\sf {A-D-E}$ singularity.  In all cases, $\zeta$ acts as a hyper-Kahler
automorphism of the singularity of type $G^*$ (this fact was first used in string theory in \cite{aspingross}) and  hence as a symmetry of the corresponding $(0,2)$ model.\footnote{As has been pointed
out by the author of \cite{tachikawa}, it is not true that all outer automorphisms of simply-laced groups act as hyper-Kahler automorphisms
of the corresponding singularity.  Rather, this is so precisely for the pairs $(G^*,\zeta)$  that are associated to groups $G^\vee$ that
are not simply-laced.   These pairs are $G^*=\sf A_{2n-1}$ with the automorphism of complex conjugation combined with a suitable inner automorphism (related to $G^\vee ={\sf C_n}=Sp(2n)$),
$G^*={\sf D_{2n}}$ with the automorphism a reflection of one variable (related to $G^\vee={\sf B_{n-1}}=SO(2n-1)$), $G^*=E_6$ with its outer
automorphism (related to $G^\vee={\sf F}_4$), and $G^*=\sf D_4$ with an outer automorphism of order 3 (related to $G^\vee={\sf G}_2$).  A concise
way to state the relation between these pairs is that (by the usual duality that exchanges long and short roots of the Dynkin diagram) the loop group of $G^\vee$ is GNO or Langlands dual to the $\zeta$-twisted loop
group of $G^*$.  The example
of an outer automorphism that does not arise as a hyper-Kahler symmetry of the appropriate singularity and is not related to a non-simply-laced
Lie group is $\sf A_{2n}$ with the automorphism of complex conjugation.}

Before generalizing to include the automorphism $\zeta$, let us recall the standard claim about compactification of the $(0,2)$ model of type $G^*$ on
a two-torus $\tilde S^1\times S^1$.  If one formulates the $(0,2)$ model on $M_4\times \tilde S^1\times S^1$ for some $M_4$, and
scales down the metric of $\tilde S^1$, then it reduces to supersymmetric gauge theory on $M_4\times S^1$.  The gauge group in this
description is a global form of the group $G^*$.  Which global form arises depends on a subtle choice one makes in defining the theory
in six dimensions; see footnote \ref{feathers}.  If instead one reduces on $S^1$, one gets a five-dimensional gauge theory based
on a possibly different global form of $G^*$ -- the Langlands or GNO dual form.  (This duality exchanges the center of $G^*$ with its
fundamental group, so for instance the adjoint form of the group is dual to the simply-connected form.)

Now let us repeat this discussion with $\zeta$ included.  We consider the $(0,2)$ model of type $G^*$ on $M_4\times \tilde S^1\times S^1$,
but now with a twist by $\zeta$ in going around one of the two circles.  Again, we consider what happens when $\tilde S^1$ is scaled down.
There are two cases:

{\it (i)} If the twist is made around $ S^1$, then the reduction on $\tilde S^1$ gives five-dimensional $G^*$ gauge
theory on $M_4\times  S^1$, just as if there were no twist.  But in this gauge theory description, one sees  a twist by $\zeta$ in going around  $S^1$.  The twist breaks $G^*$ down to $G^\vee$,
so in four dimensions one gets $G^\vee$ gauge symmetry.

{\it (ii)} If instead the twist is made  around $\tilde S^1$, one gets in five dimensions gauge theory
on $M_4\times  S^1$ with gauge group $G$, the dual of $G^\vee$. Since there is no twist around $S^1$,
 the compactification on $S^1$ does not affect the gauge group
observed in four-dimensions at scales large compared to the radius of $S^1$.

Statements {\it (i)} and {\it (ii)} are related by electric-magnetic duality in four dimensions,
since obviously exchanging the two circles (which is the basic operation of electric-magnetic duality) is equivalent to changing the circle
around which the twist is made.  Statement {\it (ii)} is used in the literature as a way to generate non-simply-laced gauge symmetry
starting from $M$-theory or Type II superstring theory.

We need to know one more fact about the $(0,2)$ model of type $G^*$, beyond the fact that it admits $\zeta$ as a global symmetry.
This model admits a half-BPS defect consisting of a codimension two submanifold around which all fields undergo the automorphism
$\zeta$.  This fact has been briefly mentioned in \cite{beauty} and exploited in \cite{tachikawa}.

\subsubsection{Two Constructions}\label{twoc}

Using these facts, we can now describe two six-dimensional constructions that are related to the two formulas  presented in eqn. (\ref{zelbow})
for the knot invariants derived from Chern-Simons theory of a simple but not simply-laced Lie group $G$.  In explaining these
formulas, as in section \ref{zorky}, $G^*$ will be a simply-laced Lie group that possesses an outer automorphism $\zeta$ that leaves fixed $G^\vee$,
the dual of $G$.  Now, however, we will also need a simply-laced Lie group $G^\diamond$ that is related to $G$ the way $G^*$ is related to $G^\vee$.
Thus, $G^\diamond$ admits an outer automorphism $\zeta'$ that leaves fixed $G$.  If $G$ is of type ${\sf G}_2$ or ${\sf F}_4$, then $G=G^\vee$ and
$G^\diamond =G^*$.  The case that $G^\diamond $ and $G^*$ are different is that $G=Sp(2n)$ and $G^\vee=SO(2n+1)$ (or vice-versa); then
$G^*=SO(2n+2)$ and $G^\diamond=SU(2n)$.

Now we consider two constructions that will lead to the two formulas in eqn. (\ref{zelbow}):

{\it (1)} The first construction is familiar. We consider the $(0,2)$ model of type $G^*$ on $M_6=\R\times W_3\times D$.  We write $\K^*$ for its space of physical ground states.
After reducing on the $U(1)_D$ orbits, $\K^*$ can be computed by solving the supersymmetric equations (\ref{torm}) in $G^*$ gauge theory.

{\it (2)} In the second construction, we start with the $(0,2)$ model of type $G^\diamond$, again on $M_6=\R\times W_3\times D$.  Now,
however, we include a defect operator associated to the outer automorphism $\zeta'$ and supported on $\R\times W_3\times p$,
where $p\in D$ is the $U(1)_D$ fixed point.  Reducing on the $U(1)_D$ orbits, we get a description by supersymmetric gauge theory on
$\R\times W_3\times \R_+$ with gauge group $G^\vee$.  This assertion reflects statement {\it (ii)} in section \ref{meld}, except that,
since we started with $G^\diamond$ instead of $G^*$, the roles of $G$ and $G^\vee$ are exchanged.
In determining the gauge symmetry in this description, it suffices to consider the situation at large
$y$, and we do not need to know what is happening at $y=0$.  However, because of
the supersymmetry of the problem, we expect the boundary condition at $y=0$ to be the usual one with the regular Nahm pole (for the five-dimensional bulk gauge group $G^\vee$, of course).

We write $\K$ for the space of physical
ground states in  construction {\it (2)}.  It can be obtained by studying the supersymmetric equations (\ref{torm}) in $G^\vee$  gauge theory.  So
in particular the spaces $\K^*$ and $\K$ that arise in our two constructions coincide with the ones that were denoted the same way in section \ref{zorky}.

Now we compactify the time direction, possibly with a global symmetry twist:

{\it (1$'$)} In case {\it (1)},  we replace $M_6$ by $S^1\times W_3\times D$, but making a twist by $\zeta$ around the $S^1$ direction.
The resulting path integral on $S^1\times W_3\times D$ can be interpreted as a trace in $\K^*$.  In the absence of the twist, the path integral  would compute $\Tr_{\K^*}\, q^\EH(-1)^\EF$, but as we have included the twist, we get instead
$\Tr_{\K^*}\, q^\EH(-1)^\EF\zeta$.  This is the right hand side of one of the two formulas in (\ref{zelbow}).

{\it (2$\,'$)} In case {\it (2)}, we again replace $M_6$ by $S^1\times W_3\times D$, but now without any twist in the $S^1$ direction.
The path integral around $S^1$ now computes $\Tr_{\K}\, q^\EH(-1)^\EF$.  This is the right hand side of the other formula in (\ref{zelbow}).

As for why  these two six-dimensional constructions agree with the left hand-side of eqn. (\ref{zelbow}) -- that is, with the path integral of Chern-Simons theory with gauge group $G$ -- we simply observe the following.  In either of the two constructions, at distances
large compared to the size of $S^1$, we get a description by $G^\vee$ gauge theory on $W_3\times \R_+$ with D3-D5 boundary conditions.
Given this, we can retrace our way through the steps of sections \ref{dualities} and \ref{csfun}, first making an $S$-duality to a description by $G$
gauge theory with D3-NS5 boundary conditions, and finally relating this to Chern-Simons theory on $W_3$ with gauge group $G$.

\section{Another Path To Six Dimensions}\label{morebranes}

\subsection{Overview}\label{moreover}

\subsubsection{Some Background}\label{someb}

In this section, we will repeat the analysis of the present paper along a different route.

For a first orientation, let us recall some of the defect operators in gauge theories.  A basic defect
operator in dimension 1 is the Wilson line operator.  In codimension 3, there are 't Hooft operators.
These are the two types of defect operator that we have considered so far.

More obvious than the 't Hooft operator is another type of defect operator that appears in codimension 2.
This is an operator associated with a prescribed monodromy.  In gauge theory with gauge group $G$ on any manifold $X$, let $U$ be a submanifold of codimension 2.  Let $\mathcal C$ be a conjugacy
class in $G$.  Then one considers
gauge theory on $X\backslash U$ with the condition that the gauge fields have a monodromy around $U$
that is in the conjugacy class $\mathcal C$.  A surface operator supported on $U$ is defined by asking in addition that the fields should have the mildest
type of singularity consistent with this monodromy or (depending on the context) by imposing additional
conditions on the singular behavior along $U$.  We will call codimension two operators of this sort
monodromy defects.  We introduce this terminology because, in comparing related theories
in different dimension, we want  a way to emphasize the codimension rather than the
dimension on which the defect is supported.

Chern-Simons theory is a theory in dimension 3, and since $3-2=1$, in this case
the defect operators
defined by monodromy are also line operators, just like the Wilson
operators.\footnote{Similarly, since $3-3=0$, an 't Hooft operator in a
three-dimensional theory is simply a local operator.  However, the
Chern-Simons function $\CS(A)$ is not gauge-invariant in the presence of the
singularity corresponding to an 't Hooft operator, and hence there are no 't Hooft
operators in pure Chern-Simons theory.  't Hooft operators -- which in this context
are often called monopole operators -- do exist in Chern-Simons theories
with matter fields \cite{KS,KBar}.}  Moreover, in Chern-Simons
theory, the two types of line operator are equivalent.  This statement is a slight
reformulation of matters explained in \cite{witten} and \cite{wilecture} and in
much more detail in \cite{Beasley}.  The basic
reason for a relation between the two types of line operator
can be seen for $G=U(1)$.  Consider $U(1)$ Chern-Simons theory on a
three-manifold $W_3$ at level $k$, coupled to a knot $K$ that
is labeled by the charge $n$ representation of $U(1)$.  The action is
\begin{equation}\label{dolfox}I=-\frac{k}{4\pi}\int_{W_3}
A\wedge \d A -n\oint_K A.            \end{equation}
The equation of motion is
\begin{equation}\label{zofox} F=-\frac{2\pi n}{k}\delta_K,\end{equation}
where $\delta_K$ is a delta function that is Poincar\'e dual to $K$.
This means that the gauge field $A$ has a singularity along $K$, the
monodromy around $K$ being $\eurm M=\exp(-2\pi i n/k)$.  It is equivalent to
consider Chern-Simons theory for ordinary $U(1)$ gauge fields on $W_3$ with a
Wilson operator of charge $n$ on the knot $K$ or Chern-Simons theory on
$W_3$ for $U(1)$ gauge fields that are required to have a singularity
along $K$ of the form (\ref{zofox}).

This construction is particularly simple for $G=U(1)$ because a representation is
one-dimensional and a Wilson operator $\exp(i n \oint_KA)$ is
constructed by exponentiating a local expression that can be included in the action.
The analog for a nonabelian gauge group $G$ with a Wilson line associated
to an irreducible representation $R$ is to include
in the microscopic description a matter system, supported on $K$, whose
quantization gives the representation $R$.  In view of the Borel-Weil-Bott theorem,
such a system is the theory of maps
$K\to G/T$, where the ``flag manifold'' $G/T$ is endowed with a homogeneous
line bundle whose first Chern
class is the highest weight $\lambda_R$ of the representation $R$.
Thus, one considers a quantum theory
of pairs $(A,\Phi)$, where $A$ is a connection on a
$G$-bundle $E\to W_3$ and $\Phi$ is a section of the
$G/T$ bundle $\mathcal E\to K$ that is associated to $E$
(if $E$ is understood as a principal $G$-bundle, one can set $\mathcal E=G/T\times_G E$).

After introducing $\Phi$, one can gauge $\Phi$ away,
since $G/T$ is a homogeneous space, and then the
equation of motion for $A$ takes the form of (\ref{zofox}) with the integer $n$
replaced by the Lie algebra element $\lambda_R$.  The monodromy
around $K$, if computed classically,
turns out to be $\eurm M=\exp(-2\pi \lambda_R^*/k)$.
($\lambda_R$ is naturally an element of $\frak t^\vee$; we have used the
usual metric in which short roots have length squared two to map $\lambda_R$ to an element
of $\frak t$ that we call $\lambda_R^*$.)  It is known, however, that many formulas take
their simplest form if $k$ is replaced by $\Psi=k+h\,\mathrm{sign}\,k$ and $\lambda_R^*$ by
$\lambda_R^*+\varrho^*$, where $\varrho$ is one-half the sum of the positive roots.
The shift from $k$ to $\Psi$ has an interpretation that was explained
in section \ref{onecite}, and this interpretation indicates that all formulas of
$\N=4$ super Yang-Mills theory should be expressed in terms of $\Psi$.   Unfortunately, we
do not have an equally clear picture of what the shift $\lambda_R\to \lambda_R+\varrho$
means in the context of $\N=4$ super Yang-Mills theory and hence we do not know
whether this shift should be included in the microscopic formulas in this description.
When we introduce the description by $\N=4$ super Yang-Mills theory, we will not
incorporate this shift, and thus we will take the monodromy to be $\eurm M=\exp(-2\pi \lambda_R^*/\Psi)$.  But this is only a provisional choice and is one of many points in the present section that
merit a more careful reconsideration.

\subsubsection{Contents Of This Section}\label{contents}

Although Wilson operators and monodromy defects are equivalent in Chern-Simons theory, they lead to two
quite different pictures when we lift to four dimensions.
A one-dimensional defect in three dimensions can be lifted to four dimensions as a one-dimensional defect.
This is what we have done in the present paper, beginning in section \ref{csfun}, in relating Wilson
operators in three dimensions to Wilson or 't Hooft operators in four-dimensional gauge theory.
Alternatively, a codimension two defect in three dimensions can be lifted to four dimensions
as a codimension two defect.  That will be our approach in the present section.  The use of codimension
two defects in four dimensions to describe Wilson operators in three dimensions is not essentially new;
this actually was done in \cite{wittentwo}.  The motivation
there was to study a semiclassical limit of Chern-Simons theory in which $k$ and $\lambda_R$
are both large, with a fixed ratio so that the monodromy $\eurm M$ remains fixed.  This semi-classical limit is
related to the volume conjecture  for Chern-Simons theory (see for instance \cite{Mura,Gu}), and related
developments.  In the present paper, we started with Wilson operators rather than monodromy defects
because this seemed to give the most direct route to Khovanov homology.  However, in the present
section we will describe at least the beginnings of an analogous story based on monodromy defects.

Monodromy defects in four dimensions are supported on a surface of dimension two and are often
called surface operators.  The appropriate ones were described in \cite{GuWRam} and will
be reviewed in
section \ref{zolg}, where we will also describe the basic four-dimensional construction that
is related to Chern-Simons theory in this perspective.  In section \ref{zelfus}, we describe
the $S$-dual construction in four dimensions, and the resulting formulas for knot
invariants, in terms of counting of solutions of elliptic differential equations.  In section \ref{liftoff},
we lift the story to five dimensions, giving a description of Chern-Simons theory in terms
of dimensions of vector spaces rather than counting of solutions, and in section \ref{liftsix}, we
make the further lift to an ultraviolet-complete description in six dimensions.  Finally, in
section \ref{trying}, we attempt to use this form of the duality to actually say something
about Chern-Simons knot invariants.
What we are able to say is quite limited.

Thus, in brief, in the rest of this paper, we aim to recapitulate what we have done
so far with
Wilson operators of Chern-Simons theory replaced by the equivalent monodromy defects.  But we
make only the barest beginnings in this direction.

\subsection{From Three Dimensions To Four}\label{zolg}

\subsubsection{Review Of Monodromy Defects}\label{olg}
\def\VV{{\mathcal V}}
Our first step is to relate Chern-Simons theory on a three-manifold $W_3$ to $\N=4$
super Yang-Mills theory on $V_4=W_3\times \R_+$, but now in the presence of a monodromy defect.
Just as in section
\ref{onecite}, in doing this, it is convenient to take the twisting parameter $t$ to be real,
so as to get a localization on the solutions of the elliptic differential equations
$\VV^+=\VV^-=\VV^0=0$.
And it is convenient to take the $Q$-invariant complex connection on the boundary of $V_4$ to
be simply $\A=A+i\phi$.

\def\SS{C}
A monodromy defect supported on a knot $K\subset W_3$ will be extended to a monodromy
defect in $V_4$.  The monodromy defect is defined by specifying
 the singularity that fields are supposed to have along a two-dimensional surface
$\SS\subset V_4$.  For our analysis, we will take $\SS=K\times \R_+$, but more generally
one may take $\SS$ to be any surface in $V_4$ whose boundary is the original knot $K\times \{0\}$.

The singularity along $\SS$ must be compatible with the localization equations $\VV^+=\VV^-=\VV^0=0$.
In fact, the relevant monodromy defects, which have been described in \cite{GuWRam},
are   half-BPS and are compatible with the localization equations for any value of the twisting
parameter $t$.

The singular solution that defines the monodromy defect operator is a solution on $\R^2$
with an isolated singularity at the origin $0\in \R^2$.  One can think of this $\R^2$ as the normal
plane to $\SS$.  The relevant solution on $\R^2$ is a solution of Hitchin's
equations
\begin{align}  F-\phi\wedge\phi & = 0\cr
               \d_A\phi & = 0 \cr
                \d_A\star\phi & = 0 \end{align}
for the pair $(A,\phi)$.  Any solution of these equations on $\R^2$, when
pulled back to $\R^4=\R^2\times \R^2$,  obeys the four-dimensional
equations $\VV^+=\VV^-=\VV^0=0$ for every value of $t$.  This is related to the fact
that Hitchin's equations are actually half-BPS, that is, they preserve one-half the supersymmetry
of $\N=4$ super Yang-Mills theory.

We will consider only the most basic monodromy defect operator considered in \cite{GuWRam}
(as opposed to refinements that depend on the choice of a non-minimal Levi subgroup of $G$).
The defect operator has parameters $(\alpha,\beta,\gamma,\eta)$.
Here $\alpha,\beta,$ and $\gamma$
are elements of the Lie algebra $\frak t$ of a maximal torus
$T\subset G$ (as described later, $\alpha$ is more precisely an element of
$\frak t/\Lambda_{\mathrm{cochar}}=T)$.
  Introducing polar coordinates
$r,\theta$ on $\R^2$, the singular solution of Hitchin's equations corresponding to
$\alpha,\beta,\gamma\in \frak t$ is
\begin{align}\label{thelx} A & = \alpha\,\d\theta\cr
                           \phi & = \beta \frac{\d r}{r}-\gamma \,\d\theta.\end{align}
The defect operator is defined by saying that one studies
$\N=4$ super Yang-Mills fields in a space of
fields that coincide with this singular solution modulo less
singular terms, that is, modulo terms with
a singularity milder than $1/r$.  As an important example of the subtlety
of this definition, let us consider
the case that $\alpha,\beta,\gamma\to 0$, or more generally, the case
that the triple $(\alpha,\beta,\gamma)$
becomes nonregular.  (We call this triple regular if the subgroup of $G$
that leaves fixed the solution
(\ref{thelx}) is only the maximal torus; more generally, we say that a collection of elements
of $\frak t$, $T$, and/or $T^\vee$ is regular if the collection is not left fixed by any nontrivial
element of the Weyl group.)
Naively, for $\alpha,\beta,\gamma\to 0$, it seems that the
singularity associated to the defect operator disappears, but the correct statement is
that the limit as $\alpha,\beta,\gamma\to 0$ is a surface
operator characterized by the fact that the singularity in the fields is milder than $1/r$.
The generic behavior of Hitchin's equations for  $\alpha,\beta,\gamma\to 0$ is given, as found
 in \cite{Simpson}, by a solution that is slightly less singular than $1/r$.  (We have seen
  a similar behavior in section \ref{anothersol}; for $\lambda\to 0$, the solution (\ref{nuphox}) does
 not become regular at $z=0$, but reduces to the solution (\ref{polzom}) that has
 a singularity that is slightly milder than $1/|z|$.)   The gauge theory surface operator with
 nonregular parameters
 must be defined to allow the same behavior, as explained in detail in \cite{GuWRam}.

The parameters $\alpha$ and $\gamma$ in (\ref{thelx}) have the following simple interpretation.
By virtue of Hitchin's
equations, the complex
connection $\A=A+i\phi$ is flat on the complement of the point $r=0$.
 Its monodromy around that singular point is
\begin{equation}\label{helx} \eurm M =\exp(-2\pi(\alpha-i\gamma)).\end{equation}
The combination $\beta+i\gamma$ also has a simple interpretation.  Write $\varphi$ for the $(1,0)$ part of
the one-form $\phi$; then away from the singularity, $\varphi$ is holomorphic by virtue of Hitchin's
equations.  It has a pole at $z=0$ with polar residue $(\beta+i\gamma)/2$:
\begin{equation}\label{omelx} \varphi=\frac{1}{2}(\beta+i\gamma)\frac{\d z}{z}.
\end{equation}
Because of the subtlety noted in the last paragraph, we have to be careful in interpreting these formulas
if the pairs $(\alpha,\gamma)$ or $(\beta,\gamma)$ are nonregular.  For example, for $\alpha=\gamma=0$,
although the model solution has monodromy $\eurm M=1$, a generic solution
that coincides with the model solution modulo terms less singular than $1/r$, and therefore
is allowed in the presence of the monodromy defect, has
nontrivial but unipotent monodromy (that is, $\eurm M-1$ is nilpotent but otherwise unconstrained).
This is relevant in the $G^\vee$ description
introduced in section \ref{zelfus}, because there the vanishing of the parameters analogous
to $\alpha$ and $\gamma$ will be natural.

The fourth parameter $\eta$ has a more quantum mechanical nature.   As long as $(\alpha,\beta,\gamma)$ is a regular triple, the presence
along a surface $\SS\subset V_4$ of a singularity of the form (\ref{thelx}) means that, along $\SS$,
the structure group of the $G$-bundle $E\to V_4$ is reduced to $T$.  For $G=SU(2)$, this means that
the structure group of $E|_\SS$ reduces to $T=U(1)$.
A $U(1)$ bundle over a two-manifold $\SS$ has a $\Z$-valued
first Chern class
$c_1$. We can introduce  a theta-angle $\eta$ and include in the path integral a factor
$\exp(2\pi  i\eta c_1)$. If $G$ is of rank greater than one, then, as explained in \cite{GuWRam},
 a $T$-bundle
over $\SS$ has a natural characteristic class $\eurm m$ that takes values in a lattice in $\frak t$ that
is known as the cocharacter lattice $\Lambda_{\mathrm{cochar}}$.  The generalization of a theta-angle is
a homomorphism from $\Lambda_{\mathrm{cochar}}$ to $U(1)$; we write this homomorphism as
$\eurm m\to \exp(2\pi i (\eta,\eurm m))$, where $\eta$ takes values in $\frak t^\vee/\Lambda_{\mathrm{char}}$.
Here $\frak t^\vee$ is the dual of $\frak t$ and $\Lambda_{\mathrm{char}}\subset\frak t^\vee$ is the
character lattice.  (More informally, $\eta$ is simply a collection of theta-angles, one for each $U(1)$
subgroup of $T$.)  Moreover, $\frak t^\vee/\Lambda_{\mathrm{char}}$ is naturally isomorphic to the maximal
torus $T^\vee$ of the GNO or Langlands dual group $G^\vee$.

Reciprocally, a gauge transformation with a singularity at $r=0$ can shift $\alpha$ by an element of
$\Lambda_{\mathrm{cochar}}$, so $\alpha$ is naturally an element of $\frak t/\Lambda_{\mathrm{cochar}}$,
which is the maximal torus $T\subset G$.  The element of $T$ corresponding to $\alpha$ is simply
$\exp(-2\pi\alpha)$.

The quadruple of parameters $(\alpha,\beta,\gamma,\eta)$ thus take values in $T\times \frak t \times \frak t
\times T^\vee$ or more precisely in the quotient of this space by the Weyl group of $G$.
Under electric-magnetic duality, $T$ and $T^\vee$ are exchanged, and $\frak t$ is mapped
to $\frak t^\vee$.  A metric on $\frak t$ gives a map from $\frak t$ to $\frak t^\vee$; we use the usual
metric in which short roots have length squared 2, and write $\beta^*$ and $\gamma^*$ for the images of $\beta$
and $\gamma$ in $\frak t^\vee$. The electric-magnetic duality transformation
$\tau \to -1/\frak n_{\frak g}\tau$ then maps the quadruple $(\alpha,\beta,\gamma,\eta)$ to the
quadruple $(\alpha^\vee,\beta^\vee,\gamma^\vee,\eta^\vee)$ defined by \cite{GuWRam}
\begin{equation}\label{zorox}(\alpha^\vee,\beta^\vee,\gamma^\vee,\eta^\vee)=
(\eta,|\tau|\beta^*,|\tau|\gamma^*,
-\alpha). \end{equation}

If the triple $(\alpha,\beta,\gamma)$ is nonregular, then our definition of
$\eta$ does not make sense.  For example, if $G=SO(3)$, the only nonregular triple is $\alpha=\beta=\gamma=0$;
this leaves $SO(3)$ unbroken and so the reduction of the structure group of $E|_\SS$ to $T$, which we
assumed in the definition of $\eta$, does not hold.  Nevertheless, there is a well-behaved surface operator
as long as the quadruple $(\alpha,\beta,\gamma,\eta)$ is regular. For example, a surface operator
with parameters $(0,0,0,\eta)$ is hard to define directly in terms of $G$ gauge theory, but in the
$S$-dual description by $G^\vee$ gauge theory, the parameters are $(\eta,0,0,0)$, and now it is obvious that
there is no problem as long as $\eta$ is regular.  An alternative description of the surface operator
which makes it clear that it behaves well as long as the quadruple $(\alpha,\beta,\gamma,\eta)$ is regular
is presented in section 3 of \cite{GuWSurf}.  In this approach, the surface operator is defined by coupling
gauge fields on the four-manifold $V_4$ to a supersymmetric sigma-model that is supported on the two-manifold
$\SS\subset V_4$.  In this description, $\alpha,\beta,\gamma$, and $\eta$ are parameters of the sigma-model.
The sigma-model becomes singular (Coulomb and Higgs branches intersect) precisely when the quadruple
$(\alpha,\beta,\gamma,\eta)$ is nonregular.

In section \ref{zelfus}, we will use $G^\vee$ gauge theory to develop a semiclassical method to calculate
in the presence of a monodromy defect.  Even though the monodromy defect makes sense as long as the quadruple
$(\alpha^\vee,\beta^\vee,\gamma^\vee,\eta^\vee)$ is regular, a semiclassical picture based on $G^\vee$
gauge theory is possible only under the stronger condition that $(\alpha^\vee,\beta^\vee,\gamma^\vee)$
is regular.  So we will usually make this assumption.

\subsubsection{Specialization To $V_4=W_3\times \R_+$}\label{special}

So far, we have considered a monodromy defect supported on an arbitrary surface $\SS$ in a general
four-manifold $V_4$.   Now let us specialize to the case that $V_4=W_3\times \R_+$ with $\SS=K\times \R_+$,
$K$ being a knot in $W_3$.  Moreover, since our interest is in Chern-Simons theory, we assume
that the boundary conditions at $y=0$ are the D3-NS5 boundary conditions discussed in section
\ref{csfun}, or their generalization discussed from a more purely topological field theory point of view
in \cite{wittenthree}.

The starting point in relating Chern-Simons theory on $W_3$ to $\N=4$ super Yang-Mills theory on $W_3\times \R_+$
is supposed to be that, given a critical point of the Chern-Simons function on $W_3$, one uses
this critical point
to define boundary conditions at $y=\infty$ for the $\N=4$ path integral.
To be more precise, in the absence of a knot, a critical point is a flat bundle $E\to W_3$, and, invoking
a theorem of Corlette \cite{corlette}, such a flat bundle (given a mild condition of semi-stability)
can be promoted to a solution of the supersymmetric equations, which in three dimensions
read $F-\phi\wedge\phi = \d_A\phi =\d_A\star\phi=0$.  In the presence of a knot $K$ labeled by parameters
$\alpha,\beta,\gamma$, these equations acquire delta function sources:
\begin{align}\label{lanko} F-\phi\wedge\phi &=2\pi \alpha\,\delta_K   \cr
                \d_A\star\phi &= 2\pi \beta \,\d s\wedge \delta_K \cr
                 \d_A \phi &= 2\pi \gamma \,\delta_K. \end{align}
In these equations, $\delta_K$ is a delta function two-form Poincar\'e dual to $K$, and $\d s$ is a one-form
defined along $K$ that measures the length element of $K$ defined using the Riemannian metric on $W_3$.
(Multiplying
it by $\delta_K$, we promote it to a closed three-form $\d s\wedge \delta_K$ on $W_3$.)
A generalization of Corlette's theorem to include such singularities is apparently not known in the context
of Riemannian geometry, though there are such results in the context of Kahler manifolds, the most basic
case being a Riemann surface \cite{Simpson}.
Given a solution of these equations, we use it to define initial conditions for the Morse theory flow equations
at $y=\infty$.  The space of solutions of the flow equations gives an integration cycle $\Gamma$ for
Chern-Simons theory on the boundary at $y=0$, in the presence of a monodromy defect. This procedure has  been described in  \cite{wittentwo}, though without the physical interpretation
by $\N=4$ super Yang-Mills theory.

The $\N=4$ path integral on $W_3\times \R_+$
with the given boundary conditions at $y=\infty$ reproduces the path integral of Chern-Simons theory
on the integration cycle $\Gamma$.  However, we do face the fact that, at least generically,
$\Gamma$ is not
equivalent to any standard integration
cycle of Chern-Simons theory.  In our earlier analysis in which knots were
associated to Wilson operators rather than monodromy defects, to partly avoid this problem,
we relied on the fact that there is an important case
in which there is only one possible integration cycle.  This was the case $W_3=\R^3$: as $\R^3$ is
simply-connected, the Chern-Simons functional for gauge fields on $\R^3$ has only one critical
point up to a gauge transformation,
and any possible integration cycle is equivalent to  the standard
one.  Hence results obtained by
the  procedure of the present paper can be compared to results of ordinary Chern-Simons theory for
expectation values of
knots in $\R^3$.  As soon as we allow a monodromy defect operator supported on some $K\subset \R^3$,
the critical point and the integration cycle are no longer unique. (This is because there typically are inequivalent
flat connections over $\R^3\backslash K$ with prescribed monodromy around $K$.)  We will try to find something almost
as convenient as we had from the Wilson loop point of view, but this will involve some assumptions and
to some extent has been included in the present paper only to orient the reader about what one
might hope for.

From the point of view of Chern-Simons theory, the natural problem involving a monodromy
defect was described in section \ref{someb}: it is a path integral in the space of gauge
fields on $W_3$ that have a singularity along $K$ with prescribed monodromy.  For simplicity, we assume that the monodromy is given by a semisimple (diagonalizable) element
$\eurm M\in G_\C$ (the more general case is discussed in \cite{wittentwo}).  Then $\eurm M$ can be conjugated
to the complex maximal torus $T_\C\subset G_\C$ and has the form $\eurm M=\exp(-2\pi(\alpha-i\gamma))$, with
$\alpha,\gamma\in \frak t$.  To describe a Chern-Simons path integral for gauge fields with monodromy
conjugate to $\eurm M$, we must use a monodromy defect operator with $\alpha$ and $\gamma$ as two of its parameters.

What about the other parameters $\beta$ and $\eta$?  We must set the parameter $\eta$ to zero for the
following reason.  What $\eta$ multiplies is supposed to be a topological invariant, which for $G=SU(2)$
would be the first Chern class of a $U(1)$ bundle over $\SS = K\times \R_+$.  To define the first Chern
class as a topological invariant on the non-compact Riemann surface $\SS$, one needs trivializations
of the $U(1)$ bundle at both $y=\infty$ and $y=0$. Although our boundary condition does allow a trivialization
at $y=\infty$, it does not allow a trivialization at $y=0$, where arbitrary fluctuations in $\A=A+i\phi$
are allowed.  More fundamentally, the integration cycle in Chern-Simons theory defined by Morse theory flow
from a critical point (or even a connected family of critical points) is connected, so there is no hope of decomposing it in components according to the values of
a generalized first Chern class.

As regards the parameter $\beta$, it has no natural meaning in Chern-Simons theory. This makes one wonder
if one should set $\beta$ to zero, but that does not seem to be the case in general.  Given a flat bundle
$E\to V_4\backslash C$, for any value of $\beta$  for which we can find a solution of eqns. (\ref{lanko}),
we can use this to give a boundary condition on $\N=4$ super Yang-Mills theory at $y=\infty$.  Since $\beta$ has no role in the Chern-Simons interpretation of the theory, one would
expect the resulting path integral on $W_3\times \R_+$ to be independent of the choice of $\beta$.
A smooth deformation of the integration cycle $\Gamma$, such as one gets by varying $\beta$,
should not change its homology class.

There is, however, one important situation in which $\beta$ must definitely be set to zero.  Suppose that
$G=U(1)$.  Then the second equation in (\ref{lanko}) reduces to $\d\star\phi = 2\pi \beta\, \d s\wedge
\delta_K$,
and this equation has no solution except for $\beta=0$.  The reason for this last statement is that the closed
three-form $\d s\wedge\delta_K$ represents a nonzero element of de Rham cohomology
(its integral is the circumference
of the knot $K$), so unless $\beta=0$, the closed form $2\pi\beta\,\d s\wedge \delta_K$ cannot be written
as $\d\star\phi$ for any $\phi$.

More generally, for any $G$, in the case of a flat bundle $E\to W_3\backslash K$ whose monodromy reduces
to an abelian subgroup of $G$, the same argument shows that we must take $\beta=0$.

It seems likely that what has just been described is essentially the only obstruction to varying $\beta$
away from zero, and that for example in the case of an irreducible flat $G_\C$-bundle $E\to W_3\backslash K$,
one may take arbitrary $\beta$.  However, as already noted, the appropriate generalization of Corlette's
theorem does not appear to be available in the literature.

Comparing the formula $\eurm M=\exp\left(-2\pi(\alpha-i\gamma)\right)$ to the discussion
at the end of section \ref{someb}, we see that
if we want to use $\N=4$ super Yang-Mills theory with a monodromy defect to generate a Chern-Simons path
integral (albeit on an unusual integration cycle) with a Wilson
loop in the representation $R$, we must relate the parameters by
\begin{equation}\label{gufilo} \frac{\lambda_R^*}{\Psi}=\alpha-i\gamma.\end{equation}
Here as usual $\Psi=k+h\,\sign(k)$, and the formula is provisional in the sense that
possibly we should replace $\lambda_R$ by $\lambda_R+\varrho$.  A notable fact is that, since
$\lambda_R^*,\,\alpha$, and $\gamma$ are all
elements of the real Lie algebra $\frak t$, in order to have  $\gamma\not=0$ we must take $\Psi$ off the real axis.
In this case, $q=\exp(2\pi i/\frak n_{\frak g}\Psi)$ does not have modulus 1, and a description
by ordinary Chern-Simons theory (in which $k$ and $\Psi$ are integers) is not possible.  In any event,
from the point of view of $\N=4$ super Yang-Mills theory, we are certainly not limited to values of $\alpha$,
$\gamma$, and $\Psi$ that obey a relation such as (\ref{gufilo}).

\subsubsection{An Important Detail}\label{zongot}

In the standard perturbative expansion of Chern-Simons theory on a three-manifold $W_3$
around a flat
connection $\A_\rho$ associated to a representation $\rho$ of the fundamental group, the leading contribution in the semiclassical limit is simply the
exponential of the classical action $\exp(-i k \CS(\A_\rho))$.  A one-loop correction converts
this to
\begin{equation}\label{moppo}Z_{\CS}\sim  \exp(-i\Psi \CS(\A_\rho)),\end{equation}
and this is the leading behavior of the Chern-Simons partition function for large $\Psi$.

In the analogous calculation in $\N=4$ super Yang-Mills on $V_4=W_3\times \R_+$, we use $\A_\rho$
to define a boundary condition at $y=\infty$.  To emphasize this, in the $\N=4$
context, we write $\A_\infty$ instead of $\A_\rho$.  Apart from an inessential $Q$-exact
term, the $\N=4$ description differs from the Chern-Simons description by an important
constant in the action -- the constant $-i\Psi \,\CS(\A_\infty)$, which can be found in eqn. (\ref{muster}).
   This means that while
the leading behavior of the Chern-Simons path integral expanded around a flat connection
$\A_\rho=\A_\infty$ is the exponential factor (\ref{moppo}), this factor is completely absent in
the corresponding $\N=4$ path integral: it cancels between $y=0$ and $y=\infty$.
The relation between them is
\begin{equation}\label{oppo} Z_{\CS}=\frak N_0 \exp(-i\Psi\CS(\A_\infty))Z_{\N=4},\end{equation}
where we allow for the possibility of a constant factor $\frak N_0$ as in (\ref{zondo}).

This is not important in studying knots in $\R^3$ via Wilson loops, because in that context
$\A_\infty$ is trivial.  However, when we study knots via monodromy defects, $\A_\infty$ has
a prescribed monodromy around $K$ and is not trivial.

In the present paper, we will consider one question for which this is important.  This
is the framing anomaly for knots.  Under a change in framing of a knot $K$, $Z_\CS$
transforms by a power of $q$ -- the framing anomaly.  But in fact, the exponential of the
classical action $\exp(-i\Psi\CS(\A_\infty))$ itself has a framing anomaly.  As we will now explain,
in a sense most of the framing anomaly is contained in the classical action and only a
quantum correction to the framing anomaly is contained in $Z_{\N=4}$.

Consider first the case $G=U(1)$.  Inserting a Wilson operator $\exp(in\oint_K A)$ in effect
adds a linear term to the action, namely the second term in eqn. (\ref{dolfox}).  Since the
action is quadratic in $A$, once we shift to a classical solution in the presence of the knot,
the linear term in the action disappears.  At this point, except for an additive constant -- the value of the action
at the classical solution -- the action coincides with what it would be in the absence
of the knot,
and the rest of the quantum computation proceeds as if the knot were
absent.  Hence, for $U(1)$ gauge theory, the framing anomaly for knots arises entirely
from the evaluation of the classical action.  For a discussion of the $U(1)$ framing anomaly
in this vein, see \cite{Marino}, section 2.4.

The result of the computation is that for $U(1)$ Chern-Simons theory, the partition function transforms under a unit
change in framing of a knot by
\begin{equation}\label{fruit}Z_\CS\to Z_\CS\, q^{n^2/2}=Z_\CS\exp(\pi i \Psi m^2)\end{equation}
where we use the fact that $\Psi=k$ for $U(1)$, and $m=n/k=n/\Psi$ is essentially the logarithm
of the monodromy around the knot (that monodromy is $\eurm M=\exp(-2\pi i m)$, as we explained in
relation to (\ref{zofox})).  We stress that this formula is purely classical in the sense that it
comes entirely from evaluating the classical action.

For a general compact Lie group, the analog is
\begin{equation}\label{ruit}Z_{\CS}\to Z_{\CS} \,q^{\ng(\lambda_R
+2\varrho,\lambda_R)/2},  \end{equation}
where $(~,~)$ is the usual inner product on $t^\vee$ in which short roots have length squared
two, and $(\lambda_R,\lambda_R+2\varrho)/2\Psi$, which reduces to $n^2/2k$ in the
abelian case, is the dimension of a chiral primary field of highest weight $\lambda_R$ in
two-dimensional current algebra at level $k$.  As usual, $q=\exp(2\pi i/\ng\Psi)$, so the factor of $\ng$
is absent if the formula is written in terms of $\Psi$.

In the same sense that the framing anomaly for knots is entirely classical in abelian gauge
theory, it is mostly classical in the nonabelian case.
If $G$ is a nonabelian group, then the flat connection $\A_\infty$ over $W_3\backslash K$
may have nonabelian monodromy.  But its restriction to a neighborhood
of $K$ in $W_3\backslash K$ is always
abelian, since the fundamental group in such a neighborhood is the abelian group $\Z\times \Z$.
The classical part of the framing anomaly comes only from the behavior of the classical
solution near $K$, and can be obtained from the abelian formula (\ref{fruit}) by replacing
$m^2$ by $(m,m)$, where $m$ is the logarithm of the monodromy.  For $m$ we will
take $\lambda_R^*/\Psi$, as explained at the end of section (\ref{someb}). But this choice
really needs more justification; it is not clear whether we should be making a shift
$\lambda_R\to\lambda_R+\varrho$.    At any rate, with our choice, we can express the
factor by which $Z$ transforms under a change in framing as
\begin{equation}\label{uit}q^{\ng(\lambda_R+2\varrho,\lambda_R)/2}
=\exp(\pi i\Psi(m,m))q^{\ng(\lambda_R,\varrho)}.\end{equation}
On the right hand side, the first factor is classical and the second, which is subleading
in the semiclassical limit (large $\Psi$ with fixed $m$), is a quantum correction.  However, there has been
some guesswork in the way we have written the formula.

The reason that we have made this decomposition is the following.  In view of the
formula (\ref{oppo}), the classical part of the framing anomaly in $Z_\CS$ is contained
in the factor $\exp(-i\Psi\CS(\A_\infty))$.  Only the quantum correction to the framing
anomaly will appear in $Z_{\N=4}$.  If therefore we accept the decomposition (\ref{uit})
at face value, then the transformation of $Z_{\N=4}$ under a unit change in the framing of a knot
will be
\begin{equation}\label{toxi}Z_{\N=4}\to Z_{\N=4}\,q^{\ng(\lambda_R,\varrho)}.\end{equation}

\subsection{The $S$-Dual In The Presence Of A Monodromy Defect}\label{zelfus}

The next step is $S$-duality.  The gauge group is transformed from $G$ to $G^\vee$, and the
boundary condition at $y=0$ becomes that of a D3-D5 system.  The partition function can be evaluated
by counting solutions of the supersymmetric equations $\VV^+=\VV^-=\VV^0=0$ with the appropriate elliptic
boundary conditions.

In particular, the boundary condition at $y=0$, away from the monodromy defect, is the familiar one
associated with a regular Nahm pole. Near the monodromy defect, the boundary condition
must be modified.  As usual the corrected boundary condition is based on a model solution.
The model solution should now be a solution on $\R^2\times \R_+$ of the three-dimensional reduction of
our supersymmetric equations. We assume that a monodromy defect is present on the ray $\ell=p\times \R_+$,
with $p$ some point in $\R^2$.  Near any point in $\R^2$ except
 $p$, the model solution should have a regular Nahm pole, and around any point of the
ray $\ell$ except the endpoint at $y=0$, it should have the singularity (\ref{thelx}) of
a monodromy defect.  The interest in the model solution is its behavior
 at the exceptional point $p\times \{y=0\}$ where $\ell$ meets the boundary; whatever this behavior is,
 we define a boundary condition by requiring  this behavior where
 a monodromy defect meets the boundary.   Happily, for $G^\vee = SO(3)$, the requisite model solutions
 have been found, though not in complete generality, in section \ref{anothersol}.  Eqn. (\ref{nuphox}) is
 the  solution with
 $\alpha^\vee=0$, $\beta^\vee,\gamma^\vee\not=0$; eqn. (\ref{omigo}) corresponds to $\alpha^\vee\not=0$,
 $\beta^\vee=\gamma^\vee=0$; and eqn. (\ref{polzom}) exhibits the subtle behavior for $\alpha^\vee,\beta^\vee,
 \gamma^\vee\to 0$.

Now let us discuss what values we should take for the parameters $(\alpha^\vee,\beta^\vee,\gamma^\vee,
\eta^\vee)$ in the context of topological field theory on $W_3\times \R_+$.  Since $\alpha^\vee$ corresponds
to $\eta$ in the description of section \ref{zolg}, and in that context we had to set $\eta=0$, we expect
that we will have to set $\alpha^\vee=0$.  Indeed, there is a simple reason for this, which can be stated
most briefly for $G^\vee=SO(3)$.  The solution (\ref{omigo}) with $\alpha^\vee\not=0$ makes perfect sense
when $W_3$ is flat, but has a monodromy $\exp(-2\pi\alpha^\vee)$ around $K$.
However, as we know from section \ref{vivisect}, away from a monodromy defect,
the $G^\vee$ bundle $E\to W_3$ is the tangent bundle to $W_3$ with its Riemannian connection.  For generic
$W_3$, the Riemannian connection is irreducible and there is no way to ``twist'' it by a monodromy $\exp(-2\pi
\alpha^\vee)$ around a knot $K\subset W_3$, while leaving it locally
unchanged up to gauge transformation on the complement
of $K$.  Hence, the boundary condition of the D3-D5 system with generic $W_3$ and a monodromy defect only
makes sense if $\alpha^\vee=0$.

As for $\beta^\vee$, its status seems to be just parallel to that of $\beta$ in the context of the
Chern-Simons like description.  At $y=\infty$,
we pick a homomorphism $\rho^\vee:\pi_1(W_3\backslash K)\to G^\vee$,
and then try to promote this to a solution of the supersymmetric equations (\ref{lanko}) in the presence of the monodromy defect, now with parameters $\alpha^\vee,\,\beta^\vee,\,\gamma^\vee$, of course.  For a given $\rho^\vee$, we may use whatever
$\beta^\vee$ is compatible with the equations.

To understand $S$-duality between the two descriptions, we need to know how the homomorphism $\rho:\pi_1(W_3\backslash K)\to G$ that is used to determine
a boundary condition at $y=\infty$ on one side of the duality is related to the homomorphism
$\rho^\vee:\pi_1(W_3\backslash K)\to G^\vee$ that is similarly used on the other side.
We get a clue from the hypothesis that the only case in which $\beta$ or $\beta^\vee$ must
vanish is an abelian representation.  The relation $\beta^\vee=|\tau|
\beta^*$ shows that
$\beta^\vee$ is constrained to vanish if and only if $\beta$ is so constrained.  So we are led
to conjecture that  $\pi_1(W_3\backslash K)$  is mapped by $\rho$ to a commutative subgroup of
$G$  if
and only if it is mapped by $\rho^\vee$ to a commutative subgroup of $G^\vee$.  This conjecture
is particularly powerful if $W_3=S^3$, for then there is precisely one choice of $\rho$ or $\rho^\vee$
with given monodromy around $K$ and with abelian image.  (This statement would not
hold if we replace the knot $K$ by a link with several components.)  So in that case, the conjecture
is that the abelian representation $\rho$ is mapped to the abelian representation $\rho^\vee$.

More generally, the number of free parameters in the choice of $\beta$ or $\beta^\vee$
is the rank of $G$ minus the rank of the automorphism group of $\rho$ or $\rho^\vee$. So a generalization of the
above argument indicates that the map from $\rho$ to $\rho^\vee$ preserves the rank of
the automorphism group.

As for the other parameters, from (\ref{zorox}) we have $\eta^\vee=-\alpha$, $\gamma^\vee=
|\tau|\gamma^*$.  In the Chern-Simons-like description, the model depends holomorphically on the
logarithm of the monodromy $\alpha-i\gamma$, so in the dual description, it depends
holomorphically on $\eta^\vee+i\gamma$.

An important detail is dual to the discussion of eqn. (\ref{omelx}).  In the $G^\vee$ description,
for  $\alpha^\vee=\gamma^\vee=0$,  the monodromy around $K$ is unipotent, but not necessarily 1.

\subsubsection{The Partition Function}\label{thepar}

  In section \ref{dualities},
solutions of the supersymmetric equations $\VV^+=\VV^-=\VV^0=0$ were labeled by
the instanton number $\EP$ (whose precise definition depended on a framing of both
$W_3$ and $K$).  The contribution of a given solution to the partition function was $(-1)^gq^\EP$
where $(-1)^g$ is the sign of the fermion determinant in expanding around the given
solution, $\EP$ is its instanton number, and $q=\exp(2\pi i/\frak n_{\frak g}\Psi)$.  In the present context, assuming $\beta^\vee$
and $\gamma^\vee$ are not both zero (we have set $\alpha^\vee=0$), there is an additional
topological invariant.  When the $G^\vee$ bundle $E\to V_4$ is restricted to a two-manifold
$\SS\subset V_4$,
its structure group reduces to $T^\vee$, so roughly speaking it has a generalized first Chern class
$\eurm m^\vee$
valued in $\Lambda_{\mathrm{char}}$.  (We postpone to section \ref{revisited} some subtleties that arise
if $\SS$ is not compact, which is the case in our application to knots.)

How does the contribution of a given classical solution to the partition function depend on
$\eta^\vee$ and $\gamma^\vee$?  The dependence on $\eta^\vee$ is a simple factor of
$\exp\bigl(2\pi i (\eta^\vee,\eurm m^\vee)\bigr)=\exp\bigl(-2\pi i(\alpha,\eurm m^\vee)\bigr)$.
Since the partition function is holomorphic in $\alpha-i\gamma$,  the full dependence on
$\alpha$ and $\gamma$ must be a factor $\exp\bigl(-2\pi i(\alpha-i\gamma,\eurm m^\vee)\bigr)$.
We will not show explicitly how to calculate the $\gamma$-dependence, but we expect
that this will involve a computation somewhat analogous to eqns.  (\ref{omox}) and (\ref{uster}): in the presence of a
monodromy defect, when one writes the action as a $Q$-exact term plus a topological invariant,
the topological invariant includes a multiple of $(\gamma,\eurm m^\vee)$.

We can now write a formula for the partition function along the lines of eqn. (\ref{unky}).  Let $S$
be the set of solutions of the supersymmetric equations.  For $s\in S$, let $n_s$,
$\eurm m^\vee_s$, and $(-1)^{g_s}$ be the values of $\EP$, $\eurm m^\vee$, and
the sign of the fermion determinant for the classical solution corresponding to $s$.
The partition function is then
\begin{equation}\label{mombo} Z(q)=\sum_{s\in S} q^{n_s}\exp(-2\pi i(\alpha-i
\gamma,\eurm m^\vee_s))(-1)^{g_s}.\end{equation}
Making use of (\ref{gufilo}) and the definition of $q$, we can write this as
\begin{equation}\label{umombo}Z(q)=\sum_{s\in S}q^{n_s-\frak n_{\frak g}(\lambda_R,\eurm m^\vee_s)}(-1)^{g_s}.\end{equation}
Alternatively, let $w_{r,\eurm c}$ be the ``number'' of solutions of $\EP=r $ and $\eurm m^\vee
=\eurm c$, where in computing this number we weight each solution with the sign of the fermion
determinant.  Then
\begin{equation}\label{dozzo} Z(q)=\sum_{r,\eurm c}w_{r,\eurm c}q^{r-\frak n_{\frak g}(\lambda_R,\eurm c)}.\end{equation}
These formulas have the usual proviso that $\lambda_R$ should possibly be replaced by
$\lambda_R+\varrho$.

To be more exact, though we have kept the notation minimal, all these formulas describe a partition function in $\N=4$ supersymmetric $G^\vee$ gauge theory with a boundary condition at $y=\infty$ set
by a suitable homomorphism $\rho^\vee:\pi_1(W_3\backslash K)\to G^\vee$, and with a monodromy
defect operator whose parameters are determined by the representation $R$ of $G$.   For some
purposes, it may be best to write these formulas in terms of the logarithm of monodromy $\alpha-i\gamma$, but as they can be elegantly written in terms of $\lambda_R$, we have done so.

\subsubsection{The Framing Anomaly Revisited}\label{revisited}

For the case $V_4=W_3\times \R_+$,
$\SS=K\times \R_+$, because $\SS$ is not compact, the definition of $\eurm m^\vee$
depends on  a trivialization of $E|_\SS$ at both ends of $\R_+$.  The dependence on a choice of
trivialization at $y=\infty$ means that the right topological data in fixing the boundary
condition at infinity is a little more than the choice of $\rho^\vee$, but we will not say
more about this.

The dependence on
the trivialization at $y=0$ leads to a framing anomaly for the $\N=4$
partition function on $W_3\times \R_+$ in the presence of a monodromy defect.   We can see
this as follows.
The restriction of the $G^\vee$ bundle $E\to W_3\times \R_+$ to the boundary
$W_3\times \{y=0\}$
is the tangent bundle $TW_3$ of $W_3$, or more exactly it is the $G^\vee$ bundle associated
to the $SO(3)$ bundle $TW_3$ by a principal embedding $\xi:\frak{su}(2)\to \frak g^\vee$.
A framing of the knot $K$ trivializes the restriction of $TW_3$ to $K$, so it trivializes the
restriction of $E|_\SS$ to $\SS\cap \{y=0\}$.  Thus a framing of $K$ (together with
whatever data was used at $y=\infty$) makes $\eurm m^\vee$ well-defined, so that we can write
the formula (\ref{dozzo}) for the $\N=4$ partition function.  Under a unit change of framing of $K$,
$\eurm m^\vee$ transforms to $\eurm m^\vee-\varrho$, and this gives the expected formula (\ref{toxi}).   The statement
about how $\eurm m^\vee$ transforms under a change in framing amounts to the following.
For $G^\vee=SO(3)$, a unit change of framing shifts $\eurm m^\vee$ by one unit, that is by
$\varrho_{SU(2)}$. (A weight of $G=SU(2)$ is an element of $\frak t^\vee=\frak t_{SO(3)}$, so
in particular $\varrho_{SU(2)}\in \frak t_{SO(3)}$.) A minus sign comes from comparing orientations. For general $G$, the homomorphism $\xi:\frak{su}(2)\to \frak{g}$ maps $\varrho_{SU(2)}\in \frak t_{\frak{so}(3)}$ to $\varrho=\varrho_{G}\in \frak t^\vee\subset\frak g^\vee$ (this is a standard fact about principal $\frak{su}(2)$
subalgebras), and this gives our result.   But since we do not really know where the
shift $\lambda_R\to\lambda_R+\varrho$ should enter in the present formalism, what we have
described is more a scenario  than a derivation of the framing anomaly.

\subsection{Lifting To Five Or Six Dimensions}\label{liftoff}

\subsubsection{Five Dimensions}\label{five}

The next step is to lift to five dimensions, following the same logic as in section \ref{tdual}.
We promote the solutions of the four-dimensional equations $\VV^+=\VV^-=\VV^0=0$ on
$V_4=W_3\times\R_+$ to time-independent solutions of the five-dimensional supersymmetric
equations (\ref{torm}) on $S^1\times V_4$. Here $S^1$ is viewed as the time direction. We lift the monodromy defect supported  on $K\times \R_+\subset V_4$ to a monodromy defect supported on $S^1\times K\times \R_+$.

The basic idea  of a monodromy defect in five-dimensional super Yang-Mills theory on a five-manifold $M_5$ is similar
to what it is in four dimensions, and can be described without specializing to the setting of
the present paper.   The support of a monodromy defect is now a three-manifold $U$,
which is of codimension two in $M_5$.
As long as the triple of parameters $(\alpha^\vee,\beta^\vee,\gamma^\vee)$ is regular,
a monodromy defect in five dimensions can be defined by postulating in the normal plane
to $U$ the same type of singularity as in eqn. (\ref{lanko}).  For $\phi$ in this formula,
we take two of the scalar fields of five-dimensional super Yang-Mills theory.  Which two
depends on the context.  In our application, $M_5=M_4\times \R_+$, $U=\SS\times
\R_+$ for some $\SS\subset M_4$, and three of the scalar fields are twisted to a field
$B\in \Omega^{2,+}(M_4)\otimes \mathrm{ad}(E)$.  Along $ \SS$,  $\Omega^{2,+}(M_4)$
has the decomposition (\ref{restc}) with a two-dimensional real subbundle corresponding to
$\mathfrak \L$,
and the part of $B$ valued in this subbundle is what appears in the five-dimensional
analog of (\ref{lanko}).

The most striking difference from four dimensions is possibly that the monodromy defect
operator has no parameter corresponding to $\eta^\vee$, because the generalized first
Chern class is now associated not to a spacetime history but to a physical state.
In other words, if the triple $(\alpha^\vee,\beta^\vee,\gamma^\vee)$ is regular, then the
bundle $E\to M_5$, when restricted to $U$, has abelian structure group $T^\vee$
and its curvature  is a $\frak t^\vee$-valued closed two-form $f$ that is defined along $U$.
Then $\star_U f$ (here $\star_U$ is the Hodge star operator for the three-manifold
$U$) is a conserved current defined on $U$.  Its integral on an initial
value surface $C\subset U$ is a conserved quantity in the sense that it
only depends on the homology class of $C$.  We call this conserved quantity
$\eurm m^\vee$.  (What $\eurm m^\vee$ means when the triple $(\alpha^\vee,\beta^\vee,\gamma^\vee)$ is
nonregular will be explained in section \ref{nonreg}.
Technical issues in the definition of $\eurm m^\vee$ involving the fact that in our application to knots,
the relevant $C$ is not compact were discussed in section \ref{revisited}.)

For our application, we take $M_5=\R\times W_3\times \R_+$, $U=\R\times K\times \R_+$,
where $K$ is a knot in the three-manifold $W_3$.  The space $\K$ of physical states defined
on the initial value surface $K\times \R_+$
is then graded by the conserved charges $\EP,\,\EF$, and $\eurm m^\vee$.

The time-independent solutions on $M_5$  supply a basis for a space $\K_0$
of approximate supersymmetric ground states.
A salient fact here -- just as in the absence of the monodromy defect -- is that from a four-dimensional perspective, a time-independent solution
has a $\Z_2$-valued invariant, the sign of the fermion determinant.  But from a five-dimensional
perspective, this $\Z_2$-valued invariant is the mod 2 reduction of a $\Z$-valued invariant,
the $R$-charge or fermion number $\EF$.  This is a large part of the reason that the lift to
five dimensions gives a richer theory than the  four-dimensional one.

$\K_0$ is an approximation to
the space $\K$ of exact supersymmetric ground states.
To determine $\K$,  one follows the standard recipe described in section \ref{york}.  One considers solutions that interpolate between different
time-independent solutions in the far past and the far future.  By  counting such solutions
in an appropriate way, one constructs the operator $Q$ of eqn. (\ref{omex}) whose
cohomology is  $\K$.

By the same reasoning as in section \ref{ondo}, we can restate (\ref{dozzo})  as a formula
for the partition function via a trace in $\K$:
\begin{equation}\label{gondo}Z(q)=\Tr_{\K}\,q^{\EP-\ng (\lambda_R,\eurm m^\vee)}(-1)^\EF. \end{equation}

More generally, we can consider knot cobordisms interpolating between two knots $K$ and $K'$
 by considering in $\R\times W_3\times \R_+$
a monodromy defect supported on $\SS\times\R_+$, where $\SS\subset \R\times W_3$
is asymptotic to $\R\times K$ in the past and $\R\times K'$ in the future.  Still more generally, we
can replace $\R\times W_3$ with any oriented four-manifold $M_4$, and $\SS$ by any
oriented two-manifold in $M_4$.

\subsubsection{The Non-Regular Case And An Action Of $G$}\label{nonreg}

The description of the monodromy defect in five dimensions via the singularity (\ref{lanko})
is adequate when the triple $(\alpha^\vee,\beta^\vee,\gamma^\vee)$ is regular.  For the general
case, one needs a more powerful point of view.

The monodromy defect can be alternatively defined by coupling the five-dimensional
$G^\vee$ gauge theory to a three-dimensional supersymmetric theory known
as $T(G^\vee)$.   ($T(G^\vee)$ was systematically discussed in \cite{gw2} for all $G^\vee$; the prototype
$T(SU(2))$ is a basic example  of three-dimensional mirror symmetry \cite{IS}.  $T(G^\vee)$
is a rather subtle theory which, for example, can be interpreted as the universal kernel
of geometric Langlands duality, as briefly explained in section 3.5 of \cite{wittentech}.)
The theory $T(G^\vee)$ has $OSp(4|4)$ superconformal symmetry; it has an action of
$G^\vee$ on its Higgs branch and $G$ on its Coulomb branch.\footnote{It is believed
that the groups that act faithfully are the adjoint forms of $G^\vee$ and $G$, so the distinction
between them is unimportant in the simply-laced case.  The mirror of
$T(G^\vee)$ is $T(G)$.  In parallel with  the Fayet-Iliopoulos parameters that are introduced
momentarily, there is a mirror triple of mass parameters that violate the $G^\vee$ symmetry;
these are not relevant in the present context as the $G^\vee$ symmetry is gauged.}
We couple $T(G^\vee)$ to $G^\vee$ gauge theory using the $G^\vee$ action on the Higgs branch.
$T(G^\vee)$ can be deformed by Fayet-Iliopoulos parameters $(\alpha^\vee,
\beta^\vee,\gamma^\vee)$; this breaks the $G$ symmetry to the maximal torus,
eliminates the Coulomb branch, and makes the Higgs branch smooth.  Once the Higgs
branch is smooth, the theory is infrared free and one can aim for a classical
description of the defect operator associated to coupling to $T(G^\vee)$.  This classical
description involves the singularity postulated in eqn. (\ref{lanko}).  The steps involved
in reducing from a description involving a coupling to a field theory on the defect to a
description involving the singularity are similar to what they are in one dimension less;
see section 3 of \cite{GuWSurf}.

Describing the defect operator by coupling the bulk gauge theory to $T(G^\vee)$ has the advantage of making sense when
the triple $(\alpha^\vee,\beta^\vee,\gamma^\vee)$ is nonregular.  Let us consider
the extreme case that these parameters  vanish.  Then the theory admits
an action of $G$, acting only on fields supported along the defect.  The conserved
quantities $\eurm m^\vee$ generate the action of the maximal torus of $G$, in the sense that the group
element corresponding to $\eta^\vee\in T$ is $\exp(2\pi i (\eta^\vee,\eurm m^\vee))$.

Naively speaking, it appears that, upon setting $\alpha^\vee,\,\beta^\vee$, and $\gamma^\vee$
to zero, since the theory has a $G$ action, the cohomology of $Q$
would also admit such an action and the trace (\ref{gondo}) would then
be a trace in a $G$-module.  This would have strong implications for the knot invariants
-- probably too strong.  An instructive problem arises here.  Precisely when the triple
$(\alpha^\vee,\beta^\vee,\gamma^\vee)$ is nonregular, the theory $T(G^\vee)$ flows
to a non-trivial CFT in the infrared.  The noncompactness of the initial value surface
$K\times \R_+$ then becomes essential and it is likely that the continuous spectrum cannot
be ignored.  Even in the nonregular case, it is possible to express
the partition function $Z(q)$ as a trace
analogous to (\ref{gondo}) in a much bigger Hilbert space -- the space of all physical
states of the $(0,2)$ model, without reducing to the cohomology of $Q$.  But it may not be
possible to reduce to a discrete spectrum of BPS states with $G$ action.  For example, trying to do so would entail setting $|q|=1$ in the expansions made for the unknot in section \ref{trying}.

\subsubsection{Lifting To Six Dimensions}\label{liftsix}

The last step of this type is the lift to an ultraviolet-complete description in six
dimensions, along the lines of section \ref{fivebranes}.  The six-dimensional geometry
is now $M_4\times D$, where $D$ is a two-manifold with $U(1)$ symmetry.

The six-dimensional theory is classified by the choice of a simply-laced Dynkin diagram,
and  the distinction between $G$ and $G^\vee$ arises from a subtle choice mentioned in
footnote \ref{feathers}.   (To relate the six-dimensional theory to
gauge theory of a Lie group that is not simply-laced, one makes one of the two constructions
described in section \ref{twoc}.)  Since the six-dimensional theory
is not infrared-free, it is not clear that a system consisting of the six-dimensional theory
with a codimension two defect can be obtained by coupling the six-dimensional theory to
a four-dimensional theory that is defined independently.  However, the combined
system consisting of the six-dimensional theory with a four-dimensional  defect does exist.
In fact, there are a family of half-BPS codimension two defects; see \cite{G,GMa,BTX}.
They parallel the corresponding half-BPS monodromy defects described in
gauge theory in \cite{GuWRam}
and associated to Levi subgroups of $G$.  We will consider here only the ``full'' defect
which in reduction to gauge theory corresponds to a monodromy defect operator
with the full set of parameters $(\alpha,\beta,\gamma,\eta)$.

The six-dimensional theory does not have a Lie group or
gauge group of symmetries, but in the presence of a codimension two defect, it does have
a global symmetry group, which  is a form  of $G$.   The
full defect corresponds after reduction on a circle
to  the monodromy defect in five-dimensional gauge theory that we have derived from
eqn. (\ref{thelx}).   In six dimensions, the full defect is characterized only by the parameters
$\beta^\vee$ and $\gamma^\vee$. (One may as well call these parameters $\beta$ and $\gamma$, as the six-dimensional description is symmetrical beween $G$ and $G^\vee$.)  $\alpha^\vee$ arises  if, in compactifying on a circle to get to five
dimensions,
one twists by the element $\exp(-2\pi\alpha^\vee)$ of the global symmetry group.\footnote{The
form of $G$ that acts as a global symmetry group in six dimensions has not been
fully analyzed and may depend on a choice as in footnote \ref{feathers}.  It appears
that after reducing on a circle, the global
symmetry group coincides with the gauge group.}  As we have already discussed,
$\eta^\vee$ is not present as a parameter in five dimensions; instead the five-dimensional theory
has a conserved current with $\eurm m^\vee$ as the conserved charge.

It is clear what to do with a codimension two defect in the context of the present paper.
We place such a defect on $\SS\times D\subset M_4\times D$,
where $\SS\subset M_4$ is an oriented
two-manifold.    Upon reducing on the $U(1)$ orbits on $D$, we return to the five-dimensional
construction that we have already analyzed.    To study a knot, we make the usual
specialization to $M_4=\R\times W_3$, $\SS=\R\times K$.

\subsection{Using The Duality}\label{trying}

In the part of this paper that was based on representing knots by Wilson operators,
there were a few technical problems in actually using the duality to learn about
Chern-Simons theory for knots in a three-manifold $W_3$.  One problem is that if
$W_3$ is compact, then gauge theory on $W_3\times  \R_+$ with a reducible flat connection
at infinity leads to infrared divergences. Their role in the duality is not yet understood.  Another problem is that in defining a boundary
condition at $y=\infty$, we have to pick a homomorphism $\rho:\pi_1(W_3)\to G_\C$;
we do not know how this is related to the homomorphism $\rho^\vee:\pi_1(W_3)\to G^\vee_\C$
that one introduces in the dual description.   Happily, in an important situation -- knots
in $\R^3$ with only gauge transformations that are trivial at infinity allowed -- these issues do not arise.

For the equivalent story with monodromy defects, we are not so fortunate.  We can still
avoid infrared divergences by taking $W_3=\R^3$.  But now to study a knot $K$, we have to consider homomorphisms
from the fundamental group of $\R^3\backslash K$ to $G_\C$ or $G^\vee_\C$, with a prescribed
monodromy around $K$.  Because of the prescribed monodromy, there is no longer a trivial flat connection, and once one only allows gauge transformations that are trivial at infinity, any
non-trivial flat connection becomes non-isolated.  So to proceed, we need to learn something
 about the relation between the Chern-Simons path integral and that of $\N=4$ super
Yang-Mills for the case that the flat connection at infinity is not isolated.    Also, for
generic $K$, there are multiple homomorphisms of $\pi_1(\R^3\backslash K)$ to $G_\C$
or $G^\vee_\C$, even when the conjugacy class of the monodromy around $K$ is prescribed.
So we cannot avoid the question of the relation under duality of the homomorphisms $\rho$ and
$\rho^\vee.$

In short, to actually use the duality based on monodromy defects, we need to learn more.
And so far we have only mentioned questions of principle.   In practice, for either
the duality based on Wilson operators or that based on monodromy defects,
to learn a lot one will need to know more about actually solving the equations.

Rather than say nothing at all, we will make a few remarks about the unknot  $K_0\subset \R^3$.
The fundamental group of $\R^3\backslash K_0$ is simply the abelian group $\Z$, so it
has up to conjugacy
only one homomorphism to $G$ or $G^\vee$ with prescribed monodromy, and
the image of this homomorphism is abelian.   So there is essentially only one possible
integration cycle in Chern-Simons theory, and the standard integration cycle must coincide
with the one we get in the $G^\vee$ description using the unique possible flat connection at
infinity.  The Chern-Simons action of an abelian
flat connection vanishes (with the canonical framing),
so we do not need to worry about a factor in the duality involving the classical action.
There might be a correction to the formula involving the fact that the abelian flat connection is
not isolated (in the context of $\R^3\backslash K_0$), or a constant $\frak N_0$, as in
(\ref{zondo}), but we will just proceed and see
what happens.

For simplicity, we consider the case of $G=SU(2)$.  The path integral for a Wilson
operator in the spin $j$ representation placed on the unknot in $\R^3$ is
\begin{equation}\label{pelfro}  J(q;K_0,j)=\frac{q^{(2j+1)/2}-q^{-(2j+1)/2}}{q^{1/2}-q^{-1/2}}.   \end{equation}

We would like to express this function in the form of (\ref{dozzo}), which for $G=SU(2)$
should become
\begin{equation}\label{elfro}J(q;K_0,j)=\sum_{r,c}w_{r,c}q^{r-cj}.\end{equation}
What sort of expansion will this be?  Actually, there are two expansions that we should make.
In general, in the $G^\vee$ description, we have $\alpha^\vee=0$, and in the present
case, we are relying on an abelian homomorphism $\rho^\vee$, so also $\beta^\vee=0$.
Hence if $\gamma^\vee=0$, then we are in the nonregular case described at the end of
section \ref{nonreg}, where the space of BPS states may not be well-defined.  So we prefer
to take $\gamma^\vee\not=0$.  In this case, as explained at the end of section \ref{special},
$q$ does not have modulus 1, so there are two cases, $|q|<1$ or $|q|>1$.  In these
two cases, we will interpret (\ref{elfro}) as a Laurent series around $q=0$ or $q=\infty$,
respectively.

There are simple expansions of this type which moreover are consistent with the fact
that in (\ref{elfro}) the coefficients $w_{r,c}$ are supposed to be independent of $j$.
We use either
\begin{equation}\label{torbo}\frac{1}{q^{1/2}-q^{-1/2}}=-q^{1/2}\sum_{t=0}^\infty q^t,~~
|q|<1 \end{equation}
or
\begin{equation}\label{porbo} \frac{1}{q^{1/2}-q^{-1/2}}=q^{-1/2}\sum_{t=0}^\infty q^{-t},~~
|q|>1.     \end{equation}
For example, the first leads to the formula
\begin{equation}\label{melfro}J(q;K_0,j)=\left(-q^{j+1}+q^{-j}\right)\sum_{t=0}^\infty
q^t,\end{equation}
in which the finite Laurent polynomial $J$ is written as the difference of two infinite Laurent series.
This expansion takes the form (\ref{elfro}); the coefficients $w_{r,c}$ are nonzero
if and only if $c=\pm 1$ and $r$ is a positive integer, or $r=0$ with $c=1$.  A similar  formula can be
written straightforwardly for $|q|>1$.  Of course, to be satisfied with the expansion (\ref{melfro})
or its cousin for $|q|>1$, one would like to know that solutions with the claimed topological
invariants actually exist.   In the present context, it is unclear why there are solutions
leading to the geometric series in (\ref{melfro}).   Possibly a hint comes from recent
approaches to related problems such as \cite{DGH}.

 The fact that one has to make two different expansions may be special to a reducible
 flat connection.   In the case of an irreducible flat connection,
one is free to take $\beta^\vee\not=0$, and this means that $\gamma^\vee$ can be varied
in an arbitrary way while avoiding nonregular triples.  This suggests that the contribution
to the  path integral of an irreducible flat $G^\vee$
connection with monodromy around $K$ will be given by a
Laurent polynomial (powers of $q$ bounded above and below) rather  than a Laurent
series (powers of $q$ bounded in only one direction).  At any rate, there is plenty to understand.

\vskip 1cm
 \noindent {\it Acknowledgments}
 Research supported in part by NSF Grant PHY-0969448.  I would
 like to thank A. Ashtekar, M. Aganagic, C. Beasley, S. Cherkis,  D. Bar-Natan,
 R. Cohen, R. Dijkgraaf, D. Gaiotto, J. Gomis, S. Gukov, J. Heckman, L. Hollands,
 J. Kamnitzer,  A. Kapustin, B. Kostant,
 S. Lewallen, R. Mazzeo, M. Mari\~no,
 G. Moore, L. Rozansky, Y. Tachikawa,
 C. Taubes,  C. Vafa, and the members of the Stanford and IAS particle
 theory groups
 for their comments and I. Frenkel and D. Bar-Natan for having introduced me to the subject.
\bibliographystyle{unsrt}

\end{document}